\documentclass[12pt]{article}
\usepackage{cite}
\usepackage{amsmath, amsthm, amssymb, graphicx,slashed}
\parskip=\baselineskip
\def\Bbb{\mathbb}
\def\tx{\textsf}
\def\Tr{{\rm Tr}}
\def\T{{\cal T}}
\def\B{{\mathcal B}}
\def\M{{\mathcal M}}
\def\16{{\bf 16}}
\def\1{{\bf 1}}
\def\2{{\bf 2}}
\def\4{{\bf 4}}
\def\bar{\overline}
\def\tilde{\widetilde}
\def\R{{\Bbb{R}}}\def\Z{{\Bbb{Z}}}
\def\N{{\mathcal N}}
\def\hat{\widehat}
\numberwithin{equation}{section}
\def\frak{\mathfrak}
\font\teneurm=eurm10 \font\seveneurm=eurm7 \font\fiveeurm=eurm5
\newfam\eurmfam
\textfont\eurmfam=\teneurm \scriptfont\eurmfam=\seveneurm
\scriptscriptfont\eurmfam=\fiveeurm
\def\eurm#1{{\fam\eurmfam\relax#1}}
 \font\teneusm=eusm10 \font\seveneusm=eusm7 \font\fiveeusm=eusm5
\newfam\eusmfam
\textfont\eusmfam=\teneusm \scriptfont\eusmfam=\seveneusm
\scriptscriptfont\eusmfam=\fiveeusm
\def\eusm#1{{\fam\eusmfam\relax#1}}
\font\tencmmib=cmmib10 \skewchar\tencmmib='177
\font\sevencmmib=cmmib7 \skewchar\sevencmmib='177
\font\fivecmmib=cmmib5 \skewchar\fivecmmib='177
\newfam\cmmibfam
\textfont\cmmibfam=\tencmmib \scriptfont\cmmibfam=\sevencmmib
\scriptscriptfont\cmmibfam=\fivecmmib
\def\cmmib#1{{\fam\cmmibfam\relax#1}}

\parskip=\baselineskip
\def\Bbb{\mathbb}
\def\BC{\mathbb C}
\def\Bbb{\mathbb}
\def\Tr{{\rm Tr}}
\def\A{{\mathcal A}}
\def\C{\Bbb{C}}
\def\1{{\bf 1}}
\def\2{{\bf 2}}
\def\3{{\bf 3}}
\def\4{{\bf 4}}
\def\bar{\overline}
\def\tilde{\widetilde}
\def\R{{\Bbb{R}}}\def\Z{{\Bbb{Z}}}
\def\N{{\mathcal N}}
\def\hat{\widehat}
\def\neg{\negthinspace}
\numberwithin{equation}{section}
\begin{document}

\begin{titlepage}
\begin{flushright}
hep-th/yymm.nnnn
\end{flushright}
\vskip 1.5in
\begin{center}
{\bf\Large{$S$-Duality Of Boundary Conditions}\vskip0cm
\bf\Large{in ${\mathcal N}=4$ Super Yang-Mills Theory} }\vskip
0.5cm {Davide Gaiotto and Edward Witten} \vskip 0.3in {\small{
\textit{School of Natural Sciences, Institute for Advanced
Study}\vskip 0cm {\textit{Einstein Drive, Princeton, NJ 08540
USA}}}}

\end{center}
\vskip 0.5in

\baselineskip 16pt

\begin{abstract}
By analyzing brane configurations in detail, and extracting
general lessons, we develop methods for analyzing $S$-duality of
supersymmetric boundary conditions in $\N=4$ super Yang-Mills
theory.  In the process, we find that $S$-duality of boundary
conditions is closely related to mirror symmetry of
three-dimensional gauge theories, and we analyze the IR behavior
of large classes of quiver gauge theories.
\end{abstract}
\end{titlepage}
\vfill\eject

\date{May, 2008}

\tableofcontents

\section{Introduction}\label{intro}

\def\fB{\frak B}

 In a recent paper \cite{Gaiotto:2008sa}, we have
described half-BPS boundary conditions in $\N=4$ super Yang-Mills
theory with gauge group $G$. The general classification of boundary
conditions is rather elaborate and depends on a triple
$(\rho,H,\fB)$.  The gauge group $G$ is explicitly broken near the
boundary to a subgroup $H$. Part of the symmetry breaking involves a
choice of homomorphism $\rho:\frak{su}(2)\to \frak g$
 from the Lie algebra of $SU(2)$ to that of $G$.  Finally, $\fB$
is a boundary field theory with $H$ symmetry.  Because of the
explicit symmetry breaking, the gauge fields on the boundary are
valued in $\frak h$, the Lie algebra of $H$, and can be naturally
coupled to $\fB$.

In this brief summary, we have omitted the role of the
four-dimensional theta-angle, which adds an extra layer of
structure as explained in \cite{Gaiotto:2008sa,Gaiotto:2008sd}. In
the present paper, we take the theta-angle to vanish until section
\ref{stheta}.

Our goal in the present paper is to understand the action of
electric-magnetic duality on this class of boundary conditions. To
gain experience, we begin with concrete examples. In section
\ref{about}, we review boundary conditions in $U(n)$ gauge theory
that can be constructed using D3-branes, D5-branes, and NS5-branes
of Type IIB superstring theory. Because our topic turns out to be
closely related to IR dynamics in three dimensions, we also
re-examine the behavior of purely three-dimensional theories
constructed from those ingredients.  We describe an important and
mirror symmetric class of three-dimensional theories, and we use
``monopole operators''
 to learn
something about their IR dynamics.  Monopole operators were
discussed qualitatively in relation to supersymmetric gauge
dynamics in \cite{Aharony:1997bx,Kapustin:1999ha} and the
formalism we use was developed in
\cite{Borokhov:2002ib,Borokhov:2002cg,Borokhov:2003yu}. In section
\ref{examples}, we analyze in detail $S$-duality for boundary
conditions constructed from branes. This involves many interesting
details but also some general lessons. One important lesson is
that $S$-duality of four-dimensional boundary conditions is
closely related to mirror symmetry of three-dimensional gauge
theories \cite{Intriligator:1996ex}. A second important lesson is
that a certain class of superconformal field theories plays an
important role. The most basic of these is a certain self-mirror
theory that we call $T(SU(n))$ that appears when one applies
$S$-duality to Dirichlet boundary conditions. (For $n=2$,
$T(SU(n))$ coincides with one of the main examples in
\cite{Intriligator:1996ex}.)

In section \ref{general}, we attempt to extract the important
lessons from our investigation of branes and formulate some general
statements that are valid for any compact gauge group $G$. The key
step is to generalize $T(SU(n))$ and its close cousins, which we do
using Janus domain walls \cite{BGH,CFKS,CK,DEG,DEG2}.  We describe
the key  properties of the theories $T_{\rho^\vee}^\rho(G)$ that we
construct this way, and show in general how they can be used to
construct the $S$-dual of a given boundary condition.

In section \ref{orthosymp}, we return to three dimensions and
analyze some important properties of quivers with orthogonal and
symplectic gauge groups. We use the results in section \ref{orient}
to analyze $S$-duality of boundary conditions in $U(n)$ gauge theory
constructed with orientifold and orbifold fiveplanes.  These give
tractable and interesting illustrations of some of the general ideas
of this paper.  In section \ref{bcos}, by using brane with O3
planes, we extend many of our results to the case that the gauge
group is $SO(n)$ or $Sp(n)$. Among other things, we describe quiver
constructions of $T(SO(n))$ and $T(Sp(n))$.

Up to this point, our analysis concerns the basic electric-magnetic
duality operation $S:\tau\to -1/\tau$, rather than the complete
duality group $SL(2,\Z)$.  Indeed, for most half-BPS boundary
conditions, only the action of $S$ can be defined. In section
\ref{stheta}, we incorporate the gauge theory $\theta$-angle, and
describe the action of  $SL(2,\Z)$ on those half-BPS boundary
conditions that admit such an action. As an application, we give a
quiver-like description of the low energy effective field theory
that describes the interaction of D3-branes with a
$(p,q)$-fivebrane. This description uses Chern-Simons couplings with
$\N=4$ supersymmetry.  Finding such a description has been a
longstanding problem.

We will often refer to the three-dimensional theory $\fB$ that is
part of the definition of a supersymmetric boundary condition as a
boundary superconformal field theory or SCFT since the conformally
invariant case tends to be particularly interesting.  Moreover,
once one understands $S$-duality of conformally invariant boundary
conditions, one can understand the general case by following the
duality under relevant perturbation.  Focussing on the IR limit
has another important advantage.  The brane configurations that we
will use for our explicit examples are most tractable if one is
free to make standard rearrangements of the fivebranes.  The
justification for these rearrangements is that they involve
deformations that are irrelevant in the IR.

\section{Brane Constructions For Unitary Groups}\label{about}

Rather than attempt an abstract explanation from the beginning, we
will start this paper by considering the case $G=U(n)$, where
everything can be described concretely via manipulations of
branes.  In the present section, we describe the necessary facts
about brane constructions of boundary conditions, and we describe
some facts about dynamics of three-dimensional supersymmetric
gauge theories that will also be important.  In section
\ref{examples}, we use these facts together with standard brane
manipulations to gain a fairly detailed understanding of the
$S$-duality of boundary conditions for $G=U(n)$.

\subsection{Brane Construction Of Boundary
Conditions}\label{branecon}
\def\N{{\mathcal N}}
Supersymmetric boundary conditions of any kind inevitably break the
$R$-symmetry group of $\N=4$ super Yang-Mills theory to a subgroup.
For half-BPS boundary conditions, we can be more specific.  The full
$R$-symmetry group, which is $SO(6)_R$ (or its cover $SU(4)_R$), is
broken to a subgroup\footnote{For example, in the conformally
invariant case, a half-BPS boundary condition breaks the conformal
group $PSU(4|4)$ to $OSp(4|4)$, whose $R$-symmetry subgroup is
$SO(4)\cong SO(3)\times SO(3)$.} $SO(3)\times SO(3)$ (or its cover
$SO(4)$). Under this subgroup, the six adjoint-valued scalar fields
of $\N=4$ super Yang-Mills theory split up into two groups of three
scalar fields, say $\vec X$ and $\vec Y$, which are rotated
respectively by the two factors of  $SO(3)\times SO(3)$.  We call
these factors $SO(3)_X$ and $SO(3)_Y$.

Since the idea of a boundary condition determined by a triple
$(\rho,H,\fB)$ (as summarized in the introduction) is daunting at
first sight, we will begin by using a concrete and familiar brane
construction to build half-BPS boundary conditions and study their
$S$-duality.  The construction \cite{Ganor,Hanany:1996zelf}
involves branes in ten-dimensional Minkowski spacetime with
coordinates $x^0,x^1,\dots,x^9$.  We make use of three types of
brane: D3-branes with worldvolume spanned by $x^0,x^1,x^2,x^3$,
D5-branes with worldvolume spanned by $x^0,x^1,x^2$ together with
$x^4,x^5,x^6$, and NS5-branes with worldvolume spanned by
$x^0,x^1,x^2$ together with $x^7,x^8,x^9$. Thus all branes share
the directions $x^0,x^1,x^2$.  The D3-branes are semi-infinite in
the $x^3$ direction, being supported on the region $x^3\geq 0$,
with a boundary at $x^3=0$. We also write $y$ for $x^3$.  The
fivebranes are located at specified values of $y$ (such as $y=0$)
and are used to provide boundary conditions (or couplings to
matter systems) for the D3-branes.

In the gauge theory on the D3-branes, fluctuations in $x^4,x^5,x^6$
correspond to the scalar fields $\vec X$ of $\N=4$ super Yang-Mills
theory and fluctuations in $x^7,x^8,x^9$ correspond to the scalar
fields $\vec Y$. Brane configurations of the type just summarized
are useful in studying three-dimensional mirror symmetry and the
methods used in that context will be very helpful in what follows.

\begin{figure}
  \begin{center}
    \includegraphics[width=3.5in]{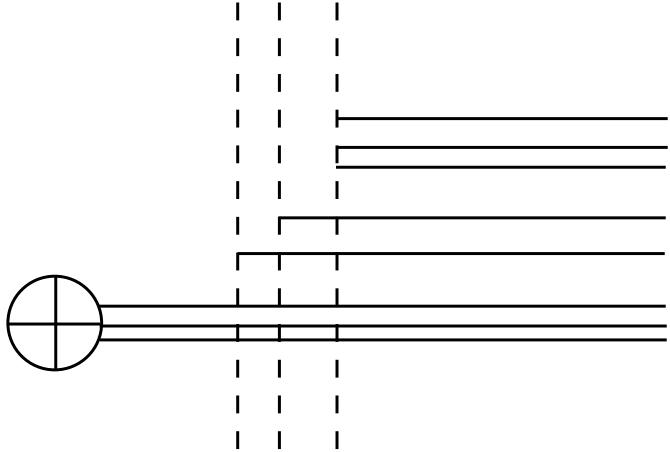}
  \end{center}
\caption{\small A brane configuration that determines a half-BPS
boundary condition in ${\mathcal N}=4$ super Yang-Mills theory. Here
and later, horizontal solid lines designate D3-branes spanning
directions 0123; vertical dotted lines designate D5-branes spanning
directions 012456.  In this example, there are eight D3-branes and
the gauge group is $U(8)$. The symbol $\bigoplus$ denotes a further
fivebrane system, of which some possible examples are sketched in
fig. \ref{Fig2}.}
  \label{Fig1}
\end{figure}

Let us recall from \cite{Gaiotto:2008sa}, section 2.5.1, the
boundary conditions obtained from such a brane configuration. In the
example sketched in fig. \ref{Fig1}, there are 8 D3-branes, so the
four-dimensional gauge group is $U(8)$. Reading the figure from
right to left, the first three D3-branes terminate on a D5-brane. At
this point, $\vec X$ develops a rank 3 pole, reducing the gauge
symmetry from $U(8)$ to $U(5)$. This pole is governed by Nahm's
equations and represents the way the D3-branes flare out into a
fuzzy funnel that joins the D5-brane
\cite{Callan:1997kz,Constable:1999ac}. A single D3-brane ends on
each of the next two D5-branes, reducing the rank of the gauge group
without a further pole.   To the left of the D5-branes, the D3-brane
gauge group is reduced to $U(3)$. The symbol $\bigoplus$ then
represents a further system of NS5-branes and D5-branes that
describes a three-dimensional matter system coupled to the $U(3)$
gauge fields.

\begin{figure}
  \begin{center}
    \includegraphics[width=5.5in]{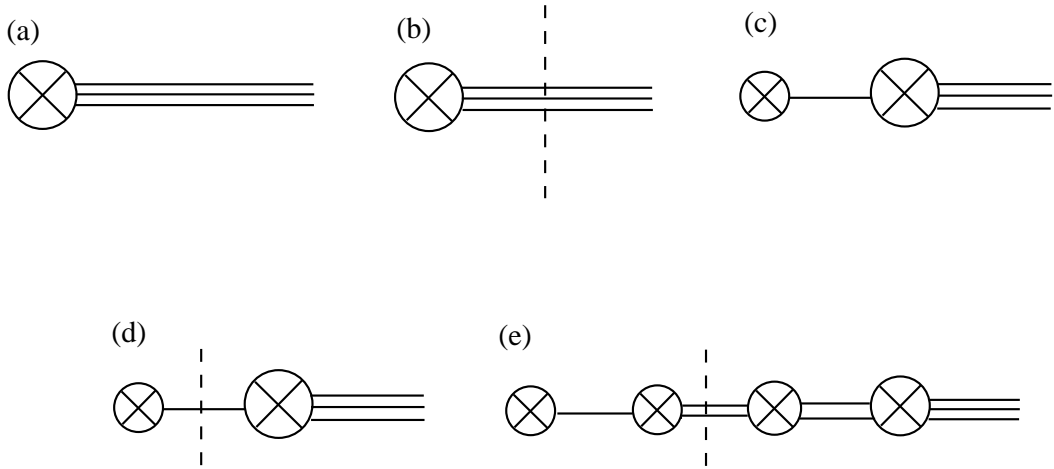}
  \end{center}
\caption{\small Some brane configurations, any one of which can
correspond to the symbol $\bigoplus$ on the left of fig. \ref{Fig1}.
 Here and later, the symbol $\bigotimes$
represents an NS5-brane spanning directions 012789. In (a), three
D3-branes end on a single NS5-brane.  This leads to Neumann
boundary conditions in $U(3)$ gauge theory. In (b), the D3-branes
intersect a D5-brane before terminating on a single NS5-brane.
This leads (in the limit that all fivebrane separations in the
$y=x^3$ direction are taken to zero) to Neumann boundary
conditions with a fundamental hypermultiplet supported on the
boundary. The hypermultiplet comes from the brane intersection. In
(c), (d), and (e), there is more than one NS5-brane. This leads to
Neumann boundary conditions modified by coupling to a non-trivial
boundary
 SCFT, as described in the text.}
  \label{Fig2}
\end{figure}

There are many possible choices of this further system. Some
illustrative examples are shown in fig. \ref{Fig2}. In fig.
\ref{Fig2}(a), the additional system consists of a single
NS5-brane, and the $U(3)$ gauge fields simply obey Neumann
boundary conditions.  In fig. \ref{Fig2}(b), a D5-brane has been
added.  As a result, the $U(3)$ gauge fields couple to a
hypermultiplet in the fundamental representation (or more briefly
a fundamental hypermultiplet) that is supported in codimension 1.

Fig. \ref{Fig2}(c) requires a more detailed explanation.  Two
NS5-branes are separated by a distance $L$ in the $x^3$ direction.
The worldvolume theory for the  D3-brane in the slab between the two
NS5-branes is a $U(1)$ gauge theory  with Neumann boundary
conditions on the two ends. If the four-dimensional gauge coupling
$g_{4d}$ is sufficiently small, the Kaluza-Klein scale on the slab,
which is $1/L$, is much larger than the scale set by the
three-dimensional gauge coupling, which is $g_{3d}^2=g_{4d}^2 /L$.
The result is that at sufficiently low energy the worldvolume theory
reduces to a three-dimensional gauge theory. The $3-3$ strings
stretched across the NS5-branes give a single bifundamental
hypermultiplet coupled to both this three-dimensional $U(1)$ gauge
theory and to the bulk gauge theory on the half line. The $U(3)$
gauge symmetry of the semi-infinite D3-branes is a global symmetry
from the point of view of the three-dimensional gauge theory. The
three-dimensional theory is actually a theory of a $U(1)$ vector
multiplet coupled to three hypermultiplets of charge 1 and in the
fundamental representation of a $U(3)$ global symmetry. In the
infrared, the three-dimensional gauge coupling becomes large. If we
simply turn off the four-dimensional gauge coupling on the
semi-infinite D3-branes of fig. \ref{Fig2}(c), the IR flow gives a
purely three-dimensional SCFT. Turning back on the four-dimensional
gauge coupling, we get a combined system consisting of
four-dimensional gauge fields on a half-space coupled to a boundary
SCFT.  The boundary theory is coupled to the bulk gauge fields by
gauging its $U(3)$ global symmetry.

Fig. \ref{Fig2}(d)  is a small modification of (c): each extra D5
brane inserted between the two NS5-branes, with no D3-branes
ending on it, adds a fundamental hypermultiplet coupled to the
three-dimensional gauge theory. So here (with one extra D5-brane
in this example) we get a boundary SCFT that has $U(4)$ global
symmetry, of which a $U(3)$ subgroup is coupled to bulk gauge
fields. Finally, in general, if the brane system $\bigoplus$
consists of several displaced NS5-branes with a variable number of
D3-brane segments stretched between them, and extra D5-branes with
no D3-branes ending on them, as in fig. \ref{Fig2}(e),  then at
low energies, the worldvolume theory consists of a certain linear
quiver of three-dimensional $U(n_i)$ gauge theories with
fundamental matter possibly coupled to each of the nodes. The
quiver for this example is sketched in fig. \ref{Fig3}. If the
brane multiplicities obey certain inequalities, which will be
explained, then the quiver system flows in the IR to an SCFT with
${\mathcal N}=4$ supersymmetry, and the overall system can be
described by gauge fields on a half-space coupled to this SCFT.

\def\n{{\bf n}}
\def\m{{\bf m}}
\begin{figure}
  \begin{center}
    \includegraphics[width=3in]{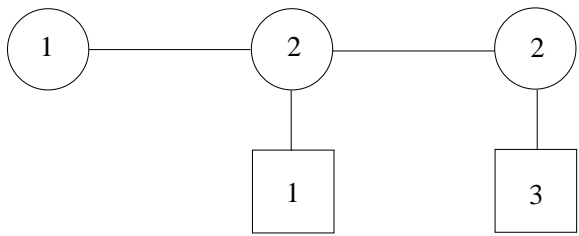}
  \end{center}
\caption{\small A quiver such as this one gives a convenient way to
summarize the construction of a gauge theory with suitable gauge
group and matter representation.  A circle containing an integer $n$
represents a $U(n)$ factor in the gauge group.  The gauge group is
the product of such factors, one for each circle.  A line joining
two circles labeled by $n$ and $m$ represents a bifundamental
hypermultiplet, that is a collection of hypermultiplets transforming
under $U(n)\times U(m)$ as $(\n,\bar\m)\oplus (\bar\n,\m)$. Finally,
if a circle labeled by $n$ is linked to square labeled $p$, this
means that there are $p$ fundamental hypermultiplets of $U(n)$. For
every square labeled by $p$, there is a  $U(p)$ global symmetry
acting on the corresponding hypermultiplets.  The specific quiver
drawn here represents the boundary SCFT that arises from the brane
configuration of fig. \ref{Fig2}(e).}
  \label{Fig3}
\end{figure}

In all cases, we have slightly separated the various branes in the
$x^3$ direction to avoid ambiguities and to make possible a
description by gauge theory.  However, the intent is always to
consider a limit in which the brane separations $L_i$ are taken to
zero and the brane configuration determines a boundary condition for
the four-dimensional gauge fields.  The boundary conditions obtained
this way are special cases of the general definition in
\cite{Gaiotto:2008sa}, involving a triple $(\rho,H,\fB)$.  Here
$\rho:\frak{su}(2)\to\frak g$ is an embedding of the Lie algebra of
$SU(2)$ in that of $G$, $H$ is a subgroup of $G$ that commutes with
$SU(2)$, and $\fB$ is a boundary superconformal field theory with
$H$ symmetry.  For the brane constructions that we have described,
$G$ is a unitary group $U(N)$ for some $N$; $\rho$ is arbitrary;
$\fB$ is constructed from a quiver gauge theory; and $H$, which is a
subgroup of $G$ of the form  $U(M)$ for some $M$, is a global
symmetry group acting at one end of the quiver.

\subsection{Ordering Of Branes}\label{ordering}

We are not interested in studying brane configurations for their
own sake, but as a tool for generating boundary conditions and
studying the action of $S$-duality.  For this purpose, it turns
out that it suffices to consider a certain subset of brane
configurations. As we will see, other configurations can be
reduced to this subset by moving branes along the lines of
\cite{Hanany:1996zelf}.

In the brane constructions that we have described, the brane
separations $L_i$ are irrelevant in the infrared, but the specific
ordering of the fivebranes along $y=x^3$ is quite important. We have
ordered the fivebranes in figs. \ref{Fig1} and \ref{Fig2} in a way
that makes the field theory interpretation of the boundary condition
understandable and the infrared limit simple.  This depends on two
constraints that will be described here.

To state these constraints, one important concept is the net number
of D3-branes ending on a fivebrane.  We define this number to be the
number of D3-branes ending on the fivebrane from the right, minus
the number ending from the left.

\subsubsection{The First Constraint}\label{firstconstraint}

Our first constraint is that any D5-brane on which a net non-zero
number of threebranes ends is to the right of all NS5-branes.  This
constraint has been incorporated in figs. \ref{Fig1} and \ref{Fig2}.
The D5-branes that are shown explicitly in fig. \ref{Fig1} are to
the right of the $\bigoplus $ symbol, in which any NS5-branes are
hidden.  In expanding out the $\bigoplus$ symbol in fig. \ref{Fig2},
there may be additional D5-branes, but the net number of D3-branes
ending on any one of them is zero.

This constraint ensures that the brane configurations of fig.
\ref{Fig2} have an interpretation in gauge theory.  If the net
number of D3-branes ending on any D5-brane is zero, it is possible
(in one phase of the theory) to detach all D3-branes from D5-branes
and let D3-branes end on NS5-branes only.  A moveable D3-brane
connecting two NS5-branes is described by a vector multiplet of a
gauge theory. The branch of the moduli space of vacua  in which all
moveable D3-branes end on NS5-branes is the Coulomb branch of this
gauge theory.  Thus our condition implies that the configuration
labeled
 $\bigoplus$ can be interpreted in gauge theory. (In fact, it is a quiver
 gauge theory, as we have already noted.) The IR limit of this gauge
theory is the boundary SCFT $\fB$ that is part of the definition of
our boundary condition. In this gauge theory, each vector multiplet
comes from a D3-brane segment that is of finite extent in the $y$
direction, so it is reasonable to hope that one can extract the zero
modes of all vector multiplets and reduce to a purely
three-dimensional gauge theory before taking the infrared limit.
(For this actually to be true depends on the additional constraint
of section \ref{secondconstraint}.)

The gauge theory associated with the $\bigoplus$ symbol has a global
symmetry group $H$ that couples to four-dimensional gauge fields. In
general $H$ is not the gauge group $G$ of the bulk four-dimensional
gauge theory, but a subgroup. For instance, in fig. \ref{Fig1},
$H=U(3)$, where 3 is the number of D3-branes near the $\bigoplus$
symbol. This results from the fact that, as one comes in from the
right in fig. \ref{Fig1}, some of the D3-branes terminate on
D5-branes before interacting with the gauge theory hidden in the
$\bigoplus$ symbol. D3-branes ending on  D5-branes can have moduli
(if there is more than one D5-brane involved, as in fig. \ref{Fig1})
since D3-brane segments that join two D5-branes can break away and
move in the $\vec X$ direction.  Modes resulting from motion of such
D3-brane segments are hypermultiplets (rather than vector
multiplets).  The resulting hypermultiplet moduli space can be
described, as we have explained in detail in \cite{Gaiotto:2008sa},
by Nahm's equations $d\vec X/dy+[\vec X,\vec X]=0$.  These equations
cannot be readily described in terms of purely three-dimensional
gauge dynamics, since they describe precisely the $y$-dependence of
$\vec X$ .

To summarize then the first constraint, it says that as one
approaches the boundary from the right, one first encounters the
part of the construction (the Nahm pole $\rho:\frak{su}(2)\to\frak
g$ and the reduction from $G$ to a subgroup $H$) that is not
naturally expressed in terms of three-dimensional field theory. Then
one meets, compressed to the symbol $\bigoplus$, the construction
via three-dimensional gauge theory of a three-dimensional boundary
theory $\fB$.  It only makes sense to describe $\fB$ once $H$ is
known (since $\fB$ must have $H$ symmetry, not $G$ symmetry), so it
is convenient to encounter $\rho$ and the reduction to $H$
``first.''

There is no loss in imposing this first constraint, since, given a
second constraint that we describe next, it can always be
implemented without changing the infrared physics by moving branes,
as we will see in section \ref{sduality}.

\subsubsection{The Second Constraint}\label{secondconstraint}

The second constraint that we want can be succinctly stated in
terms of a certain ``linking number'' invariant that was defined
in \cite{Hanany:1996zelf}.  The linking number of a fivebrane is
the D3-brane charge measured at infinity on that fivebrane. Since
a D3-brane ending on a fivebrane is a magnetic source for the
$U(1)$ gauge field on the fivebrane, the D3-brane charge on a
fivebrane can be computed by integrating the $U(1)$ field strength
over a two-sphere at infinity.  The reason that the linking number
is important is that, since it can be measured at infinity along a
brane, it is invariant under the sort of brane manipulations that
are needed to understand $S$-duality.

\begin{figure}
  \begin{center}
    \includegraphics[width=6in]{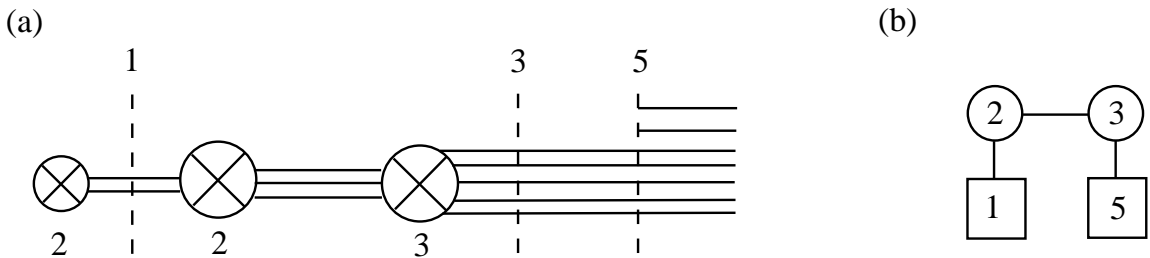}
  \end{center}
\caption{\small (a) A configuration of 3 D5-branes and  3 NS5-branes
 in $U(7)$ gauge theory.  Each fivebrane has a linking number,
defined as the number of fivebranes of the opposite kind that are to
the left of the given fivebrane, plus the net number of threebranes
ending on the right of the given fivebrane.   In the figure, the
linking number of a D5-brane (or an NS5-brane) is given by the
integer that is written just above (or below) the brane in question.
This configuration has been chosen so that the linking numbers of
fivebranes of a given type are non-decreasing if one reads the
figure from left to right. (b) In the boundary condition derived
from (a), a $U(5)$ subgroup of the gauge group is coupled at the
boundary to an SCFT with $U(5)$ symmetry. This SCFT can be obtained
as the infrared limit of the three-dimensional gauge theory
associated with the quiver indicated here (together with a free
fundamental hypermultiplet from interaction with the D5-brane of
linking number 3).}
  \label{Fig4}
\end{figure}

 Concretely, the linking number of a
fivebrane is the number of fivebranes of the opposite kind to the
left of the given fivebrane, plus the net number of D3-branes
ending on this fivebrane on the right.\footnote{This definition
differs by an inessential constant from the definition used in
\cite{Hanany:1996zelf}.} The constraint that we want on the brane
ordering is that for each kind of fivebrane -- NS or D -- the
linking numbers are nondecreasing from left to right. An example
is given in fig \ref{Fig4}.

Let us first discuss what this constraint means for D5-branes. First
consider a D5-brane that is not to the right of all NS5-branes.  The
net number of D3-branes ending on such a D5-brane is zero (by our
first constraint), so its linking number is just the number of
NS5-branes to its left.  The number of NS5-branes to one's left can
only increase (or remain constant) as one moves to the right along
the chain.  So for such D5-branes, the linking numbers are
automatically nondecreasing.

Hence for D5-branes, the linking number constraint only says
something non-trivial for those D5-branes that are to the right of
all NS5-branes -- for example, the ones drawn explicitly in fig.
\ref{Fig1} and the two on the right in fig. \ref{Fig4}(a).  Since
all such D5-branes have the same number of NS5-branes to their left,
the constraint is that the net number of D3-branes ending on a
D5-brane is nondecreasing as one moves to the right.  This
constraint is satisfied in both examples.

The meaning of the constraint was explained in section 3.5 of
\cite{Gaiotto:2008sa}. To get a boundary condition, we must take the
limit that the brane separations $L_i$ are all taken to zero. The
behavior in this limit of the moduli space of solutions of Nahm's
equations associated to a brane configuration is most simple if the
D5 linking numbers are nondecreasing. Moreover the interesting
boundary conditions all arise from configurations of this type.  If
the D5-branes are not arranged in order of increasing (or at least
nondecreasing) linking number, then the moduli space of solutions of
Nahm's equations contain extra hypermultiplets that decouple as
$L_i\to 0$.  For our goal of studying boundary conditions, it does
not add anything to consider brane configurations that generate such
decoupled hypermultiplets.

\begin{figure}
  \begin{center}
    \includegraphics[width=4in]{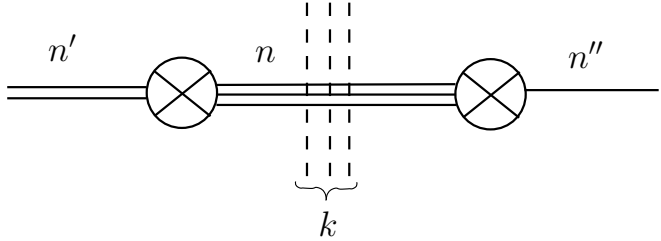}
  \end{center}
\caption{\small Two NS5-branes with $k$ D5-branes between them.  To
the left of the NS5-branes, between them, and to their right, there
are respectively $n'$, $n$, and $n''$ D3-branes.  Here $n',n,n''$
equal 2,3, and 1, respectively.}
  \label{Fig5}
\end{figure}

$S$-duality suggests that similarly, the Coulomb branch will contain
vector multiplets that decouple for $L_i\to 0$ unless the NS5-branes
are arranged with nondecreasing linking number. To see what the
condition means, let us examine in detail the condition that the
linking number for a pair of successive NS5-branes is nondecreasing.
We write $n',n,n''$ for the number of D3-branes to the left of the
two NS5-branes, between them, and to their right (fig \ref{Fig5}).
There may also be D5-branes between the two NS5-branes, but if so
(as the net number of D3-branes ending on such a D5-brane is
required to vanish) the number of D3-branes does not jump in
crossing them.  Let $k$ be the number of such D5-branes and let $t$
be the number of D5-branes to the left of both NS5-branes drawn in
the picture. The linking numbers $\ell_L$ and $\ell_R$ of the left
and right NS5-brane in fig. \ref{Fig5} are $\ell_L=t+n-n'$,
$\ell_R=t+k+n''-n$.  The condition that $\ell_R\geq \ell_L$
therefore gives
\begin{equation}\label{holfo} n'+n''+k\geq 2n.\end{equation}

This condition has a simple interpretation in gauge theory.  The
$n$ D3-branes between the two NS5-branes support a $U(n)$ gauge
theory. This $U(n)$ gauge theory is coupled to $k$ fundamental
hypermultiplets that arise from D3-D5 intersections.  In addition,
the interactions of the D3-branes that meet at the two NS5-branes
in the picture give bifundamental hypermultiplets of $U(n')\times
U(n)$ and $U(n)\times U(n'')$.  From the point of view of the
$U(n)$ theory, these are $n'+n''$ fundamental hypermultiplets.

Altogether, the $U(n)$ theory therefore interacts with a total of
$n_f=n'+n''+k$ fundamental hypermultiplets, and the condition of
eqn. (\ref{holfo}) is equivalent to
\begin{equation}\label{thelc}n_f\geq 2n.\end{equation}
This condition on the matter fields in a three-dimensional $U(n)$
gauge theory with ${\mathcal N}=4$ supersymmetry is similar to
conditions encountered in \cite{Intriligator:1996ex}.

The most direct interpretation of   (\ref{thelc}) is that it is the
condition under which complete Higgsing is possible; that is, it is
the condition under which there exists a vacuum in which the
hypermultiplets have expectation values, the scalars in the vector
multiplets do not, and the gauge symmetry is completely broken.
Consider an $\N=4$ theory with gauge group $U(n)$ and $n_f$
fundamental hypermultiplets. Viewing the $\N=4$ theory as an $\N=2$
theory, the hypermultiplets consist of an $n\times n_f$ matrix $A$
and an $n_f\times n$ matrix $B$. The scalars in the vector multiplet
are an $n\times n$ matrix $\phi$. The superpotential is $\Tr \,\phi
AB$, so the condition for a critical point of the superpotential
with $\phi=0$ is
\begin{equation}\label{dunk} AB=0.\end{equation}
For $n_f=2n$, we can satisfy this condition and completely break the
gauge symmetry with
\begin{equation}\label{ank}A= \begin{pmatrix} M & 0\end{pmatrix}, ~~B=
\begin{pmatrix} 0\\ N \end{pmatrix},\end{equation}
where here $M,N$ and 0 are all $n\times n$ blocks and $M$ and $N$
are generic.

For $n_f>2n$, we simply add more rows and columns of zeroes to $A$
and $B$.  However, for $n_f<2n$, it is not possible to completely
break the gauge symmetry while also satisfying (\ref{dunk}) and
the $D$-term condition $AA^\dagger=B^\dagger B$. That latter
condition implies that $A$ and $B$ have the same rank $r$.
(\ref{dunk}) implies that $2r\leq n_f$ so $r<n$ if $n_f<2n$. The
equation $AA^\dagger=B^\dagger B$ says that $B$ and $A^\dagger$
have the same kernel.  The dimension of the kernel is $n-r$, so if
$r<n$, the kernel is nonempty and complete Higgsing has not
occurred.

When complete Higgsing is possible, the three-dimensional gauge
theory has a critical point at the intersection of the Coulomb and
Higgs branches at which all vector multiplets are strongly coupled.
In our context, this gives the SCFT $\fB$ that is part of the
boundary conditions. When complete Higgsing is not possible, some
vector multiplets remain free in the IR.  For example, if $n_f=0$,
then the Coulomb branch is smooth and all vector multiplets are free
in the IR limit. Since we are interested in boundary SCFT's rather
than in brane configurations, we are not interested in considering
brane configurations whose Coulomb branch has degrees of freedom
that decouple in the IR.

Some brane configurations that do not obey our constraints are also
understandable, but it is not necessary to consider them. A more
precise description of the kind of infrared limit that we want in
our study of boundary conditions is given in section \ref{quivers},
along with another interpretation of the condition $n_f\geq 2n_c$.

\subsection{$S$-Duality}\label{sduality}

$S$-duality of a brane configuration can be defined in a purely
formal way. We simply replace NS5-branes with D5-branes, and
vice-versa. We also exchange $\vec X$ with $\vec Y$, or
equivalently, we make a spatial rotation transforming $x^4,x^5,x^6$
into $x^7,x^8,x^9$, and vice-versa.  The combined operation maps the
class of configurations that we have been considering back to
itself.

The only problem is that after this transformation, the branes are
incorrectly ordered; the first constraint of section \ref{ordering}
is not obeyed. The second constraint, which states that linking
numbers are nondecreasing for fivebranes of both types, remains
valid.

However, a reader familiar with three-dimensional mirror symmetry
may guess what to do. Brane configurations of the type considered
here can be manipulated without changing their infrared limit by
moving D5-branes in the $x^3$ direction.  The D5-branes may cross
NS5-branes, but every time a D5-brane is moved across an NS5-brane,
a D3-brane segment stretched between them will be created or
destroyed in such a way that the linking numbers remain constant. By
judiciously moving the D5-branes, we can get back to a configuration
that obeys the constraints.

 Though it is also
possible to move NS5-branes, this does not add anything. To make
NS5-branes cross each other, or D5-branes cross each other,
complicates the analysis of the infrared limit. The simplest type of
brane motion is to move  D5-branes relative to NS5-branes without
changing their ordering. This is enough to restore the first
constraint on ordering of branes, so it is the only operation we
need to consider.

We move any D5-brane whose linking number $l$ is smaller than the
total number of NS5-branes  to the interval between the $l^{th}$ and
$l+1^{th}$ NS5-branes.  Then the net number of D3-branes ending on
it will be zero. Any other D5-brane can be pushed  to the right of
all NS5-branes, and an appropriate number of D3-branes will end on
it from the right to give the right linking number.  Once this has
been done, our constraints are obeyed.  The number of D3-brane
segments between each consecutive pair of NS5-branes is uniquely
determined by the linking numbers of the NS5-branes.

There is only one catch: we need to show that the number of D3-brane
segments in the mirror configuration always turns out to be
positive.  For concreteness, let us denote the linking numbers of
the $P$ NS5-branes as $\ell_i,\,i=1,\dots,P$ and of the $Q$
D5-branes as $\tilde \ell_a,\,a=1,\dots,Q$. Let $n_i$ be the number
of D3-branes ending on the $i^{th}$ NS5-brane on the right. By
definition,
\begin{equation} \ell_i = n_{i}-n_{i-1} + \#\{a|\tilde \ell_a <i\}.
\end{equation} We can invert this relation as \begin{equation} n_i =
\sum_{j=1}^i \ell_j - \sum_{a | \tilde \ell_a<i} (i-\tilde \ell_a)
\end{equation} In particular, the number of D3-branes just to the right
of the rightmost NS5-brane in the original configuration is
\begin{equation} n_P = \sum_{j=1}^P \ell_j - \sum_{a | \tilde
\ell_a<P} (P-\tilde \ell_a).
\end{equation} Moving farther to the right, the number of D3-branes will
further increase when crossing the remaining D5-branes. At each D5
brane of linking number $\tilde \ell$, the number of D3-branes will
increase by $\tilde \ell -P$.  The final number of semi-infinite
D3-branes is
\begin{equation} \label{seminumber} n = \sum_{j=1}^P \ell_j - \sum_{a=1}^Q (P-\tilde \ell_a)
= \sum_{j=1}^P \ell_j + \sum_{a=1}^Q \tilde \ell_a -PQ.
\end{equation}

We can assume that in this original configuration, the numbers
$n_i$ of D3-brane segments are strictly positive; otherwise, the
system would break into decoupled subsystems which we would study
separately. Hence we have a simple inequality
\begin{equation}\label{guelfo} \sum_{j=1}^i \ell_j
> \sum_{a | \tilde \ell_a<i} (i-\tilde \ell_a).
\end{equation}

The sum on the right hand side starts from $a=1$ and ends at some
$a=b$. We can actually take $b$ to be unconstrained: if we lower
$b$ we are omitting some positive terms from the sum; if we
increase $b$, we include some extra non-positive terms in the sum.
Hence we can consider the more symmetric inequality
\begin{equation}\label{telfo} \sum_{j=1}^i \ell_j+  \sum_{a=1}^b \tilde \ell_a
> bi
\end{equation}

$S$-duality exchanges the two kinds of linking numbers $\ell$ and
$\tilde \ell$. This leaves the collection of inequalities
(\ref{telfo}) unaffected. In particular the dual number of D3-branes
\begin{equation} \tilde n_b = \sum_{a=1}^b \tilde \ell_a - \sum_{j |
\ell_j<b} (b-\ell_j)
\end{equation} is positive. Moreover, the number of semi-infinite D3-branes $n$ also
had a symmetric expression (\ref{seminumber}) in $\ell_j$ and
$\tilde \ell_a$. Thus, the class of brane configurations that obey
our constraints is closed under $S$-duality.  Starting with such a
configuration and applying $S$-duality, there is a unique way to
move D5-branes to put it back in the desired form.

We are interested in applying this result both for boundary
conditions in four dimensions and for purely three-dimensional
configurations, where no semi-infinite D3-branes are present. In
the above inequalities the two cases differ only by whether $n$ is
positive or zero. In case $n=0$, the linking numbers obey
$0<\ell_i<Q$ and $0<\tilde \ell_a <P$, ensuring that the original
and $S$-dual configurations have no D5-branes to the left or right
of all NS5-branes. For the purely three-dimensional
configurations, we have proved that every linear quiver with $n_f
\geq 2 n_c$ at each node has a mirror of the same kind.

In section \ref{domain}, we will consider a few examples of domain
walls, generated as configurations of branes with semi-infinite
D3-branes on both sides. In that case the linking numbers are still
nondecreasing from left to right, but they are not necessarily
positive. The inequalities above apply to such configurations with
very minor modifications.

\subsection{Quivers: Good, Bad and Ugly}\label{quivers}
In our study of $S$-duality of boundary conditions, we will need
some understanding of the infrared dynamics of three-dimensional
gauge theories defined by linear quivers of unitary groups. The
information we need can be extracted by supplementing what is
visible classically with the properties of monopole operators
\cite{Borokhov:2002ib,Borokhov:2002cg,Borokhov:2003yu}.

We will make extensive use of the operators constructed in those
papers, but our point of view is slightly different. We do not want
to assume properties of the infrared theory, so we define monopole
operators at short distances, using the fact that three-dimensional
gauge theory is ultraviolet-free, and then we see what deductions we
can make about the infrared behavior.

The definition of a monopole operator in three dimensions is
analogous to the definition of an 't Hooft operator in four
dimensions.  The definition is based on a codimension three magnetic
monopole singularity of gauge fields, leading to a local operator in
three dimensions, or a line operator in four dimensions.

\def\dd{\mathrm d}
\def\n{{\vec n}}
\def\O{{\mathcal O}}
Like disorder operators in statistical mechanics, monopole
operators are most easily defined by giving a recipe to calculate
in the presence of a monopole operator. If the gauge group is
$G=U(1)$, to compute in the presence of a charge $a$ monopole
operator inserted at a point $x=x_0$ in $\R^3$, we perform the
path integral over a space of fields with a suitable Dirac
monopole singularity:
\begin{equation}\label{ranc} F=\frac{a}{ 2}\star \dd\frac{1}{|\vec x-\vec
x_0|}.\end{equation} $a$ must be an integer. In supersymmetric
gauge theory, to define a BPS monopole operator, we pick one of
the scalar fields $\vec Y$ of the vector multiplet, say $Y_3$, and
require that it also should have a singularity compatible with the
Bogomolny equations $\dd Y_3=\star F$. The choice of $Y_3$ is
determined by the choice of a unit vector $\vec n$ in three-space,
and accordingly we will denote the resulting monopole operator of
charge $a$ as $\O_\n(a)$.

The special case of free $U(1)$ gauge theory (with ${\mathcal N}=4$
supersymmetry in three dimensions) is illuminating. The monopole
operator $\O_\n(a)$ can be written as $\exp(a (Y_3+i\phi)/e^2)$, or
equivalently as $\exp(a(\n\cdot \vec Y+i\phi)/e^2)$, where $\phi$ is
the dual photon. There is such an operator for any specified choice
of $\n$, with no reason to treat $\n$ as a collective coordinate.
The choice of a particular $\n$ breaks $SO(3)_Y$  to $SO(2)_Y$. This
enables us to define an $\N=2$ algebra\footnote{The supercharges of
three-dimensional $\N=4$ supersymmetry transform as $(2,2)$ under
$SO(3)_X\times SO(3)_Y$. The $R$-symmetry of an $\N=2$ subalgebra is
actually a diagonal subgroup of $SO(2)_X\times SO(2)_Y$, where
$SO(2)_X$ is an arbitrarily chosen $SO(2)$ subgroup of $SO(3)_X$.
$SO(3)_X$ leaves invariant the monopole operators of interest (since
it acts trivially on the fields whose singularity characterizes
them), as a result of which the $R$-symmetry acts on monopole
operators via $SO(2)_Y$.} whose global $R$-symmetry acts on monopole
operators via $SO(2)_Y$. The operator $\O_\n(a)$ is a chiral
operator (the lowest component of a chiral multiplet) from the point
of view of this $\N=2$ algebra. This multiplet has the unusual
property that its $R$-charge is zero, since $\exp(a (\n\cdot \vec
Y+i\phi)/e^2)$ is certainly invariant under rotations around $\vec
n$.  There are two important comments to make about this:

(1) If we want to place the operator $\O_\n(a)$ in a multiplet of
the microscopic global $\N=4$ supersymmetry, we have to let $\n$
vary. When we do this, we get an infinite-dimensional $\N=4$
multiplet, since the operator $\O_\n(a)$ has no simple dependence on
$\n$.

(2) In conformal field theory, the dimension of a chiral operator
is at least the $R$-charge, so $R$-charges of chiral operators
other than the identity are positive.  The fact that this theory
has a chiral operator of $R$-charge zero means that, although it
has global $\N=4$ supersymmetry, it cannot be given the structure
of a superconformal field theory in which the $R$-symmetry is the
microscopic $SO(3)_X\times SO(3)_Y$ symmetry.  (This can also be
shown more directly by observing that the field $Y_1+iY_2$ is a
chiral operator with $R$-charge 1 and dimension $1/2$.)  After
dualizing the photon, the model describes four free scalars and
four free spinors, so it can be given an $\N=4$ superconformal
structure, but the $R$-symmetry is not the one that one sees in
the ultraviolet.\footnote{\label{sasaki}A necessary condition
\cite{deWit:1998zg,Gibbons:1998xa} for an $\N=4$ superconformal
structure is that, near some chosen vacuum, the Coulomb branch
should look like a tri-Sasakian cone.  This means that the metric
must be conical, with a scaling symmetry generated by a vector
field $V$ that obeys $D_\nu V^\mu=\delta^\mu_\nu$, and moreover
the generators of the $SO(3)$ $R$-symmetry must be the vector
fields $W^k=I^k V$, where $I^k$, $k=1,2,3$, are the three complex
structures.  For a free vector multiplet, the Coulomb branch is a
smooth manifold $\R^3\times S^1$, so it looks conical near any
point, but the microscopic $SO(3)_Y$ $R$-symmetry (which acts by
rotating $\R^3$) does not have the required form, regardless of
which vacuum we choose in taking the infrared limit.}

\subsubsection{$U(1)$ Examples With Hypermultiplets}\label{hyperex}

The next step, still with gauge group $U(1)$, is to add
hypermultiplets of charges $q_1,\dots,q_s$.  The $q_i$ are all
non-negative integers.  (A hypermultiplet contains fields of equal
and opposite charge, and by convention we take the positive sign for
the hypermultiplet charge.)  The monopole operator of charge $a$ is
defined in the same way, and is still a chiral operator, but as
shown in \cite{Borokhov:2002cg}, it now carries a non-zero
$R$-charge
\begin{equation}\label{rcharge}
q_R=\frac{1}{2}\sum_{i=1}^s|aq_i|,\end{equation} due to an asymmetry
in the fermionic spectrum.\footnote{The computation in
\cite{Borokhov:2002cg} is justified by using a large $n_f$ limit to
suppress fluctuations. Here, we use the fact that the gauge theory
is free in the ultraviolet to justify the computation for all $n_f$.
The Dirac equation in an external field is conformally invariant,
justifying the conformal mapping to $\R\times S^2$ that is used in
\cite{Borokhov:2002cg}.}

Since the monopole operators $\O_\n(a)$ now have positive
$R$-charges,  it is conceivable that such a theory might flow to an
infrared critical point in which the $R$-symmetry is the microscopic
$SO(3)_X\times SO(3)_Y$. We will refer to any $\N=4$ critical point
with this $R$-symmetry as a standard critical point.

Since (by definition of a Higgs or Coulomb branch) $SO(3)_Y$ acts
trivially on a Higgs branch of vacua and $SO(3)_X$ acts trivially on
a Coulomb branch, a vacuum at the intersection of the two branches
is automatically $SO(3)_X\times SO(3)_Y$-invariant.  So this is a
candidate for the locale of a standard critical point. A necessary
condition for this \cite{deWit:1998zg,Gibbons:1998xa} is that the
Higgs and Coulomb branches must both be tri-Sasakian cones near
their intersection, with the microscopic  $SO(3)_X$ and $SO(3)_Y$
$R$-symmetries entering in the tri-Sasakian structures. In $\N=4$
supersymmetric gauge theories (without bare masses, FI terms, or
Chern-Simons couplings) the Higgs branch always has the appropriate
tri-Sasakian structure. Mirror symmetry sometimes makes it possible
to show that the Coulomb branch also has the right structure, near
its intersection with the Higgs branch. For instance, this is true
for models derived from linear quivers with $n_f\geq 2n_c$ at every
node, since we have shown in section \ref{sduality} that such models
have mirrors of the same type. It is reasonable to expect that when
both branches have the appropriate structure near their
intersection, a standard critical point does indeed arise at this
intersection.

\def\OO{\Bbb O}

If a $U(1)$ theory coupled to hypermultiplets flows in the
infrared to a standard critical point, the structure of the
superconformal algebra $OSp(4|4)$ implies that in this limit the
operators $\O_\n(a)$ transform in a finite-dimensional multiplet,
even though this is not true in the ultraviolet.  Indeed, in the
infrared theory, the operator $\O_\n(a)$ has dimension $q_R$, and
is part of an irreducible $\frak{so}(3)_Y$ representation of
dimension $2q_R+1$. (In an appropriate formalism, $\O_\n(a)$
varies with $\n$ as a holomorphic section of the line bundle
$\OO(q_R)\to \Bbb{CP}^1$. The relevant multiplet is an $\OO(q_R)$
multiplet, in the language of
\cite{Lindstrom:1987ks,Lindstrom:2008gs}.) Otherwise, $\O_\n(a)$
would be related by repeated action of raising or lowering
operators in $\frak{so}(3)_Y$ to an operator of dimension $q_R$
but with $R$-charge greater than $q_R$ in absolute value. This
would contradict unitarity of the IR fixed point.

Let us consider a few special cases.  If there is only a single
hypermultiplet of charge 1, the $R$-charge is $q_R=|a|/2$. Setting
$a=\pm 1$, the basic monopole operators have $R$-charge $\pm 1/2$.
In three dimensions, unitarity implies that a chiral superfield of
$R$-charge and dimension 1/2 is actually part of a free
hypermultiplet.  In the present case, we actually have a twisted
hypermultiplet, in the sense that it transforms non-trivially under
$SO(3)_Y$ and trivially under $SO(3)_X$, like the scalar fields
$\vec Y$ of the vector multiplet and in contrast to the bosonic
fields of an ordinary electrically charged hypermultiplet.  The
existence of this field shows \cite{Borokhov:2002cg} that the
Coulomb branch of the $U(1)$ theory with a single charge 1
hypermultiplet is equivalent to $\Bbb{R}^4$, parametrized by a free
twisted hypermultiplet.

A free hypermultiplet has a global symmetry group $SU(2)$, commuting
with the superconformal group.  Let us try to find this symmetry.
It helps to know that in a theory with $\N=4$ superconformal
symmetry, a conserved current $J$ appears in a multiplet whose
lowest component $\mu$ is an $\N=2$ chiral superfield with dimension
and $R$-charge 1.  (In free field theory, $\mu$ is the moment map
for the symmetry associated with the conserved current $J$.)

One symmetry of the Coulomb branch of this $U(1)$ theory with one
hypermultiplet is visible classically.   This is the group
$U(1)_\phi$ of translations of the dual photon, $\phi\to\phi+{\rm
constant}$. To find additional symmetries, we need monopole
operators of $R$-charge 1.  Such operators arise precisely for
$a=\pm 2$, and their presence extends the classical symmetry of the
Coulomb branch from $U(1)$ to $SU(2)$.

Under shifts of the dual photon, the monopole operator
$\O_\n(a)=\exp(a(Y_3+i\phi)/e^2)$ transforms by a phase that is
clearly proportional to $a$.  The conserved currents associated
with operators  $\O_\n(\pm 2)$ therefore do not commute with
$U(1)_\phi$.  Together they generate a three-dimensional Lie
algebra which must be $SU(2)$. This is the expected symmetry of
the Coulomb branch.

The fact that the hypermultiplet fields arise for $a=\pm 1$ and the
symmetry currents for $a=\pm 2$ has a simple interpretation: the
hypermultiplet fields transform in the two-dimensional
representation of $SU(2)$, whose weights are one-half the nonzero
weights of the adjoint representation.

For our next example, consider the case of two hypermultiplets of
charges $q_1=q_2=1$.  The $R$-charge of a monopole operator is
$q_R=|a|$.  The smallest possible value is 1, for $a=\pm 1$. Again,
the existence of these chiral operators of $q_R=1$ means that if the
theory flows to a standard IR critical point, the classical $U(1)$
symmetry of the Coulomb branch is extended to $SU(2)$.  In contrast
to the case of $n_f=1$, where the Coulomb branch has an $SU(2)$
symmetry that follows from something more fundamental (existence of
a free twisted hypermultiplet), for $n_f=2$ the $SU(2)$ symmetry of
the Coulomb branch seems to be an irreducible statement.

The $U(1)$ theory with $n_f=2$ hypermultiplets of charge 1 has a
Higgs branch\footnote{This is the $n_c=1$ case of what is
described in eqn. (\ref{ank}).} and is believed to flow to a
standard critical point. Indeed, this model was one of the
original examples of three-dimensional mirror symmetry
\cite{Intriligator:1996ex}. (For an explanation of its mirror
symmetry, see fig. \ref{Fig10}.) The Higgs branch of the model has
a classical $SU(2)$ global symmetry, rotating the two
hypermultiplets.  Mirror symmetry exchanging the Higgs and Coulomb
branches  implies that the Coulomb branch must also have an
$SU(2)$ symmetry in the infrared, as we have just argued in
another way.

The IR critical point of $U(1)$ coupled to two hypermultiplets will
turn out to be an important example for understanding $S$-duality of
boundary conditions. We will call it $T(SU(2))$. As we have just
seen, this model has $SU(2)\times SU(2)$ global symmetry, with one
factor acting on the Higgs branch one and on the Coulomb branch.
(The group that acts faithfully is really $SO(3)\times SO(3)$.)

Continuing in this vein, consider $U(1)$ coupled to $n_f>2$
hypermultiplets of charge 1.  Monopole operators have $R$-charges
$|an_f|/2$.  As these numbers are greater than 1 (for $|a|\geq
1$), the Coulomb branch has no symmetries beyond its classical
$U(1)$ symmetry.  Likewise, there are no free
hypermultiplets.

What happens if we couple $U(1)$ gauge theory to hypermultiplets of
charge greater than 1?  The only case that leads to a monopole
operator of $|q_R|\leq 1$ is the case that we couple to one
hypermultiplet of charge 2, leading for $a=\pm 1$ to $|q_R|=1$. This
theory is simply an orbifold of the $n_f=1$ theory. The orbifolding
operation is a shift in the dual photon, changing the quantum of
charge.  In the orbifolded theory, the Coulomb branch is $\R^4/\Z_2$
rather than $\R^4$.  The monopole operators of $q_R=1/2$ are
projected out, but those of $q_R=1$ persist.

Returning to the case of $n_f$ hypermultiplets of charge 1, let us
summarize some properties of these models:

(1) For $n_f\geq 2$, we get what we will call ``good'' theories.
They have Higgs branches, and flow to standard IR critical points
at the intersection of the Higgs and Coulomb branches. These are
non-Gaussian critical points, as is clear from the singularity of
the moduli space of vacua, and no more elementary description of
them is evident. (They have equally good mirror descriptions.) The
borderline case $n_f=2$ is what we will call a ``balanced''
theory, with $n_f=2n_c$ (here $n_c=1$). In the balanced theory,
the classical $U(1)$ symmetry of the Coulomb branch is extended in
the IR to $SU(2)$.

(2) The theory with $n_f=1$, which has no Higgs branch, still
flows in the IR to a standard critical point.  However, this
critical point is Gaussian and has a more economical description
in terms of a free twisted hypermultiplet.  We regard the $U(1)$
theory with $n_f=1$ as an ``ugly'' description of a theory that
actually is Gaussian.

(3) Finally, the theory with $n_f=0$ is ``bad'' in that, because
it has chiral operators of $R$-charge 0, it cannot flow in the IR
to a standard critical point.

\subsubsection{Monopole Operators In Nonabelian Gauge
Theory}\label{nonamon}

To describe monopole operators in a three-dimensional theory with
any gauge group $G$, we first pick a homomorphism
$\rho:\frak{u}(1)\to\frak g$.  $\rho$ plays the role of the monopole
charge $a$ in the $U(1)$ case.  Then we modify the ansatz
(\ref{ranc}) for the singularity characterizing the monopole
operator to
\begin{equation}\label{tanc}F=\frac{\rho(1)}{ 2}\star \dd\frac{1}{|\vec x-\vec
x_0|},\end{equation} where 1 is a generator of $\frak{u}(1)$, and
$\rho(1)$ is its image in $\frak g$.  After requiring that $X_3 $
should have a singularity compatible with the Bogomolny equations,
we arrive at the definition of a monopole operator in the nonabelian
case.

The $R$-charges of these operators were computed in
\cite{Borokhov:2002cg}.  Let $h_i$ and $v_i$ be the charges of
vector multiplets and hypermultiplets under the $U(1)$ subgroup of
$G$ that is defined by $\rho$.  (The quantities $h_i$ correspond to
$aq_i$ in the notation we used for $G=U(1)$.)  The $R$-charge of the
monopole operator defined by $\rho$ is then
\begin{equation}\label{zanc}q_R=\frac{1}{2}\left(\sum_i|h_i| - \sum_j
|v_j|\right).\end{equation}

The fact that vector multiplets and hypermultiplets make
contributions to $q_R$ of equal magnitude and opposite sign is
clear if one considers the special case that the hypermultiplets
and vector multiplets transform in the same representation of $G$.
This gives a theory with enhanced supersymmetry ($\N=8$ in three
dimensions) and $R$-symmetry.  The unbroken $SO(2)_Y$ of the
monopole operator is extended to $SO(6)_Y$.  The monopole operator
for given $\rho$ now furnishes a one-dimensional representation of
$SO(6)_Y$, and as this group has no non-trivial one-dimensional
representation, $q_R$ must vanish, along with its $SO(6)_Y$
completion.

To orient ourselves to the implications of (\ref{zanc}), we will
consider a basic example: $U(n_c)$ gauge theory with $n_f$
fundamental hypermultiplets.  We define $\rho:\frak{u}(1)\to
\frak{u}(n_c)$ by giving a diagonal matrix of integers
$a_1,a_2,\dots,a_n$. (The $a_i$ are only defined up to
permutation.) With a little group theory, we find that
\begin{equation}\label{danc}q_R=\frac{n_f}{2}\sum_{i=1}^{n_c}|a_i|-\sum_{1\leq
i<j\leq n_c}|a_i-a_j|.\end{equation} Alternatively, this can be
written
\begin{equation}\label{obanc}q_R=\frac{n_f-2n_c+2}{2}\sum_{i=1}^{n_c}|a_i|+\sum_{1\leq
i<j\leq n_c}\bigl(|a_i|+|a_j|-|a_i-a_j|\bigr).\end{equation} This
formula is useful, since $|a_i|+|a_j|-|a_i-a_j|\geq 0$ for all
$i,j$.

We can draw the following conclusions:

(1) First consider the ``good'' theories with Higgs branches, the
ones with $n_f\geq 2n_c$.  If $n_f>2n_c$, then $|q_R|>1$ for all
monopole operators. This follows immediately from (\ref{obanc}).
There are no free hypermultiplets and no enhanced symmetries of
the Coulomb branch. The last good theory is the ``balanced''
theory with $n_f=2n_c$. In this theory, the monopole operators in
which the $a_i$ are $(\pm 1,0,\dots,0)$ have $q_R=1$.  So, just as
we saw earlier for $n_c=1$, the classical $U(1)$ symmetry of the
Coulomb branch is extended to $SU(2)$.

(2) Now consider the next case, $n_f=2n_c-1$, which we consider
``ugly.'' Here we get $q_R=1/2$ if the $a_i$ are $(\pm
1,0,\dots,0)$, so again there is  a free twisted hypermultiplet. In
addition, $q_R=1$ arises from $(\pm 2,0,\dots,0)$ and also from
$(1,-1,0,\dots,0)$. (The first of these we have already seen for
$n_c=1$.)  Combining this together, the Coulomb branch has a {\it
four}-dimensional group of symmetries. By looking at how the
operators transform under $U(1)_\phi$, one can see that the group is
$SU(2)\times U(1)$.

\def\C{{\mathcal C}}
\def\H{{\mathcal H}}
What does this mean?  The existence of a free twisted hypermultiplet
is consistent with the fact that, since $n_f<2n_c$, the model cannot
be completely Higgsed.  Instead, adapting the logic of eqn.
(\ref{ank}), one finds that it can be Higgsed to $U(1)$, so it has a
branch of vacua of the form $\C\times \H$, where $\C$ parametrizes a
$U(1)$ vector multiplet and $\H$ parametrizes the expectation values
of the hypermultiplets.  The factor $\C$ is associated with  the
free twisted hypermultiplet. The $U(n_c)$ theory with $2n_c-1$
flavors must be equivalent in its standard IR fixed point to a
theory of a free twisted hypermultiplet times some other theory that
can be completely Higgsed and whose Higgs branch is $\H$. To find
this second theory, move on the Coulomb branch of the $U(n_c)$
theory to the locus where $U(n_c)$ is broken to $U(1)\times
U(n_c-1)$ and the $U(1)$-invariant hypermultiplets are massless.  It
is then possible to give expectation values to those
hypermultiplets.  The result is a component of the moduli space of
vacua that is of the form $\C\times \H$, where $\C$ has hyper-Kahler
dimension 1, and $\H$ is the Higgs branch of a $U(n_c-1)$ theory
with $n_f=2n_c-1$.

(3) Finally, the models with $n_f=2n_c-2$ have monopole operators of
$q_R=0$, and those of $n_f<2n_c-2$ have monopole operators of
negative $q_R$.  So these ``bad'' models do not have standard IR
critical points.

\subsubsection{Quiver Theories}\label{quivertheories}

We now consider a gauge theory derived from a general linear
quiver with $P-1$ nodes.\footnote{A linear array of $P-1$ nodes
can be viewed as the Dynkin diagram of the Lie group
$\textsf{A}_{P-1}$. In section \ref{bifurc}, we will obtain
results for Dynkin diagrams of type $\textsf{D}$ that are
analogous to what we will explain here for type $\textsf{A}$.  We
also will give partial results for quivers of type
$\textsf{E}_6,\textsf{E}_7,$ and $\textsf{E}_8$.} At the $i^{th}$
node there is a $U(n_i)$ gauge theory, coupled to $m_i$
fundamental hypermultiplets. There is also a bifundamental
hypermultiplet of $U(n_i)\times U(n_j)$ if and only if $j=i\pm 1$.

We define a ``good'' quiver to be one for which $n_f\geq 2n_c$ at
each node.  Explicitly, this means that the quantities
\begin{equation}\label{elf}e_i=m_i+n_{i-1}+n_{i+1}-2n_i\end{equation}
are non-negative.  We call $e_i$ the
``excess'' at the $i^{th}$ node, and say that a node is
``balanced'' if it has zero excess.  Our first goal is to show
that in a gauge theory derived from a ``good'' quiver, every
monopole operator has $q_R\geq 1$.  We also want to determine for
every good quiver precisely which monopole operators have $q_R=1$.
This will give us the symmetry of the Coulomb branch.

Consider a monopole operator whose  magnetic charges at the
$i^{th}$ node are integers $a_{i,k}$,  for  $1\leq k \leq n_i$.
The $R$-charge $q_R$ of the monopole operator receives a
contribution
\begin{equation} \frac{m_i}{2} \sum_{k=1}^{n_i} |a_{i,k}|
\end{equation} from the $m_i$ flavors at each node, a contribution
\begin{equation}-\frac{1}{2} \sum_{k=1}^{n_i} \sum_{t=1}^{n_i} |a_{i,k} -
a_{i,t}|\end{equation}
 from the vector multiplets at that node, and a contribution
\begin{equation} \frac{1}{2} \sum_{k=1}^{n_i} \sum_{t=1}^{n_{i+1}} |a_{i,k} -
a_{i+1,t}|\end{equation} from the bifundamental matter between the
nodes $i$ and $i+1$.

If we plug the definition of $e_i$ into the $R$-charge formula,
and make substitutions like $n_i = \sum_{k=1}^{n_i}1$ judiciously,
we find that $q_R=\sum_i\left(\Delta_i+A_i+B_i\right)$, where
\begin{equation}\Delta_i=\frac{ e_i}{2} \sum_{k=1}^{n_i} |a_{i,k}| \end{equation}
is nonnegative, and the other contributions are
\begin{align}A_i&=\frac{1}{2} \sum_{k=1}^{n_i} \sum_{t=1}^{n_i} \left( |a_{i,k}| +
|a_{i,t}|-|a_{i,k} - a_{i,t}|\right) \cr B_i&=- \frac{1}{2}
\sum_{k=1}^{n_i} \sum_{t=1}^{n_{i+1}} \left( |a_{i,k}| +
|a_{i+1,t}|- |a_{i,k} - a_{i+1,t}|\right). \end{align} Each term
in these sums is of the form $|x|+|y|-|x-y|$, and is zero if $x$
and $y$ have opposite signs. If $x$ and $y$ are both of the same
sign, $|x|+|y|-|x-y| = 2\,\mathrm{min}(|x|,|y|)$.

These formulas make clear that the total $R$-charge is the sum of
a contribution from those charges $a_{i,k}$ that are positive,
plus a contribution from those charges $a_{i,k}$ that are
negative. The contribution from positive charges can be computed
by setting the negative charges to zero, and vice-versa.
 Without essential loss of generality, we
are then free to consider a configuration with non-negative
charges only.  A further simplification is as follows. If all
$a_{i,k}=0$ for some node $i$, we can erase that node from the
quiver and add its rank $n_i$ to the number of flavors of the
neighboring nodes without changing $q_R$. We can then treat each
disconnected component of the reduced quiver separately. This
means that we can restrict the analysis to monopoles with some
nonzero charge at each node.

Let us also order the charges at each node in a nondecreasing
fashion $a_{i,k+1} \geq a_{i,k}$. Then we can write
\begin{equation}\label{owb}A_i=\frac{1}{2} \sum_{k=1}^{n_i} \sum_{t=1}^{n_i} \,2\,\,\mathrm{min}(
a_{i,k},a_{i,t}) = \sum_{k=1}^{n_i} a_{i,k} (2 n_i - 2k +1)
\end{equation}
and similarly
\begin{equation}\label{otowb} B_i=- \frac{1}{2} \sum_{k=1}^{n_i} \sum_{t=1}^{n_{i+1}}
\,2\,\,\mathrm{min}(a_{i,k},a_{i+1,t}) \end{equation} Now we want
a lower bound
\begin{equation}\label{lowb} B_i\geq  -\sum_{k=1}^{n_i} a_{i,k} (n_i - k ) -
\sum_{t=1}^{n_{i+1}} a_{i+1,t} (n_{i+1} - t +1), \end{equation}
which will give a lower bound on $q_R$. We can get a lower bound on
$B_i$ by replacing $\mathrm{min}(a_{i,k},a_{i+1,t})$ with either
$a_{i,k}$ or $a_{i+1,t}$.  We pick $a_{i,k}$ if $t>n_{i+1}-n_i+k$,
which for given $k$ happens for at most $n_i-k$ values of $t$; and
we pick $a_{i+1,t}$ if $t\leq n_{i+1}-n_i+k$, which for given $t$
happens for at most $n_{i+1}-t+1$ values of $k$.  Adding up the
possibilities now yields the lower bound (\ref{lowb}).

The formula (\ref{owb}) and lower bound (\ref{lowb}) lead to a very
useful result when we sum over $i$.  Everything cancels except for
contributions from the ends of the chain, and we get simply
\begin{equation}\label{howb} \sum_{i=1}^{P-1}(A_i+B_i)\geq
\sum_{k=1}^{n_1}
a_{1,k}(n_1-k+1)+\sum_{k=1}^{n_{P-1}}a_{P-1,k}(n_{P-1}-k).\end{equation}
So
\begin{equation}\label{powb} q_R\geq \sum_{i=1}^{P-1}\frac{ e_i}{2} \sum_{k=1}^{n_i} a_{i,k}
+\sum_{k=1}^{n_1}
a_{1,k}(n_1-k+1)+\sum_{k=1}^{n_{P-1}}a_{P-1,k}(n_{P-1}-k).\end{equation}

In particular, since $e_i\geq 0$ and we have performed this
computation in a sub-quiver in which $a_{1,k}\not=0$ for some $k$,
we get the desired result that $q_R\geq 1$. The inequality
$q_R\geq 1$ actually holds separately for the contributions to
$q_R$ from positive monopole charges $a_{i,k}$ as well as the
contribution from negative ones.  So a monopole operator of
$q_R=1$ has all $a_{i,k}$ non-negative or all $a_{i,k}$
non-positive.

We further see that in order to get $q_R=1$, we must have all
$a_{i,k}=0$ unless the $i^{th}$ node is balanced, that is, unless
$e_i=0$.  So to study monopole operators of $q_R=1$, we can
restrict ourselves to quivers in which every node is balanced.
Moreover, we can assume that the set of nodes with nonzero charge
is connected.  Otherwise, each component would contribute at least
1 to $q_R$.

So now we want to consider a connected linear quiver with every
node balanced.  We want to determine exactly which monopole
operators with nonzero charge at each node have $q_R=1$. For this,
a slight generalization of the above inequalities is useful.
Obviously, the inequality (\ref{lowb}) has a ``mirror image,''
\begin{equation}\label{ozowb} B_i\geq  -\sum_{k=1}^{n_i} a_{i,k} (n_i - k+1 ) -
\sum_{t=1}^{n_{i+1}} a_{i+1,t} (n_{i+1} - t ). \end{equation} Now
after picking a node $s$, to get a lower bound on $\sum_iB_i$, we
use the mirror image inequality for $i<s$ and the original one for
$i\geq s$.  The result is a slightly modified inequality for
$q_R$:
\begin{equation}\label{plowb} q_R\geq \sum_{i=1}^{P-1}\frac{ e_i}{2} \sum_{k=1}^{n_i} a_{i,k}
+\sum_{k=1}^{n_1}
a_{1,k}(n_1-k)+\sum_{k=1}^{n_s}a_{s,k}+\sum_{k=1}^{n_{P-1}}a_{P-1,k}(n_{P-1}-k).\end{equation}
Now we see to get $q_R=1$, it must be that at each value of $s$ in
the reduced quiver, the monopole charges take the form
$a_{s,k}=(0,0,\dots,0,1)$.

The complete set of monopole operators with $q_R=1$ is therefore
easy to describe.  Let $b_i=\sum_k a_{i,k}$. A monopole operator of
$q_R=1$ is completely determined by the $b_i$. The reduced quiver is
supported in a range $i_0\leq i\leq i_1$ for some $i_0,i_1$. The
$b_i$ (and $a_{i,k}$) vanish outside this range, and in the range
the $b_i$ are all 1, or all $-1$.

We want to argue that for a linear quiver with $P-1$ consecutive
balanced nodes, the classical symmetries of the Coulomb branch
combine with the monopole operators to generate an $SU(P)$
symmetry group. The classical symmetries of the Coulomb branch,
acting by translations of dual photons, are an abelian group that
we will call $U(1)_\phi^{P-1}$.

We identify the group $U(1)_\phi^{P-1}$ with the maximal torus of
$SU(P)$.  Its action on a monopole operator can be read off from
the charges $b_i$.  The simple roots of $SU(P)$ correspond to
monopole operators with only a single $b_i$ equal to 1, and the
rest vanishing.  The other monopole operators of $q_R=1$ furnish
the other roots of $SU(P)$.

So a string of $P-1$ balanced nodes in any quiver with a standard
IR limit leads to an $SU(P)$ symmetry of the Coulomb branch. More
generally, the symmetry of the Coulomb branch for any good linear
quiver is as follows. Every unbalanced node with $e_i>0$
contributes a $U(1)$ factor.  Every sequence of $P-1$ balanced
nodes contributes a factor $SU(P)$.

These results can also be obtained from mirror symmetry. To have
$P-1$ successive balanced nodes in a linear quiver means that $P$
consecutive NS5-branes have the same linking number.  In the
mirror, the $P$ dual D5-branes, since they have the same linking
number, are located in the same D3-brane segment. They can be
taken to be located at the same point in space and therefore
generate a $U(P)$ symmetry, of which the center may act trivially,
depending on the details of the quiver.  We factor $U(P)$ as
$SU(P)\times U(1)$.  Every cluster of D5-branes at the same
location gives a $U(1)$ factor in the symmetry group (one overall
diagonal $U(1)$ decouples), and in addition every cluster of $P>1$
D5-branes gives an $SU(P)$ factor.

Now we would like to analyze ``ugly'' and ``bad'' quivers.  We call
a quiver ugly if the smallest value of $q_R$ is 1/2, so that there
can be a standard infrared limit, but it must have free twisted
hypermultiplets.  And we call a quiver bad if there are monopole
operators of $q_R\leq 0$, so that a standard infrared limit is not
possible.  For example, any quiver with a node of $e_i\leq -2$ is
bad, since there exist monopole operators with charges only at that
node and $q_R\leq 0$.  So an ugly quiver has $e_i\geq -1$  for all
nodes.  An ugly linear quiver has at least one of the $e_i$ equal to
$-1$.  We call a node with $e_i=-1$ minimally unbalanced. We will
mainly be concerned with linear quivers with a single minimally
unbalanced node, and $e_i\geq 0$ for all other nodes.  These quivers
turn out to be always ugly.\footnote{Quivers with more than one
minimally unbalanced node can be either ugly or bad.  For example, a
linear quiver with two nodes of $e=-1$ connected by a chain of
balanced nodes is bad, since one can explicitly exhibit a monopole
operator of $q_R=0$. A linear quiver in which all nodes have
$e_i\geq -1$ and every two nodes of $e=-1$ are separated by a node
of $e>0$ is ugly. This can be shown by further use of the
inequalities (\ref{lowb}) and (\ref{ozowb}).}

Let the minimally unbalanced node of such a quiver be at position
$s_0$. Consider the inequality (\ref{plowb}) for
$s=s_0$:\begin{equation}\label{uplowb} q_R\geq \sum_{i \not=
s_0}\frac{ e_i}{2} \sum_{k=1}^{n_i} a_{i,k} +\sum_{k=1}^{n_1}
a_{1,k}(n_1-k)+ \frac{1}{2}
\sum_{k=1}^{n_{s_0}}a_{s_0,k}+\sum_{k=1}^{n_{P-1}}a_{P-1,k}(n_{P-1}-k).\end{equation}
$q_R$ may be 1/2 only if $a_{s_0,k}$ is of the usual form
$(0,0,\dots,0,1)$ and any node with $e_i >0$ has charge zero.
Suppose that there are $P_1-1$ consecutive balanced nodes to the
left of $s_0$ and $P_2-1$ consecutive balanced nodes on the right.
To get $q_R=1/2$, the inequality (\ref{plowb}) for each balanced
node forces the charges at that node to be of the usual form
$a_{i,k}=(0,0,\dots,0,b_i)$ with $b_i=0,1$. The nodes with
non-zero monopole charge must form a connected set for the same
reason as before (or we will get $q_R\geq 3/2$). There are $P_1
P_2$ such monopoles, associated with all possible reduced quivers
supported in a range $i_0\leq i\leq i_1$ for some $i_0,i_1$ with
$i_0\leq s_0\leq i_1$. All these monopoles actually have
$q_R=1/2$.

There are several ways to construct monopoles with $q_R=1$. The
ones with zero charge at the minimally unbalanced node combine
with the classical symmetries at the balanced nodes to give a
$SU(P_1) \times SU(P_2)$ symmetry group. The $P_1 P_2$ monopoles
of charge $q_R=1/2$ carry the weights of a bifundamental
representation of this $SU(P_1) \times SU(P_2)$ symmetry group. As
the monopoles of charge $q_R=1/2$ are expected to flow to free
twisted hypermultiplets in the infrared, we expect a full $Sp(2
P_1 P_2)$ symmetry group acting only on them. Apart from this,
there may be symmetries which act trivially on the free twisted
hypermultiplets. Indeed many more monopoles of charge $q_R=1$ can
be found which have nonzero charges at the unbalanced node. The
full analysis of the symmetry group is complicated and depends on
the $n_i$.

\section{ $S$-Duality for $U(n)$ Boundary
Conditions}\label{examples}

We will now use the brane constructions reviewed and analyzed in the
last section to study $S$-duality of boundary conditions in $U(n)$
gauge theory.  Before considering examples, we make a few general
remarks.

The duality transformation $S:\tau\to -1/\tau$ exchanges D5-branes
and NS5-branes.  To ensure that the class of brane configurations we
consider is $S$-invariant (rather than being mapped by $S$ to a
different but equivalent class), we accompany $S$ with a rotation
that exchanges $\vec X$ and $\vec Y$.  The $R$-symmetries of Higgs
and Coulomb branches are $SO(3)_X$ and $SO(3)_Y$, respectively, so
$S$ exchanges Higgs and Coulomb branches.

The analyses of $S$-duality of brane configurations will almost
always depend on the freedom to move D5-branes in the $y$
direction. The precise positions of the D5-branes are irrelevant
in, but only in, the infrared limit. For this reason, brane
methods are natural for studying $S$-duality of infrared critical
points.

Focusing on critical points is not a real limitation. Once one
establishes $S$-duality between two conformally invariant boundary
conditions, one can expect to follow the duality after turning on
relevant operators on the boundary.

The IR limits that are naturally studied via branes are standard IR
limits in the sense of section \ref{quivers} -- superconformal
critical points at which the $R$-symmetry is the one seen in the
ultraviolet.  The ultraviolet $R$-symmetry is the $R$-symmetry that
is visible in a brane configuration and whose behavior under
$S$-duality is known. However, some constructions depend on global
symmetries (as opposed to $R$-symmetries) that only exist in the IR
limit.

Some final remarks mainly concern notation.   We will generically
write $\B$ for a boundary condition, and $\fB$  for a boundary SCFT.
An important special case is that $\B$ might be constructed from
Neumann boundary conditions coupled to some boundary SCFT $\fB$.
Then we say that $\B$ is the boundary condition associated to $\fB$.
(It is not true that every boundary condition is associated in this
way to a boundary theory, since other ingredients -- Nahm poles and
reduction of gauge symmetry -- can also enter.)  If $G$ is a compact
group, we write $G^\vee$ for its dual group.  If $\B$ is a boundary
condition in $G$ gauge theory, then the $S$-dual of $\B$ is a
boundary condition in $G^\vee$ gauge theory; we denote this $S$-dual
as $\B^\vee$.  If $\B^\vee$ is obtained by coupling Neumann boundary
conditions to a boundary SCFT, then we denote this SCFT as
$\fB^\vee$.

An important point is that $\fB^\vee$ is {\it not} the $S$-dual of
$\fB$.  Such a statement would not even make sense, since
$S$-duality is an operation on four-dimensional field theories,
while $\fB$ and $\fB^\vee$ (when they exist) are three-dimensional
SCFT's.  There is, however, an operation of mirror symmetry for
three-dimensional SCFT's \cite{Intriligator:1996ex} that is closely
related to four-dimensional $S$-duality.  The mirror of a
three-dimensional SCFT $\fB$ is a theory that we will call $\tilde
\fB$, obtained by exchanging the Higgs and Coulomb branches of
$\fB$.  It turns out that when $\fB^\vee$ exists, it is possible to
construct its mirror $\tilde\fB^\vee$ directly from $\fB$.
Explaining this will be one of our main goals.  But first we will
work out a number of examples.

\subsection{$U(1)$ Examples}\label{u1}

We begin with the case of a single D3-brane.  This means that the
bulk gauge group is $G=U(1)$, and that electric-magnetic duality
in bulk can be explicitly understood.  It maps the field strength
$F$ to a multiple of $\star F$, where $\star$ is the Hodge star.

\subsubsection{Dirichlet And Neumann}\label{dirn}

The simplest statement of all is that Dirichlet boundary conditions
are dual to Neumann boundary conditions. Dirichlet boundary
conditions assert that $F_{\mu\nu}=0$ on the boundary, where
$\mu,\nu=0,1,2$ are tangent to the boundary, or more succinctly
$F|=0$.  Neumann boundary conditions assert that $F_{\mu 3}=0$ on
the boundary, or more succinctly $\star F|=0$. These two conditions
are exchanged via $F\leftrightarrow \star F$.

\begin{figure}
  \begin{center}
    \includegraphics[width=4.5in]{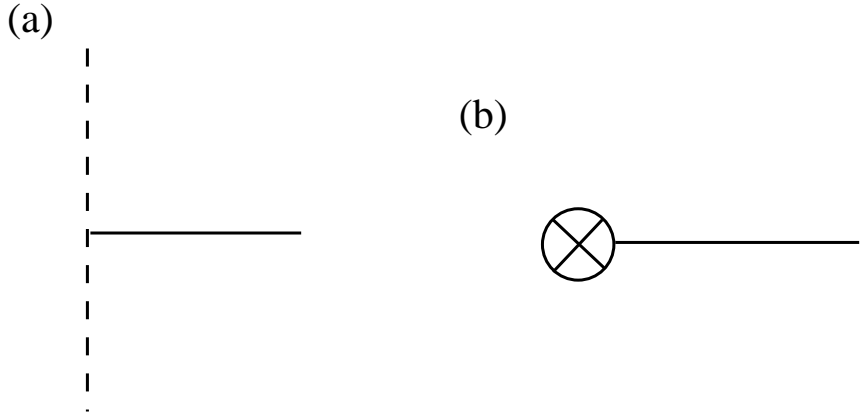}
  \end{center}
\caption{\small The most basic boundary conditions in $U(1)$ gauge
theory are Dirichlet and Neumann.  These arise for a single D3-brane
ending on a D5-brane, as in (a), or an NS5-brane, as in (b).}
  \label{Fig6}
\end{figure}

The corresponding brane picture is simple (fig. \ref{Fig6}).
Dirichlet boundary conditions arise for a D3-brane ending on a
D5-brane, and Neumann boundary conditions arise for a D3-brane
ending on an NS5-brane.  For a direct path integral explanation of
the duality of Dirichlet and Neumann for $U(1)$, see section
\ref{tuone}.

What we have just described involves practically the only example
of a half-BPS boundary condition $\B$ in $U(1)$ gauge theory that
is {\it not} associated with coupling to a boundary SCFT $\fB$.
For the abelian group $G=U(1)$, there is no room for a Nahm pole.
A reduction of gauge symmetry would necessarily reduce $U(1)$ to a
finite group.  The Dirichlet boundary conditions that we have just
analyzed are the ones that reduce $U(1)$ to the trivial subgroup
consisting only of the identity.  Reduction to a finite subgroup
is equivalent locally to Dirichlet (and our concerns in this paper
are purely local).  So in our further study of $U(1)$ gauge
theory, we can assume that  the boundary condition $\B$ and its
$S$-dual $\B^\vee$ are associated to boundary SCFT's $\fB$ and
$\fB^\vee$.

\subsubsection{Coupling To A Boundary Hypermultiplet}\label{coupbou}

\begin{figure}
  \begin{center}
    \includegraphics[width=5.5in]{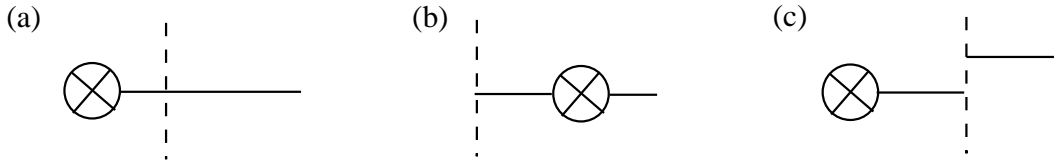}
  \end{center}
\caption{\small (a) A brane configuration, made from a D3-brane
interacting with one D5-brane and one NS5-brane, that leads to
$U(1)$ gauge theory coupled to a boundary hypermultiplet of charge
1. This boundary condition is selfdual, because the fivebranes both
have linking number 1.  (b) A naive application of $S$-duality turns
the configuration of (a) into the one sketched here.  Upon moving
the D5-brane to the right of the NS5-brane, we get back to (a), an
outcome that is ensured because the two fivebranes have equal
linking numbers.  (c) Starting with (a) and displacing the D3-brane
at infinity in the $\vec X$ direction, the D3-brane ``breaks'' and
we arrive at the configuration indicated here.}
  \label{Fig7}
\end{figure}

The next example is more interesting.  We consider a $U(1)$ gauge
field with Neumann boundary conditions coupled to a charged
hypermultiplet on the boundary.  We take the hypermultiplet to
have charge 1 as that is the only value that we can conveniently
get from branes.  As in fig. \ref{Fig7}, we realize this boundary
condition by letting a D3-brane pass through a D5-brane at $y=L$
and end on an NS5-brane at $y=0$.  The boundary condition arises
in the limit $L\to 0$, but the advantages of starting with $L>0$
will become clear, especially when we get to the nonabelian case.

In fig. \ref{Fig7}(a), the D5-brane and NS5-brane both have
linking number 1.  So the dual will be precisely the same
configuration.  To see in more detail how this comes about, we
first make naive $S$-duality, turning the D5-brane into an
NS5-brane and vice-versa, while exchanging $\vec X$ and $\vec Y$.
This gives the configuration of fig. \ref{Fig7}(b).  A D3-brane
passes through the NS5-brane and ends on the D5-brane.  Then we
move the D5-brane to the right of the NS5-brane.  When it crosses
the NS5-brane, a new D3 segment connecting the two fivebranes is
created, according to \cite{Hanany:1996zelf}.  So we get back to
the configuration of fig. \ref{Fig7}(a).

Thus, a boundary condition $\B$ consisting of Neumann boundary
conditions coupled to a charge 1 hypermultiplet is selfdual. One
might naively argue that this is so simply because $\B$ was
constructed from one D5-brane and one NS5-brane, which are exchanged
by duality. But  the crucial point is really that the linking
numbers were the same.  This will be particularly clear when we get
to nonabelian examples.

To elucidate the physical meaning of the $S$-duality of $\B$, it
is useful to move away from the conformal fixed point by giving
expectation values to the scalar fields $\vec X$ or $\vec Y$ along
the D3-brane. In the original configuration of fig. \ref{Fig7}(a),
if we give an expectation value to $\vec Y$, this causes the
D3-brane to move along the NS5-brane so that it no longer
intersects the D5-brane. Hence, the electrically charged
hypermultiplet acquires a mass. Therefore, $S$-duality implies
that if we give an expectation value to $\vec X$, we should see a
massive hypermultiplet of magnetic charge 1.

If we give an expectation value to $\vec X$, then the brane
configuration of fig. \ref{Fig7}(a) is deformed to the
configuration sketched in (c).  The D3-brane splits at the
D5-brane; the two fivebranes are connected by a finite D5-brane
segment.  From a field theory point of view, what is happening is
the following.  The charged hypermultiplet at $y=0$ has a
hyper-Kahler moment map $\vec \mu$.  The boundary condition for a
vector multiplet with Neumann boundary conditions coupled to
boundary hypermultiplets is
\begin{equation}\label{bound} \vec X(0)+\vec \mu=0,\end{equation}
according to eqn. (2.33) of \cite{Gaiotto:2008sa}.  Therefore,
when we give $\vec X$ an expectation value, we force a Higgsing of
the $U(1)$ gauge symmetry, so as to get $\vec\mu\not=0$. This
Higgsing occurs only on the boundary, because that is where the
charged fields are.  A spontaneously broken $U(1)$ gauge theory in
two space dimensions has vortices; in the field of a vortex, there
is a unit of magnetic flux integrated over a spatial section of
the boundary.  By flux conservation in abelian gauge theory, the
magnetic flux measured at spatial infinity is the same, and
therefore in $3+1$ dimensions this configuration looks like a
magnetic monopole of charge 1, localized at the boundary.

Supersymmetry is broken if we give expectation values to both $\vec
X$ and $\vec Y$, because turning on $\vec Y$ gives a mass to the
boundary hypermultiplet, and turning on $\vec X$ forces this
hypermultiplet to have an expectation value.  From the point of view
of branes, for a supersymmetric configuration, the D3-brane must end
on one of the fivebranes, and so must be located at $\vec X=0$ or
$\vec Y=0$.

\subsubsection{Two Boundary Hypermultiplets}\label{twoofthem}

\begin{figure}
  \begin{center}
    \includegraphics[width=2in]{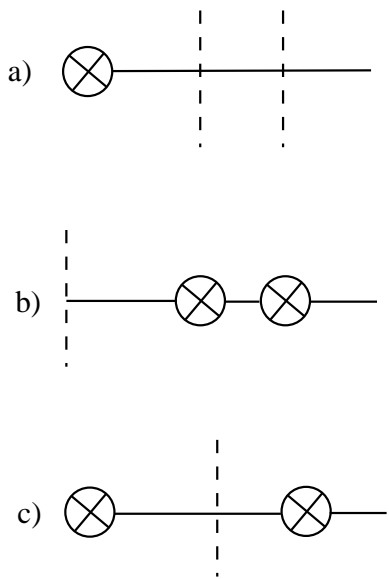}
  \end{center}
\caption{\small (a) A brane configuration that leads to a $U(1)$
theory coupled to two boundary hypermultiplets. (b) The naive
$S$-dual.  (c) A standard form of the $S$-dual after some
rearrangement.}
  \label{Fig8}
\end{figure}

Our next example will be a $U(1)$ theory coupled to two boundary
hypermultiplets of charge 1.  The brane configuration and the
steps in understanding its $S$-duality are sketched in fig.
\ref{Fig8}.  Each fivebrane has linking number 1.  Starting with
configuration (a), the naive $S$-dual is (b), and a brane
rearrangement that preserves the linking numbers brings us to (c),
which is more easily interpreted.

This gives us our first example in which the $S$-dual $\B^\vee$ of
a boundary condition $\B$ involves a coupling of four-dimensional
gauge theory to a non-trivial three-dimensional boundary SCFT
$\fB^\vee$. In fig. \ref{Fig8}(c), there is a $U(1)$ gauge theory
in the D3-brane segment connecting the two NS5-branes.  From a
three-dimensional point of view, the $U(1)$ gauge theory is
coupled to two hypermultiplets of charge 1. One comes from the
intersection of the D3-brane with the D5-brane and one from the
interaction with the semi-infinite D3-brane on the right.  The
second hypermultiplet is also charged under the bulk $U(1)$ gauge
theory.   $U(1)$ gauge theory with two hypermultiplets was
discussed in section \ref{hyperex}.  It has a standard IR limit
consisting of a nontrivial
 SCFT, which we have called $T(SU(2))$.   So the dual of $U(1)$ theory coupled to two boundary hypermultiplets is
 $U(1)$ theory coupled to $\fB^\vee=T(SU(2))$.  To make this
 statement precise, one must describe how $U(1)$ is coupled to
 $T(SU(2))$.  The relevant coupling is simply derived from
 the fact that before the infrared flow, one of the two
hypermultiplets of fig. \ref{Fig8}(c) has charge 1 under the bulk
gauge group and one is neutral.

\begin{figure}
  \begin{center}
    \includegraphics[width=2.5in]{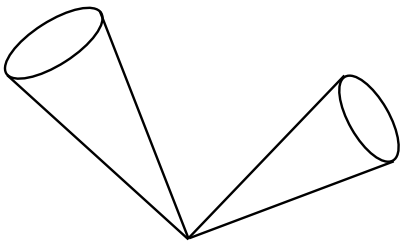}
  \end{center}
\caption{\small In a superconformal field theory with ${\mathcal
N}=4$ supersymmetry, the Higgs and Coulomb branches are both
conical, and meet at their common origin.  In general, there may
also be mixed Higgs-Coulomb branches.}
  \label{Fig9}
\end{figure}

\begin{figure}
  \begin{center}
    \includegraphics[width=4.5in]{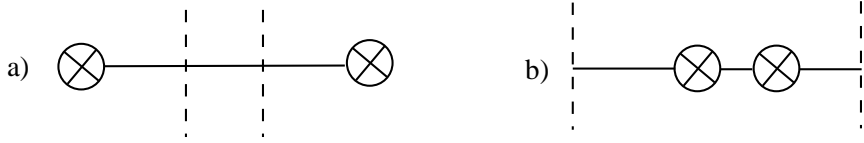}
  \end{center}
\caption{\small The conformal field theory $T(SU(2))$ can be
realized by the brane configuration of (a).  All fivebranes have
linking number 1, which implies the self-mirror property.  Naive
$S$-duality leads to (b), and brane rearrangement leads back to the
mirror theory (a).}
  \label{Fig10}\end{figure}

The theory $T(SU(2))$, which will play an important role in this
paper,  has\footnote{The dimension of the Higgs branch is the
difference between the numbers of hypermultiplets and vector
multiplets; the dimension of the Coulomb branch is the rank of the
gauge group.} Higgs and Coulomb branches that are both of
hyper-Kahler dimension 1. And there is a mirror symmetry
\cite{Intriligator:1996ex} that exchanges the two branches. As
summarized in fig. \ref{Fig10}, the mirror symmetry can be
established by the same sort of brane manipulations that we are
using to analyze boundary conditions. Some key aspects of this
mirror symmetry picture are as follows.  The Higgs branch has an
$SU(2)$ flavor symmetry that is obvious classically; the Coulomb
branch, as described in section \ref{hyperex}, has a manifest $U(1)$
global symmetry (translations of the dual photon) that is extended
to $SU(2)$ in the infrared.  So the global symmetry is really
$SU(2)\times SU(2)$, with one factor acting on each branch. In the
absence of hypermultiplet bare masses and Fayet-Iliopoulos (FI)
parameters, each branch is a copy of $\R^4/\Z_2$, with an $A_1$
singularity at the origin; the two branches meet at their common
singularity, which is a superconformal critical point (fig.
\ref{Fig9}). If FI parameters are turned on, the Coulomb branch
disappears and the singularity of the Higgs branch is resolved. If
bare masses are turned on, the Higgs branch disappears and the
singularity of the Coulomb branch is resolved. Mirror symmetry
exchanges the two sets of parameters.

Going back to fig. \ref{Fig8}(a), even when the vacuum is fixed at
infinity by requiring that $\vec X$ and $\vec Y$ vanish, the theory
still has a moduli space of vacua that depends on the hypermultiplet
expectation values.  From the point of view of the brane picture,
this is so because the D3-brane segment between the two D5-branes is
free to move in the $\vec X$ direction.  In the description by
$U(1)$ gauge theory coupled to two boundary hypermultiplets, we can
reason as follows. In this description, supersymmetry requires that
$\vec X$ is independent of $y$, and so $\vec X(0)$ equals the value
of $\vec X$ at infinity. (In nonabelian gauge theory, $\vec X$ would
obey Nahm's equations, as described in \cite{Gaiotto:2008sa}, but
for $G=U(1)$, these equations merely say that $\vec X$ is constant.)
In addition, we have the boundary condition $\vec X(0)+\vec \mu=0$,
where $\vec \mu$ is the moment map of the fundamental
hypermultiplets. After also dividing by $G$ the space $Z\cong \R^8$
that parametrizes the two hypermultiplets, the moduli space of vacua
is the hyper-Kahler quotient $Z/\neg/\neg/U(1)$. This quotient is
the $A_1$ singularity $\R^4/\Z_2$ if $\vec X(0)=0$; otherwise, it is
a resolution of this singularity.  If we take $\vec Y$ to be nonzero
at infinity, the D3-brane is displaced in the $\vec Y$ direction,
the hypermultiplets become massive, and this branch of vacua
disappears.

Of course, these statements must have an analog in the $S$-dual
theory of fig. \ref{Fig8}(c).   First we might look for a Higgs
branch in the $S$-dual theory.  Just as in the last paragraph,
such a Higgs branch would come by taking the hyper-Kahler quotient
by $G=U(1)$ of the Higgs branch of the conformal field theory
$T(SU(2))$. However, the Higgs branch of that SCFT is $\R^4/\Z_2$,
and its hyper-Kahler quotient by $U(1)$ is trivial. On the other
hand, the Coulomb branch of the SCFT does not couple directly to
the bulk gauge theory and survives the coupling of the $U(1)$
symmetry of the SCFT to the bulk gauge fields. This is visible in
fig. \ref{Fig8}(c); the D3-brane segment that connects the two
NS5-branes is free to move independently of the semi-infinite
D3-brane.   The Coulomb branch disappears if we give an
expectation value to $\vec X$ at infinity, because those are FI
parameters that force the hypermultiplets to have expectation
values.  (An expectation value of $\vec X$ causes a brane
reconnection similar to that in fig. \ref{Fig8}(c), and the brane
modulus disappears.) An expectation value for $\vec Y$ at infinity
gives a hypermultiplet bare mass and modifies the geometry of the
Coulomb branch.  These statements are $S$-dual to the statements
in the last paragraph.

\begin{figure}
  \begin{center}
    \includegraphics[width=5in]{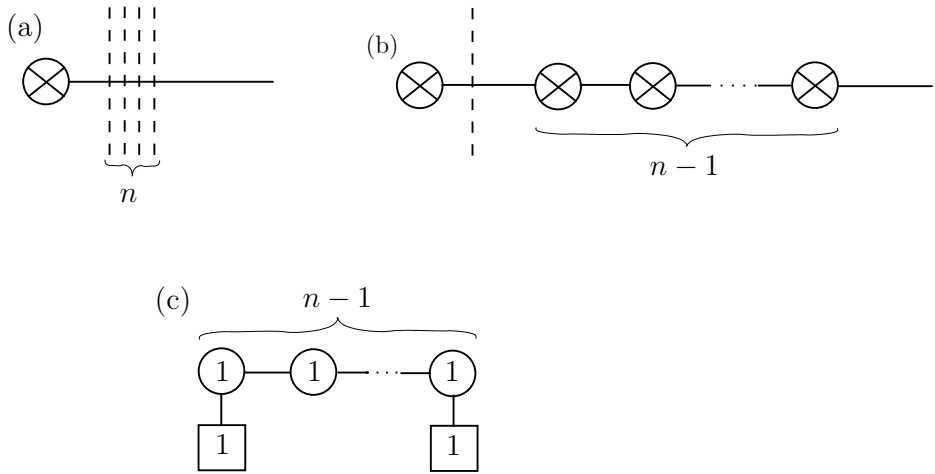}
  \end{center}
\caption{\small (a) A brane configuration representing a $U(1)$
theory with $n$ boundary hypermultiplets of charge 1.  All linking
numbers are 1.  (b) The $S$-dual configuration, after a standard
rearrangement. All linking numbers are again 1.  (c) The quiver
corresponding to the SCFT that arises from the infrared limit of the
dual configuration. The hypermultiplet at one end comes from the
D3-D5 intersection, and the hypermultiplet at the other end comes
from the interaction of the rightmost D3-brane segment with the
semi-infinite D3-brane.}
  \label{Fig11}\end{figure}
\subsubsection{Generalization}\label{generalization}

A natural generalization of the previous example is to introduce
$n$ boundary hypermultiplets of charge 1.  We represent this via a
single D3-brane passing through $n$ D5-branes and ending on an
NS5-brane (fig. \ref{Fig11}(a)).  All linking numbers equal 1. The
$S$-dual brane configuration has $n$ NS5-branes and 1 D5-brane.
Since the D5-brane has linking number 1, in the dual configuration
(fig. \ref{Fig11}(b)), it is to the right of just one NS5-brane.
The first NS5-brane has linking number 1, and no D5-branes to the
left; hence a single D3-brane must end on it. All the other
NS5-branes have a D5-brane to the left; hence to make the linking
numbers equal 1, no net D3-branes end on them.

\def\B{{\mathcal B}}
 In the infrared limit, the dual configuration corresponds to
$U(1)$ gauge theory coupled to a boundary SCFT $\fB^\vee$ that is
the IR limit of a well-known quiver, shown in fig. \ref{Fig11}(c).
(This is sometimes called a quiver of type $\textsf{A}_{n-1}$.) It
has a Higgs branch of hyper-Kahler dimension 1 and a Coulomb
branch of hyper-Kahler dimension $n-1$. The Higgs branch can be
computed classically, and is the $\textsf{A}_{n-1}$ singularity
$\R^4/\Z_n$, according to a well-known result of Kronheimer.  The
Coulomb branch is most easily determined by mirror symmetry. In
fact, the theory $\fB^\vee$ is part of one of the original
examples of a mirror pair in three-dimensions
\cite{Intriligator:1996ex}. Its mirror is another theory
$\tilde\fB^\vee$ that arises from the infrared limit of $U(1)$
coupled to $n$ hypermultiplets of charge 1. The Higgs branch of
$\tilde\fB^\vee$ and therefore Coulomb branch of $\fB^\vee$ have
an $SU(n)$ flavor symmetry, which is $S$-dual to the flavor
symmetry of the $n$ hypermultiplets in the original boundary
condition.  (In fact, it can be shown that as a complex manifold,
this Higgs or Coulomb branch is isomorphic to the minimal
nilpotent orbit of $SL(n,\BC)$.)

\begin{figure}
  \begin{center}
    \includegraphics[width=4.5in]{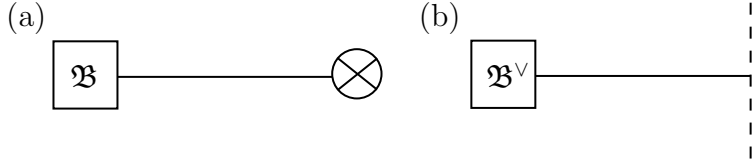}
  \end{center}
\caption{\small (a) This four-dimensional picture, with Neumann
boundary conditions coupled to a three-dimensional theory $\fB$ at
one end and pure Neumann boundary conditions at the other end,
flows in the IR to the three-dimensional theory $\fB'$.  (b) The
$S$-dual configuration has Dirichlet boundary conditions on the
right and on the left a coupling to a boundary theory $\fB^\vee$
that provides the boundary condition that is $S$-dual to the one
determined by $\fB$.
 Dirichlet boundary conditions have the effect of turning off the gauge couplings on the interval,
 and the infrared limit is simply the three-dimensional theory $\fB^\vee$.  Hence $\fB'$ is mirror to
 $\fB^\vee$.}
  \label{Fig12}\end{figure}

The reader may notice a resemblance between the three-dimensional
theory $\tilde\fB^\vee$ and the original boundary condition that we
started with.  They involve the same $n$ hypermultiplets coupled to
four-dimensional or three-dimensional gauge fields. Explaining this
resemblance will lead to one of the main ideas of this paper.

To formalize what is going on, we start with a boundary condition in
four-dimensional $U(1)$ gauge theory obtained by coupling to some
boundary theory $\fB$. In our example, $\fB$ is the theory of $n$
free hypermultiplets.

$\fB$ has a $U(1)$ global symmetry that is used in coupling it to
four-dimensional gauge fields.  By means of the same $U(1)$ global
symmetry, we could instead couple $\fB$ to {\it three}-dimensional
gauge fields.  If the coupled theory has a standard IR limit, this
limit is a new three-dimensional theory $\fB'$.  We claim that in
fact, $\fB'$ is the mirror of $\fB^\vee$, the SCFT that defines
the $S$-dual boundary condition $\B^\vee$.

This can be deduced as follows. The three-dimensional theory $\fB'$
can be obtained from the boundary condition $\B$ by taking the
D3-brane that generates the $U(1)$ gauge symmetry to be of finite
extent in the $y=x^3$ direction. We take boundary conditions $\B$ at
$y=0$, and Neumann boundary conditions at $y=L$ (fig.
\ref{Fig12}(a)). At low energies or for $L\to 0$, the gauge fields
become effectively three-dimensional and we get the
three-dimensional theory $\fB'$.

On the other hand, we can apply  $S$-duality to this configuration.
The boundary condition $\B$ is replaced by its $S$-dual $\B^\vee$,
and the Neumann boundary conditions at $y=L$ are replaced by
Dirichlet (fig. \ref{Fig12}(b)).

Because of the Dirichlet boundary conditions at one end, the gauge
fields on the segment are massive. So although the theory $\fB^\vee$
appears to have been coupled to gauge fields, at low energies it is
effectively ungauged.  Hence the configuration of fig.
\ref{Fig12}(b) just leads to the three-dimensional theory
$\fB^\vee$.

The $S$-duality operation that relates figs. \ref{Fig12}(a) and (b)
is mirror symmetry from a three-dimensional point of view,
 since this operation exchanges $\vec X$ and $\vec
Y$ and therefore exchanges Higgs and Coulomb branches. So the
conclusion is that the relation between $\fB$ and $\fB^\vee$ is that
$\fB'$ is the mirror of $\fB^\vee$.

In the case that $\B$ and $\B^\vee$ are realized by branes, the
brane manipulations that  relate them are equivalent, after adding
an extra fivebrane at $y=L$ (to give Dirichlet or Neumann boundary
conditions in fig. \ref{Fig12}), to the brane manipulations used to
show mirror symmetry between $\fB'$ and $\fB^\vee$.

Now let us assess what we actually learn from this construction.
 To
understand $S$-duality of a boundary condition $\B$, what we
really want is to describe the boundary SCFT $\fB^\vee$ associated
to the dual boundary condition $\B^\vee$.  By coupling $\fB$ to a
three-dimensional $U(1)$ gauge field $B$, we have an explicit way
to construct $\tilde\fB^\vee$, the mirror of $\fB^\vee$, starting
directly with $\B$. From an abstract point of view, constructing
the mirror $\tilde\fB^\vee$ is just as good as constructing
$\fB^\vee$. We simply declare that $\fB^\vee$ is the same as
$\tilde\fB^\vee$ but with Higgs and Coulomb branches exchanged.

To make this answer useful, we want to make concrete how to couple
a four-dimensional $U(1)$ gauge theory, with gauge field $C$, to
$\fB^\vee$. We must couple $A$ to the $U(1)$ symmetry of the Higgs
branch of $\fB^\vee$, but by definition the Higgs branch of
$\fB^\vee$ is the Coulomb branch of $\tilde\fB^\vee$. The $U(1)$
isometry of the Coulomb branch of $\tilde\fB^\vee$ comes from a
shift in the scalar dual to the three-dimensional gauge field $B$.
The conserved current generating this symmetry is simply $J=\star
dB$, and the coupling $A_\mu J^\mu$ of another gauge field $A$ to
this current is simply a Chern-Simons-like coupling $A\wedge dB$.
In the present case, since $A$ is a four-dimensional gauge field,
while $B$ is defined only on the boundary, the appropriate
interaction is a coupling between $B$ and the restriction of $A$
to the boundary:
\begin{equation}\label{cslike}\frac{1}{2\pi}\int_{y=0} A\wedge
dB.\end{equation} This coupling has an extension that has $\N=4$
supersymmetry in the three-dimensional sense.

To give this description of the dual boundary theory $\fB^\vee$,
we did not have to come to grips with mirror symmetry in any
serious way. The reason for this was that in abelian gauge theory,
the symmetries of the Coulomb branch are visible classically and
one can write the explicit classical coupling (\ref{cslike}).
Nonabelian gauge theory is completely different; there is a
procedure similar to what we have just described for finding the
mirror of $\fB^\vee$, but the relevant symmetries of its Coulomb
branch are hard to see without coming to grips with mirror
symmetry.

With some care about the definitions, the exchange of Dirichlet and
Neumann boundary conditions can also be understood as a special case
of the above procedure.  We will say more about this in the
nonabelian case, where the exceptions are more numerous.

As a final comment, we want to explain why the above description
of the action of $S$-duality on boundary conditions is consistent
with the fact that $S^2$ should be the identity. The operation
that goes from $\fB$ to $\fB'$ can be described as follows.  $\fB$
by definition is a three-dimensional theory with a $U(1)$
symmetry, generated by a current that we may call $J$. To define
$\fB'$, we gauge the $U(1)$ symmetry of $\fB$ by coupling $J$ to a
$U(1)$ gauge field $B$.  This gives a new theory $\fB'$ with a new
conserved current $J'=\star dB$ and therefore also a $U(1)$
symmetry. The operation that goes from $(\fB,J)$ to $(\fB',J')$
has been considered before \cite{Witten:2003ya}; its square is
indeed the identity.

\subsection{$U(2)$ Examples}\label{utwo}

For $G=U(2)$, we can construct three kinds of examples from branes.
$U(2)$ may be completely broken by the interaction with D5-branes,
it may be broken to a $U(1)$ subgroup that then couples to a
boundary theory, or it may couple directly to a boundary theory.  We
will analyze all three cases.

\subsubsection{Complete Breaking By D5-Branes}\label{complete}

\begin{figure}
  \begin{center}
    \includegraphics[width=4in]{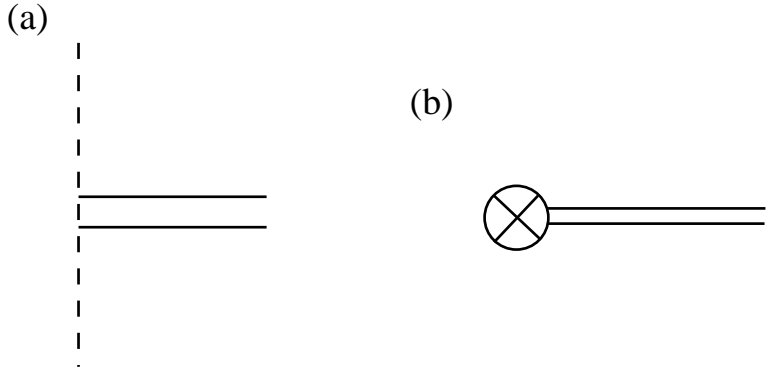}
  \end{center}
\caption{\small  (a) Two D3-branes ending on a single D5-brane,
leading to a boundary condition involving a pole in $\vec X$.  (b)
The dual configuration: two D3-branes ending on a single NS5-brane,
leading to Neumann boundary conditions.}
  \label{Fig13}\end{figure}

As reviewed in \cite{Gaiotto:2008sa}, a configuration of two
D3-branes ending on the same D5-brane (fig. \ref{Fig13}(a)) leads to
a boundary condition for $U(2)$ gauge theory in which the scalar
fields $\vec X(y)$ have a singularity at $y=0$ of the form $\vec
X\sim \vec t/y$, where $\vec t$ are the images of standard
$\frak{su}(2)$ generators under the obvious embedding
$\rho:\frak{su}(2)\to \frak{u}(2)$. This breaks $U(2)$ to its center
$U(1)$ and the $U(1)$ gauge fields obey Dirichlet boundary
conditions.  The $S$-dual of this corresponds to Neumann boundary
conditions (fig. \ref{Fig13}(b)), with two D3-branes ending on a
single NS5-brane. Both of these boundary conditions respect the
factorization of the $U(2)$ theory as a local product of $SU(2)$ and
$U(1)$ theories. The boundary conditions for $U(1)$ are the usual
Dirichlet/Neumann pair, while for $SU(2)$, the boundary condition on
one side is determined by the Nahm pole, and on the other side is
Neumann.

\begin{figure}
  \begin{center}
    \includegraphics[width=5in]{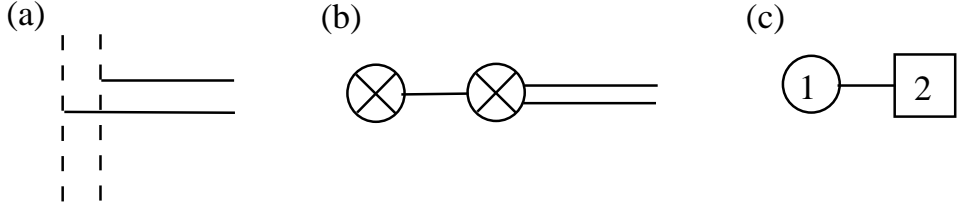}
  \end{center}
\caption{\small  (a) Two D3-branes ending on two D5-branes. (b)
The $S$-dual configuration, which describes $U(2)$ gauge theory
coupled to a non-trivial boundary SCFT.  To establish the
$S$-duality, one observes that in each picture, all linking
numbers are 1. (c) The boundary SCFT is $T(SU(2))$, the infrared
limit of $U(1)$ coupled to two charged hypermultiplets (which in
(b) arise from interaction of the D3-brane segment with the
semi-infinite D3-branes).}
  \label{Fig14}\end{figure}
To get Dirichlet boundary conditions for the full $U(2)$ group, we
need to consider a configuration in which two D3-branes end on two
D5-branes (fig. \ref{Fig14}(a)).  As usual, we displace the
D5-branes slightly in the $y$ direction, and then the $S$-dual
configuration is straightforward to describe (fig.
\ref{Fig14}(b)). It does not consist of standard Neumann boundary
conditions for $U(2)$ gauge theory, which we already encountered
in the last paragraph.  Rather, we get Neumann boundary conditions
modified by coupling to a certain boundary SCFT.  The relevant
SCFT can be read off from the figure.  It is the infrared limit of
$U(1)$ coupled to two charge 1 hypermultiplets, and thus is
actually the self-mirror theory $T(SU(2))$ that we have already
encountered more than once.

Several remarks are in order:

(1) The flavor symmetry of $U(1)$ with two hypermultiplets of
charge 1 is $SU(2)$, not $U(2)$ (a $U(1)$ flavor rotation is
equivalent to a gauge transformation).  Accordingly, the flavor
symmetry of the Higgs branch of $T(SU(2))$ is $SU(2)$ rather than
$U(2)$, as is also clear from the fact that this branch is
$\R^4/\Z_2$.  As the theory is self-mirror, $SU(2)$ is also the
symmetry of the Coulomb branch, which is also $\R^4/\Z_2$, so the
full global symmetry is $SU(2)\times SU(2)$. So the $U(1)$ part of
$U(2)=SU(2)\times U(1)$ does not couple to $T(SU(2))$ and in the
above construction it simply obeys Neumann boundary conditions.
The coupling of the $SU(2)$ gauge theory to $T(SU(2))$ removes one
of the $SU(2)$ symmetries of that theory. The other one matches
with the global symmetry of $SU(2)$ gauge theory with Dirichlet
boundary conditions.

(2) We recall from \cite{Gaiotto:2008sa} that the $U(2)$ gauge
theory with Dirichlet boundary conditions, even after one fixes
$\vec X$ and $\vec Y$ to vanish at infinity, has a moduli space of
vacua given by the space of solutions of Nahm's equations.
Moreover this space is the nilpotent cone of $SL(2,\Bbb{C})$, or
equivalently is the $A_1$ singularity $\R^4/\Z_2$.  In the present
construction, that moduli space matches the Coulomb branch of the
boundary theory $T(SU(2))$.  Before $S$-duality, $\R^4/\Z_2$ is
the moduli space of solutions of Nahm's equations, and after
$S$-duality, it is the Coulomb branch of the boundary theory.

\subsubsection{Breaking To $U(1)$}\label{breaking}

\begin{figure}
  \begin{center}
    \includegraphics[width=5in]{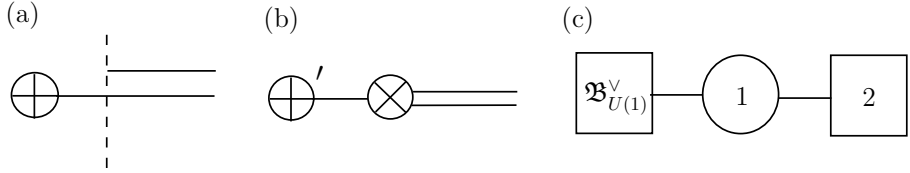}
  \end{center}
\caption{\small (a) Two D3-branes, one of which ends on a
D5-brane, while the second then ends with Neumann boundary
conditions coupled to some matter system.  This boundary condition
is indicated by the symbol $\bigoplus$. (b) The dual
configuration; $\bigoplus'$ is the $S$-dual of the boundary
condition $\bigoplus$ in $U(1)$ gauge theory. (c) A schematic
representation of the dual boundary condition as a quiver.
$\fB_{U(1)}^\vee$ is the boundary theory associated with the
boundary condition $\bigoplus'$ in $U(1)$ gauge theory.}
  \label{Fig15}\end{figure}

To construct a boundary condition $\B$ reduces the gauge group from
$U(2)$ to $U(1)$, we end one D3-brane on a D5-brane and let the
second D3-brane end on a brane system which defines Neumann boundary
conditions, possibly coupled to a boundary theory $\fB_{U(1)}$ with
$U(1)$ symmetry.  In fig. \ref{Fig15}, we denote this generic
boundary system by the symbol $\bigoplus$, and we write $\bigoplus'$
for its $S$-dual. We assume that $\bigoplus'$ is $U(1)$ coupled to
some boundary theory $\fB^\vee_{U(1)}$ with $U(1)$
symmetry.\footnote{\label{try} As explained at the end of section
\ref{dirn}, this is so except in one case: if $\bigoplus$
corresponds to pure Neumann boundary conditions with no boundary
matter fields, then its dual is Dirichlet.  We will discuss this
exceptional case in section \ref{full}.}  To get the $S$-dual of
$\B$, we apply $S$-duality to the whole picture, arriving at fig.
\ref{Fig15}(b). Now the boundary theory is a composite, which we
will call $\fB_{U(2)}^\vee$; it is obtained (fig. \ref{Fig15}(c)) by
coupling a three-dimensional $U(1)$ gauge theory both to
$\fB_{U(1)}^\vee$ and to two hypermultiplets of charge 1, which also
have a $U(2)$ flavor symmetry.  In this discussion, we did not
really need to know that $\bigoplus$ can be realized by branes; it
could be any boundary condition in $U(1)$ gauge theory, with
$\bigoplus'$ as the dual boundary condition.

\begin{figure}
  \begin{center}
    \includegraphics[width=6in]{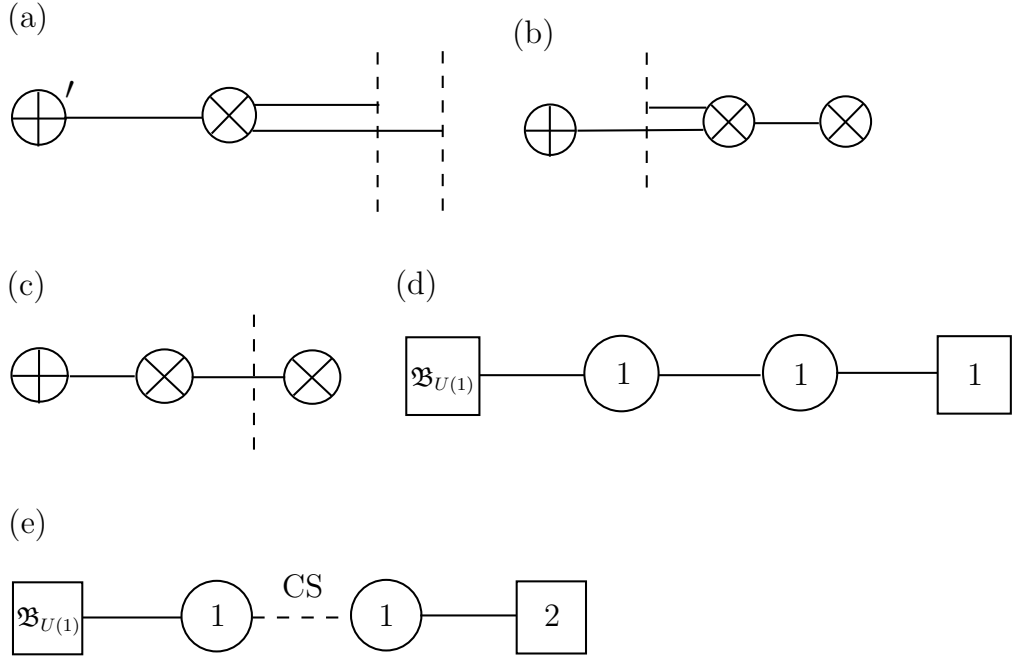}
  \end{center}
\caption{\small (a) The three-dimensional theory $\fB^\vee_{U(2)}$
can be recovered by ungauging the gauge symmetry with the help of
Neumann boundary conditions. (b) The naive $S$-dual configuration.
(c) A well-ordered rearrangement; the D5-brane has been moved to
the right to satisfy the usual conditions. (d) A quiver that
schematically represents the configuration of (c); its  infrared
limit is  the mirror of $\fB^\vee_{U(2)}$. The symbol $\bigoplus$
has been replaced by a coupling to the corresponding boundary
theory $\fB_{U(1)}$. (e) A variant of the quiver that makes the
global symmetries visible. The dotted line indicates a
Chern-Simons coupling between the two $U(1)$ theories represented
by the circles. (This coupling is the supersymmetric completion of
eqn. \ref{cslike}.)}
  \label{Fig16}\end{figure}
As in our analysis of $U(1)$ theories, a convenient way to study the
theory $\fB^\vee_{U(2)}$ is to ungauge the $U(2)$ symmetry, by
terminating the two D3-branes on the right on a pair of D5-branes,
giving Dirichlet boundary conditions (fig. \ref{Fig16}(a)).  Since
we have ungauged the symmetry, the infrared limit of this
configuration is simply the three-dimensional theory
$\fB^\vee_{U(2)}$. The $S$-dual configuration is shown in fig.
\ref{Fig16}(b) and after a brane rearrangement to obey our rules, we
arrive at fig. \ref{Fig16}(c). The infrared limit of this
configuration gives the mirror $\tilde\fB^\vee_{U(2)}$ of
$\fB^\vee_{U(2)}$. It is again a composite theory, sketched in fig.
\ref{Fig16}(d). A simple interpretation of this theory is that we
have coupled a $U(1)$ gauge field $C$ to $\fB_{U(1)}$ (the original
boundary theory associated with the configuration $\bigoplus$) and
to a $U(1)$ subgroup of the flavor symmetry of $T(SU(2))$. The
Coulomb branch of the composite theory has a $U(1)$ global symmetry
(translations of the scalar field dual to $C$) and an $SU(2)$ global
symmetry (acting on the Coulomb branch of $T(SU(2))$).  These
combine to the $U(2)$ symmetry of $\fB^\vee_{U(2)}$.

The theory  $\tilde\fB^\vee_{U(2)}$ is explicitly constructed and
the desired boundary condition is constructed from its mirror
$\fB^\vee_{U(2)}$.  We could simply define $\fB^\vee_{U(2)}$ as
$\tilde\fB^\vee_{U(2)}$ with the Higgs and Coulomb branches
exchanged.  However, to explicitly construct a boundary condition in
$U(2)$ gauge theory, we need to be able to see the $U(2)$ currents
that act on the Coulomb branch of $\fB^\vee_{U(2)}$. Unfortunately,
in the construction of $\fB^\vee_{U(2)}$, only the currents of a
Cartan subalgebra are realized as classical symmetries; the other
currents are monopole or vortex operators
\cite{Borokhov:2002cg,Borokhov:2003yu}, as described in section
\ref{quivers}. Though later in the paper, this will be a problem in
making some of the constructions explicit, in the present example,
we can circumvent the problem with a trick. In fig. \ref{Fig16}(d),
we have gauged a $U(1)$ symmetry of the Higgs branch of $T(SU(2))$.
Since $T(SU(2))$ is self-mirror, we could equivalently have gauged a
$U(1)$ symmetry of its Coulomb branch. This we can do explicitly,
since the abelian symmetries of the Coulomb branch are visible
classically, and when we do that, we are left with the visible
$SU(2)$ symmetries of the Higgs branch. The resulting description of
$\fB^\vee_{U(2)}$ is sketched in fig. \ref{Fig16}(e).

For $G=U(1)$, we had an analogous description of the dual boundary
condition involving mirror symmetry.  There was no problem because
the symmetries of the Coulomb branch were visible classically.

\subsubsection{Full $U(2)$ Gauge Symmetry At The
Boundary}\label{full}

The most basic boundary condition with full $U(2)$ gauge symmetry
at the boundary is Neumann.  We have already encountered this
boundary condition as the dual of the Nahm pole.  We want to
explore what happens if we add matter at the boundary.

\begin{figure}
  \begin{center}
    \includegraphics[width=4.5in]{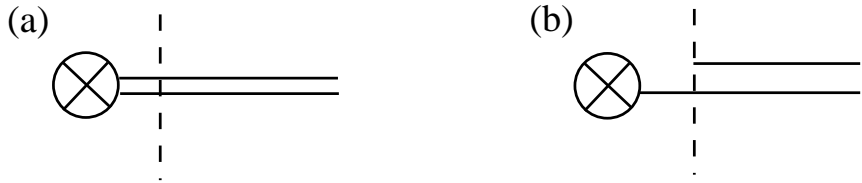}
  \end{center}
\caption{\small (a) This configuration leads to $U(2)$ gauge theory
coupled to a fundamental hypermultiplet at the boundary. (b) The
$S$-dual configuration, in which one D3-brane ends on a D5-brane,
reducing the gauge symmetry to $U(1)$, and then the second ends on
an NS5-brane, leading to Neumann boundary conditions. }
  \label{Fig17}\end{figure}
As in our treatment of $G=U(1)$, we begin by adding a fundamental
hypermultiplet at the boundary.  The $S$-dual can be found in the
standard way (fig. \ref{Fig17}) and turns out to be the
exceptional case mentioned in footnote \ref{try} of a theory in
which the gauge symmetry is reduced to $U(1)$ at the boundary and
the boundary theory $\fB_{U(1)}^\vee$ is trivial (so that the
$U(1)$ vector multiplet simply obeys Neumann boundary conditions).

In fig. \ref{Fig16}(d), we gave a general recipe for analyzing the
$S$-dual of any $U(2)$ boundary condition that involves reduction
of gauge symmetry to $U(1)$ and coupling to a {\it non-trivial}
boundary theory $\fB_{U(1)}$ with $U(1)$ symmetry.  (The role of
non-triviality is to avoid the exceptional behavior mentioned in
footnote \ref{try}.)  Now that we know what the $S$-dual is for
the case that $\fB_{U(1)}$ is trivial, it is interesting to
compare the answer to what we would get from the general recipe.
When $\fB_{U(1)}$ is trivial, fig. \ref{Fig16}(d) reduces to an
ugly quiver theory in the sense of section \ref{quivers}, and as
explained there, it is equivalent to a theory of two free twisted
hypermultiplets, which parametrize its Coulomb branch. These free
twisted hypermultiplets have $U(2)$ (and in fact $Sp(4)$)
symmetry, and can be identified with the fundamental
hypermultiplets that we started with in fig. \ref{Fig17}(a). So,
though it involves grappling with an ugly quiver, the general
recipe works even in the exceptional case that does not quite fit
the original discussion.

\begin{figure}
  \begin{center}
    \includegraphics[width=3.5in]{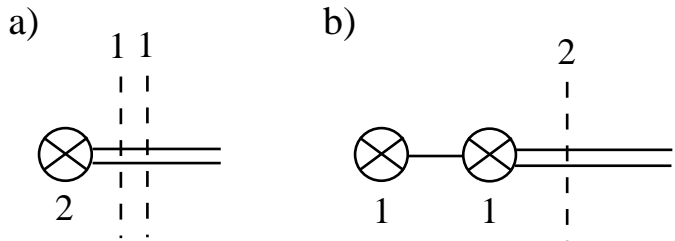}
  \end{center}
\caption{\small (a) $U(2)$ gauge theory with two fundamental
hypermultiplets at the boundary.  Linking numbers are indicated. (b)
The dual configuration. The boundary SCFT is the decoupled sum of
$T(SU(2))$ coming from the left of the picture, and a fundamental
hypermultiplet of the bulk $U(2)$, coming from the D3-D5
intersection. }
  \label{Fig18}\end{figure}
Another exceptional case is $U(2)$ coupled to two fundamental
hypermultiplets at the boundary (fig. \ref{Fig18}). The $S$-dual
configuration, shown in (b), corresponds to Neumann boundary
conditions for $U(2)$ coupled to a boundary SCFT $\fB^\vee$ that
is a product of two factors.  One factor is $T(SU(2))$, and the
other consists of a fundamental hypermultiplet.

\begin{figure}
  \begin{center}
    \includegraphics[width=5.5in]{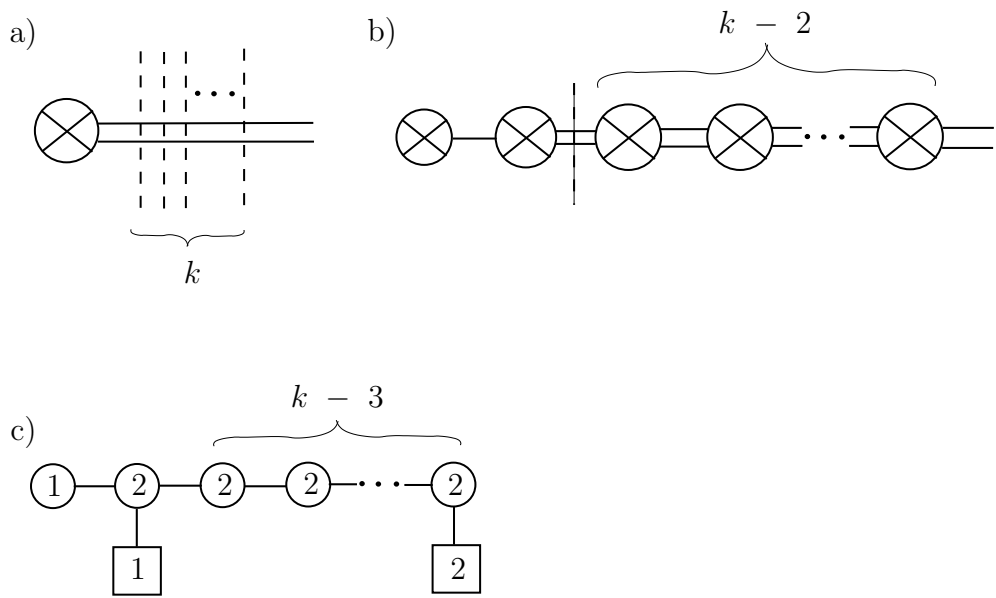}
  \end{center}
\caption{\small (a) $U(2)$ gauge theory coupled to $k$ fundamental
hypermultiplets at the boundary.  (b) The dual configuration,
which is related to a boundary SCFT $\fB^\vee$. (c) The quiver
corresponding to $\fB^\vee$.}
  \label{Fig19}\end{figure}
Before completing the analysis of those two exceptional cases, let
us look at a more generic case, such as a boundary coupling to
$k>2$ fundamental hypermultiplets (fig. \ref{Fig19}). The dual
boundary condition can be found in the usual way and corresponds
to a boundary SCFT $\fB^\vee$ that can be represented by the
quiver of fig. \ref{Fig19}(c).

\begin{figure}
  \begin{center}
    \includegraphics[width=5.5in]{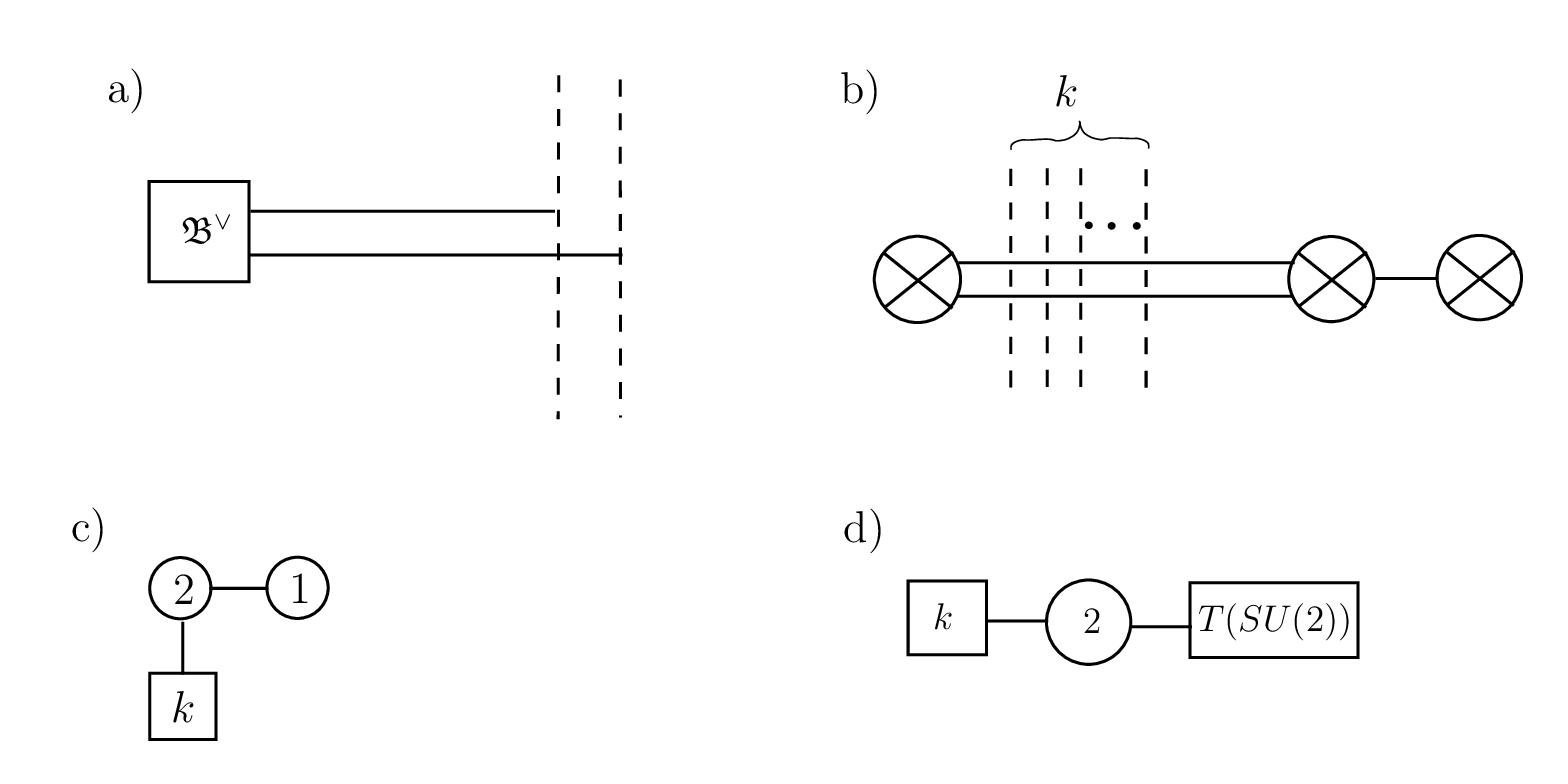}
  \end{center}
\caption{\small  To find the mirror to the dual boundary condition
$\fB^\vee$, we ungauge the symmetry, using Dirichlet boundary
conditions (a), then take the $S$-dual, which is depicted in (b)
after a brane rearrangement, and can be represented as a quiver (c).
This quiver describes the mirror of
 $\fB^\vee$. It is the mirror of the quiver in fig. \ref{Fig19}(c), as
one can verify by computing the linking numbers.  As is shown
schematically in (d), this quiver couples $U(2)$ gauge theory to
the product of $T(SU(2))$ and $k$ hypermultiplets.}
  \label{Fig20}\end{figure}
The mirror to $\fB^\vee$ can also be found in the familiar way.
Starting with $\fB^\vee$ coupled to bulk $U(2)$ gauge fields (as in
fig. \ref{Fig19}(b)), we ungauge the $U(2)$ gauge symmetry by
introducing a second boundary with Dirichlet boundary conditions
(fig. \ref{Fig20}(a)), and take the $S$-dual (fig. \ref{Fig20}(b)).
We arrive at a remarkably simple quiver description of the mirror
$\tilde \fB^\vee$ (fig. \ref{Fig20}(c)). It has a very simple
interpretation: it is described by a three-dimensional $U(2)$ gauge
theory coupled to the original theory $\fB$ of $k$ free
hypermultiplets, and to $T(SU(2))$.  (The node labeled 2 in the
quiver represents a $U(2)$ gauge field which couples to a
bifundamental hypermultiplet of $U(2)\times U(1)$ -- part of the
definition of $T(SU(2))$ -- and to $k$ fundamental hypermultiplets.)

So there is a simple general prescription to go from the starting
theory $\fB$ to $\tilde\fB^\vee$, the mirror of the desired boundary
theory $\fB^\vee$. We simply couple $\fB$ to $T(SU(2))$ via $U(2)$
gauge fields, as in fig. \ref{Fig20}(c). We call the theory made
this way the composite gauge theory and denote it as
$\fB\times_{U(2)}T(SU(2))$:
\begin{equation}\tilde\fB^\vee=\fB\times_{U(2)}T(SU(2)).\end{equation}

Then the ``answer'' $\fB^\vee$, which determines the $S$-dual
boundary condition, is the mirror of $\tilde\fB^\vee$.  The only
trouble with this answer is that to understand $\fB^\vee$ as a
theory with $U(2)$ global symmetry (so that we can couple it to
$U(2)$ gauge fields), we need to be able to see the $U(2)$
symmetries of the Coulomb branch of $\fB^\vee$. As usual, only a
Cartan subalgebra is visible classically. (In any three-dimensional
gauge theory in which the center of the gauge group has rank $r$, a
$U(1)^r$ symmetry of the Coulomb branch is visible classically,
acting by shifts of the dual photons.  In the present example, $r=2$
as the relevant quiver has two nodes. For more on this, see section
\ref{quivers}.)

If we could find a realization of $T(SU(2))$ with manifest
$SU(2)\times SU(2)$ global symmetry, then using this in fig.
\ref{Fig20}(c) would give an explicit way to construct the dual
boundary condition. This is not available at the moment.  However,
if the original boundary theory $\fB$ is constructed via branes,
then the usual $D$-brane manipulations give a construction of the
mirror to $\fB^\vee$ with the relevant symmetries visible, as in
fig. \ref{Fig19}(c) for the case that the starting theory consists
of $k$ fundamental hypermultiplets.

As long as $k\geq 3$, the quiver that we have arrived at in fig.
\ref{Fig20}(c) obeys the constraint $n_f\geq 2n_c$ at each node
and is a good quiver in the sense of section \ref{quivers}. This
is the condition under which a quiver gauge theory gives the most
economical possible description of whatever infrared critical
point it describes.  For $k=2$, it is an ugly quiver with one
minimally unbalanced node and therefore describes free twisted
hypermultiplets, times an additional SCFT. This is reflected in
the fact that for $k=2$, the $S$-dual configuration of fig.
\ref{Fig18}(b) involves coupling to a free fundamental
hypermultiplet, times $T(SU(2))$.  (The appearance of $T(SU(2))$
can be argued by further analysis of the ugly quiver, but we will
omit this.)

For $k<2$, the quiver is a bad one, with operators of $q_R\leq 0$
according to section \ref{quivers}.  Not coincidentally, this is
also the case that the $S$-dual boundary condition has reduced
gauge symmetry, possibly with a Nahm pole. When this is so, our
derivation of the quiver is not valid; the starting point of this
derivation was to assume that the dual boundary condition has full
$U(2)$ gauge symmetry. Indeed, the first step in the derivation
was to ungauge the $U(2)$ symmetry by means of Dirichlet boundary
conditions. But for $k<2$, the dual boundary condition has reduced
gauge symmetry, as we have seen.

In sections \ref{recipe} and \ref{symbr}, we will give another
derivation of the recipe using $T(SU(2))$ which works for any
starting boundary condition. The details  will be more subtle. The
main point is that as we flow to the infrared, the $U(2)$ isometries
of $\fB\times_{U(2)}T(SU(2))$ may be spontaneously
broken.\footnote{Such spontaneous breaking cannot occur in a theory
that has a standard infrared critical point, as such a point is
always invariant under continuous global symmetries.  The coupling
of a conserved current to a Goldstone boson would violate conformal
invariance.} As a result, the $S$-dual $U(2)$ gauge symmetry will be
Higgsed at the boundary, and reduced to a subgroup in the infrared.
Moreover, if the $U(2)$ isometry at the boundary is spontaneously
broken, the hyper-Kahler moment maps of the isometry group may
acquire non-zero expectation values, which are dimensionful. The
boundary conditions $\vec X| + \vec\mu=0$ will then force the scalar
fields $\vec X$ to acquire nonzero boundary values which diverge as
one flows to the infrared, leading to a Nahm pole.

\begin{figure}
  \begin{center}
    \includegraphics[width=5.5in]{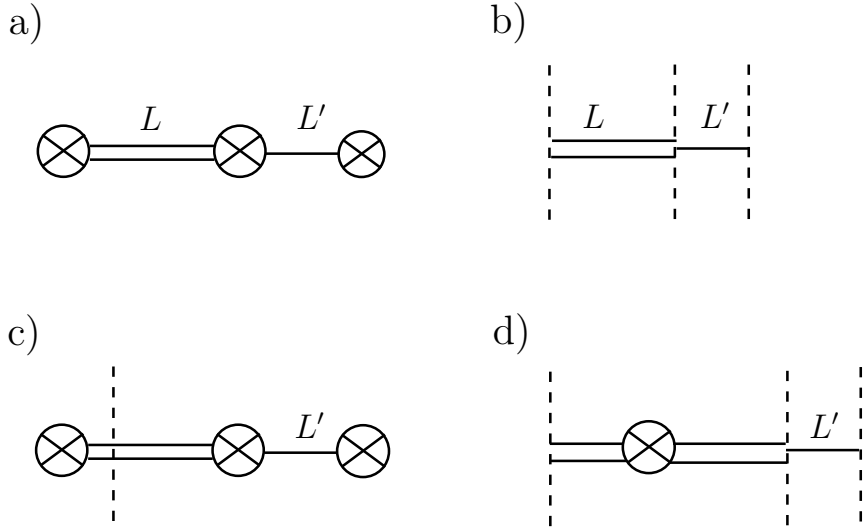}
  \end{center}
\caption{\small (a) The $k=0$ case of the bad quiver discussed in
the text can be studied via this brane configuration.  The full
$U(2)$ symmetry of the Coulomb branch arises when  $L'\to 0$ with
fixed $L$. (b) The $S$-dual of (a). The expected $U(2)$ global
symmetry arises when  $L'\to 0$ and the two D5-branes are
coincident. For $L'\to 0$, the desired Coulomb branch of (a) is the
moduli space of solutions of Nahm's equations on an interval, with a
rank two Nahm pole at $y=0$ (where two D3-branes end on one
D5-brane) and Dirichlet boundary conditions at $y=L$ (where two
D3-branes end on two D5-branes). The existence of a Nahm pole at
$y=0$ means that $\vec X$ cannot vanish at $y=L$, and breaks the
global $U(2)$ symmetry (which acts on $\vec X$ by conjugation) to
its center. (c) The $k=1$ bad quiver corresponds to this brane
configuration. (d) The Coulomb branch in (c), for $L'\to 0$, can
again be obtained by solving Nahm's equations with a Nahm pole in
the left and Dirichlet boundary conditions on the right.  But now
$\vec X$ can jump in crossing the NS5-brane, leaving unbroken a
$U(1)$ subgroup of $U(2)$ consisting of elements of the form ${\rm
diag}(*,1)$. }
  \label{Fig21}\end{figure}

As a preview of this, we will discuss this symmetry breaking
scenario for our bad quivers with $k<2$.   These quivers have no
Higgs branch (not even a mixed Coulomb-Higgs branch). They each have
a Coulomb branch, of hyper-Kahler dimension 3. For the question of
whether the $U(2)$ global symmetries of the Coulomb branch are
spontaneously broken to make any sense, these symmetries have to be
present.  This is the case only when one takes the strong coupling
limit on the $U(1)$ node of the quiver, so as to generate
$T(SU(2))$.  So we take that limit, and then analyze the $U(2)$
gauge dynamics at the remaining node.  We claim that this $U(2)$
gauge dynamics leads to a spontaneous breaking of the $U(2)$ global
symmetry of the Coulomb branch.  If the claim is correct, it is best
to analyze the gauge dynamics at finite gauge coupling, since a
spontaneously broken global symmetry is simply lost in an IR limit.
The Coulomb branch of the bad quivers with finite $U(2)$ gauge
coupling can be analyzed by  Nahm's equations, as explained in fig.
\ref{Fig21}, and the expected breaking of global symmetries does
occur.

Indeed, the bad quiver gauge theories with $k=0$ and $k=1$ can be
represented by the brane configurations of fig. \ref{Fig21}(a) and
(c).  Their Coulomb branches are most conveniently determined by
solving Nahm's equations in the $S$-dual brane configurations of
fig. \ref{Fig21}(b) and (d).  In solving Nahm's equations, we
require $\vec X(y)$ to have a Nahm pole at $y=0$, where two
D3-branes end on a single D5-brane.  In the limit of strong
coupling on the $U(1)$ node, $\vec X(y)$ obeys  Dirichlet boundary
conditions at the other endpoint $y=L$.  The existence of the pole
at $y=0$ makes it impossible for $\vec X$ to vanish at $y=L$, so
the $U(2)$ global symmetry is spontaneously broken.  For $k=0$,
the Nahm pole breaks the symmetry to its center, but for $k=1$,
because $\vec X$ jumps in crossing the NS5-brane (see section
3.9.3 of \cite{Gaiotto:2008sa}, at the end of which the relevant
solution is described), $\vec X(L)$ can have rank 1, giving a
solution invariant under a noncentral $U(1)$ subgroup of $U(2)$
(consisting of matrices of the form ${\rm diag}(*,1)$). This is
indeed the unbroken symmetry of the $S$-dual boundary condition
for $k=1$, as we recall from fig. \ref{Fig17}. For $k\geq 2$,
$\vec X$ would jump in crossing two successive NS5-branes, and can
vanish at $y=L$, leaving the global symmetry unbroken, despite the
Nahm pole at $y=0$.

\subsection{$U(n)$ Examples}\label{unex}

Examples with $n$ D3-branes leading to $U(n)$ gauge theory can be
studied in much the same way, but the details are  richer.

\subsubsection{The Dual of Dirichlet Boundary Conditions} \label{unexdir}
\begin{figure}
  \begin{center}
    \includegraphics[width=5in]{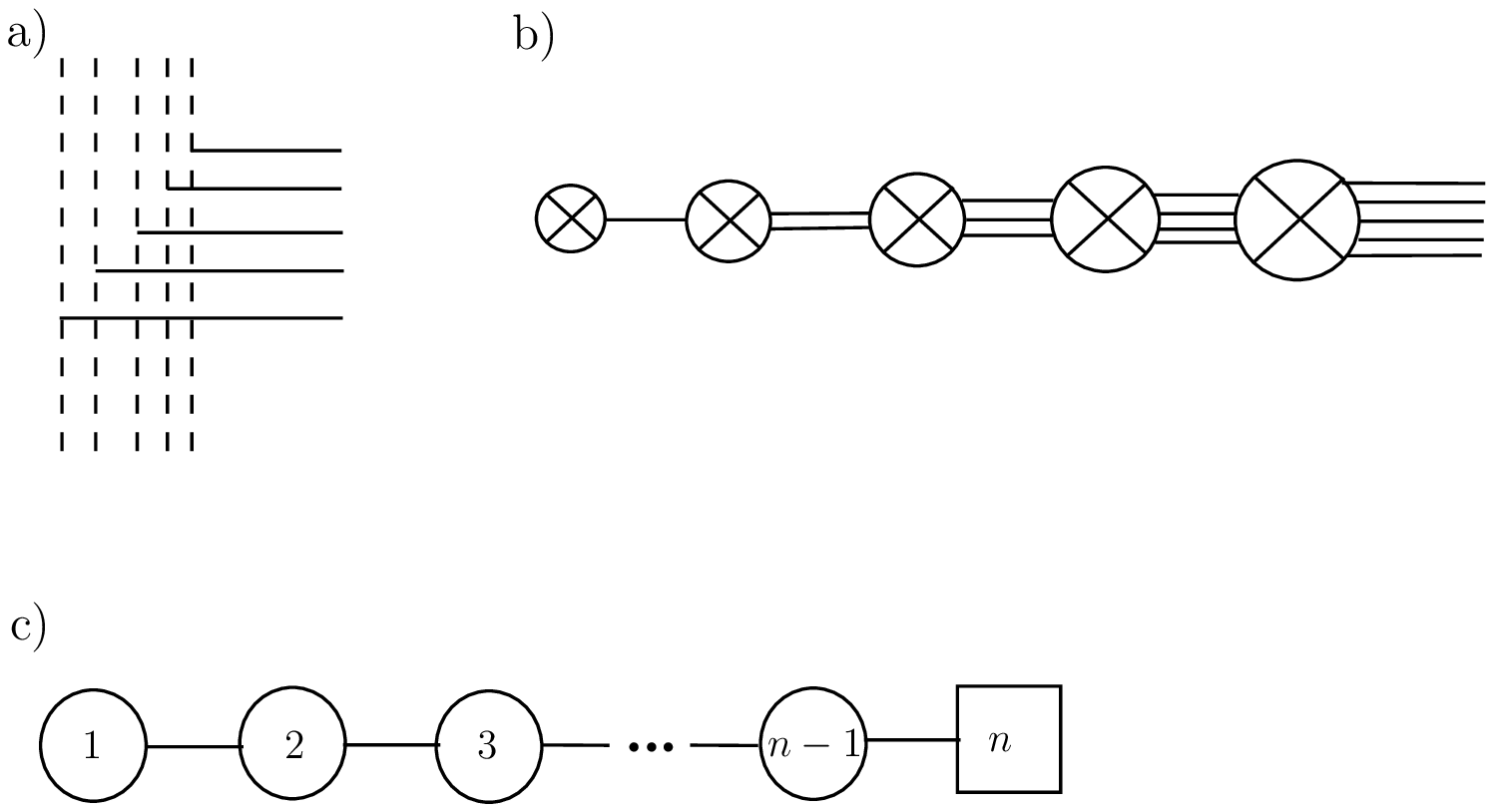}
  \end{center}
\caption{\small (a) Dirichlet boundary conditions in $U(n)$ gauge
theory arise from $n$ D3-branes ending one at a time on $n$
D5-branes, as sketched here for $n=5$.  (b) The $S$-dual is obtained
by simply converting D5-branes to NS5-branes.  There is no need for
any rearrangement. (c) This leads to Neumann boundary conditions
coupled to the SCFT that is the IR limit of this quiver.}
  \label{Fig22}\end{figure}

Our first example in $U(n)$ gauge theory will be a simple Dirichlet
boundary condition. This is realized by ending each of $n$ D3-branes
on a distinct D5-brane. The steps to the $S$-dual configuration are
shown in fig. \ref{Fig22}. Each D5-brane has linking number 1. The
$S$-dual boundary condition couples the $U(n)$ gauge fields to a
special boundary theory $T(SU(n))$ which generalizes  our friend
$T(SU(2))$. $T(SU(n))$ is given by the IR limit of the quiver gauge
theory in fig. \ref{Fig22}(c).

\begin{figure}
  \begin{center}
    \includegraphics[width=4.5in]{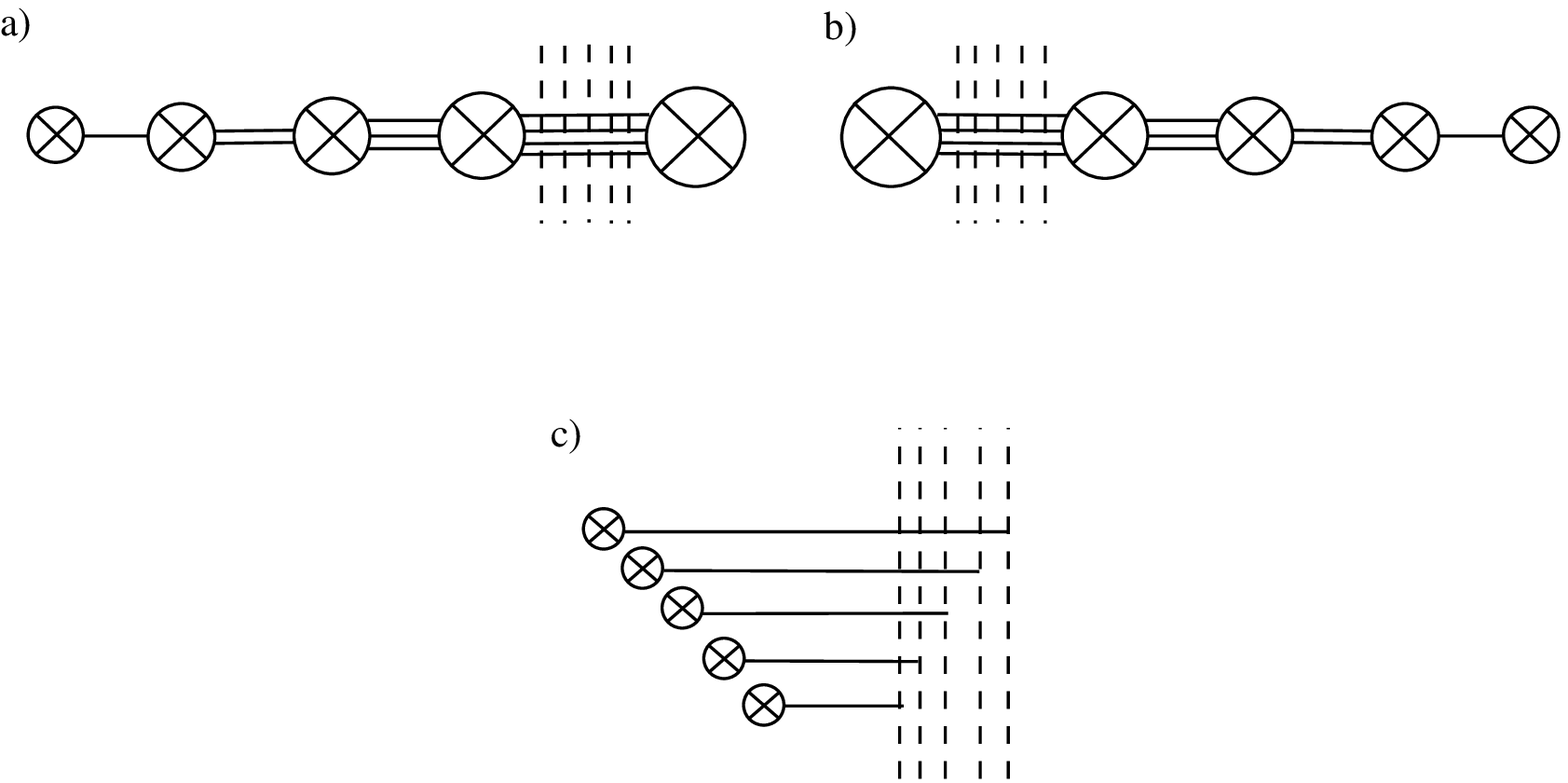}
  \end{center}
\caption{\small (a) A brane configuration bounded in the $y$
direction that leads to the theory $T(SU(n))$, sketched here for
$n=5$. D5-branes all have linking number $n-1$ and NS5-branes all
have linking number 1. (b) The mirror is the same quiver written
backwards. (c) The self-mirror property becomes manifest if we
move all NS5-branes to the left and all D5-branes to the right. To
make the picture visible, the NS5-branes and D5-branes have been
displaced; a $U(n)\times U(n)$ global symmetry (whose center acts
trivially) appears when they become coincident. But this
configuration has no interpretation in gauge theory.}
  \label{Fig23}\end{figure}

Like $T(SU(2))$, the theory $T(SU(n))$ is self-mirror.  To show
this, we observe that
 $T(SU(n))$ can be derived from the purely three-dimensional
configuration shown in fig. \ref{Fig23}(a). The linking numbers of
the D5-branes are all $n-1$, while the linking numbers of the
NS5-branes are all $1$. As a consequence, the mirror is given by the
same brane configuration written backwards, as in fig.
\ref{Fig23}(b). This establishes the mirror symmetry.  Finally, by
separating all the NS5-branes from the D5-branes, it is possible to
make the self-duality of the configuration more manifest as in fig.
\ref{Fig23}(c), at the price of obscuring the physical content of
the theory, since this configuration has no gauge theory
interpretation.

Because of the self-mirror property of $T(SU(n))$, its Higgs and
Coulomb branches are isomorphic. By $S$-duality, the Coulomb branch
of $T(SU(n))$ is the same as the moduli space of solutions of the
Nahm equations on the half line $y\geq 0$ with Dirichlet boundary
conditions at $y=0$ and prescribed behavior of $\vec X$ at infinity.
This moduli space is a hyper-Kahler manifold first introduced by
Kronheimer \cite{Kronheimer}. In any one of its complex structures,
the manifold is isomorphic to the nilpotent cone $\N$ of
$SL(n,\BC)$. $\N$ is defined as the subspace of nilpotent elements
of the $SL(n,\BC)$ Lie algebra. It is a complex symplectic manifold
of dimension $n(n-1)$. For an explanation of these matters, and many
other topics that are relevant here (including Nahm's equations and
Slodowy slices), see section $3$ of \cite{Gaiotto:2008sa}.

\begin{figure}
  \begin{center}
    \includegraphics[width=3.5in]{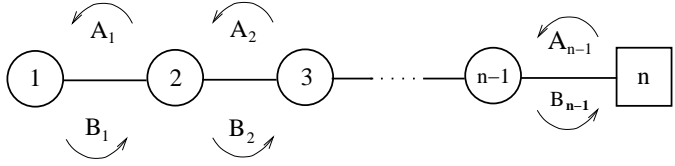}
  \end{center}
\caption{\small The quiver that leads to $T(SU(n))$, with chiral
multiplets labeled.}
  \label{Fig24}\end{figure}

It is instructive to see directly that the Higgs branch of the
quiver in fig. \ref{Fig22}(c) coincides with $\N$ as a complex
manifold. Picking a specific complex structure, each
hypermultiplet splits into a pair of chiral multiplets in
conjugate representations of the gauge group. Labeling these
chiral multiplets as in fig. \ref{Fig24}, they consist of $i
\times (i+1)$ matrices $A_i$, $i=1,\dots,n-1$, transforming as
$\overline{\bf i}\otimes {\bf (i+1)}$ under $U(i)\times U(i+1)$,
and $(i+1) \times i$ matrices $B_i$  transforming in the conjugate
representation. Here $U(i),$ $i<n$ are the gauge groups at the
nodes of the quiver, and $U(n)$ is the global symmetry group
acting at the end of the quiver.  The center of $U(n)$ acts
trivially, so the flavor symmetry of the quiver theory is really
$SU(n)$. The composite field $M=B_{n-1}A_{n-1}$ is an $n \times n$
matrix of rank at most $n-1$. The traceless part of $M$ is the
complex moment map for the $SU(n)$ flavor symmetry, and will
parametrize the Higgs branch.

Using the $F$-term constraints \begin{align}\notag  A_{i+1} B_{i+1}
&= B_i
A_i,~i=1,\dots,n-2\\
A_1B_1&=0,\end{align} we can compute
\begin{equation} M^2 = B_{n-1}A_{n-1} B_{n-1}A_{n-1} =
B_{n-1}B_{n-2}A_{n-2}A_{n-1}\end{equation} and discover that $M^2$
has rank at most $n-2$. Iteratively, \begin{equation}M^a =
\prod_{i=n-1}^{n-a} B_i \prod_{j=n-a}^{n-1} A_j,\end{equation} and
has rank at most $n-a$. Finally, $M^n=0$, so $M$ is nilpotent. Any
nilpotent $n\times n$ matrix will satisfy the rank constraints
$\mathrm{rk}\, M^a \leq n-a$, and it is not hard to show that for
any nilpotent $M$, a set of matrices $A_i, B_i$, unique up to
gauge transformations, can be found satisfying the above
conditions. So the Higgs branch coincides with the nilpotent cone
$\N$.

We can also introduce $n-1$ possible FI terms $t_i$ for the center
of the gauge group. This deforms the $F$-term constraints to
\begin{align}\notag A_{i+1} B_{i+1} &= B_i
A_i + t_{i+1},~i=1,\dots,n-2\\
A_1B_1&=t_1 .\end{align} The above analysis changes only slightly.
Starting with
\begin{equation}M^2 = B_{n-1}A_{n-1} B_{n-1}A_{n-1} =
B_{n-1}B_{n-2}A_{n-2}A_{n-1} + t_{n-1} M,\end{equation}  we derive
iteratively \begin{equation}M (M-t_{n-1}) (M-t_{n-1}-t_{n-2}) \cdots
(M- \sum_{i=n-a}^{n-1}t_i)= \prod_{i=n-1}^{n-a} B_i
\prod_{j=n-a}^{n-1} A_j.\end{equation} Finally,
\begin{equation}\label{ortuf}M(M-t_{n-1})(M-t_{n-1}-t_{n-2}) \cdots (M-
\sum_{i=1}^{n-1} t_i) =0\end{equation} expresses the characteristic
polynomial of $M$ in terms of the FI parameters. The Higgs branch as
a complex manifold is the set of Lie algebra elements with the
eigenvalues indicated by the characteristic polynomial.

In the brane realization in fig. \ref{Fig23}(c), the FI parameters
of the quiver gauge theory correspond to the positions of the
NS5-branes in the $\vec X$ directions. The positions of the
D5-branes in the $\vec Y$ directions correspond to mass parameters
for the fundamental hypermultiplets of the quiver. In the original
$D5$ brane boundary condition, the positions of the D3-branes at
infinity in the $\vec X$ direction are the FI parameters, as
explained in section 3 of \cite{Gaiotto:2008sa}, where it is also
shown that turning on those parameters deforms the nilpotent cone to
a more generic conjugacy class.

\subsubsection{The Dual of a Nahm Pole}\label{dualnahm}
\begin{figure}
  \begin{center}
    \includegraphics[width=3in]{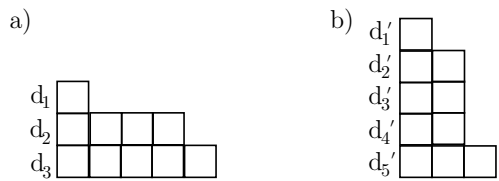}
  \end{center}
\caption{\small (a) A Young diagram with $i^{th}$ row of length
$d_i$.  (b) The dual Young diagram, with rows and columns
exchanged.}
  \label{Fig25}\end{figure}
 Dirichlet boundary conditions can be generalized by requiring a pole for the scalar fields $\vec
X$  at the boundary. Supersymmetry requires that the residue of the
pole should be the images of a standard set of $\frak{su}(2)$
generators $\vec t$ under a homomorphism $\rho:\frak{su}(2) \to
\frak{u}(n)$.

Let us recall that for every positive integer $n$, the Lie algebra
$\frak{su}(2)$ has up to isomorphism a unique irreducible
representation of that dimension. Therefore, the choice of $\rho$
is determined by a decomposition $n=\sum_{i=1}^kd_i$, where the
$d_i$ are positive integers that we can assume to be arranged in
nondecreasing order. The information contained in this
decomposition is conveniently displayed in a Young diagram whose
$i^{th}$ row, counting from top to bottom, has length $d_i$, as in
fig. \ref{Fig25}(a).

The choice of $\rho$ has another interpretation.  The image under
$\rho$ of the raising operator in the $\frak{su}(2)$ Lie algebra is
a nilpotent element $\rho_+$ of $\frak{gl}(n)$, the complexification
of $\frak{u}(n)$. It is the direct sum of nilpotent Jordan blocks of
dimension $d_i$. The existence of a Jordan canonical form for every
matrix means that any nilpotent element of $\frak{gl}(n)$ is
conjugate to this form for some $d_i$.  Thus, the choice of $\rho$
is equivalent to the choice of a nilpotent conjugacy class in the
complexified Lie algebra. (According to the Jacobson-Morozov
Theorem, this has an analog for every semi-simple Lie algebra, not
just for $\frak{u}(n)$.) One more remark will be helpful.  There is
a natural duality of Young diagrams in which they are reflected
along the main diagonal (fig. \ref{Fig25}(b)).  This gives a duality
operation on homomorphisms $\rho:\frak{su}(2)\to \frak{u}(n)$. We
write $\rho_D$ for the dual of $\rho$ in this sense.

\begin{figure}
  \begin{center}
    \includegraphics[width=2.5in]{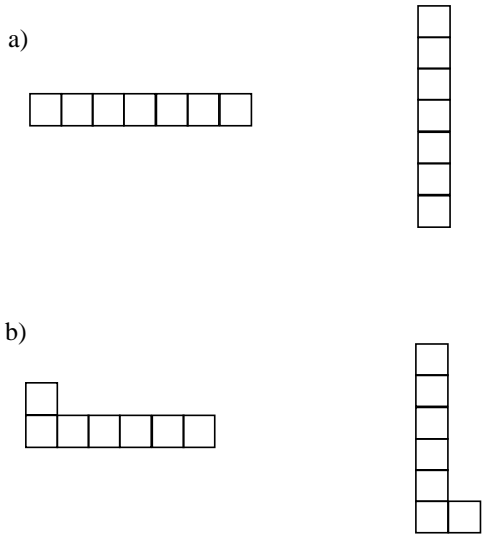}
  \end{center}
\caption{\small (a) The Young diagram associated with the regular
or irreducible representation $\rho:\frak{su}(2)\to\frak{u}(n)$
(left), and its dual (right), which is associated with the trivial
representation. (b) The Young diagram associated with the
subregular representation, which corresponds to the decomposition
$n=(n-1)+1$, and its dual, which corresponds to
$n=2+1+1+\dots+1$.}
  \label{Fig26a}\end{figure}
We pause to give a few examples (fig. \ref{Fig26a}) because dual
pairs will enter our story momentarily, though not in a symmetrical
way. The regular or irreducible representation $\rho$, which
corresponds to the case $k=1$ and $d=n$, is dual to the trivial
representation $\rho=0$, with $k=n$ and all $d_i=1$. If $\rho$ is
the regular representation, then $\rho_+$ is known as a regular
nilpotent element; if $\rho=0$, then $\rho_+=0$. Similarly, the
subregular representation, corresponding to the decomposition
$n=1+(n-1)$, is dual to the decomposition $n=1+1+\dots 1+2$, related
to the most obvious embedding of $SU(2)$ in $SU(n)$.  The nilpotent
element corresponding to the subregular representation is called a
subregular nilpotent element, and the nilpotent element
corresponding to the dual representation is called a minimal
nilpotent.  For more background on these subjects and many other
matters that will appear below, see section 3 of
\cite{Gaiotto:2008sa}.

\begin{figure}
  \begin{center}
    \includegraphics[width=4.5in]{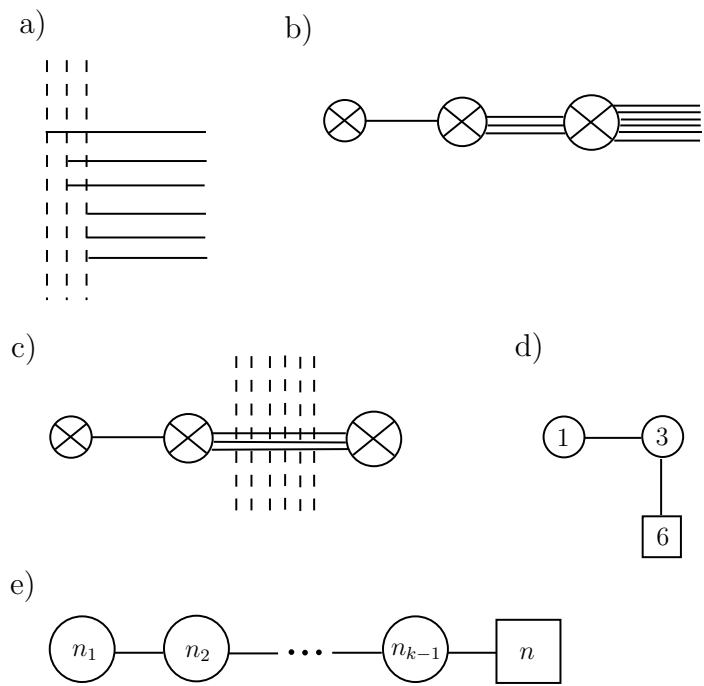}
  \end{center}
\caption{\small (a) Numbering the $k$ D5-branes in this picture from
left to right, $d_i$ D3-branes end on the $i^{th}$ D5-brane. In this
example, $k=3$ and the $d_i$ are $1,2,3$. (b) The $S$-dual
configuration, and (c) a macroscopically three-dimensional
configuration that leads to the same boundary SCFT.
 (d) The quiver gauge
theory that flows in the IR to this boundary SCFT. The gauge group
is $U(n_1)\times U(n_2)$, where $n_1=d_1=1$ and $n_2=d_1+d_2=3$.
(e) The general quiver of this type, representing the theory
$T_\rho(SU(n))$.}
  \label{Fig27a}\end{figure}

An embedding which decomposes the fundamental representation of
$U(n)$ via $n=\sum_{i=1}^k d_i$  can be realized by a boundary
condition in which $n$ D3-branes end on $k$ D5-branes (fig.
\ref{Fig27a}(a)). Numbering the D5-branes from left to right, we
take $d_i$ D3-branes to end on the $i^{th}$ D5-brane. The linking
numbers of the D5-branes are then equal to the $d_i$. Since we
take the $d_i$ to be non-increasing, the D5-branes can be
separated  in the $y$ direction without introducing extra degrees
of freedom, as we have done in the figure.

In fig. \ref{Fig27a}(d,e), we indicate the quiver gauge theory which
will flow in the infrared to the SCFT  that determines the dual
boundary condition. It is a linear quiver of $k-1$ unitary groups of
rank $n_i = \sum_{j=1}^i d_j$. We will denote this SCFT as
$T_{\rho}(SU(n))$. If $\rho$ is trivial, then $T_{\rho}(SU(n))$ is
the same as $T(SU(n))$.

The Higgs branch of this quiver has an $SU(n)$ flavor symmetry, and
can be analyzed in the same fashion as the Higgs branch for the
quiver of $T(SU(n))$.  The matrices $A_i$ and $B_i$ can be defined
as before, but now have size $n_i \times n_{i+1}$ and $n_{i+1}
\times n_i$ respectively. We can define again $M = B_{k-1}A_{k-1}$,
whose traceless part is the moment map for the $SU(n)$ action. The
$F$-term constraints are
\begin{align} A_{i+1} B_{i+1} &= B_i
A_i + t_{i+1},~i=1,\dots,k-2\\
A_1B_1&=t_1 .\end{align} The same proof as before shows that $M$
is nilpotent, but now the rank of $M^i$ is at most $n_{k-i}$. This
constraint is not satisfied by an arbitrary nilpotent matrix. If
we consider a nilpotent matrix $M$ with Jordan blocks of sizes
$j_1,j_2,\dots,j_t$ (which we arrange in nonincreasing order),
 then the rank of $M^i$ is $\sum_{a|j_a\geq i} (j_a-i)$. An $M$
which saturates the rank constraints will have $\sum_{a|j_a\geq i}
(j_a-i) = n_{k-i}$.  The orbit of nilpotent matrices with this
Jordan structure is a dense open subset of the Higgs moduli space
$\mathcal H$.  $\mathcal H$ is actually the {\it closure} of this
orbit, since it also includes points at which the rank of $M^i$ is
lower for some $i$.

\begin{figure}
  \begin{center}
    \includegraphics[width=1.5in]{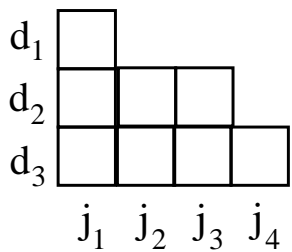}
  \end{center}
\caption{\small The Young diagram associated to $\rho$. The
lengths of the rows are the sizes $d_i$ of the Jordan blocks of
$\rho_+$ (arranged from top to bottom), and the heights of the
columns are the sizes $j_a$ of the Jordan blocks of $M$ (arranged
from left to right). The formula $n_{k-i} =
\sum_{j=1}^{k-i}d_j=\sum_{a|j_a\geq i} (j_a-i)$ asserts that the
number of blocks above the bottom $i$ rows can be computed by
summing over either rows or columns.}
  \label{Fig28}\end{figure}
We started with one nilpotent element $\rho_+$, and we have
arrived at another nilpotent element $M$.  The relation between
them appears obscure, but it has a simple combinatoric
interpretation, as in fig. \ref{Fig28}: $j_a$ is the height of the
$a^{th}$ column in the Young diagram of $\rho$, reading from left
to right. Indeed the number of boxes above the bottom $i$ rows is
 $n_{k-i} = \sum_{j=1}^{k-i} d_j$, and the
 formula  $n_{k-i}=\sum_{a|j_a\geq i}
(j_a-i)$ gives a way to compute this number by summing over
columns. This means that, letting $\rho_D$ denote the dual of the
$\frak{su}(2)$ embedding in the sense of fig. \ref{Fig25}(b), the
Higgs branch of the SCFT $T_\rho(SU(n))$ is the closure of the
orbit of the nilpotent element $\rho_{D\,+}$.

\begin{figure}
  \begin{center}
    \includegraphics[width=5in]{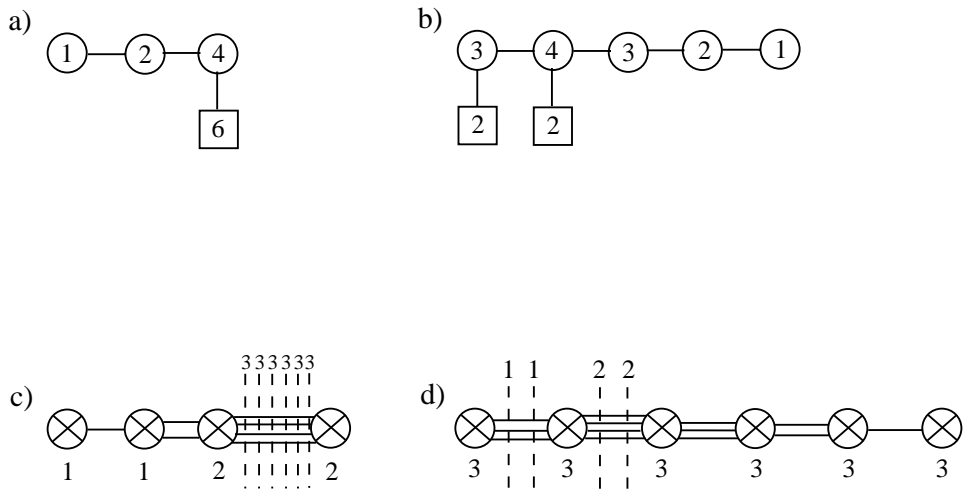}
  \end{center}
\caption{\small (a) The quiver representing $T_\rho(SU(n))$ for
$n=6$ and a particular $\rho$.  (b) The mirror quiver.  It
contains a chain of five balanced nodes, leading to an $SU(6)$
global symmetry of the Coulomb branch, which matches the flavor
symmetry of the Higgs branch in (a).   The mirror symmetry between
the two quivers in (a) and (b)  arises by comparing the two brane
configurations of (c) and (d), in which linking numbers are
indicated.}
  \label{Fig29}\end{figure}

To understand the Coulomb branch of $T_{\rho}(SU(n))$, we need to
build the mirror quiver, which describes the mirror theory that we
will call $T^{\rho}(SU(n))$. The linking numbers of the NS5-branes
in the brane realization of $T_{\rho}(SU(n))$ in fig. \ref{Fig27a}
coincide with the $d_i$, while the linking numbers of the $n$
D5-branes are all $k-1$. The mirror quiver has $n-1$ nodes, and the
$i^{th}$ fundamental flavor sits at the node number $d_i$ from left
to right. The ranks of the gauge groups are computed from the
linking numbers $k-1$ of the NS5-branes. For example the rank of the
leftmost gauge group is $k-1$ and the rank of the rightmost gauge
group is $1$.  For an illustration of this procedure, see fig.
\ref{Fig29}.

\def\S{\mathcal S}
Unfortunately, we do not know how to describe the Higgs moduli space
of this mirror quiver directly by solving the $F$-term constraints.
According to $S$-duality, this Higgs moduli space should be the same
as the moduli space of solutions of the Nahm equations with the
original boundary condition determined by $\rho$ and with $\vec X\to
0$ at infinity. As explained in section 3 of \cite{Gaiotto:2008sa},
this moduli space is isomorphic as a complex manifold to the Slodowy
slice $\S_\rho$ transverse to the nilpotent orbit associated with
$\rho$ intersected with the nilpotent cone $\N$. An indirect
argument given by Nakajima \cite{Nakajima} using the ADHM transform
of instantons confirms the identity between the Higgs moduli space
of $T^{\rho}(SU(n))$ with this intersection $\S_\rho\cap \N$.

\subsubsection{A Simple Application of $S$-Duality For Nahm
Poles}\label{application}

\begin{figure}
  \begin{center}
    \includegraphics[width=4in]{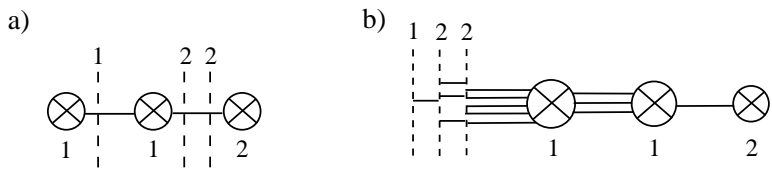}
  \end{center}
\caption{\small (a) A typical good linear quiver related to a
three-dimensional gauge theory.  All D3-brane segments end on
NS5-branes and all linking numbers are nondecreasing from left to
right. (b) Moving all D5-branes to the left, so that all
NS5-branes are on the right, we get an IR equivalent configuration
that treats D5-branes and NS5-branes symmetrically, but has no
direct interpretation in gauge theory. The D5 configuration on the
left determines an $\frak{su}(2)$ embedding $\rho$, and the NS5
configuration on the right determines the $S$-dual of another
$\frak{su}(2)$ embedding $\rho'$.  The example given here is
self-mirror, since in (b), exchange of the two types of fivebrane
is equivalent to a reflection $y\leftrightarrow -y$.}
  \label{Fig30}\end{figure}

Consider a three-dimensional SCFT built from a general good linear
quiver gauge theory, realized on a set of D3-brane segments
stretched between NS5-branes and interacting with D5-branes.   The
linking numbers are non-decreasing from left to right, and there
are no semi-infinite D3-branes (fig. \ref{Fig30}). A very natural
rearrangement of the fivebranes is to push all the D5-branes to
the left of all the NS5-branes. The D5-branes can be brought
together to define a boundary condition at $y=0$ given by a
certain $SU(2)$ embedding $\rho$. The sizes of the irreducible
blocks of $\rho$ coincide with the linking numbers of the
D5-branes. Similarly the NS5-branes can be brought together to
give a boundary condition at $y=L$, which is $S$-dual to the
boundary condition given by another $SU(2)$ embedding $\rho'$. The
sizes of the irreducible blocks of $\rho'$ can be read off from
the linking numbers of the NS5-branes.

Although this last configuration has no direct relation to a
three-dimensional gauge theory, it is certainly a four-dimensional
$U(n)$ gauge theory on the segment $0<y<L$ with boundary
conditions associated with $\rho$ and $\rho'$. Here $n$ is the
number of D3-brane segments in the space between the D5's and
NS5's. The Higgs moduli space of vacua of this theory is described
by Nahm's equations on the segment with appropriate conditions at
the two ends.  At $y=0$, $\vec X$ must have a pole with residue
determined by $\rho$.  The appropriate condition at $y=L$ is that
$\vec X$ must equal the moment map $\vec\mu$ of $T_{\rho'}$. It is
now straightforward to describe this moduli space as a complex
manifold. Because of the boundary condition at $y=0$, ${\mathcal
X}(L)=X_1(L)+iX_2(L)$ lies in the Slodowy slice $\S_\rho$
transverse to the nilpotent orbit $\O_{\rho}$ associated to
$\rho$. On the other hand, the complex moment map of the Higgs
branch of $T_{\rho'}$ takes values in the closure of the dual
orbit $\O_{\rho'_D}$ related to $\rho'_D$. The boundary condition
at $y=L$ therefore gives the intersection  $\S_\rho \cap
\O_{\rho'_D}$. This intersection is the Higgs branch.
Reciprocally, the Coulomb branch is the intersection
$\S_{\rho'}\cap \O_{\rho_D}$.

 We denote an SCFT constructed this
way using two homomorphisms $\rho,\rho':\frak{su}(2)\to
\frak{u}(n)$ as $T_{\rho'}^{\rho}(SU(n))$. Clearly $T(SU(n))$,
$T_{\rho}(SU(n)$, and $T^{\rho}(SU(n))$ are all examples of this
class, with $\rho$ and/or $\rho'$ taken to be trivial. The mirror
of $T^\rho_{\rho'}$ is $T^{\rho'}_\rho$.

One limitation of what we have said is that starting from a
suitable quiver gauge theory, we identified a pair $\rho, \rho'$,
but we gave no indication of what pairs can be produced this way.
Alternatively, for any given pair, we can generate a
three-dimensional theory as the infrared limit of the
four-dimensional $U(n)$ gauge theory on a segment with boundary
conditions given by that pair, as in fig. \ref{Fig30}(b). But this
theory may spontaneously break supersymmetry.

For example (assuming $n>1$), if $\rho$ is regular and $\rho'$ is
nonzero (which implies that $\rho'_D$ is not regular), the Slodowy
slice transverse to $\rho$ does not intersect the orbit of
$\rho'_D$, so supersymmetry is broken. If $\rho$ and $\rho'$ are
both regular (so that $\rho'_D=0$), this statement is equivalent
to the mysterious ``$s$-rule'' from \cite{Hanany:1996zelf}. A
borderline example is $\rho=\rho'_D$: the Slodowy slice transverse
to $\rho$ intersects the orbit of $\rho'_D=\rho$ in a single
point. The theory has a single massive supersymmetric vacuum. In
this case, $T_{\rho'}^\rho$ is a trivial SCFT.

\subsubsection{Reductions of the Gauge Symmetry to
$U(1)$}\label{reduone}

Just as for $U(2)$,  it is instructive to consider configurations
in which the gauge group at the boundary is reduced from $U(n)$ to
$U(1)$. There are several distinct ways to do that; the partial
Dirichlet boundary conditions which implement the reduction from
$U(n)$ to $U(1)$ may be accompanied by a Nahm pole. Brane
configurations do not allow us a generic choice of a $U(1)$
subgroup of  $U(n)$. The gauge symmetry at the boundary will have
to be carried by a single D3-brane and so will correspond to the
first factor $U(1)$ of a subgroup $U(1)\times U(n-1)\subset U(n)$.

\begin{figure}
  \begin{center}
    \includegraphics[width=5in]{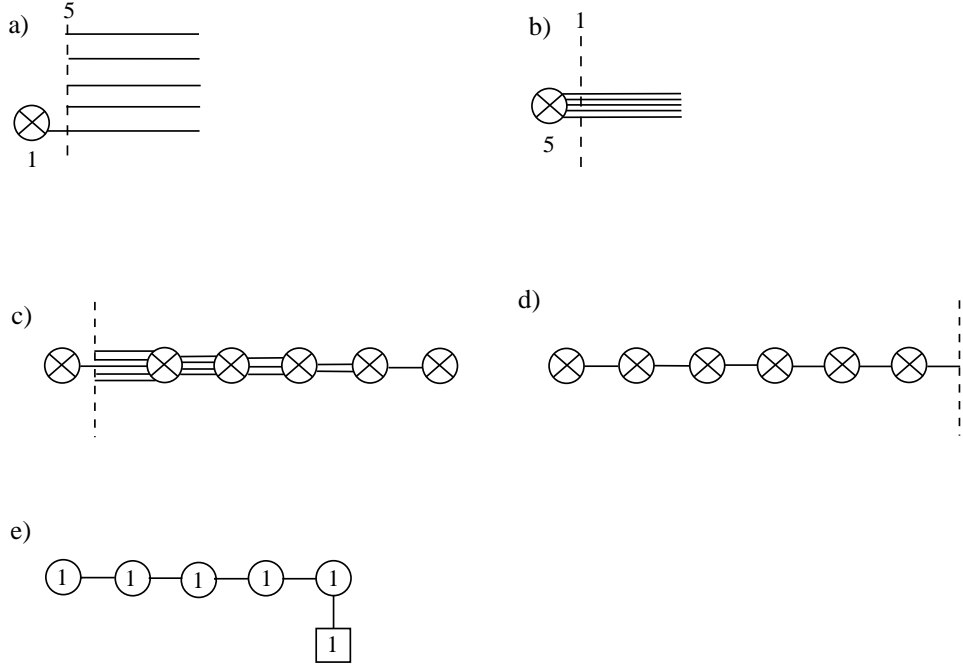}
  \end{center}
\caption{\small (a) Reduction of $U(n)$ gauge theory to $U(1)$ by
a subregular Nahm pole (shown for $n=5$).  The $U(1)$ obeys
Neumann boundary conditions. (b) The $S$-dual boundary condition,
found by the usual brane procedure. It arises by coupling to a
SCFT $\fB^\vee$ which in this case consists of a free
hypermultiplet. (c) As usual, one can try to find $\fB^\vee$ by
the ungauging procedure. Alternatively, the mirror
$\tilde\fB^\vee$ of $\fB^\vee$ can be found by coupling the
original boundary condition to the dual of Dirichlet, represented
here by the brane configuration related to $T(SU(n))$. (d) A more
convenient rearrangement of (c), which is equivalent to the quiver
in (e). All nodes but the leftmost one are balanced, leading to an
$SU(n)$ symmetry of the Coulomb branch. The leftmost node is
minimally unbalanced, so the Coulomb branch describes twisted
hypermultiplets in the fundamental representation of $SU(n)$, in
agreement with (b).}
  \label{Fig31}\end{figure}
An example in which the $U(1)$ subgroup has pure Neumann boundary
conditions is shown in fig. \ref{Fig31}(a). In this example, the
reduction in gauge symmetry is accomplished by ending the other
$n-1$ D3-branes on a single D5-brane, whose linking number is $n$.
As the NS5-brane in the figure has linking number 1, in the dual
configuration it will appear as a D5-brane just to the right of a
single NS5-brane of linking number $n$, with no D3-brane ending on
it (fig. \ref{Fig31}(b)). The dual boundary condition then will have
full $U(n)$ symmetry at the boundary, coupled to a single
fundamental hypermultiplet.

It is interesting to compare this to the result of the usual
ungauging strategy, by ending the D3-branes in the dual
configuration on a Dirichlet boundary condition at $y=L$ and
taking the $S$-dual and IR limit. The resulting brane
configuration is shown in fig. \ref{Fig31}(c) and (d) before and
after rearrangement.

The final result is that the mirror $\tilde\fB^\vee$ of the SCFT
that defines the $S$-dual boundary conditions is given by the linear
quiver gauge theory of fig. \ref{Fig31}(e). In the language of
section \ref{quivers}, this is an ugly quiver with a single
minimally unbalanced node (at the extreme left) and a chain of $n-1$
balanced nodes.  The balanced nodes generate a $U(n)$ symmetry of
the Coulomb branch, and because of the minimally unbalanced node,
there are free twisted hypermultiplets in the fundamental
representation of $U(n)$.  Counting dimensions, we see that the
Coulomb branch should have hyper-Kahler dimension $n$ and the Higgs
branch is absent.  So in fact, this quiver theory is precisely
equivalent in the IR to a free theory of $n$ twisted
hypermultiplets.  Thus, the $S$-dual of the original boundary
condition is equivalent to the coupling to a free SCFT $\fB^\vee$
consisting of $n$ free hypermultiplets in the fundamental
representation.  This is in perfect accord with what we found from a
direct brane construction in fig. \ref{Fig31}(b).

\begin{figure}
  \begin{center}
    \includegraphics[width=4.5in]{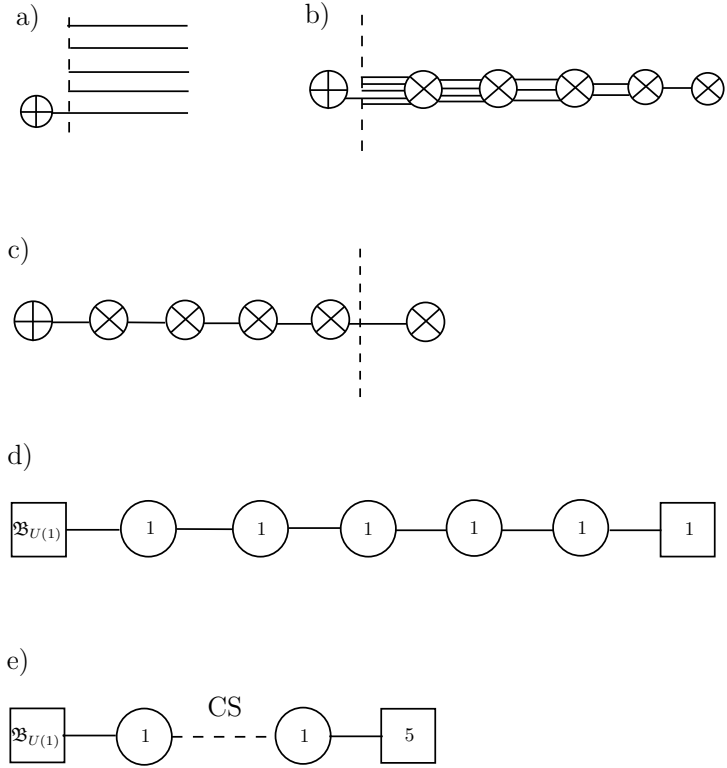}
  \end{center}
\caption{\small (a) The analog of fig. \ref{Fig31}(a) with Neumann
for the unbroken $U(1)$ altered by coupling to a boundary SCFT
$\fB_{U(1)}$, represented by a brane system $\bigoplus$.  The
gauge group is here $G=U(5)$. Assuming that the $S$-dual boundary
condition is obtained by coupling to an SCFT $\fB^\vee$, the
mirror $\tilde\fB^\vee$ to $\fB^\vee$ can be found via the usual
steps of (b), (c), and (d). (e) depicts a more exotic realization,
analogous to fig. \ref{Fig16}(e) for $U(2)$, that makes the
$SU(5)$ flavor symmetry manifest.}
  \label{Fig32}\end{figure}

It is straightforward to generalize this to the case that the
$U(1)$ gauge field at the boundary is not simply free but is
coupled to some
 SCFT
$\fB_{U(1)}$ coupled to the $U(1)$ gauge group at the boundary. The
final result from the ungauging procedure for $\tilde\fB^\vee$ (the
mirror of the SCFT $\fB^\vee$ that defines the $S$-dual boundary
condition) is depicted in fig. \ref{Fig32}(d): it is given by the
same chain of $n$ $U(1)$ groups coupled to $\fB_{U(1)}$ at one end.
As the reader may suspect, this result is a simple generalization of
the proposal involving $T(SU(2))$ in the $n=2$ case.

The chain of $n-1$ $U(1)$ gauge groups on the right of fig.
\ref{Fig32}(d) is something we have seen before, in fig.
\ref{Fig11}(c). It is the mirror of a $U(1)$ gauge theory coupled
to $n$ hypermultiplets of charge $1$.

This has another interpretation.  In the examples considered here,
we reduce the gauge symmetry from $U(n)$ to $U(1)$ by letting
$n-1$ D3-branes end on a single D5-brane, or equivalently by means
of a subregular Nahm pole $\rho:\frak{su}(2)\to\frak{u}(n)$. For
this choice of $\rho$, the theory $T_\rho(SU(n))$ is the IR limit
of $U(1)$ coupled to $n$ hypermultiplets of charge 1.  Its mirror
$T^\rho(SU(n))$ is the IR limit of the quiver theory of fig.
\ref{Fig11}(c).

We can therefore reformulate the prescription of fig. \ref{Fig32}(d)
in a more intrinsic fashion: the mirror $\tilde\fB^\vee$ to the SCFT
$\fB^\vee$ that defines the dual boundary condition is built by
coupling a three-dimensional $U(1)$ gauge theory to the product of
the input SCFT  $\fB_{U(1)}$ and the theory $T^{\rho}(SU(n))$, and
then flowing to the IR.  In brief, $\tilde
\fB^\vee=\fB_{U(1)}\times_{U(1)}T^\rho(SU(n))$.

Thus, at least for this class of examples, if we start with a
boundary condition that contains a Nahm pole $\rho$, then
$T^\rho(SU(n))$ plays the same role that $T(SU(2))$ played, in the
absence of the Nahm pole, in section \ref{utwo}. In section
\ref{genpres}, we will give a more systematic explanation of this.
We did not see this role of $T^\rho(SU(n))$ in section \ref{utwo},
because if the gauge group is $U(2)$, the only way to get a Nahm
pole is to end both D3-branes together on a single D5-brane,
leaving no analog of the input SCFT $\fB_{U(1)}$.

The importance of using $T^\rho(SU(n))$ rather than $T(SU(n))$ in
constructing $\tilde\fB^\vee$ is that, although we can see the
$T(SU(n))$ brane configuration on the right of fig.
\ref{Fig31}(c), this configuration does not have a gauge theory
interpretation; instead, fig. \ref{Fig31}(d) does have such an
interpretation, but involves $T^\rho(SU(n))$.

As in fig. \ref{Fig16}(e), we can alternatively use the CS-like
coupling to the $U(1)$ symmetry of the Coulomb branch of the
mirror $T_{\rho}(SU(n))$ to make the $U(n)$ flavor symmetry
manifest. We depict this in fig. \ref{Fig32}(e).

\subsubsection{Reductions of the Gauge Symmetry to
$U(2)$}\label{redutwo} For further practice, we will consider the
case that the gauge symmetry at the boundary is reduced to $U(2)$
by ending $n-2$ D3-branes on a single D5-brane.
\begin{figure}
  \begin{center}
    \includegraphics[width=5.5in]{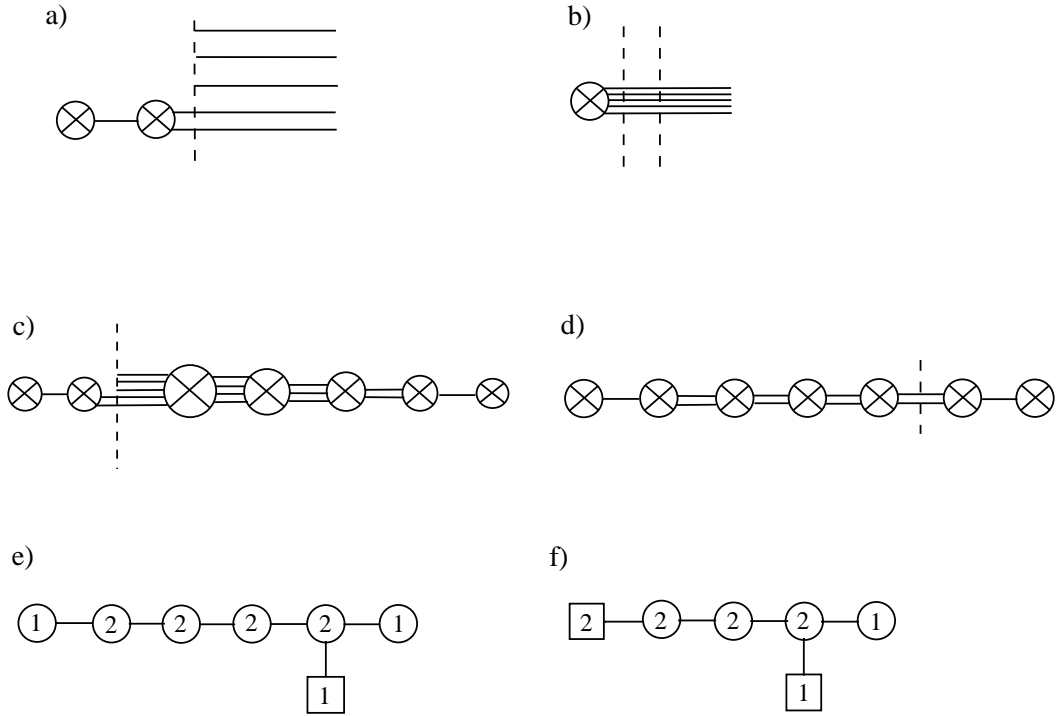}
  \end{center}
\caption{\small (a) Reduction of $U(n)$ gauge theory (in this
figure, $n=5$) to $U(2)$ by a Nahm pole; the remaining two D3-branes
end on two NS5-branes. (b) The dual has a single NS5-brane of
linking number $n$, with two D5-branes to its right. (c) If one
terminates the D3-branes of (b) at $y=L$ using Dirichlet boundary
conditions, and applies $S$-duality, one arrives here.  (d) The
result of a standard brane rearrangement applied to (c).  (e) The
quiver representing the configuration of (d). This is an ugly
quiver, in the language of section \ref{quivertheories}, as the
second node from the left has $e=-1$. (f) The relevant
$T^\rho(SU(n))$ is associated with this quiver.}
  \label{Fig33}\end{figure}

As the first example, we will let the two remaining D3-branes end on
two NS5-branes, as in fig. \ref{Fig33}(a). The corresponding theory
$\fB_{U(2)}$ coupled to the $U(2)$ gauge symmetry at the boundary is
$T(SU(2))$. Keeping track of the linking numbers, we construct the
$S$-dual configuration of  fig. \ref{Fig33}(b). The dual boundary
condition consists of a pair of fundamental hypermultiplets coupled
to the $U(n)$ gauge theory. Lets see how the standard ungauging
prescription reproduces such a result is an ``uneconomical''
fashion.

In the dual description of (b), if we end the D3-branes at $y=L$
with Dirichlet boundary conditions and take the $S$-dual, we
obtain the configuration in fig. \ref{Fig33}(c), with the initial
boundary condition on the left and $T(SU(n))$ (realized via
branes) on the right.  Rearranging the branes to get something
with a gauge theory interpretation, and keeping track of the
linking numbers, we arrive at
 fig. \ref{Fig33}(d), from which we get  in fig. \ref{Fig33}(e) a quiver whose
IR limit should be the mirror $\tilde\fB^\vee$ of the theory
$\fB^\vee$ that defines the dual boundary condition.

That quiver is interpreted directly as the coupling of the input
SCFT $\fB_{U(2)} = T(SU(2))$ to a quiver that represents
$T^{\rho}(SU(n))$ for a certain $\rho$. Differently put, the
relevant theory is a three-dimensional $U(2)$ gauge theory coupled
to the product of $T(SU(2))$ and $T^\rho(SU(n))$, both of which (for
the relevant $\rho$) have $SU(2)$ global symmetry. This is what we
call the composite gauge theory and denote
$\fB_{U(2)}\times_{U(2)}T^\rho(SU(n))$. The relevant $\rho$ is
associated with the decomposition $n=(n-2)+1+1$, corresponding to
the Nahm pole that we started with in \ref{Fig33}(a).

This result is quite like what we found in section \ref{reduone}:
the theory $T^\rho(SU(n))$ plays the same role in the presence of
a Nahm pole that $T(SU(n))$ plays without one.  As usual, we
expect this prescription to work well for every choice of the
input theory $\fB_{U(2)}$, as long as the dual boundary condition
has full $U(n)$ gauge symmetry.

\begin{figure}
  \begin{center}
    \includegraphics[width=4in]{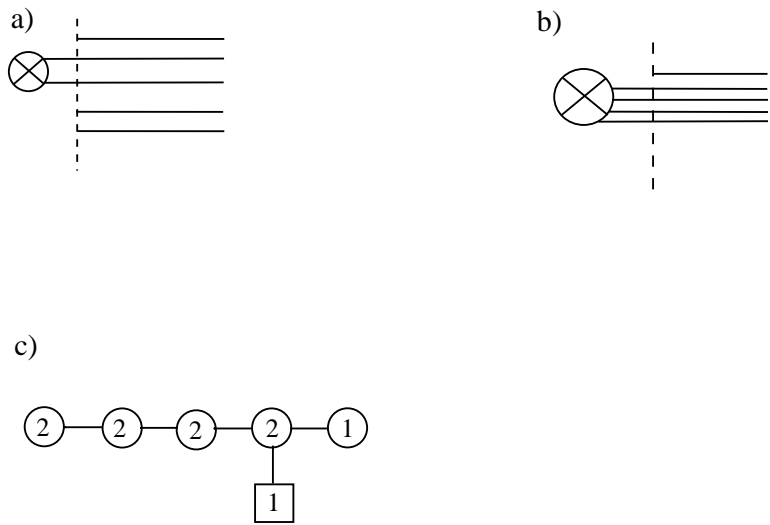}
  \end{center}
\caption{\small (a) $n-2$ D3-branes end on a single D5-brane, and
the remaining ones then end on a single NS5-brane. (b) The dual
brane configuration, as determined from the linking numbers. (c)
The dual quiver, constructed in the same way as fig.
\ref{Fig33}(e). It describes a $U(2)$ gauge theory (acting at the
left-most node of the quiver) coupled to $T^\rho(SU(n))$, for the
choices of $n$ and $\rho$ that were made in (a).  This is a bad
quiver, in the language of section \ref{quivertheories}, as the
leftmost node has $e=-2$.  }
  \label{Fig34}\end{figure}

Finally, another interesting example is given by taking $\fB_{U(2)}$
to be trivial.  In terms of branes, we do this by ending the two
leftmost D3-branes on a single NS5-brane, as in fig. \ref{Fig34}(a),
so as to get a pure Neumann boundary condition for the surviving
gauge group $U(2)$. In the dual brane configuration in fig.
\ref{Fig34}(b), which is found as usual by matching the linking
numbers, the gauge group is reduced to $U(n-1)$, which obeys Neumann
conditions at the boundary. The reduction of the dual gauge group
invalidates our simple derivation of the mirror $\tilde \fB^\vee$ of
the SCFT $\fB^\vee$ living at the $S$-dual boundary condition. If we
nevertheless follow this recipe, by coupling the initial boundary
condition to $T(SU(n))$ and moving branes to get something with a
gauge theory configuration, we arrive at the quiver of fig.
\ref{Fig34}(c). (This can be obtained from the quiver of fig.
\ref{Fig33}(e) by deleting the leftmost node, which in that example
generates $\fB_{U(2)}=T(SU(2))$.  In the present case, we take
$\fB_{U(2)}$ to be trivial so we replace the leftmost node of fig.
\ref{Fig33}(e) by nothing.)  This is a bad quiver, in the sense of
section \ref{quivers}, as the leftmost node has $e=-2$.  By
reasoning similar to that of fig. \ref{Fig21}, one can show that the
symmetries of the Coulomb branch of this bad quiver gauge theory are
spontaneously broken.  As we will explain more fully later, this is
related by $S$-duality to the fact that the dual boundary condition
has reduced gauge symmetry.

\subsubsection{Full Gauge Symmetry With Matter At The Boundary
}\label{neumfun}

Now we consider boundary conditions with full $U(n)$ symmetry
coupled to a boundary theory $\fB$.   A simple example is to use
D5-branes to generate $k$ fundamental hypermultiplets at the
boundary. We will get a result similar to what we found in section
\ref{utwo} for $n=2$:
 if the number $k$ of
fundamental hypermultiplets at the boundary is large enough, the
dual boundary condition is described by coupling to a dual SCFT
$\fB^\vee$ whose mirror $\tilde\fB^\vee$ admits a construction in
terms of $T(SU(n))$. This standard construction will give an
economical description of $\tilde\fB^\vee$ if $k>n$ and a less
economical one if $k=n$. Conversely, if $k<n$, the $S$-dual
boundary condition is not obtained by coupling to an SCFT; rather,
the gauge group at the boundary is reduced to $U(k)$, possibly in
the presence of a Nahm pole.
\begin{figure}
  \begin{center}
    \includegraphics[width=4.5in]{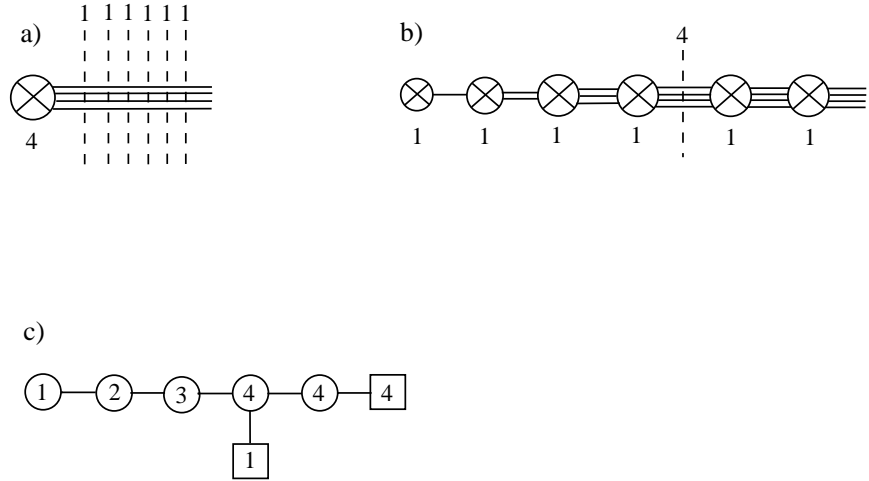}
  \end{center}
\caption{\small (a) $U(n)$ gauge theory coupled to $k$ fundamental
hypermultiplets at the boundary, as shown here for $n=4$ and
$k=6$. Linking numbers have been labeled.  (b) The dual brane
configuration, as determined from the linking numbers. (c) The
quiver that represents the dual brane configuration. All nodes are
balanced; the chain of five consecutive balanced nodes leads to an
$SU(6)$ global symmetry of the Coulomb branch, matching the global
symmetry of the six coincident fivebranes in (a).}
  \label{Fig35}\end{figure}

\begin{figure}
  \begin{center}
    \includegraphics[width=3.5in]{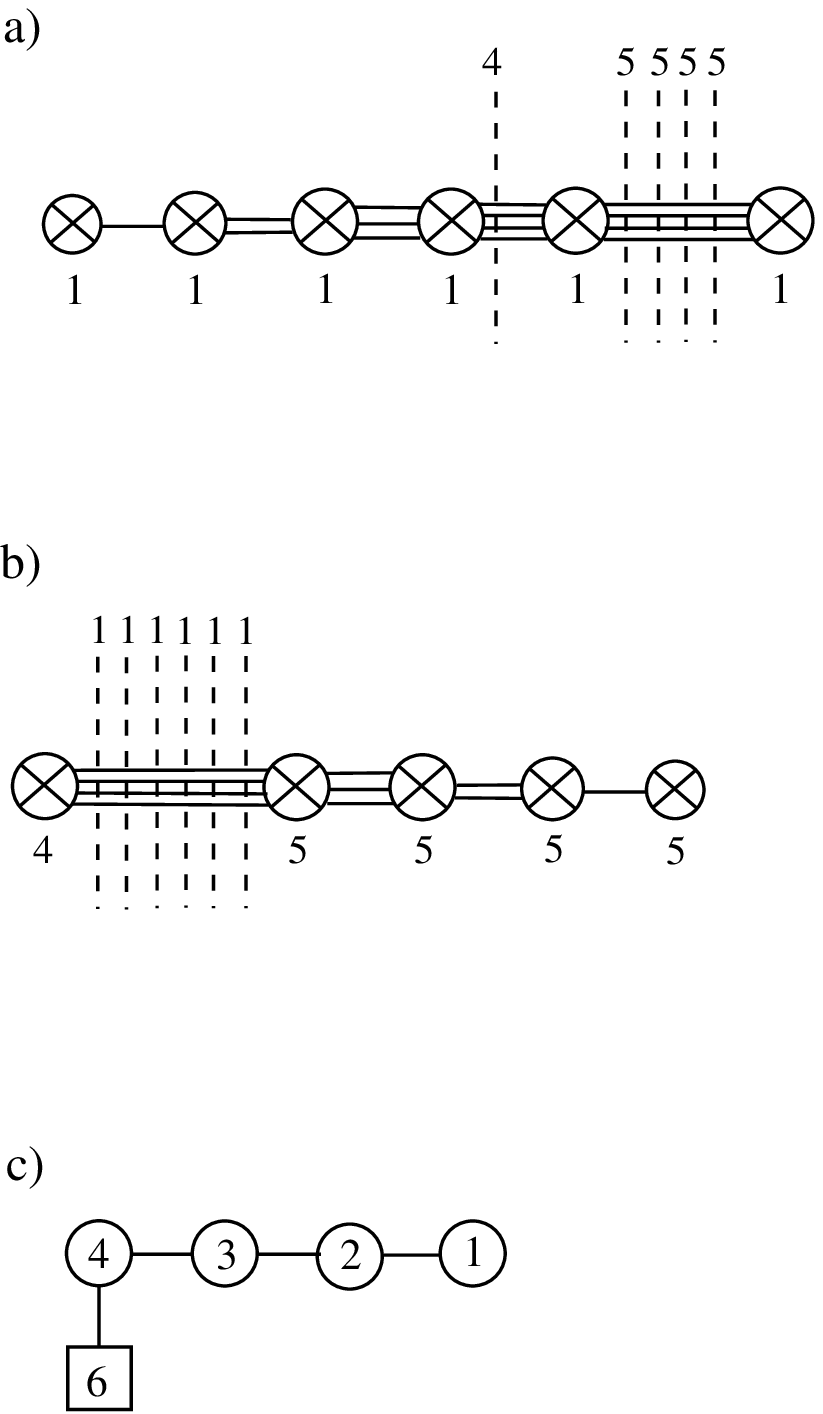}
  \end{center}
\caption{\small (a) The ``flavors'' coming from interaction with
semi-infinite D3-branes on the right of fig. \ref{Fig35}(b) have
been replaced by flavors coming from intersection with D5-branes.
 This will facilitate a purely three-dimensional construction of the relevant
 SCFT. Linking numbers are labeled. (b) The mirror arrangement, found
from the linking numbers. (c) The quiver corresponding to (b). The
original boundary condition (six fundamental hypermultiplets of
$U(4)$) has been coupled to $T(SU(4))$ by gauging their common
$U(4)$ symmetry, which acts at the leftmost node.  The other three
nodes are balanced, giving the Coulomb branch an $SU(4)$ global
symmetry.}
  \label{Fig36}\end{figure}

The brane manipulations and quivers for $k>n$ are shown in figs.
\ref{Fig35} and \ref{Fig36}. In fig. \ref{Fig35}, we construct a
quiver for the SCFT $\fB^\vee$ that appears in the dual boundary
condition, and in fig. \ref{Fig36}, we construct the mirror quiver
associated to the mirror SCFT $\tilde\fB^\vee$. This mirror quiver
 has a simple interpretation as $T(SU(4))$
coupled to the original boundary condition ($k=6$ fundamental
hypermultiplets) by a three-dimensional $U(4)$ gauge theory.

In general, for any $k$ and $n$, the original boundary condition has
a single NS5-brane of linking number $n$ and $k$ D5-branes of
linking number 1. The dual configuration has a single D5-brane of
linking number $n$ and $k$ NS5-branes of linking number 1. If $k>n$,
the dual configuration has the D5-brane to the left of the $n^{th}$
NS5-brane. This ensures that the dual consists of an SCFT $\fB^\vee$
coupled to Neumann boundary conditions.
\begin{figure}
  \begin{center}
    \includegraphics[width=3.5in]{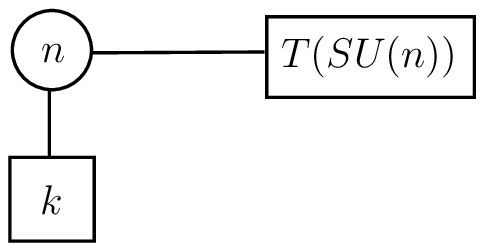}
  \end{center}
\caption{\small A schematic representation of the dual boundary
condition for $U(n)$ coupled to $k$ fundamental hypermultiplets. A
$U(n)$ gauge theory is coupled both to $T(SU(n))$ and to the
hypermultiplets.  The IR limit  gives a three-dimensional SCFT
$\tilde\fB^\vee$ that is the mirror of the SCFT $\fB^\vee$ that
defines the dual boundary condition.}
  \label{Fig37}\end{figure}
This is the condition under which the usual ungauging procedure is
a good way to determine the dual boundary condition.  To extract
$\fB^\vee$, we follow the familiar steps.  First we end the
D3-branes with Dirichlet boundary conditions.  Applying
$S$-duality, we get a representation of the mirror
$\tilde\fB^\vee$ of $\fB^\vee$ in terms of $T(SU(n))$ coupled to
the original SCFT $\fB$ (which in the present example consists of
$k$ fundamental hypermultiplets), as in fig. \ref{Fig37}.  This
strategy, which accounts for the result of fig. \ref{Fig36},
should work whenever the dual boundary condition has the full
$U(n)$ gauge symmetry at the boundary.

If $k>n$, the quiver description of $\tilde\fB^\vee$  by
$T(SU(n))$ coupled to the original boundary conditions satisfies
our constraints from section \ref{ordering}, and gives a
straightforward description of $\tilde\fB^\vee$.

If $k=n$, the dual brane configuration has a D5-brane of linking
number $n$ sitting to the right of $n$ NS branes. The resulting
 SCFT $\fB^\vee$ consists of two factors: a single fundamental
hypermultiplet (from the D3-D5 intersection) and $T(SU(n))$. In this
case, the quiver of fig. \ref{Fig37} that arises from the ungauging
procedure is ugly, accounting for the existence of the free
hypermultiplets.

If $k=n-1$, the D5-brane in the dual configuration is to the right
of $n-1$ NS5-branes but has linking number $n$; hence one D3-brane
will end on it and the dual boundary condition involves a reduction
of the gauge group down to $U(n-1)$ together with a coupling to
$T(SU(n-1))$. For smaller values of $k$, a Nahm pole appears in the
dual boundary condition, as $n-k$ D3-branes end on the single
D5-brane.  In these cases, the ungauging recipe leads to a bad
quiver.

We have accumulated by now enough examples to guess how to build the
dual of essentially any boundary conditions as long as the dual
gauge symmetry is unbroken, in terms of coupling of
$T^{\rho}(SU(n))$ to $\fB$. We will now give a general derivation of
this fact.

\subsection{A General Recipe}\label{recipe}

Let $\B$ be any half-BPS boundary condition in $U(n)$ gauge
theory, possibly but not necessarily constructed by coupling
Neumann boundary conditions to a boundary SCFT $\fB$.  Let us
further assume that the $S$-dual boundary condition $\B^\vee$ has
full gauge symmetry, and so is obtained by coupling Neumann
boundary conditions to a boundary theory $\fB^\vee$. We have
gained enough experience by now to formulate a general recipe for
construction of $\fB^\vee$.

\begin{figure}
  \begin{center}
    \includegraphics[width=4in]{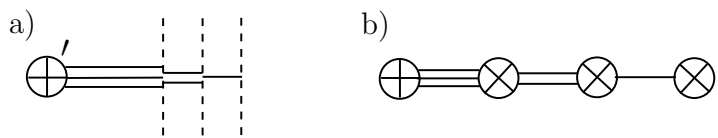}
  \end{center}
\caption{\small (a)  Let $\bigoplus$ symbolize an arbitrary given
boundary condition, with dual $\bigoplus'$, and assume that
$\bigoplus'$ is defined by coupling to an SCFT $\fB^\vee$.  Then
$\fB^\vee$ can be found by ungauging, that is by terminating the
four-dimensional gauge theory at $y=L$ using Dirichlet boundary
conditions.  This is achieved, for $G=U(n)$, by letting $n$
D3-branes end on $n$ D5-branes, as shown here for $n=3$. (b) The
mirror of $\fB^\vee$ is represented by the $S$-dual configuration.
Here $\bigoplus'$, which we do not know, is replaced by the given
boundary condition denoted $\bigoplus$, and the Dirichlet boundary
conditions are replaced by a coupling to $T(SU(n))$.}
  \label{Fig38}\end{figure}
Let us first explain what we regard as a satisfactory answer.
$\fB^\vee$ is supposed to be a three-dimensional conformal field
theory, and we are satisfied if we can give a purely
three-dimensional description of it.   Roughly speaking, if we can
do so, this means that we have reduced the understanding of
$S$-duality of boundary conditions in four-dimensional gauge
theory to a problem only involving the boundary.

We begin in a simple and by now familiar fashion (fig. \ref{Fig38}).
$\fB^\vee$, if it exists, is obtained from the $S$-dual boundary
condition $\B^\vee$ by terminating the gauge theory at $y=L$ with a
Dirichlet boundary condition and then flowing to the IR. The
$S$-dual of this is a gauge theory on a slab $\R^3\times I$ ($I$ is
the interval $0\leq y\leq L$) with the original boundary condition
$\B$ at one boundary, and a coupling to $T(SU(n))$, the dual of the
Dirichlet boundary condition, at the other.

This gives a description of $\tilde \fB^\vee$, the mirror of
$\fB^\vee$, in terms of four-dimensional gauge theory on $\R^3\times
I$, with specified boundary conditions at the two ends. To get a
satisfactory answer, we should reduce this to gauge theory on
$\R^3$.
 If the original boundary condition $\B$ has no Nahm pole, there is no problem in
doing this.  Suppose that $\B$ is defined by reducing the gauge
group $G=U(n)$ to some subgroup $H$ and then coupling to an SCFT
$\fB_H$ with $H$ symmetry. Then the low energy limit can be obtained
by simply restricting all four-dimensional fields to their zero
modes in the fourth direction -- the modes that are independent of
$y$. These modes make up the vector multiplet of $H$ gauge symmetry
in three dimensions.  So at low energies, the four-dimensional
configuration of fig. (\ref{Fig38}(b)) merely reduces to a
three-dimensional gauge theory with gauge group $H$ coupled to the
product $\fB_H\times T(SU(n))$. (In the figure, $H=G=U(3)$.)  We
denote the theory obtained by gauging the diagonal $H$ symmetry of
$\fB_H\times T(SU(n))$ as $\fB_H\times_H T(SU(n))$ and we call it
the composite gauge theory.  This gives the required
three-dimensional description of $\tilde\fB^\vee$.

\begin{figure}
  \begin{center}
    \includegraphics[width=5.5in]{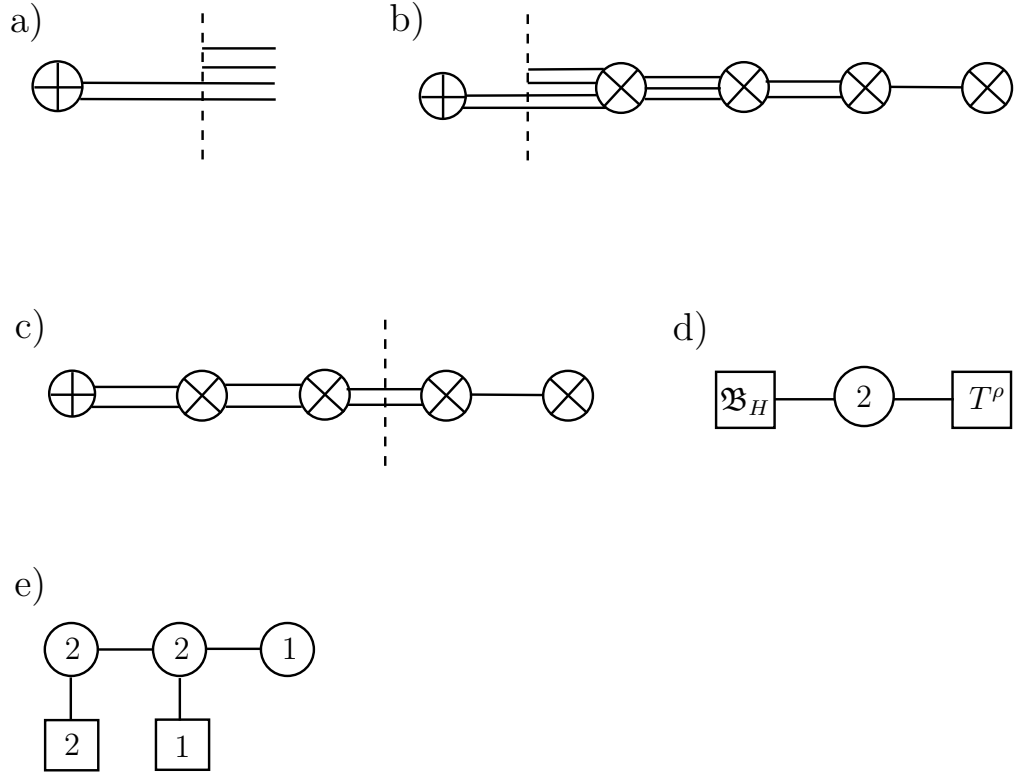}
  \end{center}
\caption{\small (a) A boundary condition  with a Nahm pole or
reduction of the gauge group. (b) A direct application of the
ungauging procedure leads to this result, with branes ordered
incorrectly. (c) A D5-brane has been moved to the right to restore
correct ordering. Now the IR limit is straightforward; for example,
if $\bigoplus$ is constructed from properly ordered fivebranes, then
this configuration corresponds to a good quiver. (d) The result can
be described as a generalized quiver, with  $H=U(2)$ coupled both to
$\fB_H$  and to another theory, which in the example given can be
represented by the quiver in (e) and is in fact our friend
$T^\rho(SU(n))$, where here $n=4$ and $\rho$ corresponds to the
decomposition $4=2+1+1$. $H$ acts at the leftmost node in (e).}
  \label{Fig39} \end{figure}

If the starting boundary condition $\B$ has a Nahm pole $\rho$
(fig. \ref{Fig39}), we begin as before and represent
$\tilde\fB^\vee$ in exactly the same way in terms of
four-dimensional gauge theory on $\R^3\times I$  (fig.
\ref{Fig39}(b)).  This gives a gauge theory in {\it four}
dimensions, with three non-compact dimensions, whose IR limit is
the desired three-dimensional theory $\tilde\fB^\vee$.  Before
declaring success, we are supposed to reduce this to a {\it
three}-dimensional description. The Nahm pole on the left of fig.
(\ref{Fig39}(b)) forces the field $\vec X$ to be $y$-dependent. So
we cannot extract a three-dimensional description by simply taking
all four-dimensional fields to be independent of $y$.

Instead, before trying to extract a low energy limit, we take the
D5-branes that create the Nahm pole in the original boundary
condition $\B$, as well as those that reduce the gauge
symmetry,\footnote{We recall that a D5-brane on which several
D3-branes end creates a Nahm pole, while one on which a single
D3-brane ends reduces the gauge symmetry. Both kinds of D5-brane
are depicted in fig. \ref{Fig1}, while for simplicity only a
D5-brane creating a Nahm pole is depicted in fig. \ref{Fig39}.}
and move them  to the right  (fig. \ref{Fig39}(c)), positioning
them in the usual way among the NS5-branes, so that no net
D3-branes end on any of these D5-branes.

At this stage, we have gauge fields of a subgroup $H$ of $G$,
coupled on the left to some boundary theory $\fB_H$ (represented in
fig. \ref{Fig39}(c) as $\bigoplus$) and on the right to a system of
D5-branes and NS5-branes that are shown explicitly in the figure.
Collapsing all the separations among these branes, we arrive at a
boundary SCFT that for the moment we call $\fB^\rho$. A
three-dimensional description of the desired SCFT $\tilde\fB^\vee$
is now at hand; it is what we will call $\fB_H\times_H \fB^\rho$,
the low energy limit of three-dimensional $H$ gauge theory coupled
to $\fB_H\times \fB^\rho$.

But in fact, $\fB^\rho$  is our friend $T^\rho(SU(n))$. The
D3-branes that produce Nahm poles depend only on $\rho$ and are
present in the description  of $T^\rho(SU(n))$ that treats the two
kinds of fivebrane symmetrically (see the example in fig.
\ref{Fig30}(b)). The process of moving them to the right to get a
well-ordered arrangement is the key step in getting a gauge theory
interpretation of $T^\rho(SU(n))$ (such as comes from fig.
\ref{Fig30}(a)).  If $H$ is trivial, so that the symbol $\bigoplus$
and the D3-branes ending on it are absent in fig. \ref{Fig39}(b)
(the leftmost D3-branes would then all end on D5-branes), then fig.
\ref{Fig39}(b) would simply be the definition of $T^\rho(SU(n))$ for
a particular $\rho$.
  In general, $\fB^\rho$ depends only on
$\rho$ and so coincides with $T^\rho(SU(n))$.  If $H$ is
non-trivial, this means that some of the global symmetries of
$T^\rho(SU(n))$ have been gauged and also coupled to another theory
$\fB_H$.  This produces what we call the composite gauge theory and
denote $\fB_H\times_H T^\rho(SU(n))$. So
\begin{equation}\label{honey}\tilde\fB^\vee=\fB_H\times_H
T^\rho(SU(n)).\end{equation} This is the single most important
conclusion of the present paper.

In this construction, the $SU(n)$ isometries of the Coulomb branch
of $T^{\rho}(SU(n))$  survive in the IR as symmetries of the full
theory $\tilde\fB^\vee$. This allows $\tilde\fB^\vee$ to be coupled
to the dual bulk gauge theory and produce the boundary condition
$\B^\vee$ dual to $\B$. The dual gauge group has to be coupled to
the Coulomb branch of $\tilde\fB^\vee$, rather than the Higgs
branch, as would be more standard, because $\tilde\fB^\vee$ is the
mirror of the theory $\fB^\vee$ that more directly defines the dual
boundary condition. As usual, to make the result useful in practice,
it is very helpful to have a representation of $\tilde\fB^\vee$ that
makes its Coulomb branch symmetries manifest.  This is what we
sometimes get from mirror symmetry.

\def\BB{\mathfrak B}

Our construction produces a $U(n)$-invariant SCFT $\fB^\vee$ that
defines the dual boundary condition whenever it exists, as the
mirror of a standard IR limit of the composite $H$ gauge theory.
However, it may be that the dual boundary condition $\B^\vee$
involves a reduction of the gauge symmetry, possibly with a Nahm
pole; if so, $\fB^\vee$ does not exist so the ungauging procedure
cannot construct it.  One would like a criterion for determining,
given $\B$, whether the dual boundary condition $\B^\vee$ has
reduced gauge symmetry.

In fact, this will occur precisely when the ``ungauging''
configuration of fig. \ref{Fig38}(a) admits three-dimensional
chiral operators of zero or negative $R$-charge $q_R$.  (Here as
in section \ref{quivers} we mean chiral operators for an $\N=2$
subalgebra of $\N=4$.) Dually, the composite $H$ gauge theory will
also have such bad chiral operators and will not have a standard
IR limit. The problematical chiral operator in fig. \ref{Fig38}(a)
will be constructed from Wilson line operators stretched between
the two boundaries.  So dually, the chiral operators of the
composite gauge theory are constructed from stretched 't Hooft
operators; they are the monopole operators reviewed in section
\ref{quivers}.

If the boundary condition $\B^\vee$ comes from coupling $U(n)$ gauge
fields to an SCFT $\fB^\vee$, then chiral operators of $\fB^\vee$
have strictly positive $q_R$, and coupling to four-dimensional gauge
fields on $\R^3\times I$ with Dirichlet boundary conditions on the
right does not change this fact.  The situation is different if
$\B^\vee$ involves a reduction of the gauge symmetry or a Nahm pole.
In this case, we will construct chiral operators of $q_R\leq 0$
using supersymmetric Wilson operators that end on the boundaries of
$\R^3\times I$.  As in section \ref{quivers}, we select a unit
vector $\vec n$ and form a linear combination of scalar fields
$X_3=\vec n\cdot \vec X$.  And we let $A_y$ be the component of the
gauge field $A$ in the $I$ direction.  Then the combination
$\A=A_y+iX_3$ is a chiral superfield for a suitable $\N=2$
subalgebra of $\N=4$. Let $S=p\times I$, with $p$ a point in $\R^3$.
Now we consider the path-ordered exponential
\begin{equation}\label{nowc}W(p)=P\exp\int_S\A.\end{equation}  Any gauge-invariant matrix
elements of $W(p)$ are chiral superfields with $q_R=0$. Whether
there are gauge-invariant matrix elements depends on the boundary
conditions.  Since $S$ stretches from $y=0$ to $y=L$, the pertinent
boundary conditions are at $y=L$ and $y=0$.  At $y=L$, we have
Dirichlet boundary conditions and gauge transformations are trivial.
If in addition the boundary condition $\B^\vee$ has reduced gauge
symmetry without a Nahm pole, then $W(p)$  has gauge-invariant
matrix elements in some representation of $G$.

If the dual boundary condition $\B^\vee$ involves a Nahm pole
associated with a homomorphism $\rho':\frak{su}(2)\to G$, then a
similar construction actually gives operators of $q_R<0$. In this
case, we have to be careful with the definition of $SO(3)_X$ and
of the Wilson operator $W(p)$.  The boundary condition forces
$\vec X$ to be nonzero near the boundary, and is not invariant
under naive rotation of $\vec X$. But it is invariant under  a
rotation of $\vec X$ combined with a gauge transformation
determined by $\rho'$.  So the $SO(3)_X$ symmetry of the boundary
condition is the combination of an ordinary rotation of $\vec X$
and a gauge transformation. As a result, the matrix elements of
$W(p)$ -- once we define them -- transform non-trivially under
$SO(3)_X$, and some of them are negatively charged under
$SO(2)_X$.  This will lead to $q_R<0$.

The Nahm pole leads to a subtlety in defining the operator $W(p)$.
By definition, the Nahm pole means that $\vec X\sim \vec t/y$, near
$y=0$, where $\vec t$ is the image under $\rho'$ of  a standard set
of $\frak{su}(2)$ generators.  So $\A\sim i t_3/y$, with $t_3=\vec
n\cdot\vec t$.  The pole in $\A$ causes a problem in defining
$W(p)$.  We regularize the resulting divergences by letting
$S_\epsilon$ be the restriction of $S$ to $y\geq \epsilon$ and
defining $W_\epsilon(p)=P\exp\int_{S_\epsilon}\A$.  Taking
$\epsilon\to 0$, we define the regularized Wilson
operator\footnote{Our gauge fields $A$ and scalar fields $\vec X$
are antihermitian, so $t_3$ is antihermitian and $it_3$ has real
eigenvalues.}
 \begin{equation}\label{renwilson}\hat W=\lim_{\epsilon\to
0}\epsilon^{-it_3}W_\epsilon(p).\end{equation}   Gauge-invariant
matrix elements of $\hat W$ are chiral operators of dimension and
$R$-charge $q_R=it_3$.

Hence we predict that the $S$-dual of a boundary condition $\B$
will break the dual gauge theory at the boundary if and only if
monopole operators of non-positive charges can be found in the
composite $H$ gauge theory.

Even if the dual boundary condition breaks the dual gauge theory at
the boundary, the D-brane realization of the system suggests that
the composite $H$ gauge theory still holds the information about the
dual boundary condition $\B^\vee$. One may start with the
``ungauging'' configuration, and move away in the $\vec Y$ direction
the D5-branes which define the Nahm pole.  The D3-segments attached
to those D5-branes will need to move away as well, and so will those
D5-branes at the Dirichlet boundary condition that are attached to
them. What is left is a four-dimensional gauge theory on $\R^3
\times I$ with a reduced gauge group $H^\vee$, with Dirichlet
boundary conditions at one end, and coupled to some boundary theory
$\fB^\vee_{H^\vee}$ at the other end. The system will flow smoothly
in the IR to $\fB_{H^\vee}^\vee$.

To carry out an $S$-dual of this process in the composite $H$
gauge theory, and thereby construct the mirror of
$\fB_{H^\vee}^\vee$, we will need to identify the FI parameters in
the composite gauge theory which correspond to the motion of the
D5-branes, and make them large. It is difficult to give a general
prescription on how to do that without knowing anything about
$\B$. The monopole operators with negative $R$-charges dual to the
regularized Wilson line operators (\ref{renwilson}) will transform
non-trivially under the Coulomb branch isometries which correspond
to the appropriate FI parameters. One might be able to collect
further information on $\B^\vee$ by exploring modifications of the
Dirichlet boundary condition: adding a Nahm pole, coupling some
judiciously chosen SCFT, etc. We will not pursue this matter
further.

\subsection{Domain Walls}\label{domain}

\def\D{\mathcal D}
\def\fD{\mathfrak D}
In general, domain walls between gauge theories with one gauge group
$G_1$ on one side and another gauge group $G_2$ on the other side
are equivalent, after a folding trick, to boundary conditions in
$G_1\times G_2$ gauge theory.  (For more on this, see section 2.6 of
\cite{Gaiotto:2008sa}.)  Hence all methods for studying boundary
conditions can be adapted to domain walls.

\begin{figure}
  \begin{center}
    \includegraphics[width=4.5in]{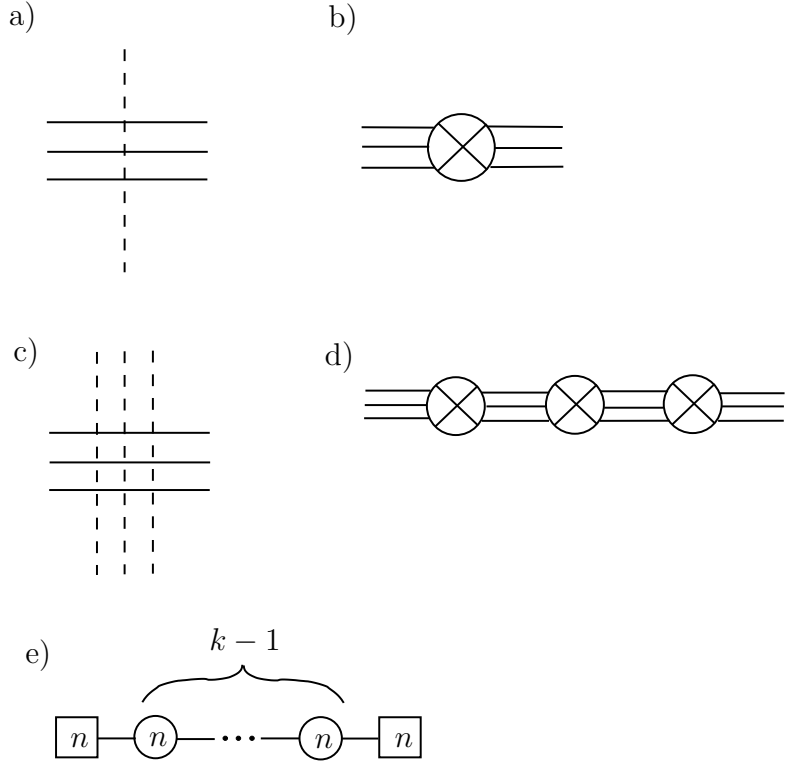}
  \end{center}
\caption{\small (a) A simple supersymmetric domain wall in $U(n)$
gauge theory (drawn for $n=3$); intersection with a D5-brane
creates a fundamental hypermultiplet supported at $y=0$.  (b) The
$S$-dual domain wall is constructed from the intersection with a
single NS5-brane.  (c) Intersection with $k$ D5-branes creates a
defect supporting $k$ fundamental hypermultiplets (sketched here
for $k=3$). (d) The $S$-dual of (c). (e) For $k>1$, the $S$-dual
involves coupling to a non-trivial SCFT, which can be represented
by a quiver.  The quiver has been drawn for the general case of
$n$ D3-branes intersecting $k$ D5-branes.}
  \label{Fig40}\end{figure}

If $G_+=U(n_+)$ and $G_-=U(n_-)$ are unitary gauge groups realized
by D3-branes for $y>0$ and $y<0$, and the domain walls are
constructed via fivebranes, we can study $S$-duality of a domain
wall by the usual manipulations.  It is also possible to apply an
ungauging procedure similar to what we have just used in discussing
boundary conditions.

Consider a gauge theory with a domain wall $\D$ at $y=0$.  The
domain wall is constructed in general by a two-sided version of the
construction used for boundary conditions.  $G_+$ and $G_-$ are
broken to subgroups $H_\pm$  by two Nahm poles ($\vec X\sim \vec
t_\pm/y$ for $y\to 0^\pm$).  The product $H_+\times H_-$ is then
further broken at $y=0$ to a subgroup $H$, which is then coupled to
a three-dimensional defect theory $\fD$.

As long as we consider only domain walls constructed from
fivebranes, there is no problem determining the $S$-duals. A few
examples are given in fig. \ref{Fig40}. In (a) we consider a
single D5-brane crossing $n$ D3-branes.  There is then a
fundamental hypermultiplet supported on the defect or domain wall
at $y=0$ and coupled to the bulk $U(n)$ gauge fields. In the
language of the last paragraph, $G_+=G_-=U(n)$, there are no Nahm
poles, $H$ is a diagonal subgroup of $G_+\times G_-$, and $\fD$
describes a free fundamental hypermultiplet.  The $S$-dual in (b)
clearly corresponds to interaction with a single NS5-brane. In
this case, there are distinct $U(n)$ gauge theories on the two
half-lines. There is no reduction of gauge symmetry at the
interface and the full $U(n)\times U(n)$ couples to a
bifundamental hypermultiplet. We can generalize this example to
have different numbers $n_\pm$ of D3-branes on the two sides.  In
(a), there would  then be a Nahm pole of rank $|n_+-n_-|$ on the
side with more D3-branes, and in (b), there would be a
bifundamental hypermultiplet coupled to $U(n_+)\times U(n_-)$.

Returning to the case $n_+=n_-$, in (c), we generalize (a) to a
defect made from $k$ D5-branes and so supporting $k$ fundamental
hypermultiplets. The $S$-dual involves a chain of $k$ NS5-branes
leading to the balanced quiver gauge theory in (d) and (e).  As
usual, the chain of balanced nodes in the quiver leads to a global
symmetry of the Coulomb branch, matching the global symmetry of
the hypermultiplets of (c).

\begin{figure}
  \begin{center}
    \includegraphics[width=4in]{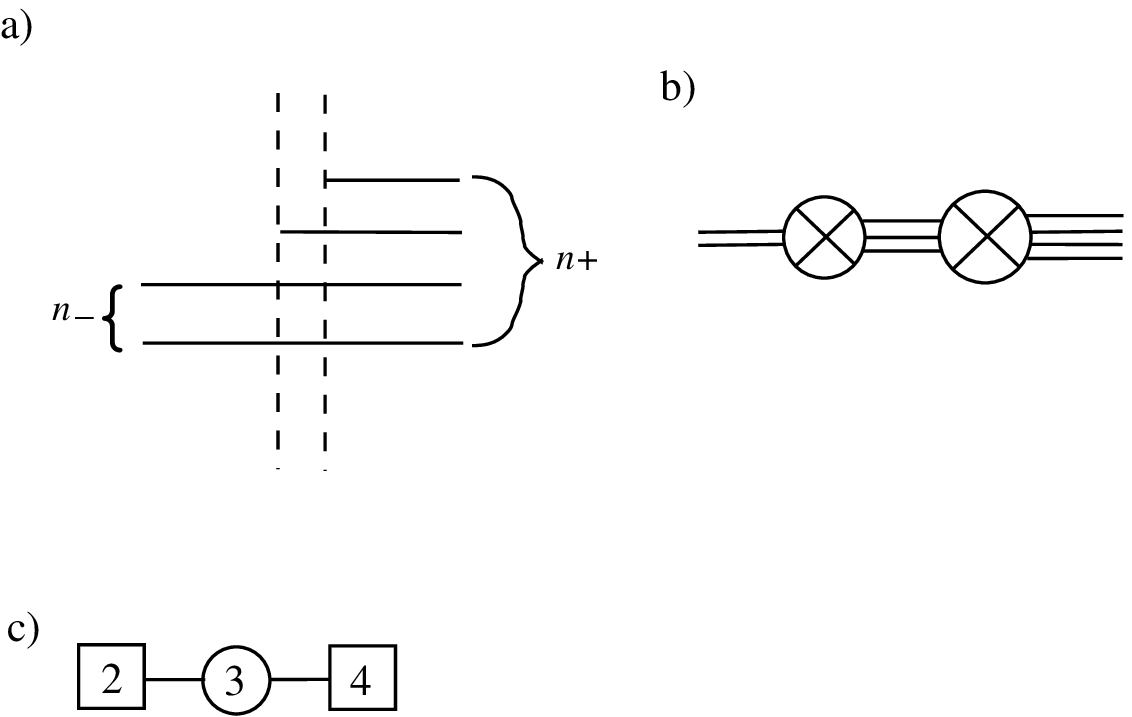}
  \end{center}
\caption{\small (a) $U(n_+)$ gauge theory for $y>0$, reduced to
$U(n_-)$ for $y<0$, by letting D3-branes end one by one on
$n_+-n_-$ D5-branes.  Sketched here is the case $n_+=4$, $n_-=2$.
(b) The $S$-dual, which as in  (c)  can be represented by a
balanced quiver gauge theory. The boxes represent global
symmetries that can be coupled to four-dimensional gauge fields in
the right or left half-spaces.}
  \label{Fig41}\end{figure}

For a slightly different example,  in fig. \ref{Fig41}(a), the
gauge group $U(n_+)$ is reduced to $U(n_-)$ in crossing a domain
wall by interaction with $n_+-n_-$ D5-branes; one D3-brane ends on
each D5-brane.  The $S$-dual is an analogous picture with
NS5-branes (fig. \ref{Fig41}(b)), which can be represented by a
balanced quiver gauge theory with $U(n_+)\times U(n_-)$ symmetry,
as in (c).  The examples we have given are particularly simple
because no brane rearrangement is required, but in general,
starting with any domain wall with fivebranes arranged to satisfy
our rules, the $S$-dual can be rearranged in the usual way to also
satisfy them, giving another domain wall with a simple gauge
theory description.

\begin{figure}
  \begin{center}
    \includegraphics[width=4.5in]{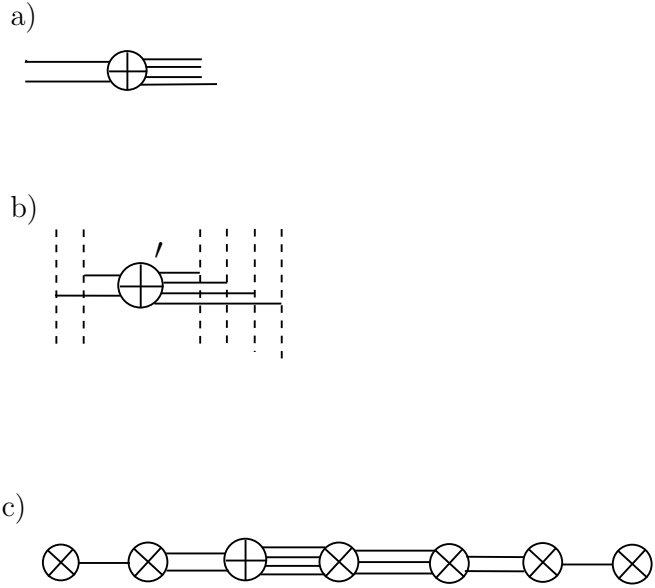}
  \end{center}
\caption{\small (a) The symbol $\bigoplus$ represents a general
domain wall between $U(n_-)$ and $U(n_+)$ gauge theory, shown here
for $n_-=2$, $n_+=4$. (b) The symbol $\bigoplus'$ represents the
dual domain wall.  If it arises by coupling to an SCFT $\frak D$
with $U(n_-)\times U(n_+)$ symmetry, then this SCFT can be
extracted by using Dirichlet boundary conditions on both sides to
ungauge the gauge symmetry.  As usual, Dirichlet boundary
conditions are constructed using a chain of D5-branes. (c) Taking
the $S$-dual, we get a quiver representation of the mirror $\frak
D^\vee$ to $\frak D$, in terms of the original domain wall coupled
by $U(n_+)\times U(n_-)$ gauge fields to  $T(SU(n_+))\times
T(SU(n_-))$, which are here represented by quivers.}
  \label{Fig42}\end{figure}
It is also possible to develop a general recipe using $T(SU(n))$.
Let $\D$ be a domain wall and $\D^\vee$ its $S$-dual. Suppose that
$\D^\vee$ arises by coupling  to four-dimensional gauge fields a
three-dimensional theory $\fD^\vee$ with $G_+\times G_-$ symmetry.
We can recover the theory $\fD^\vee$ by terminating the
four-dimensional gauge theory with Dirichlet boundary conditions at
$y=\pm L$.  The $S$-dual of this gives the mirror of the theory
$\fD^\vee$ as the infrared limit of a composite configuration
described in fig. \ref{Fig42}.  In this configuration, boundary
conditions at the two ends are provided by coupling to $T(SU(n_+))$
and $T(SU(n_-))$, and the original domain wall $\D$ appears in the
center.  The low energy limit, assuming that there are no Nahm poles
in the definition of $\D$, is a three-dimensional gauge theory with
gauge group $H$ coupled to a product $T(SU(n_+))\times
T(SU(n_-))\times \fD$.  If there are Nahm poles, then just as in
section \ref{recipe}, the low energy theory is a three-dimensional
gauge theory with $H$ coupled to $T^{\rho_+}(SU(n_+))\times
T^{\rho_-}(SU(n_-))\times \fD$.  If this three-dimensional gauge
theory has a standard IR limit, that limit will be the mirror of the
desired theory $\fD^\vee$.

\begin{figure}
  \begin{center}
    \includegraphics[width=4.5in]{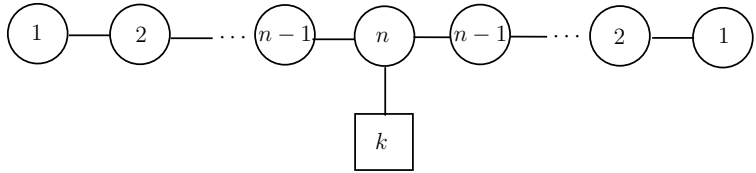}
  \end{center}
\caption{\small The result of applying the general recipe to the
domain wall in $U(n)$ gauge theory that consists of a coupling to
$k$ fundamental hypermultiplets (as sketched in fig.
\ref{Fig40}(c)). This quiver is mirror to the one in fig.
\ref{Fig40}(e).}
  \label{Fig43}\end{figure}
A simple example is to use this procedure to study the configuration
of fig. \ref{Fig40}(a) or (c) with a defect supporting $k$
fundamental hypermultiplets.  The composite gauge theory is a $U(n)$
gauge theory coupled to two copies of $T(SU(n))$ and to $k$
fundamental hypermultiplets.  This theory is described by the quiver
of fig. \ref{Fig43}.  For $k>1$, this is a good quiver, whose mirror
is the quiver that we found in fig. \ref{Fig40}(e) by direct brane
manipulations.

For $k=1$, the quiver of fig. \ref{Fig43} is an ugly quiver with a
single minimally unbalanced node in the center, and chains of
$n-1$ balanced nodes on each side. As explained in section
\ref{quivertheories}, this will generate in the infrared an
$SU(n)\times SU(n)\times U(1)$ symmetry acting on a free
bifundamental hypermultiplet.  For this particular example, the
dimension of the Coulomb branch is $n^2$, the same as the number
of free hypermultiplets, so the infrared limit of the quiver
theory describes the free hypermultiplets only, with no additional
degrees of freedom.  This is the expected answer of fig.
\ref{Fig40}(b).

For $k=0$, which describes an empty or trivial domain wall, we get a
bad quiver.  This is in accord with the fact that the $S$-dual,
which is also the trivial domain wall, does not have the full
$U(n)\times U(n)$ gauge symmetry on the two sides, but only the
diagonal $U(n)$.  We will return to this example in section
\ref{genpres}, but for now we simply  note that the hyper-Kahler
dimension of the Coulomb branch is $n^2$, which is what one needs if
the Coulomb branch is to break $U(n)\times U(n)$ to the diagonal
$U(n)$.

\begin{figure}
  \begin{center}
    \includegraphics[width=5in]{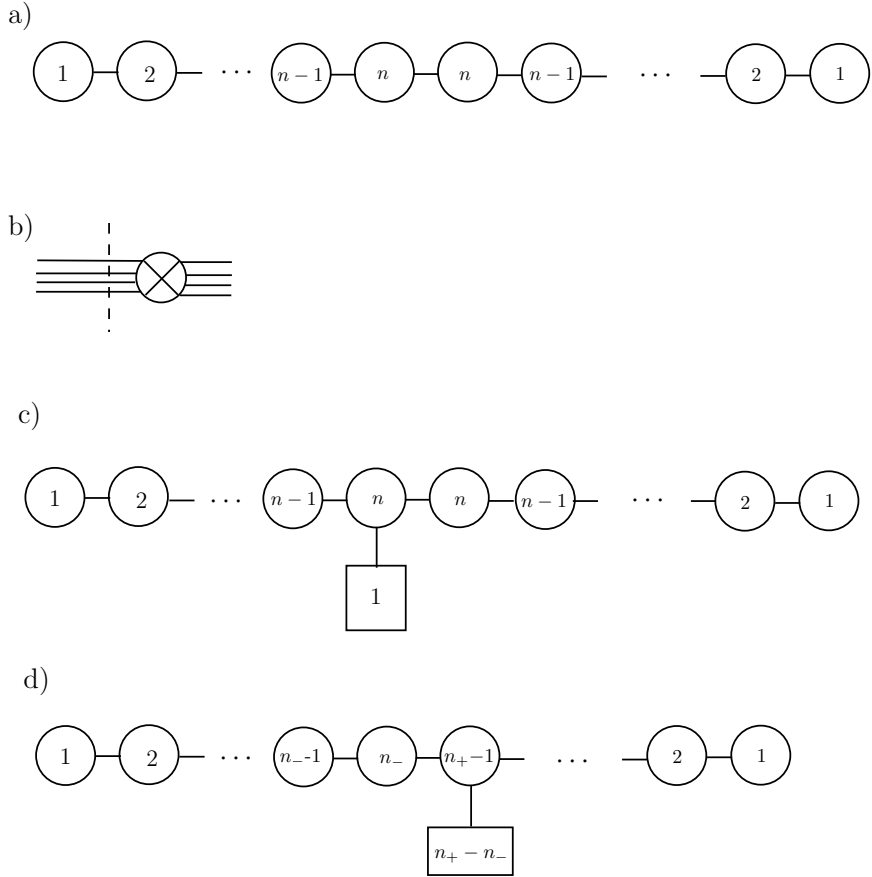}
  \end{center}
\caption{\small (a)  The ungauging recipe, applied to the problem of
generating an $S$-dual of the domain wall of fig. \ref{Fig40}(b),
produces a composite $U(n)\times U(n)$ gauge theory coupled to two
copies of $T(SU(n))$ and a bifundamental hypermultiplet. This is a
bad quiver with two adjacent minimally unbalanced nodes, so the
infrared analysis is not straightforward. (b) A domain wall
constructed from a D5-brane and an NS5-brane (as opposed to the
single NS5-brane of fig. \ref{Fig40}(b)). $S$-duality merely
exchanges the two. (c) The associated quiver gauge theory, which
differs from that in (a) by adding a fundamental hypermultiplet at
one node.  This is an ugly quiver with a single minimally unbalanced
node, reflecting the fact that the mirror domain wall (which is that
of (b) with the two fivebranes exchanged) has full $U(n)\times U(n)$
gauge symmetry with  free fundamental and bifundamental
hypermultiplets. (d) The ungauging procedure applied to the domain
wall of fig. \ref{Fig41}(a) leads to this quiver, after a
rearrangement of branes. }
  \label{Fig44}\end{figure}
Similarly, we can apply this recipe to seek a dual of fig.
\ref{Fig40}(b), where the domain wall, constructed from a single
NS5-brane, has the full $U(n)\times U(n)$ symmetry coupled to a
single bifundamental hypermultiplet. The associated composite gauge
theory is a $U(n)\times U(n)$ gauge theory coupled to two copies of
$T(SU(n))$ and a bifundamental hypermultiplet, described by the
quiver of fig. \ref{Fig44}(a). This is a bad quiver with two
adjacent minimally unbalanced nodes, in keeping with the fact that
the dual domain wall of fig. \ref{Fig40}(a) has reduced gauge
symmetry.

There is an amusing modification of this problem for the domain
wall in fig. \ref{Fig44}(b), which is constructed from a D5-brane
to the left of an NS5-brane. The $S$-dual configuration is a
mirror image, with a single D5-brane to the right of an NS5-brane.
The composite gauge theory is described by the quiver in fig.
\ref{Fig44}(c), which has a single minimally unbalanced node with
$n$ balanced nodes to the left, $n-1$ to the right. Again the
dimension of the Coulomb branch, $n(n+1)$, agrees with the number
of free hypermultiplets realized as monopole operators. These are
the expected bifundamental and fundamental free hypermultiplets at
the dual boundary.

Finally we want to apply our construction to fig. \ref{Fig41}(a).
The resulting composite gauge theory is described by the quiver
depicted in fig. \ref{Fig44}(d), which is good as long as $n_+ -
n_->1$. The quiver is mirror to the expected result in fig.
\ref{Fig41}(b). If $n_+ - n_-=1$, the quiver is ugly (and equivalent
to that of fig. \ref{Fig44}(c)); its Coulomb branch flows in the
infrared to the expected $n \times (n+1)$ bifundamental free twisted
hypermultiplets.

To give an example of analyzing in detail the IR behavior of a bad
quiver, we will consider the
 quiver in fig. \ref{Fig43} at $k=0$.  This quiver gauge theory is
supposed to be $S$-dual to a trivial domain wall.  The trivial
domain wall breaks the product of the $U(n)$ gauge groups on the
left and  right to a  diagonal $U(n)$. So the Coulomb branch of
the bad quiver in question should break $U(n)\times U(n)$ to the
diagonal subgroup. The easiest way to find the Coulomb branch
$\mathcal C$ of the quiver is to use the $S$-dual description,
which is simply a trivial domain wall cut off at both ends. In
other words, the $S$-dual description is by  four-dimensional
$\N=4$ gauge theory on a slab $\R^3\times I$ with Dirichlet
boundary conditions on both boundaries.
  The Coulomb branch $\C$ is simply the moduli space of
solutions of Nahm's equations on the interval $I$. Here Nahm's
equations are the equations $D\vec X/Dy+\vec X\times \vec X=0$,
for fields $\vec X$ and $A=A_y$; $\mathcal C$ is the moduli space
of solutions of these equations, modulo gauge transformations that
equal 1 at both ends of $I$. This is an important hyper-Kahler
manifold described in \cite{KronheimerOld} and reviewed in
\cite{Gaiotto:2008sa}, section 3.9.1. In the analysis of this
manifold, the quantity $W(p)$ defined in eqn. (\ref{nowc}) plays
an important role.  It is holomorphic on $\C$ in one of the
complex structures; in physical terms, its expectation value
$\langle W(p)\rangle$ is a $GL(n,\Bbb{C})$-valued  function on the
moduli space $\C$ of vacua that because of the chiral nature of
the operator $W(p)$ is holomorphic in one complex structure. From
a holomorphic point of view, this expectation value breaks the
product of the left and right action of $GL(n,\Bbb{C})$ to a
diagonal subgroup. In gauge theory terms, the symmetry is
$U(n)\times U(n)$ broken to a diagonal $U(n)$ (or a subgroup
thereof, depending on the choice of a point in $\C$). $\C$ is
smooth, so the quiver gauge theory certainly has no interesting
critical point.

\section{$S$-Duality And Janus}\label{general}

An important ingredient in the last section was a self-mirror
conformal field theory $T(SU(n))$, which we found most directly in
studying the $S$-dual of Dirichlet boundary conditions.  We used
this, and its generalization $T^\rho(SU(n))$, to describe the
$S$-dual of a very wide class of boundary conditions.

In this section, we will extend our results to any compact gauge
group $G$. We will give an intrinsic definition of a
three-dimensional conformal field theory $T(G)$ analogous to
$T(SU(n))$ and use it, and a generalization $T^\rho(G)$, to
formulate a recipe analogous to that of section \ref{examples}.  As
in section \ref{examples}, we will also consider a further
generalization $T_{\rho^\vee}^\rho(G)$.

\subsection{$T(G)$}\label{tg}

One of the important properties of $T(G)$ will be that it has
global symmetry $G\times G^\vee$,  where $G^\vee$ is the dual
group to $G$.  (The groups acting faithfully are the adjoint forms
of $G$ and $G^\vee$, so the distinction between them is only
important when they have different Lie algebras.  That is why we
did not encounter this distinction in section \ref{about}.)  The
mirror of $T(G)$ is $T(G^\vee)$.
 $T(G)$ will
appear as the dual of Dirichlet boundary conditions in $G$ gauge
theory.

$G$ acts on the Higgs branch of $T(G)$, and $G^\vee$ acts on its
Coulomb branch. The Higgs and Coulomb branches of $T(G)$, in any
of their complex structures, are the nilpotent cones $\N$ and
$\N^\vee$ of $G$ and $G^\vee$, respectively. As reviewed more
fully in section 3 of \cite{Gaiotto:2008sa}, $\N$ is the space of
all nilpotent elements of the Lie algebra $\frak g_\Bbb{C}$ of the
complexification $G_\Bbb{C}$ of $G$.  It is a union of finitely
many nilpotent $G_\Bbb{C}$ orbits.  Each such orbit is the orbit
of a nilpotent element $\rho_+\in \frak g_\Bbb{C}$, which is the
image of the raising operator  of $\frak{su}(2)$ under some
homomorphism $\rho:\frak{su}(2)\to\frak g_\Bbb{C}$.  $\N$ is
actually the closure of a single nilpotent orbit $\O$ associated
to a regular $\frak{su}(2)$ subalgebra.  The other orbits are of
positive codimension.

The moduli space $\M_{T(G)}$ of vacua of $T(G)$ does not just
consist of Higgs and Coulomb branches, as there are also mixed
Higgs-Coulomb branches.  The full structure of the moduli space is
a union of components
\begin{equation}\label{delm} \M_{T(G)}=\bigcup_{\alpha\in S} \C_\alpha
\times \H_\alpha,\end{equation} where $S$ is the set of components
and we call $\C_\alpha$ and $\H_\alpha$ the Coulomb and Higgs
factors of the $\alpha^{th}$ component.  $G$ and $SO(3)_X$ act
nontrivially on $\H_\alpha$ and trivially on $\C_\alpha$, and
reciprocally $G^\vee$ and $SO(3)_Y$ act non-trivially on
$\C_\alpha$ and trivially on $\H_\alpha$.  The Higgs branch  is a
component with $\C_\alpha$ equal to a point and $\H_\alpha$ equal
to $\N$, and the Coulomb branch has $\H_\alpha$ equal to a point
and $\C_\alpha=\N^\vee$. We write simply $\H$ and $\C$ for the
Higgs and Coulomb branches.

The reason that mixed branches exist is that, by adjusting
parameters  on $\C$, one can  go to a locus at which a Higgs
branch opens up.  Let $\C_\alpha$ be an irreducible component of
the locus in $\C$ at which this happens, and let $\H_\alpha$ be
the corresponding Higgs branch. Then the moduli space of vacua
contains a component $\C_\alpha\times \H_\alpha$. $\C_\alpha$ is a
$G^\vee$-invariant  hyper-Kahler subspace of $\C$. These
properties imply that (as a complex manifold in one of its complex
structures) $\C_\alpha$ is a union of nilpotent orbits of
$G^\vee_\BC$. Since we have assumed $\C_\alpha$ to be irreducible,
it is actually the closure of a single such orbit, associated with
some homomorphism\footnote{The $\rho_\alpha$ of distinct $\alpha$
are not necessarily inequivalent.}
$\rho_\alpha:\frak{su}(2)\to\frak g^\vee_\BC$. Applying the same
argument starting on $\H$, we learn that each $\H_\alpha$ is
similarly the closure of a nilpotent orbit of $G_\BC$.

Since each $\H_\alpha$ or $\C_\alpha$ is the closure of a
nilpotent orbit $\O_{\rho_\alpha}$ or $\O_{\rho^\vee_\alpha}$, the
general form of the moduli space is
\begin{equation}\label{genform}\M_{T(G)}=\bigcup_{\alpha\in S
}\overline\O_{\rho_\alpha}\times\overline\O_{\rho^\vee_\alpha}.
\end{equation}
 The union in
(\ref{genform}) is definitely not a disjoint union, as the various
components meet on subspaces.   All this can be made more explicit
for $G=SU(n)$, using the representation of $T(SU(n))$  as a
quiver.  In that case, every $\rho:\frak{su}(2)\to \frak{su}(n)$
appears in the sum exactly once, and is paired with its dual
$\rho_D$.

\subsubsection{Janus And The $S$-Dual Of Dirichlet}\label{jansd}

The strategy that we will follow to construct $T(G)$ in general is
as follows.  We start with a half-BPS domain wall with $G$ gauge
theory for $y<0$ and $G^\vee$ gauge theory for $y>0$. There are many
such domain walls, but there is a minimal one that we call the Janus
domain wall.

\begin{figure}
  \begin{center}
    \includegraphics[width=4.5in]{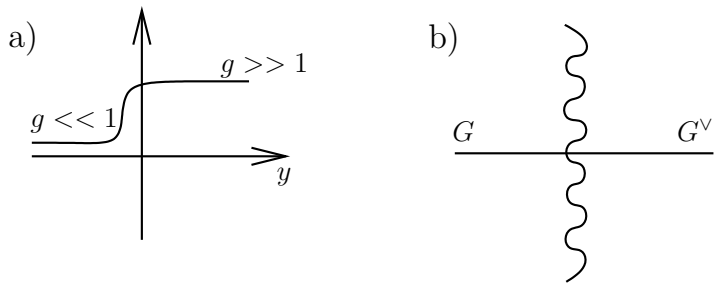}
  \end{center}
\caption{\small (a) A half-BPS configuration in $G$ gauge theory in
which the coupling $g(y)$ depends on $y$; it is small for $y<0$ and
large for $y>0$. (b) Applying $S$-duality for $y>0$, we get an
equivalent configuration described by weakly coupled $G$ gauge
theory on the left and $G^\vee$ gauge theory on the right, coupled
to a superconformal field theory $T(G)$ with $G\times G^\vee$ global
symmetry. This SCFT is schematically denoted by the vertical wiggly
line.}
  \label{Fig45}\end{figure}

The general half-BPS Janus configuration \cite{BGH,CFKS,CK,DEG,DEG2}
is a configuration in which the gauge coupling $g$ is a general
function $g(y)$.  No additional degrees of freedom are added; one
just considers $\N=4$ super Yang-Mills with $y$-dependent coupling.
This configuration admits a smooth limit to a domain wall -- a
configuration in which $g(y)$ is constant for $y<0$ and for $y>0$,
with a jump at $y=0$. This limiting configuration is half-BPS and
superconformally invariant, and we call it the Janus domain wall.

\begin{figure}
  \begin{center}
    \includegraphics[width=3.5in]{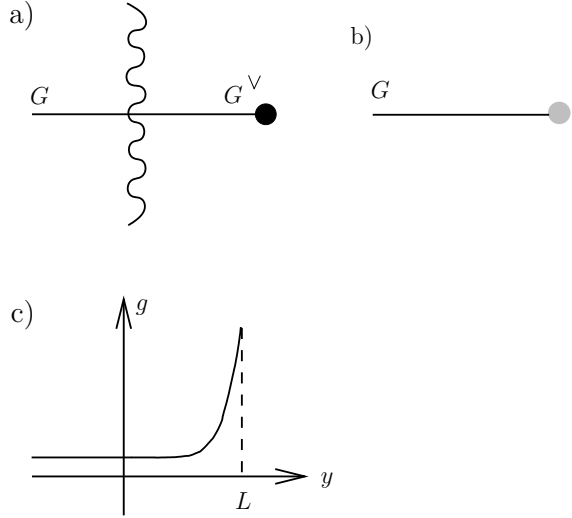}
  \end{center}
\caption{\small (a) The $G^\vee$ symmetry of fig. \ref{Fig45}(b) can
be ungauged -- and converted to a global symmetry -- by terminating
the figure on the right with Dirichlet boundary conditions, which
are schematically indicated by the black dot. (b) Applying
$S$-duality on the right of the figure, we get $G$ gauge theory
coupled to the $S$-dual of Dirichlet boundary conditions --
schematically indicated here with the shaded dot -- and with a
non-trivial coupling function $g(y)$.  (c) The profile of $g(y)$
after a deformation. The coupling is weak except very near the
boundary at $y=L$. The IR limit is obtained by shrinking away the
strongly coupled region, leaving weakly coupled $G$ gauge theory on
a half-space with boundary conditions that are the $S$-dual in $G$
gauge theory of Dirichlet boundary conditions in $G^\vee$ gauge
theory. So the $S$-dual of Dirichlet is coupling to the SCFT
$T(G)$.}
  \label{Fig46}\end{figure}

We consider a Janus domain wall with a coupling $g(y)$ that is very
small for $y<0$ and very large for $y>0$ (fig. \ref{Fig45}(a)).
Making $S$-duality in the region $y>0$, we reduce to a configuration
with weak coupling on both sides.  However, the gauge group is $G$
to the left and $G^\vee$ to the right.  In the limit that the
coupling is extremely small on both sides, we are left with some
sort of superconformal field theory weakly coupled to $G$ gauge
fields in one half space and to $G^\vee$ gauge fields in the other.
We call this superconformal field theory $T(G)$.

We can investigate $T(G)$ by ungauging the gauge fields on either or
both sides. For instance, let us introduce Dirichlet boundary
conditions for $G^\vee$ at $y=L$. (See fig. \ref{Fig46}.)  After
making $S$-duality in the region $y>0$, we have a Janus
configuration (with $y$-dependent coupling $g(y)$) that terminates
at $y=L$ with the $S$-dual of Dirichlet boundary conditions. If we
regularize the Janus domain wall by choosing a smooth function
$g(y)$ and we flow to the IR, the non-trivial profile of $g$ will
just flow away, and we will be left with the dual of Dirichlet
boundary conditions, and a constant, small $g$ everywhere. This
shows that $T(G)$ is the boundary SCFT which defines the $S$-dual of
a Dirichlet boundary condition for $G^\vee$.

Now consider $G$ gauge theory on a half-space coupled to $T(G)$ on
the boundary.  The gauging kills the Higgs branch of $T(G)$, but
leaves the Coulomb branch, which we would like to identify.  The
easiest way to do this is to apply $S$-duality, which converts the
gauge group into $G^\vee$ and turns the boundary coupling to $T(G)$
into Dirichlet boundary conditions.  So we are simply left with
$G^\vee$ gauge theory on a half-space with Dirichlet boundary
conditions.  The moduli space of vacua of this theory (with fields
vanishing at infinity) can be found by solving Nahm's equations and
is equal to the nilpotent cone $\N^\vee$ of $G^\vee$, as explained
in section 3 of \cite{Gaiotto:2008sa}.  This therefore is also the
Coulomb branch of $T(G)$.

If we introduce Dirichlet boundary conditions for both gauge groups,
at $y=L$ and $y=-L$ respectively, this will have the effect of
ungauging both gauge groups. Moreover we will be left with $G$
global symmetry acting at $y=-L$ and $G^\vee$ at $y=L$. In the
infrared, we will recover $T(G)$.

If we exchange the two ends (and $\vec X$ and $\vec Y$, as well), we
see the same configuration with $G$ and $G^\vee$ exchanged.  So
$T(G)$ and $T(G^\vee)$ are a pair of mirror SCFTs.  In particular,
the mirror symmetry implies that since $\N^\vee$ is the Coulomb
branch of $T(G)$, its Higgs branch is $\N$, the nilpotent cone of
$G$.

\subsection{Including The Nahm Pole}\label{incnahm}

The ungauging procedure can be generalized to include Nahm poles at
the two ends $y=-L$ and $y=L$, associated respectively with
homomorphisms $\rho:\frak{su}(2)\to\frak{g}$ and
$\rho^\vee:\frak{su}(2)\to\frak g^\vee$.  If this configuration has
a standard IR limit, we denote the resulting SCFT as
$T_{\rho^\vee}^\rho(G)$.  Its mirror, arrived at by exchanging the
two ends of the picture, is $T_\rho^{\rho^\vee}(G^\vee)$.

We would like to determine the Higgs and Coulomb branches of these
theories (the results are summarized in  Table \ref{firstable}).
In doing this, it is convenient to start with the case that $\rho$
or $\rho^\vee$ is trivial.  In any event, this will be the most
important case in the present paper.

\begin{table}
\caption{The Higgs and Coulomb branches $\H$ and $\C$ of the
conformal field theory $T(G)$ and its generalizations.  Here
$\S_\rho$ denotes the Slodowy slice transverse to the raising
operator $\rho_+$ of $\rho$, $\O_\rho$ is the orbit of $\rho_+$,
and $\overline\O_\rho$ is the closure of this orbit. Finally,
$C_{\rho^\vee}$ is the set of all $\alpha$ such that
$\rho_\alpha^\vee=\rho^\vee$, and $C_\rho$ is the set of $\alpha$
such that $\rho_\alpha=\rho$.}
\begin{center}
\begin{tabular}{c|c|c}
&$\H$&$\C$ \\\hline
 $T(G)$&$\N$&$\N^\vee$\\
$T_{\rho^\vee}(G)$&$~~\cup_{\alpha\in
C_{\rho^\vee}}\overline\O_{\rho_\alpha} ~~$& ${\cal
S}_{\rho^\vee}\cap\N^\vee$\\ $T^\rho(G)$&${\cal S}_\rho\cap\N$&
$~~\cup_{\alpha\in C_{\rho}}\overline\O_{\rho^\vee_\alpha}
~~$\\
 $T^{\rho}_{\rho^\vee}(G)$& $~~{\cal S}_\rho\cap\bigl(\cup_{\alpha\in
C_{\rho^\vee}}\overline\O_{\rho_\alpha}\bigr) ~~$  &$~~{\cal
S}_{\rho^\vee}\cap\bigl(\cup_{\alpha\in
C_\rho}\overline\O_{\rho^\vee_\alpha}\bigr) ~~$ \\
\end{tabular}
\end{center}\label{firstable}
\end{table}

\begin{figure}
  \begin{center}
    \includegraphics[width=3.5in]{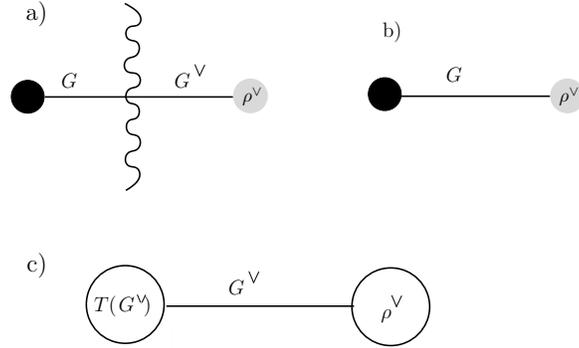}
  \end{center}
\caption{\small (a) $G$ and $G^\vee$ gauge theories joined by a
Janus domain wall (wiggly line).  On the left, we take Dirichlet
boundary conditions for $G$ (symbolized by the black dot), and on
the right, Dirichlet boundary conditions for $G^\vee$ modified by
$\rho^\vee$ (gray dot labeled $\rho^\vee$). (b) Moving Janus to
the right gives a $G$ gauge theory with Dirichlet boundary
conditions on the left and $T_{\rho^\vee}(G)$, the $S$-dual of
$\rho^\vee$ (that is, the $S$-dual of Dirichlet modified by
$\rho^\vee$) on the right. (c) Moving Janus to the left gives a
$G^\vee $ gauge theory, coupled on the left to $T(G^\vee)$ (the
$S$-dual of Dirichlet for $G$) and to $\rho^\vee$ on the right.}
  \label{Fig47}\end{figure}
Consider first $T_{\rho^\vee}(G)$. (The corresponding analysis of
$T^\rho(G)$ is made by simply exchanging $G$ and $G^\vee$ and
using mirror symmetry.)  By definition, it is obtained from a
configuration with Dirichlet boundary conditions of $G$ at the
left of an interval, Janus in the center of the interval, and
Dirichlet boundary conditions modified by $\rho^\vee$ at the right
of the interval (fig. \ref{Fig47}(a)). If we move Janus to the
right (fig. \ref{Fig47}(b)), we get a configuration with gauge
group $G$, the $S$-dual of $\rho^\vee$ on the right, and Dirichlet
boundary conditions on the left.  As usual, Dirichlet boundary
conditions just ungauge $G$, leaving it as a global symmetry.  The
 SCFT represented by fig. \ref{Fig47}(b) is therefore the $S$-dual
of $\rho^\vee$, and this then is $T_{\rho^\vee}(G)$.

On the other hand, to describe the moduli space of vacua of
$T_{\rho^\vee}(G)$, it is more convenient to move Janus to the
left (fig. \ref{Fig45}(c)).  Then we get a $G^\vee$ gauge theory,
coupled on the left to $T(G^\vee)$, and with boundary conditions
set by $\rho^\vee$ on the right.  The Coulomb branch of vacua must
be found by solving the $G^\vee$ Nahm equations $D\vec X + \vec
X\times \vec X=0$ with suitable conditions at the endpoints $y=\pm
L$:

(1)  At $y=L$, $\vec X$ must have a Nahm pole of type $\rho^\vee$,
that is $\vec X\sim \rho^\vee(\vec t)/(y-L)$.

(2) And at $y=-L$, we require that $\vec X(-L)+\vec \mu=0$, where
$\vec \mu$ is the moment map for the action of $G^\vee$ on the
Higgs branch of $T(G^\vee)$, which is the nilpotent cone $\N^\vee$
of $G^\vee$.

It is convenient to describe the result as a complex symplectic
manifold in one of its complex structures. As is explained in
section 3.3 of \cite{Gaiotto:2008sa}, condition (1) gives the
Slodowy slice transverse to the $G^\vee_\BC$  orbit of the
nilpotent element $\rho^\vee(t_+)\in \frak{g^\vee_\BC}$ (here
$t_+$ is the ``raising'' operator in $\frak{su}(2)$). Condition
(2) gives the intersection of this Slodowy slice with the
nilpotent cone $\N^\vee$ (since $\vec\mu$ takes values in this
cone). We write $\S_{\rho^\vee}$ for the Slodowy slice transverse
to $\rho^\vee(t_+)$. So the Coulomb branch of $T_{\rho^\vee}(G)$
is the intersection $\S_{\rho^\vee}\cap \N^\vee$, accounting for
one of the entries in Table \ref{firstable}. Mirror symmetry then
gives also the Higgs branch of $T^\rho(G)$.

\def\M{\mathcal M}
\def\N{\mathcal N}
\def\U{\mathcal U}
\def\V{\mathcal V}
What we have determined so far is the component of the moduli
space of vacua of $T_{\rho^\vee}(G)$ on which $SO(3)_X$ acts
trivially. To describe fully the moduli space
$\M_{T_{\rho^\vee}}(G)$ of vacua of $T_{\rho^\vee}(G)$, it helps
to be more systematic.  One ingredient is the moduli space of
vacua of $T(G)$ (or its mirror $T(G^\vee)$), whose general form
was described in (\ref{genform}).

Once a particular vacuum is picked for $T(G)$ at the end of the
interval, to get a full description, we need to consider the
behavior of $\vec X^\vee$ and $\vec Y^\vee$.  $\vec Y^\vee$ will
vanish, since the
 $\rho^\vee$ boundary conditions set it to zero at
$y=L$.  However, $\vec X^\vee$ can obey Nahm's equations.  In
solving Nahm's equations, the boundary condition (1) above is
unchanged, but (2) is modified:

(2$'$) At $y=-L$, we require that $\vec X^\vee(-L)+\vec \mu=0$,
where $\vec \mu$ is the moment map for the action of $G^\vee$ on
$\M_{T(G)}$.

The full moduli space is therefore the intersection of $\M_{T(G)}$
with the Slodowy slice $\S_{\rho^\vee}$.  (We intersect
$\S_{\rho^\vee}$ with each Coulomb branch factor of $\M_{T(G)}$ or
Higgs branch factor of $\M_{T(G^\vee)}$.) The moduli space of
vacua of $T_{\rho^\vee}(G)$ is therefore
\begin{equation}\label{fullmod}\M_{T_{\rho^\vee}(G)}=\bigcup_{\alpha\in
S} \overline\O_{\rho_\alpha}\times \left( \S_{\rho^\vee}\cap
\overline \O_{\rho^\vee_\alpha}\right).\end{equation}

For some $\alpha$, the intersection $ \S_{\rho^\vee}\cap \overline
\O_{\rho^\vee_\alpha}$ may be empty, as the Slodowy slice
transverse to $\rho^\vee$ will not intersect an orbit that is too
small.  For other $\alpha$, this intersection has positive
dimension, giving a branch of the moduli space that has a
non-trivial Coulomb factor.  To get a component of the Higgs
branch of $\M_{T_{\rho^\vee}}(G)$, the intersection
$\S_{\rho^\vee}\cap \overline \O_{\rho^\vee_\alpha}$ should have
dimension zero, which  happens precisely if
$\rho^\vee_\alpha=\rho^\vee$, in which case the intersection is a
single point.  Let $C_{\rho^\vee}$ be the set of all $\alpha$ such
that $\rho_\alpha^\vee=\rho^\vee$.  The Higgs branch of
$T_{\rho^\vee}(G)$ is then $\bigcup_{\alpha\in
C_{\rho^\vee}}\overline\O_{\rho_\alpha}$.  This accounts for
another entry in the table, and of course mirror symmetry gives
also the Coulomb branch of $T^\rho(G)$.

\begin{figure}
  \begin{center}
    \includegraphics[width=4.5in]{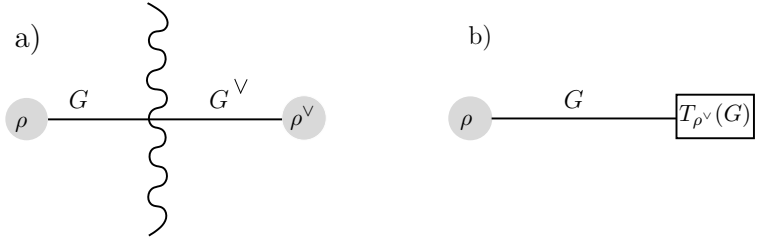}
  \end{center}
\caption{\small (a) The definition of the theory
$T_{\rho^\vee}^\rho(G)$.  On the left is $G$ gauge theory with
boundary conditions set by $\rho$.  On the right is $G^\vee$ with
boundary conditions set by $\rho^\vee$.  Between is the Janus
domain wall. The infrared limit of this configuration gives
$T_{\rho^\vee}^\rho(G)$. (b) Upon moving the Janus domain wall to
the right, we get a $G$ gauge theory with $\rho$ on the left and
$T_{\rho^\vee}(G)$ on the right.  The IR limit is the same.}
  \label{Fig48}\end{figure}

Finally, we would like to explain the last row in the Table
\ref{firstable}, which describes the Higgs and Coulomb branches of
$T^\rho_{\rho^\vee}(G)$. We start (fig. \ref{Fig48}) with the
definition of this theory in terms of gauge theory on a slab with
boundary conditions set by $\rho$ at one end and by $\rho^\vee$ at
the other end, with a Janus domain wall in the middle.  Moving the
Janus domain wall to the right, we get a configuration in $G$ gauge
theory with boundary conditions set by $\rho$ on the left and by a
coupling to $T_{\rho^\vee}(G)$ on the right.  To find a vacuum of
$T^\rho_{\rho^\vee}(G)$, we start with a vacuum of
$T_{\rho^\vee}(G)$ and solve the $G$ Nahm equations on an interval
with the obvious modification of the above conditions: $\vec X$ has
a pole of type $\rho$ at the left, and equals the moment map of
$T_{\rho^\vee}(G)$ on the right.  As a complex manifold, the result
is the intersection of the Slodowy slice $\S_\rho$ transverse to
$\rho$ with the moduli space $\M_{T_{\rho^\vee}}(G)$. Using the
description (\ref{fullmod}) of that moduli space, we arrive at a
pleasantly symmetric description of the moduli space for
$T^\rho_{\rho^\vee}(G)$:
\begin{equation}\label{symmod}\M_{T^\rho_{\rho^\vee}(G)}=\bigcup_{\alpha\in
S} \left(\S_\rho\cap \overline\O_{\rho_\alpha}\right)\times \left(
\S_{\rho^\vee}\cap \overline
\O_{\rho^\vee_\alpha}\right).\end{equation} Picking $\alpha$ so
that one factor or the other is a point, we arrive at the last row
of the table.

For $G=SU(n)$, the theories $T^\rho_{\rho^\vee}(G)$ are the general
good linear quiver theories, made from a chain of unitary gauge
groups, as we have explained in section \ref{application}.  We have
therefore described the moduli space of vacua of the general such
theory, which would have been very hard to get directly.

\subsection{A General Duality Prescription }\label{genpres}

\begin{figure}
  \begin{center}
    \includegraphics[width=5in]{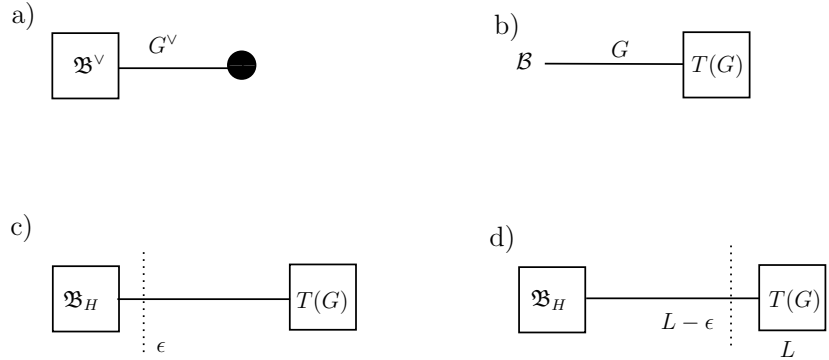}
  \end{center}
\caption{\small (a)The 4d configuration which ungauges $\fB^\vee$.
$\fB^\vee$ is the $S$-dual of a general boundary $\B$ given by a
triple $(\rho,H,\fB_H)$, where $H$ is a subgroup of the commutant
of a Nahm pole $\rho$, and $\fB_H$ is an SCFT with $H$ symmetry.
As usual the black dot represents Dirichlet boundary conditions.
(b) The $S$ dual of (a), involving a gauge theory with boundary
condition $\B$ on the left and coupling to $T(G)$ on the right.
(c) A split version of $\B$: the Nahm pole and reduction to $H$
(indicated by the vertical dotted line) have been displaced to
$y=\epsilon>0$, leaving the SCFT $\fB_H$ at the endpoint $y=0$.
(d) The Nahm pole is moved to the right. In the slab between
$y=L-\epsilon$ and $y=L$, $G$-valued gauge fields are coupled to a
Nahm pole at $y=L-\epsilon$ and to $T(G)$ at $y=L$. This generates
$T^\rho(G)$. Between $y=0$ and $y=L-\epsilon$ are $H$-valued gauge
fields, coupled at $y=0$ to $\fB_H$, which was part of the
original boundary condition, and at $y=L-\epsilon$ to the global
symmetry $H$ of $T^\rho(G)$. Thus the mirror $\fB^\vee$ is given
by a composite gauge theory with $H$ coupled to $\fB_H\times
T^\rho(G)$.}
  \label{Fig49}\end{figure}

Extending our arguments of section \ref{examples} to an arbitrary
gauge group $G$, we will now give a general recipe for understanding
the $S$-dual of a  boundary condition $\B$ associated with a triple
$(\rho,H,\fB_H)$, provided only that this $S$-dual has full $G^\vee$
gauge symmetry, and therefore can be described by coupling to an
SCFT $\fB^\vee$ with $G^\vee$ symmetry. The steps are depicted in
fig. \ref{Fig49}.  $\fB^\vee$ can be extracted via the usual
ungauging technique,  introducing a Dirichlet boundary condition at
$y=L$ and flowing to the IR. The $S$-dual of that configuration is
given as a $G$ gauge theory on a slab $\R^3\times I$, coupled to the
original boundary condition $\B=(\rho,H,\fB_H)$ at the left end, and
to $T(G)$ at the other end. When $\rho$ is trivial and $H=G$, this
gives at low energies a construction of the mirror $\tilde\fB^\vee$
to $\fB^\vee$ in terms of a three-dimensional gauge theory with
gauge group $G$ coupled to the product $\fB\times T(G)$.

In the general case (fig. \ref{Fig49}(c)), a boundary condition
$(\rho,H,\fB_H)$ can be ``split'' in space: as we approach the
boundary from $y>0$ we may first encounter the Nahm pole $\rho$,
followed by the reduction of the gauge symmetry to a subgroup $H$
(which must commute with $\rho$), and only then the coupling of $H$
to a boundary theory $\fB_H$.  This three stage nature of the
boundary condition is illustrated in fig. \ref{Fig1} (where a
description of the first two stages by branes is assumed).  The
precise positions in $y$ at which the Nahm pole and the reduction of
the gauge group are located are not important when we flow to the
infrared.

 The usefulness of this splitting is that the domain
wall that carries the Nahm pole and reduction of gauge symmetry
can then be moved to the right as in fig. \ref{Fig49}(d), towards
the $T(G)$ boundary.  Let us divide the configuration of fig.
\ref{Fig49}(d) into two slabs.  From the domain wall at
$y=L-\epsilon$ to the boundary at $y=L$, we have $G$ gauge fields
interacting with a Nahm pole at $y=L-\epsilon$ and with $T(G)$ at
$y=L$.  This (in the limit $\epsilon\to 0$) is the definition of
$T^\rho(G)$. Between $y=0$ and $y=L-\epsilon$, we have $H$ gauge
fields.  These gauge fields interact at $y=0$ with the SCFT
$\fB_H$ that was part of the original boundary condition.  At
$y=L-\epsilon$, they couple to the fields in the other slab.

The slab in fig. \ref{Fig49}(d) between the domain wall at
$y=L-\epsilon$ and the boundary at $y=L$ is the definition of
$T^\rho(G)$: gauge fields of $G$  with the Nahm pole at the left
boundary and coupling to $T(G)$ on the right.  Gauge fields of $H$
propagate to the left of $y=L-\epsilon$.  Of course, the slabs to
the left and right of the domain wall meet at their common boundary
$y=L-\epsilon$.  This means that the $H$-valued gauge fields on the
left slab are coupled to a global $H$-symmetry of the matter system
defined by the right slab.  Indeed, since $H$ commutes with $\rho$,
it is a global symmetry of $T^\rho(G)$, acting on the Higgs branch.

Hence we can formulate a prescription for $\tilde\fB^\vee$,
whenever it exists.  It is the IR limit of a composite $H$ gauge
theory depicted in fig. \ref{Fig49}(d).  This  theory, which we
call $\fB_H\times_H T^\rho(G)$, is a theory with gauge group $H$
coupled to the product $\fB_H\times T^\rho(G)$. In flowing to the
infrared, one can take zero modes of all vector multiplets in the
$y$ direction, so that the composite gauge theory is purely
three-dimensional.

In using  $\fB_H\times_H T^\rho(G)$ to define a boundary condition
in $G^\vee$ gauge theory, the four-dimensional $G^\vee$ gauge fields
couple to symmetries of the Coulomb branch of $T^\rho(G)$. As usual,
this construction  is most useful if one can find a description in
which the $G^\vee$ symmetry of the Coulomb branch is visible.

\subsubsection{Symmetry Breaking}\label{symbr}

\begin{figure}
  \begin{center}
    \includegraphics[width=4.5in]{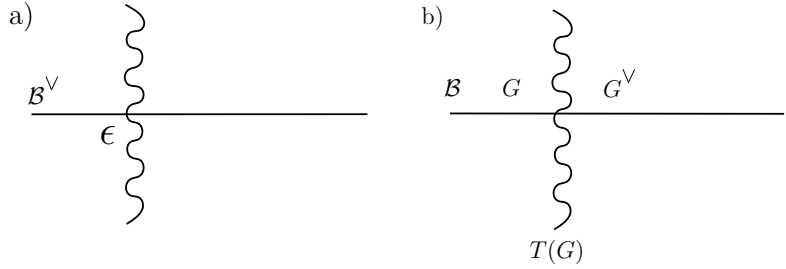}
  \end{center}
\caption{\small  (a) $G^\vee$ gauge theory on a half-space, with
boundary condition $\B^\vee$.  The gauge coupling is small except
within a distance $\epsilon$ from the boundary, to the left of the
wiggly line. (b) $S$-duality to the left of $y=\epsilon$ makes the
gauge coupling small everywhere, but now the gauge group is $G$
for $y<\epsilon$ and there is a coupling to $T(G)$ at
$y=\epsilon$.}
  \label{Fig50}\end{figure}

If we are given a boundary condition $\B$, we can construct the
composite gauge theory as above. If this theory has a standard
infrared limit, with the usual $R$-symmetry and global $G^\vee$
symmetry unbroken, then the dual boundary condition has full gauge
symmetry and is obtained by coupling to the SCFT that emerges from
the composite gauge theory.

What can we say when the composite gauge theory does not have a
standard IR limit?  To get some insight, we will describe a more
conservative variant of the above procedure that is always valid.

Our basic procedure so far has been to ungauge the dual $G^\vee$
gauge symmetry by imposing Dirichlet boundary conditions on the
$S$-dual theory at $y=L$.  This gives a quasi three-dimensional
theory, since the fourth dimension is compact, to which we then
apply $S$-duality, after which we flow to the infrared. Only after
completing the infrared flow and constructing a boundary SCFT
$\tilde \fB^\vee$ do we ``regauge'' the theory, coupling to bulk
$G^\vee$ gauge fields  so as to describe the dual of the original
$G$ theory on a half-space.

When the quasi three-dimensional theory does  not have a standard
IR limit, the reduction at an intermediate step to a quasi
three-dimensional description is not helpful.  Instead, it is
better to formulate a completely four-dimensional procedure that
is always valid (fig. \ref{Fig50}).  For this, we introduce the
theory $T(G)$ in a slightly different way.  Starting with the dual
$G^\vee$ configuration, we increase the gauge coupling in the
region $y<\epsilon$, something that does not affect the infrared
behavior.  When the $G^\vee$ gauge coupling is strong for
$y<\epsilon$ and weak for $y>\epsilon$, we make $S$-duality in the
region $y<\epsilon$.  This gives a description in which the
coupling is weak everywhere.  For $y<\epsilon$, we have weakly
coupled $G$ gauge theory, say with coupling $g_G$, coupled to the
original boundary condition $\B$.  For $y>\epsilon$, we have
weakly coupled $G^\vee$ gauge theory.  On the interface between
the two theories at $y=\epsilon$ lives the theory $T(G)$.

This description is always valid, and the question is what we can
learn from it. There are two scales in the problem: $\epsilon$ sets
the scale of the Kaluza-Klein modes of the $G$ vector multiplets,
while $\epsilon/g_{G}^2$ sets the scale of the three-dimensional
gauge coupling.  To reduce to a boundary condition, we want to take
$\epsilon\to 0$.  To flow to the infrared, we also want to take
$\epsilon/g_G^2$ to zero.  When the composite gauge theory does not
have a good IR limit, the second operation is not straightforward.
However, there is no trouble in reducing to a three-dimensional
boundary theory by taking $\epsilon\to 0$ with
$1/g_3^2=\epsilon/g_G^2$ fixed.
 The result is that
the four-dimensional $G^\vee$ gauge theory is coupled to a
composite three-dimensional field theory at the boundary, given by
the usual prescription of fig. \ref{Fig49}(d), but with finite
gauge coupling for the  three-dimensional gauge theory.

At this stage, we have $G^\vee$ gauge theory in the half-space
$y\geq 0$ coupled to a boundary theory that has full supersymmetry
and $R$-symmetry but is not superconformal.  We still want to take
the IR limit $1/g_3^2\to 0$. By hypothesis, the boundary theory
alone does not behave well in this limit.
 The final step of taking the infrared limit $1/g_3^2\to 0$ has to
 be taken for the combined theory on the half-space.  This can
 produce a boundary condition for the $G^\vee$ gauge fields that
 involves reduced gauge symmetry, possibly with a Nahm pole.  (The
 case that it produces a boundary condition with full gauge
 symmetry at the boundary is precisely the case that the IR flow
 of the composite gauge theory could have been carried out in purely
 three-dimensional terms.)

If the IR flow of the composite gauge theory spontaneously breaks
its $G^\vee$ global symmetries, then the boundary condition will
have reduced gauge symmetry.  Moreover, if the moment map
operators $\vec \mu$ for the $G^\vee$ isometries of the composite
gauge theory receive expectation values in the IR, the boundary
condition $\vec X = \vec \mu$  forces the  scalar fields $\vec X$
to acquire  expectation values at the boundary. Supersymmetry will
then require that $\vec X(y)$ should obey  the Nahm equations. By
dimensional analysis, the expectation value of $\vec X(0)$ is
proportional to the cutoff $\epsilon^{-1}$, so in this situation a
Nahm pole will emerge when we remove the cutoff.

\subsubsection{ Examples}

We will try to provide some simple examples of these phenomena,
which involve the strong coupling dynamics of the
three-dimensional gauge theory.

\bigskip\noindent{\it A  Trivial Domain Wall}

The simplest example is a trivial domain wall for the gauge group
$G$, which can be interpreted by the reflection trick as a
boundary condition for a $G\times G$ theory, broken to the
diagonal $G$ at the boundary. This example is the $k=0$ case of
fig. \ref{Fig40}(c).  Clearly the $S$-dual of a trivial $G$ domain
wall is a trivial $G^\vee$ domain wall.  We want to see how this
result arises from the ungauging procedure.  Naive application of
this procedure leads to a bad quiver which is the $k=0$ case of
 fig. \ref{Fig43}.

\begin{figure}
  \begin{center}
    \includegraphics[width=4.5in]{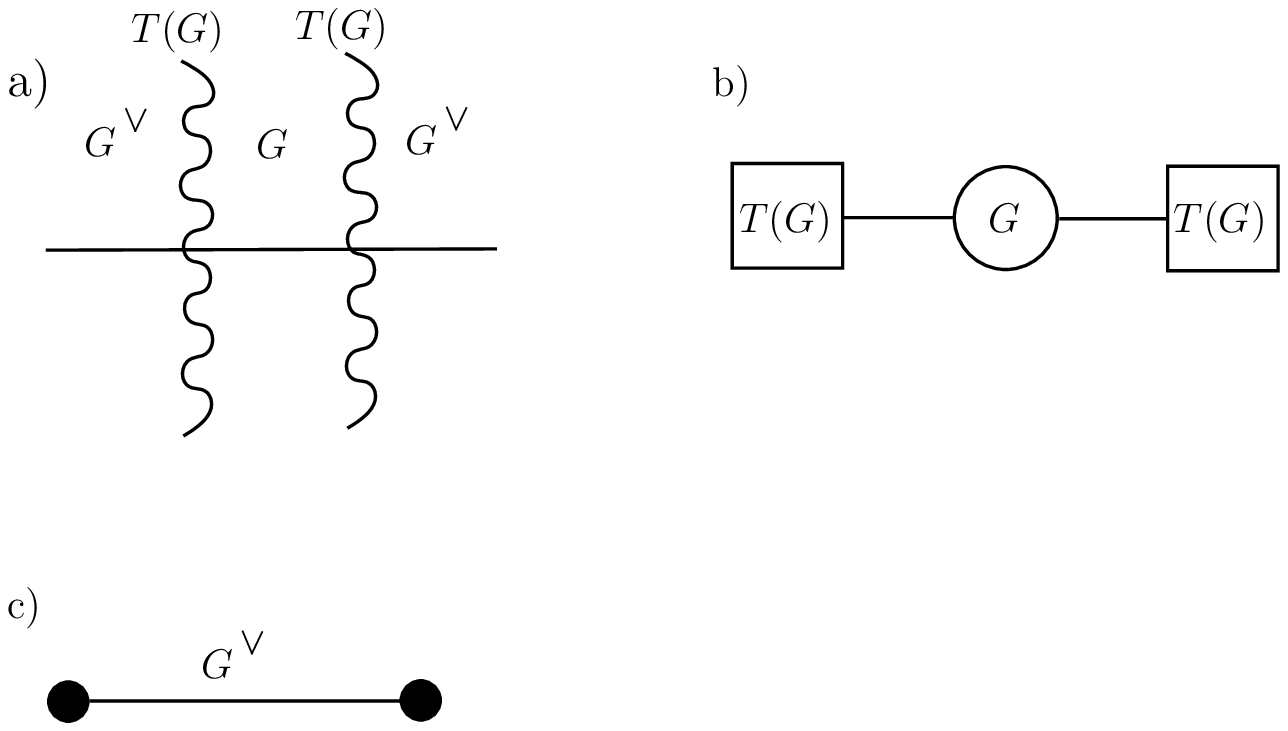}
  \end{center}
\caption{\small}
  \label{Fig51} (a) The $S$-dual of a trivial domain wall in $G$ gauge theory is a trivial
  domain wall in $G^\vee$ gauge theory.  Trying to show this using
  the ungauging procedure
  produces this configuration
  with two $T(G)$ domain walls.  The gauge group is everywhere $G^\vee$ except in
  the central slab,
  where it is $G$. (b) The three-dimensional composite gauge theory has gauge group $G$ coupled
  to two copies of $T(G)$.
  (c) A convenient way to compute its Coulomb branch is to describe it via $G^\vee$ gauge
  theory on $\R^3\times I$ with Dirichlet boundary conditions at both ends.
   The Coulomb branch is the moduli space of
  solutions of Nahm's equations. \end{figure}

We want to apply the general procedure of fig. \ref{Fig50}, but in
the two-sided case of a domain wall, this procedure must be
applied on both sides.  The analog of fig. \ref{Fig50} is
therefore a two-sided configuration in $G^\vee$ gauge theory with
two Janus domain walls separating an interval in which the gauge
group is $G$ (fig. \ref{Fig51}). Clearly as we flow to the
infrared, the two Janus domain walls will essentially meet and
cancel each other, but we would like to understand how this
happens when the Janus domain walls are represented by coupling to
$T(G)$.

For this, we need to understand the Coulomb branch of $G$ gauge
theory on $\R^3\times I$ with coupling to $T(G)$ at both ends.
(Alternatively, we can consider a $G$ gauge theory with finite
coupling, coupled to two copies of $T(G)$.) The simplest way to do
this is to use $S$-duality and the  fact that $T(G)$  is the
$S$-dual of a Dirichlet boundary condition in $G^\vee$ gauge
theory. So we want the Coulomb branch of $G^\vee$ gauge theory on
the slab with Dirichlet boundary conditions at both ends.  This is
given by the moduli space of solutions of Nahm's equations $D\vec
X/Dy+\vec X\times \vec X=0$ on $I$.  Dirichlet boundary conditions
simply mean that $\vec X$ is arbitrary at both ends and that one
divides only by gauge transformations that are trivial at both
ends.

The hyper-Kahler manifold that arises from Nahm's equations in
this situation was first studied by Kronheimer
\cite{KronheimerOld} and is described in \cite{Gaiotto:2008sa},
section 3.9.1. (This manifold also entered at the end of section
\ref{domain} in relation to the same problem for $SU(n)$.) As a
complex manifold in any of its complex structures, it is the
cotangent bundle of $G^\vee_\BC$. The $G^\vee \times G^\vee$
symmetry acts by left and right multiplication on $G^\vee_\BC$. A
maximal unbroken subgroup is a diagonal $G^\vee$ subgroup of
$G^\vee\times G^\vee$.

In other words, on the Coulomb branch, $G^\vee \times G^\vee$ is
broken down to $G^\vee$, as one would expect for the trivial
$G^\vee$ domain wall, which is the $S$-dual of the trivial $G$
domain wall with which we started.

\begin{figure}
  \begin{center}
    \includegraphics[width=3.5in]{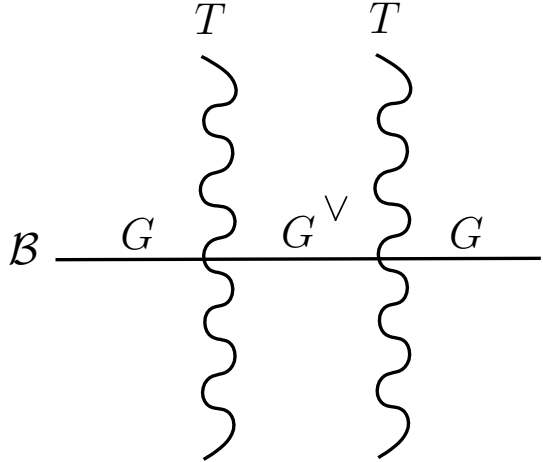}
  \end{center}
\caption{\small  The square of an $S$-duality transformation of a
boundary condition $\B$. At each step the original boundary
condition is coupled to the Higgs branch of an appropriate theory
$T(G)$ or $T(G^\vee)$.  Probing the IR dynamics of the $G^\vee$
gauge theory in the central slab leads back to the original
boundary condition.  Alternatively, one can represent the two
$T$'s by Janus domain walls and let the coupling constant profile
flow away in the infrared.}
  \label{Fig52}\end{figure}

\bigskip\noindent{\it The Square Of $S$-Duality}

We can use this construction to argue that our $S$-duality
prescription defined using $T(G)$ properly squares to the
identity. Consider any $G$ boundary condition $\B$, and apply our
$S$-duality prescription twice, as in fig. \ref{Fig52}.

The result is essentially $\B$ coupled through a $G$ gauge theory
to the Higgs branch of $T(G)$, whose Coulomb branch is in turn
coupled through a $G^\vee$ gauge theory to the Higgs branch of
$T(G^\vee)$. Finally, the bulk four-dimensional $G$ gauge theory
is coupled to the Coulomb branch of $T(G^\vee)$.

It is convenient to analyze the $G^\vee$ dynamics first. $G^\vee$
couples to two copies of $T(G^\vee)$.  This is the $S$-dual of the
system that we have just analyzed. We have learned that in the
infrared the $G \times G$ isometries of the Coulomb branch of this
composite theory are broken to a diagonal $G$. As a result, we get
a direct coupling of $\B$ to the four-dimensional $G$ gauge
theory, as desired.

In effect, what we have just analyzed is very similar to the
situation in fig. \ref{Fig51}, except that it is cut off on the
left by the boundary condition $\B$.

There is a subtlety here, unrelated to the main ideas of this
paper, which we mention only in the hope of avoiding some
confusion. Our definition of $S$ requires making a duality
transformation in $G$ gauge theory in a half-space and identifying
the result, in that half-space, with $G^\vee$ gauge theory.  If
$G$ is a group such as $SU(n)$ that has complex representations,
then $G$ and $G^\vee$ gauge theory admit a non-trivial classical
automorphism $C$ of complex conjugation; in our definition of $S$,
we did not pick an isomorphism with $G^\vee$ gauge theory, so we
did not distinguish $S$ from the product $S C$.  In defining the
two  $S$ operations from $G$ to $G^\vee$ and from $G^\vee$ to $G$,
choices can be made such that $S^2=1$. However, taking account of
the fact that in the relevant cases, $G$ and $G^\vee$ have the
same Lie algebra, it is somewhat unnatural to define the two $S$
operations independently, and a more natural set of choices
actually leads to $S^2=C$.  $C$ corresponds to the central element
$-1\in SL(2,\Z)$.

\bigskip\noindent{\it Examples With Symmetry Breaking and Nahm Pole}
\label{sympole}

The simplest  example of a boundary condition $\B$ whose dual has
a Nahm pole is a boundary condition given by a coupling to
$T_{\rho^\vee}(G)$.  Almost by the definition of
$T_{\rho^\vee}(G)$, the $S$-dual is the boundary condition
$\B^\vee$ in $G^\vee$ gauge theory given by the $\rho^\vee$ pole.

We want to understand how the general $S$-duality recipe can
reproduce this fact. That prescription produces a composite gauge
theory in which the three-dimensional  gauge group $G$ is coupled
to the product $T_{\rho^\vee}(G)\times T(G)$. This theory does not
have a good IR limit (as one can see from the quiver description
if $G=SU(n)$), so one should study it at finite gauge coupling.
More conveniently, one can study $G$ gauge theory on the slab
$\R^3\times I$ with boundary conditions at the left and right set
by coupling to $T(G)$ and $T_{\rho^\vee}(G)$, respectively.

$S$-duality converts this to $G^\vee$ gauge theory on the slab with
a $\rho^\vee$ Nahm pole at one end and Dirichlet boundary conditions
at the other end. The Coulomb branch is the moduli space of $G^\vee$
Nahm equations on an interval, with $\rho^\vee$ boundary conditions
at the right end, and Dirichlet at the left end. This moduli space
was described in \cite{Biel} and reviewed in \cite{Gaiotto:2008sa},
section 3.9.2. It has of course a $G^\vee$ isometry. The moment map
for this isometry is the initial value $\vec X(0)$ of the scalar
fields at the Dirichlet endpoint of $I$. The presence of the Nahm
pole at the other endpoint forces this moment map to be nonzero.
Indeed any complex null combination such as $\mu^1 + i \mu^2$ will
be conjugate to the raising operator $\rho^\vee(t_+)$.

As a result, the coupling of the four-dimensional $G^\vee$ gauge
fields to the Coulomb branch of the composite gauge theory forces
the  bulk scalar fields $\vec X$ to have non-zero expectation
values at the boundary, in order to obey the boundary condition
$\vec X + \vec \mu=0$.  By dimensional analysis, $\vec\mu$ and
therefore the boundary value of $\vec X$ is proportional to $1/L$,
the width of the slab. For $L\to 0$, $\vec X$ acquires a Nahm pole
at the boundary, of type $\rho^\vee$.

\subsection{$T(U(1))$}\label{tuone}

Here we will describe $T(G)$ in the abelian case $G=U(1)$.

For simple $G$, $T(G)$ is a non-trivial SCFT whose Higgs and
Coulomb branches are copies of the nilpotent cones $\N$ and
$\N^\vee$. This suggests that $T(U(1))$ might be trivial, since
the nilpotent cone for the complexification $GL(1)$ of $U(1)$
consists of only one point (a nonzero element of the Lie algebra
$\frak{gl}(1)$ is not nilpotent).

In a sense, $T(U(1))$ is trivial as an SCFT but nontrivial as a
recipe for coupling external vector multiplets.  This means the
following.  Given the SCFT $T(G)$, we can couple it to $G$-valued
and $G^\vee$-valued external gauge fields.  The ability to do this
is important in the way we use $T(G)$ in constructing the $S$-dual
of a given boundary condition.

For $G=U(1)$, $T(G)$ is trivial as an SCFT, and consists only of a
prescription for coupling external vector fields.  This turns out
to be the following.  Let $V$ and $W$ be two $\N=4$ vector
multiplets that contain $U(1)$ gauge fields $B$ and $C$.  Then the
appropriate recipe is to couple them by the supersymmetric
completion of the Chern-Simons coupling:
\begin{equation}\label{kimono} \frac{1}{2\pi}\int C\wedge
dB.\end{equation} An extension of this interaction with $\N=4$
supersymmetry does exist \cite{Brooks:1994nn,Kapustin:1999ha}.
Eqn. (\ref{kimono}) is equivalent to saying that $C$ is coupled to
the current
\begin{equation}\label{imono}J=\frac{\star
F_B}{2\pi},\end{equation} where $F_B$ is the curvature of $B$.

Assuming that this is $T(U(1))$, let us describe the appropriate
recipe to find the dual of a general boundary condition $\B$ in
$U(1)$ gauge theory. For brevity we suppose that $\B$ is given by
coupling to an SCFT $\frak B$ with $U(1)$ symmetry.  First, recall
the case of simple $G$.  In that case, $T(G)$ has $G\times G^\vee $
symmetry. We are supposed to introduce a $\frak g$-valued vector
multiplet $V$ and couple it to the diagonal $G$ symmetry of the
product $\frak B\times T(G)$. This then gives a theory $\tilde{\frak
B}^\vee=\fB\times_GT(G)$ which is the mirror of the SCFT that
defines the dual boundary condition.

As $\tilde {\frak B}^\vee$ has $G^\vee$ symmetry (since $T(G)$
does), we can couple it to a $\frak g^\vee$-valued background vector
multiplet $W$. If we do this, then at this stage, we have coupled
gauge fields $V$ and $W$ of $G\times G^\vee$ to the product $\frak
B\times T(G)$.  $V$ is dynamical and $W$ is a background field.

For $G=U(1)$, we do exactly the same thing, but the meaning is a
little different because $T(U(1))$ is ``trivial.''  Coupling $V$
and $W$ to $\frak B\times T(G)$ means that we couple $V$ to $\frak
B$ in the usual sense, and we couple $V$ and $W$ to each other by
means of the supersymmetric completion of (\ref{kimono}).

In general, for any $G$, coupling $W$ is just a way of formalizing
what is the $G^\vee$ symmetry of $\tilde{\frak B}^\vee$.  For $G$
a simple group, this step is hardly necessary\footnote{If $G$ and
$G^\vee$ have outer automorphisms, it is best to regard the choice
of $G$ and $G^\vee$ action as part of the definition of $T(G)$.}
because the only possible $G^\vee$ symmetry is the one that comes
from the $G^\vee$ symmetry of the SCFT $T(G)$. For $G=U(1)$, since
$T(U(1))$ is trivial as an SCFT, explicitly spelling out the
coupling of the two vector multiplets is the simplest way to
describe the $U(1)$ symmetry of $\tilde{\frak B}^\vee$.

The duality procedure we have just described for $U(1)$ boundary
conditions is the same as the one that we arrived at by a
different method in section \ref{generalization}. Indeed, the
coupling (\ref{kimono}) appeared previously in eqn.
(\ref{cslike}).  (The three-dimensional gauge field $C$ can be
understood as the boundary value of the gauge field $A$ of
(\ref{cslike}), which is defined on a half-space.)

We can similarly compare the general $T(G)$ procedure to the recipe
of section \ref{recipe} for the case $G=U(n)$.  The general recipe
says to construct $\tilde\fB^\vee$ by coupling $U(n)$ gauge fields
to $\frak B\times T(U(n))$, while in section \ref{recipe}, the
prescription was to couple $U(n)$ gauge fields to $\frak B\times
T(SU(n))$.  Either way, we then want to find a $U(n)^\vee=U(n)$
symmetry of the Coulomb branch of $\tilde\fB^\vee$.  This symmetry
was constructed  in section \ref{recipe} using the $SU(n)^\vee$
Coulomb branch symmetry of $T(SU(n))$ and a Chern-Simons coupling
(between $U(1)$ gauge fields that gauge the centers of $U(n)$ and
$U(n)^\vee$) to exhibit the action of the center of $U(n)^\vee$. The
local factorization $U(n)\cong SU(n)\times U(1)$ means that
$T(U(n))$ is a product $T(SU(n))\times T(U(1))$. The two procedures
are equivalent by virtue of the above description of $T(U(1))$.

\subsubsection{Computation}\label{computation}

\def\II{\cmmib I}\def\k{\cmmib k}\def\F{{\mathcal F}}
So far we have shown that everything is consistent if $T(U(1))$ is
simply a recipe for a Chern-Simons coupling of two external $U(1)$
vector multiplets.  On the other hand, in section \ref{jansd}, we
gave a general recipe for defining $T(G)$ by making a duality
transformation in a half-space. Here we aim to show that the two
approaches coincide for $G=U(1)$.

First we review electric-magnetic duality for $U(1)$ in the
absence of a boundary.  Supersymmetry plays no essential role
(since in the abelian case the additional fields  required by
supersymmetry are not coupled to the gauge field), so we focus on
pure $U(1)$ gauge theory. We follow \cite{Witten:1995gf} and
section 2.4 of \cite{Gukov:2006jk} (see
\cite{Buscher:1987qj,Rocek:1991ps} for the analog in two
dimensions), but for brevity we take the $\theta$-angle to vanish.
The action of the free $U(1)$ gauge field $B$, whose curvature we
call $F_B$, on a four-manifold $M$ is
\begin{equation}\label{zex}\II=\frac{1}{e^2}\int_M F_B\wedge \star
F_B.\end{equation}  To establish electric-magnetic duality, we
introduce a two-form field $\k$  and require the extended gauge
symmetry
\begin{align}\label{vex}\notag B&\to B+b \\
                               \k & \to \k+db,\end{align}
where $b$ is any connection on a principal line bundle $\T$ over
spacetime, and $db$ is its curvature. $\k$ should  really be
regarded as a gerbe connection (the analog of the two-form field
in string theory) with curvature ${\cmmib h}=d\k$; the periods of
$\cmmib h$ are integer multiples of $2\pi$. The extended gauge
symmetry reduces to ordinary Maxwell gauge symmetry for $B$  if
$b=d\epsilon$ for some zero-form $\epsilon$. To get the extended
gauge symmetry, it suffices to introduce $\F=F_B-\k$ and replace
$F_B$ by $\F$  in the action $\II$. The resulting theory, however,
is trivial.  To get something nontrivial, we introduce another
abelian gauge field $C$, and add to the action a coupling
$\frac{1}{2\pi}\int_M C\wedge d
\k=\frac{1}{2\pi}\int_MF_C\wedge\F$, with $F_C=dC$ the curvature
of $C$. At this stage, then, we have an extended action
\begin{equation}\label{pex}\hat\II=\frac{1}{2\pi}\int_M F_C\wedge \F+
\frac{1}{e^2}\int_M \F\wedge \star \F.\end{equation}
Electric-magnetic duality is established by comparing two ways to
study this theory. One approach is to first integrate over $C$,
which leads to a delta function in the path integral by means of
which one can set $\k=0$ modulo an extended gauge transformation
of the form (\ref{vex}). (See the discussion of eqn. (2.20) in
\cite{Witten:1995gf}. The argument depends on the precise
coefficient $1/2\pi$ in the first term of the extended action.)
This shows that the extended theory is actually equivalent to the
original theory with action (\ref{zex}). On the other hand, one
can use the extended gauge symmetry to set $B=0$, after which the
integral over $\k$ is Gaussian.  Performing this integral, we
arrive at an action for $C$
\begin{equation}\label{cex}\II_C=\frac{e^2}{16\pi^2}\int_M F_C\wedge \star
F_C\end{equation} that has the same form as the original action
(\ref{zex}) except that $\tau=4\pi i/e^2$ has been replaced by
$-1/\tau$. This is electric-magnetic duality.

Now let us consider the case that $M$ has a boundary $\partial M$.
Let us analyze the dual of Dirichlet boundary conditions.  This
means that we require that $B$ and hence also its curvature $F_B$
vanish when restricted to $\partial M$.  This being so, in the
definition of the extended gauge symmetry (\ref{vex}), we likewise
have to require that $b$ vanishes when restricted to $\partial M$.
Now we have to ask what sort of boundary conditions on $C$ and $G$
will enable the above argument to work.  The answer is that it works
if we place Neumann boundary conditions on $C$ and Dirichlet on $G$.
(For example, we cannot place Neumann boundary conditions on both or
Dirichlet on both as this does not give a well-posed boundary
problem.  If we place Dirichlet boundary conditions on $C$ and
Neumann on $G$, then the delta function from the $C$ integral does
not suffice to set $G$ to zero modulo an extended gauge
transformation.) For another explanation of the boundary condition
on $C$, see eqn. (\ref{hex}) below.

So the dual gauge theory, in which the gauge field is $C$, obeys
Neumann boundary conditions.  We have shown that the $S$-dual of
Dirichlet is Neumann.

Now we want to modify this by introducing a gauge field $A$ on
$\partial M$, with curvature $F_A$, and shifting the boundary
condition on $B$ from $B|_{\partial M}=0$ to $B|_{\partial M}=A$. We
continue to require that $b$ and $\k$ vanish on $\partial M$. The
extended theory with action $\hat\II$ still makes sense, and by
doing first the integral over $C$, one can still show that this
theory is equivalent to the original Maxwell theory (\ref{zex}) with
the shifted boundary condition. What happens if we proceed in the
opposite order?  We cannot gauge fix $B$ to 0, since this does not
obey the boundary conditions, but we can pick an arbitrary gauge
field $B_0$ on $M$ that obeys the boundary conditions, and impose
the gauge condition $B|=B_0$. Next we try to perform the integral
over $\k$.  The condition for the action to have  a critical point
as a  function of $\k$ that also obeys the boundary condition
$\k|_{\partial M}=0$ is that
\begin{equation}\label{hex} \frac{2}{e^2}\star F_C|_{\partial M} =
\frac{1}{2\pi}F_A.\end{equation} Once this boundary condition is
imposed on $C$, one can perform the Gaussian integral over $\k$,
leading back to the same bulk action (\ref{cex}) as before. The
conclusion is that the shifted boundary condition on $B$ is dual
to the deformation (\ref{hex}) away from Neumann boundary
conditions for $C$.  The correction to the boundary condition is
equivalent to the addition to the action of a boundary term. The
total action for $C$, including the boundary term, is
\begin{equation}\II_C=\frac{e^2}{16\pi^2}\int_M F_C\wedge \star
F_C+\frac{1}{2\pi}\int_{\partial M}C\wedge dA.\end{equation}

In other words, a shifted boundary condition $B|_{\partial M}=A$
with specified $A$ is dual to the boundary interaction just
indicated.  Another way to explain the result we have obtained is
that the condition $B|_{\partial M}=A$ implies that
$F_B|_{\partial M}=F_A$.  Under duality, $F_B$ maps to
$\frac{4\pi}{e^2}\star F_C$, so the boundary condition becomes
$\frac{4\pi}{e^2}\star F_C|_{\partial M}=F_A$, as in (\ref{hex}).

Now it is straightforward to perform duality in a half-space and
explain the claim we have made about $T(U(1))$. We start with a
$U(1)$ gauge field $\hat A$ on a four-manifold $M$. We select a
three-dimensional submanifold $N$ of $M$ that divides $M$ into two
pieces $M_1$ and $M_2$.  We write $A$ and $B$ for the restrictions
of $\hat A$ to $M_1$ and $M_2$, respectively. Of course they agree
on the boundary:
\begin{equation}\label{bexo}B|_{N}=A|_{N}.\end{equation}
Now we carry out the above duality procedure on $M_2$.  In the
process, $B$ is replaced by another $U(1)$ gauge field $C$ on
$M_2$, and the boundary condition (\ref{bexo}) is replaced by a
boundary interaction
\begin{equation}\label{exo}\frac{1}{2\pi}\int_{N}C\wedge dA.\end{equation}
We have justified the claim that after carrying out duality in
half of spacetime, the gauge fields $A$ and $C$ on the two sides
are coupled in this fashion. This justifies our proposal for
$T(U(1))$, and hopefully also makes more tangible the idea of
defining $T(G)$ by duality in a half-space.

\subsection{Massive Deformations Of $T(G)$}\label{northosymp}

Here we will briefly discuss the massive deformations of $T(G)$
and related theories.  These are of interest physically, and also
relevant to some aspects of recent mathematical work
\cite{BLPW,BLP,BLPW2} on classical geometry related to
three-dimensional mirror symmetry.

We begin by restating some standard facts. Let us start with a
general superconformal $\N=4$ theory $W$ in three dimensions.  The
moduli space of vacua has the general form
\begin{equation}\label{genf}\M=\cup_{\alpha\in S} \C_\alpha\times
\H_\alpha,\end{equation} where $S$ is the set of components, and
each component is the product of Coulomb and Higgs factors
$\C_\alpha$ and $\H_\alpha$. Each factor is a conical hyper-Kahler
manifold; with a scaling symmetry that leaves fixed only the
apexes of the cones, and $R$-symmetry  groups $SO(3)_X$ and
$SO(3)_Y$ that rotate respectively the complex structures of the
$\H_\alpha$ and of the $\C_\alpha$. The union in (\ref{genf}) is
not a disjoint union; the components meet on conical hyper-Kahler
subvarieties, and in particular there is one point  that they all
share in common -- the fixed point of the scaling symmetry. The
Coulomb branch $\C$ is the union of $\C_\alpha$ for which
$\H_\alpha$ is a point, and the Higgs branch $\H$ is the union of
$\H_\alpha$ for which $\C_\alpha$ is a point.

It is not true that there is always a Higgs branch or a Coulomb
branch, since there may be no component for which $\C_\alpha$ or
$\H_\alpha$ is a point.  In general, we define the maximal Higgs
branch $\H'$ as the union of $\H_\alpha$ for which $\C_\alpha$ is
minimal, and the maximal Coulomb branch $\C'$ as the union of
$\C_\alpha$ for which $\H_\alpha$ is minimal.  All $\H_\alpha$ are
subvarieties of $\H'$ (on which Coulomb directions $\C_\alpha$
appear) and all $\C_\alpha$ are subvarieties of $\C'$ (on which
Higgs directions $\H_\alpha$ appear).

Continuous symmetries act on either $\C'$ or $\H'$, but not both.
Let $\tilde H$ and $H$ be the groups that act on $\C'$ and $\H'$,
respectively. Let $\tilde T$ and $T$ be maximal tori in $\tilde H$
and $H$, and write $\tilde{\frak t}$ and $\frak t$ for their Lie
algebras. The actions of $\tilde T$ and $T$ preserve the conical
hyper-Kahler structures of the $\C'$ and $\H'$, and so always
leave fixed the apexes of these cones. Massive deformations exist
precisely if $\tilde T$ and $T$ have no other fixed points.

The massive deformations are constructed using Fayet-Iliopoulos
(FI) parameters and mass parameters.  These take values
respectively in $\tilde{\frak t}\otimes \R^3$ and ${\frak
t}\otimes \R^3$, where one can think of $\R^3$ as the space of
imaginary quaternions. The effect of turning on FI parameters is
to reduce the dimension of the $\C_\alpha$, possibly to zero, and
to resolve/deform singularities of the $\H_\alpha$, possibly
making them smooth. Indeed, each component of an FI parameter
$\vec\zeta\in \tilde{\frak t}\otimes \R^3$ corresponds to a vector
field on $\C'$ and hence on each $\C_\alpha$. We refer to the
common zeroes of these three vector fields as the zeroes or fixed
points of $\vec\zeta$. The effect of perturbing the action by
$\vec \zeta$ is to reduce the moduli space $\M$ to the zeroes of
$\vec\zeta$. In the theory $W$ perturbed by $\vec\zeta$,  the
$\C_\alpha$ are replaced by the fixed point sets
$\C_\alpha^{\vec\zeta}$, and the singularities of $\H^\alpha$ are
partially or completely deformed/resolved.

An important special case is that if $\tilde T$ acting on  $\C'$
has no fixed point except for the apex, and we choose $\vec\zeta$
generically, then the only zero of $\vec\zeta$ is the apex.  In
this situation,  $\H'$ becomes smooth and becomes the moduli space
of vacua. (Its singularities would be points of intersection with
Coulomb branches, but these are absent, since the only zero of
$\vec\zeta$ is the apex.)

We say that $\vec\zeta$ is regular if it leaves fixed only the
apex of $\C'$.  Otherwise we say that $\vec\zeta$ is non-regular.
If the set of regular $\vec\zeta$ is non-empty, then the set of
non-regular $\vec\zeta$ has real codimension at least three.  (To
be non-regular, $\vec\zeta$ must take values in ${\frak p}\otimes
\R^3$ where $\frak p$ is a proper subspace of $\tilde{\frak t}$;
the codimension of ${\frak p}\otimes \R^3$ in $\tilde{\frak t}
\otimes \R^3$ is at least three.) This has the important
consequence that the set of regular $\vec\zeta$ is connected and
simply-connected.

Similarly, we call $\vec{\eusm m}$ regular if it leaves fixed only
the apex of $\H'$, and otherwise nonregular.  Again, the set of
regular $\vec{\eusm m}$ is connected and simply-connected (though
possibly empty).

Conversely to what has been stated above, a perturbation by masses
$\vec{\eusm m}\in {\frak t}\otimes \R^3$ reduces $\H_\alpha$ to
the fixed point set $\H_\alpha^{\vec{\eusm m}}$ (defined again as
the zero set of the vector fields corresponding to $\vec{\eusm
m}$), and deforms/resolves the singularities of $\C_\alpha$.  If
the only fixed point of $T$ acting on $\H'$ is at the apex, then
after turning on a generic $\vec{\eusm m}$, $\C'$ becomes smooth
and is the moduli space of vacua.  We write $\hat \H'$ and $\hat
\C'$ for generic smoothings of $\H'$ and $\C'$ resulting from
$\vec{\eusm m}$ or $\vec\zeta,$ respectively.

Now let us focus on the situation that both $T$ and $\tilde T$
have only the apexes of the cones as fixed points. In this
situation, if both $\vec\zeta$ and $\vec{\eusm m}$ are regular,
the theory becomes massive.  We can describe the massive vacua in
two different limits, $|\vec\zeta|>>|\vec{\eusm m}|$ or
$|\vec{\eusm m}|>>|\vec\zeta|$.

If $|\vec\zeta|>>|\vec{\eusm m}|$, then we first consider the
effects of $\vec\zeta$.  The moduli space of vacua reduces to the
smooth space $\hat\H'$. Now turn on a regular $\vec{\eusm m}$. By
hypothesis,  $\vec{\eusm m}$ has only a single fixed point on the
conical space $\H'$ -- the apex of the cone. After the smoothing
to $\hat\H'$, there are finitely many fixed points. (The apex of
the cone is generically a singular point, and must be counted with
an integer multiplicity, which becomes the number of fixed points
once $\H'$ is made smooth.) These are the massive vacua of the
theory. So if $|\vec\zeta|>>|\vec{\eusm m}|$, the set $\frak M$ of
massive vacua has a natural correspondence\footnote{It does not
matter which smoothing $\hat\H'$ we use here.   To compare fixed
points for two different regular choices of $\vec\zeta$, we simply
pick an interpolating path of regular $\vec\zeta$'s, and follow
the fixed points along this path.  The choice of path does not
matter, since the space of regular $\vec\zeta$'s is
simply-connected.} with the fixed points of $T$ acting on
$\hat\H'$.

If $|\vec{\eusm m}|>>|\vec\zeta|$, we carry out these steps in the
opposite order, and establish a correspondence of $\frak M$ with
the set of fixed points of $\tilde T$ acting on $\hat\C'$. Since
we can smoothly interpolate from one limit to the other with the
theory remaining massive (as long as $\vec\zeta$ and $\vec{\eusm
m}$ both remain regular) there must be a natural correspondence
between $\tilde T$ fixed points on $\hat\C'$ and $T$ fixed points
on $\hat \H'$. Such a correspondence has been found mathematically
\cite{BLPW,BLP,BLPW2}.

\def\W{{\mathcal W}}
Now let $\W$ and $\tilde \W$ be the Weyl groups of $H$ and $\tilde
H$, respectively.  There is always a natural action of the product
$\W\times \tilde \W$ on $\frak M$.  The action of $\W$ is
clear\footnote{If $h\in H$  normalizes $T$, then $v\to hvh^{-1}$,
$v\in \frak t$, is a Weyl transformation $w_h$.  Given $w\in \W$, we
pick $h$ such that $w_h=w$.  Then the action of $w_h$ on $\hat\H'$
permutes the fixed points of $T$, in a fashion that is independent
of the choice of $h$; moreover, this gives an action of $\W$ on the
set of fixed points.} if we interpret $\frak M$ as the space of $T$
fixed points on $\hat\H'$, and the action of $\C$ is clear if we
interpret $\frak M$ as the space of $\tilde T$ fixed points on
$\hat\C'$.

\def\AA{A_{\rm reg}}
\def\BB{B_{\rm reg}}
To show that the actions of $\W$ and $\tilde\W$ commute, we look
at the problem more symmetrically.  We write $\AA$ for the space
of regular $\vec\zeta$ and $\BB$ for the space of regular
$\vec{\eusm m}$.  Conjugation of $\vec \zeta$ or $\vec{\eusm m}$
by $\tilde\W$ or $\W$ gives an equivalent theory.  Hence the
parameter space of massive deformations is $P=\AA/\tilde W\times
\BB/\W$.  Since $\AA$ and $\BB$ are connected and
simply-connected, the fundamental group of $P$ is
$\pi_1(P)=\W\times\tilde \W$.  The set $\frak M$ of massive vacua
maps to $P$, and as $\frak M$ is a finite set, this map is a
locally trivial fibration.  The global monodromy action on the
fiber gives the desired action of $\W\times \tilde\W$ on $\frak
M$.

\subsubsection{Application To $T(G)$}\label{appl}

The theory $T(G)$ gives an excellent illustration of the ideas
just summarized.  The groups $H$ and $\tilde H$ are $G$ and the
dual group $G^\vee$.  The Higgs and Coulomb branches are the
nilpotent cones $\N$ and $\N^\vee$.  The Weyl groups $\W$ and
$\W^\vee$ of $G$ and $G^\vee$ naturally coincide.

The maximal tori $T$ and $T^\vee$ of $G$ and $G^\vee$ act on $\N$
and $\N^\vee$ with fixed points only at the origin.  (In order for
$x\in \frak g$ to be $T$-invariant, it must be an element of
$\frak t$; but the intersection $\frak t\cap \N$ consists only of
the point 0, since an element of $\frak t$ that is nilpotent must
vanish.)  The condition for $\vec \zeta$ to be regular is that the
subalgebra of $\frak g^\vee$ that commutes with all three
components of $\vec \zeta$ is precisely $\frak t^\vee$; similarly
$\vec{\eusm m}$ is regular if it commutes precisely with $\frak
t$.

A regular FI perturbation eliminates the Coulomb branch $\N^\vee$
and deforms the nilpotent cone $\N$ to the orbit of a semi-simple
element, or a smooth resolution thereof.  For a particular choice of
FI perturbation\footnote{One chooses $\vec \zeta=(0,0,b)$, where
$b\in \frak t$ is regular in the usual sense, so that in one complex
structure $\vec\zeta$ generates a resolution, not a deformation, of
$\N$. The resolution in question is known as the Springer
resolution.}, $\N$ is deformed to $T^*(G/T)$, the cotangent bundle
of the compact flag manifold $G/T$.  The fixed points of the $T$
action on $G/T$ correspond precisely to elements of the Weyl group
$\W$. Indeed, let $\pi:G\to G/T$ be the projection.  For $g\in G$,
the condition for $\pi(g)$   to be a fixed point in the action of
$T$ on $G/T$ is that for any $t\in T$, we should have $tg=gt'$ for
some $t'\in T$. In other words, $g^{-1}tg=t'$, so $g$ normalizes $T$
and generates a Weyl transformation. The action of $\W\times \W$ on
$\frak M=\W$ is simply the left and right action of $\W$ on itself,
as one can see by slightly extending this analysis.

We will give two additional ways to understand these statements.
First, for $G=SU(n)$, we use the brane realization of $T(G)$ that is
sketched in fig. \ref{Fig23}(c) with $n$ D3-branes stretched between
$n$ NS5-branes on the left and $n$ D5-branes on the right. To
realize $T(SU(n))$, the $n$ NS5-branes should be coincident and
likewise  the $n$  D5-branes should be coincident. The massive
deformation is achieved by making generic displacements of the
NS5-branes in $\vec X$ and generic displacements of the D5-branes in
$\vec Y$. (This is what is actually drawn in the figure, simply
because it is easier to draw.) After this deformation, a
supersymmetric vacuum corresponds to a situation, as drawn in the
figure, in which each D3-brane ends at one end on an NS5-brane and
at the other end on a D5-brane.  The resulting vacuum is massive
since a D3-brane with opposite endpoints of this kind supports no
massless degrees of freedom.  Because of the $s$-rule, no more than
one D3-brane can end on any fivebrane, so as there are $n$ branes of
each kind, each fivebrane has precisely one D3-brane ending on it.
Connecting the $n$ NS5-branes with $n$ D5-branes gives a map from
one set of $n$ objects to another. There are $n!$ such maps, making
up the set $\frak M$ of massive vacua.

The Weyl group of $SU(n)$ is the group of permutations of $n$
objects.  The action of $\W\times \W$ is visible in the brane
picture: one factor acts by permuting the $n$ NS5-branes, and one
by permuting the $n$ D5-branes.  Clearly, we can think of $\frak
M$ as a copy of $\W$,  on which  $\W\times \W$ acts by left and
right multiplication.

An alternative approach is valid for any $G$.  We can realize $T(G)$
via gauge theory on $\R^3\times I$ (where $I$ is the interval $0\leq
y\leq L$), with $G$ gauge theory with Dirichlet boundary conditions
on the left, $G^\vee$ gauge theory with Dirichlet boundary
conditions on the right, and a Janus domain wall in between.  (This
is what we get by ``ungauging'' the configuration of fig.
\ref{Fig45} with Dirichlet boundary conditions at each end; it is
also the special case of fig. \ref{Fig47}(a) with $\rho^\vee$
trivial.)  In that language, a massive deformation is made by
shifting the Dirichlet boundary conditions at the two ends.  On the
left, instead of taking $\vec Y(0)=0$, we take $\vec Y(0)=\vec
\zeta$, and on the right, instead of taking $\vec Y^\vee(L)=0$, we
take $\vec Y^\vee(L)=\vec{\eusm m}$.  (See section 2.2.3 of
\cite{Gaiotto:2008sa} for this type of deformation.)  Once we make a
massive deformation, the Janus coupling profile $g(y)$ in fig.
\ref{Fig45} is not important.  We can deform to the case that $g$ is
constant, and small in the $G$ description. It is then useful to use
the $G$ description everywhere. The duality transformation from
$G^\vee$ to $G$ on the right part of the slab maps $\vec Y^\vee$ to
$\vec X$, and the boundary conditions become
\begin{align}\notag \vec Y(0)& = \vec\zeta \\ \vec X(L) & =
\vec{\eusm m}.\end{align} (Also, we impose Neumann boundary
conditions on $\vec X$ at $y=0$, and on $\vec Y$ at $y=L$. The
gauge field $A$ obeys Dirichlet boundary conditions at $y=0$ and
Neumann at $y=L$.) Finally, we divide by gauge transformations
that equal the identity at $y=0$.

With these boundary conditions, the theory is massive.  For example,
having opposite boundary conditions on $A$ at the two ends ensures
that it has no massive modes, while the boundary conditions on $\vec
X$ and $\vec Y$ ensure the same for them. To find a supersymmetric
vacuum, $\vec X$ and $\vec Y$ must be covariantly constant, and all
components of $\vec X$ and $\vec Y$ must commute with each other. In
addition, the curvature $F=dA+A\wedge A$ must vanish. Hence, by a
gauge transformation that is trivial at $y=0$, we can set $A=0$.  In
this gauge, $\vec X$ and $\vec Y$ are simply constant.  The boundary
condition now tells us that $\vec Y(y)=\vec \zeta$ for all $y$. As
for $\vec X$, it is also constant, but we can no longer claim that
it equals $\vec{\eusm m}$, since we have made a gauge transformation
that may be nontrivial at $y=L$.  However, $\vec X$ must be a
constant that commutes with $\vec Y=\vec \zeta$, so (as we assume
$\vec\zeta$ to be regular) $\vec X$ must be $\frak t$-valued.
Moreover, $\vec X$ must be gauge equivalent to $\vec{\eusm m}$.
These conditions imply that in this gauge, $\vec X$ is the conjugate
of $\vec{\eusm m}$ by some Weyl transformation $w$. Moreover, any
$w$ is possible.  So again we see that the set $\frak M$ of massive
vacua is a copy of $\W$. Using the definition of the $\W\times \W$
action in terms of monodromy over the parameter space $P$, one can
verify that this action is the left and right action of $\W$ on
itself.

\section{Quivers With Orthogonal And Symplectic Gauge Groups}\label{orthosymp}

In the remainder of this paper, we will extend some of the explicit
constructions to orthogonal and symplectic gauge groups.  As a
preliminary, in this section we extend to the orthogonal and
symplectic case the analysis of good and bad gauge theories and
quivers in section \ref{quivers}. This will also enable us to
describe quivers that are candidates for $T(SO(n))$ and $T(Sp(n))$.
We apply our results to $S$-duality of boundary conditions in
sections \ref{orient} and \ref{bcos}.

\subsection{Orthogonal And Symplectic Gauge Theory}\label{osymp}

The starting point is the general formula (\ref{zanc}) for the
$R$-charge of a monopole operator.  The monopole operator is
defined by specifying a $U(1)$ embedding in the gauge group $G$.
If $h_i$ and $v_i$ are the $U(1)$ charges of hypermultiplets and
vector multiplets, then the monopole operator has $R$-charge
\begin{equation}\label{zzanc}q_R=\frac{1}{2}\left(\sum_i|h_i| - \sum_j
|v_j|\right).\end{equation}

To implement this for $SO(k)$, we associate the $U(1)$ embedding
to a sequence of integer ``eigenvalues'' $a_1,\dots,a_k$; the
nonzero eigenvalues come in pairs of equal magnitude and opposite
sign, since a generator of $\frak{so}(k)$ is conjugate to a sum of
traceless
  $2\times 2$  blocks
\begin{equation}\label{lanc}\begin{pmatrix} 0 & \alpha \\ -\alpha &
0\end{pmatrix}.\end{equation}  Let us couple $SO(k)$ gauge theory
to fundamental hypermultiplets with flavor symmetry $Sp(2n_f)$.
The evaluation of (\ref{zanc}) gives
\begin{equation}\label{planc} q_R= \frac{n_f}{2}\sum_i|a_i|
-\frac{1}{2}\left(\sum_{1\leq i<j\leq k} |a_i-a_j| -\sum_{1\leq
i\leq k} |a_i|\right) .\end{equation} With the $a_i$ coming in
equal and opposite pairs, this is equivalent to
\begin{equation}\label{lanco} q_R=
\frac{n_f+2-k}{2}\sum_i|a_i|+\frac{1}{2}\sum_{i<j}\bigl(|a_i|+|a_j|-|a_i-a_j|\bigr).\end{equation}
This formula has an obvious similarity to (\ref{obanc}), with $k$
playing the role of $2n_c$.  The condition for a good theory is
\begin{equation}\label{kano}n_f\geq k-1.\end{equation}
We call an $SO(k)$ theory balanced if $n_f=k-1$.  In the balanced
case there is a single monopole operator with $q_R=1$, the
sequence of charges being $(1,-1,0,0,\dots,0)$; this leads to an
$SO(2)$ symmetry of the Coulomb branch. An important point is that
because the $a_i$ come in pairs, $q_R$ is always an integer and
there are never free hypermultiplets. We define the excess $e$ of
an $SO(k)$ gauge theory coupled to fundamentals by
\begin{equation}\label{Defex}e=n_f-k+1.\end{equation}

We can make a similar analysis for gauge group $Sp(2t)$ with
fundamental hypermultiplets of flavor symmetry\footnote{The full
flavor symmetry is actually $O(2n_f)$, not just $SO(2n_f)$, but we
primarily consider the connected component except in section
\ref{irflow}.  Classically, we could take flavor symmetry of the
form $O(2m+1)$. An anomaly would then force us to incorporate a
half-integral Chern-Simons interaction for the gauge fields,
modifying the properties of the monopole operators.} $SO(2n_f)$. The
analog of (\ref{planc}) is
\begin{equation}\label{plancop} q_R= \frac{n_f}{2}\sum_i|a_i|
-\frac{1}{2}\left(\sum_{1\leq i<j\leq 2t} |a_i-a_j| +\sum_{1\leq
i\leq 2t} |a_i|\right), \end{equation} leading to
\begin{equation}\label{olanco} q_R=
\frac{n_f-2t}{2}\sum_i|a_i|+\frac{1}{2}\sum_{i<j}\left(|a_i|+|a_j|-|a_i-a_j|\right).\end{equation}
The condition for a good theory is now
\begin{equation}\label{zano} n_f\geq 2t+1,\end{equation}
and we call a theory balanced if $n_f=2t+1$.  Again, a balanced
theory has a single monopole operator with $q_R=1$, with the same
sequence of charges  as before, leading to an $SO(2)$ symmetry of
the Coulomb branch. For the same reason as for orthogonal gauge
groups, there is no value of $n_f$ at which a free hypermultiplet
appears. We define the excess $e$ of an $Sp(2t)$ gauge theory with
fundamentals as
\begin{equation}\label{anc}e=n_f-2t-1.\end{equation}

\subsection{Orthosymplectic Quivers}\label{orquiv}
\begin{figure}
  \begin{center}
    \includegraphics[width=4.5in]{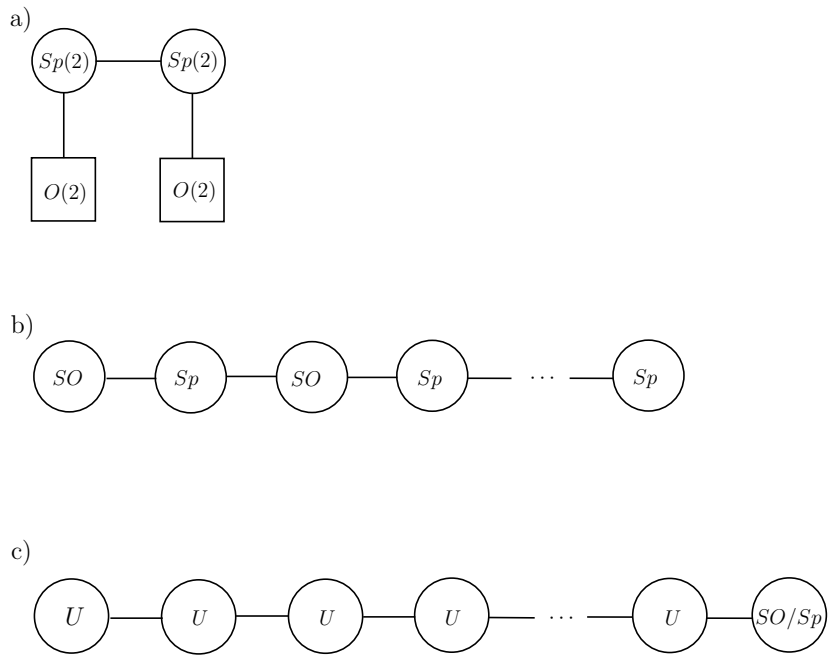}
  \end{center}
\caption{\small (a) A quiver with two adjacent balanced symplectic
nodes is bad, as explained in the text. (b) An orthosymplectic
quiver, with alternating orthogonal and symplectic nodes, each of
which has $e_i\geq 0$, is good. (c) A chain of unitary nodes
terminating at one end with an orthogonal or symplectic node is
good provided each node has $e_i\geq 0$. }
  \label{Fig53}\end{figure}

We now want to study linear quivers with unitary, orthogonal and
symplectic gauge groups. (For some examples, see fig.
\ref{Fig53}.) To each link in such a quiver, we attach
bifundamental hypermultiplets as usual; in the case of adjacent
$SO$ and $Sp$ nodes,  we place a reality condition on the
hypermultiplets, halving  the number of components.

We have already defined a notion of excess $e_i$ for each kind of
node. We also define $\epsilon_i=1$ for unitary nodes, and
$\epsilon_i=1/2$ for orthogonal and symplectic nodes. Also,
because of the reality condition on the bifundamentals, we define
$\epsilon_{i,i+1}$ to be $1/2$ for a link connecting $SO$ and $Sp$
nodes, and $1$ for any other link.

The general formula for the $R$-charge of a monopole operator is
\begin{equation} q_R = \sum_i \left(\Delta_i + \epsilon_i A_i +
\epsilon_{i,i+1} B_i\right),
\end{equation} where as in section \ref{quivers},
\begin{align}\Delta_i&=\frac{ e_i}{2} \sum_{k=1}^{n_i} |a_{i,k}|
\cr A_i&=\frac{1}{2} \sum_{k=1}^{n_i} \sum_{t=1}^{n_i} \left(
|a_{i,k}| + |a_{i,t}|-|a_{i,k} - a_{i,t}|\right) \cr B_i&=-
\frac{1}{2} \sum_{k=1}^{n_i} \sum_{t=1}^{n_{i+1}} \left( |a_{i,k}|
+ |a_{i+1,t}|- |a_{i,k} - a_{i+1,t}|\right). \end{align}

Because of the $\epsilon$ factors, it is not so that a linear quiver
with all nodes obeying $e_i \geq 0$ is automatically good. A simple
counterexample is the quiver in fig. \ref{Fig53}(a), which has two
adjacent balanced symplectic nodes. The two $e_i$ vanish, the two
$A_i$ equal $2$ and the single $B_i$ equals $-2$. Because of the
$\epsilon_i$ factors, $q_R=0$ if the charges at the two nodes are
equal, and the quiver is bad.

Nevertheless, certain classes of quivers do have the property that
if $e_i\geq 0$ for each node, then $q_R\geq 1$.  In the rest of
this paper, two classes of such good quivers will play an
important role.  These classes, which are suggested by brane
constructions with orientifolds, are as follows.

One is what we will call an ``orthosymplectic'' quiver, that is a
linear quiver of alternating orthogonal and symplectic nodes (fig.
\ref{Fig53}(b)).  The second is the case of a linear quiver of
unitary groups terminating at one end (or at each end) with an
orthogonal or symplectic group (fig. \ref{Fig53}(c)).

First we consider orthosymplectic quivers.  In this case, the
$\epsilon_i$ and $\epsilon_{i,i+1}$ all equal 1/2.  This factor of
1/2 multiplies the sum $\sum_i(A_i+B_i)$ considered in section
\ref{quivers}.  So the key inequalities of that section, such as
(\ref{powb}), have immediate analogs:
\begin{equation}\label{lopowb} q_R\geq \sum_{i=1}^{P-1}\frac{ e_i}{2} \sum_{m=1}^{n_i}
|a_{i,m}| +\frac{1}{2}\sum_{m=1}^{n_1}
|a_{1,m}|(n_1-m+1)+\frac{1}{2}\sum_{m=1}^{n_{P-1}}|a_{P-1,m}|(n_{P-1}-m)
\end{equation}
(There are $P-1$ nodes in the quiver. The group at the $i^{th}$
node is assumed to be $SO(n_i)$ or $Sp(n_i)$; $a_{i,s}$ are the
charges at the $i^{th}$ node.) There is also an analog of
(\ref{plowb}) in which one of the nodes, say the $s^{th}$ one, is
singled out:
\begin{equation}\label{odzowb}q_R\geq \sum_{i=1}^{P-1}\frac{ e_i}{2} \sum_{m=1}^{n_i}
|a_{i,m}| +\frac{1}{2}\sum_{m=1}^{n_1}
|a_{1,m}|(n_1-m)+\frac{1}{2}\sum_{m=1}^{n_s}|a_{s,m}|+\frac{1}{2}\sum_{m=1}^{n_{P-1}}|a_{P-1,m}|(n_{P-1}-m)
\end{equation}
The close relation of these formulas to the corresponding formulas
for unitary quivers reflects the fact that an orthosymplectic
quiver can be obtained as a $\Z_2$ orbifold of a quiver of unitary
groups. One divides by a $\Z_2$ that reduces each $U(n)$ gauge
group to $SO(n)$ or $Sp(n)$. The orthogonal and symplectic groups
must alternate along the chain in order that the $\Z_2$ action can
be defined for the bifundamental hypermultiplets.

It immediately follows from these formulas that for a quiver of
this kind with all $e_i\geq 0$, non-trivial monopole operators
have $q_R>0$. Allowing for the fact that the non-zero $a_{i,s}$
are paired, the bound is $q_R\geq 1$, so there are no free
hypermultiplets.

Furthermore, we can describe all of the operators of $q_R=1$ and
therefore the symmetries of the Coulomb branch.  From
(\ref{odzowb}), it follows that the charges vanish at any node
with $e_i>0$.  So it suffices to consider a chain of $p$ balanced
nodes. From (\ref{odzowb}), to get $q_R\leq 1$, at each node, the
monopole charges either vanish or are $(1,-1,0,\dots,0)$.  Just as
in section \ref{quivers}, if the subquiver of nodes with nonzero
charges is disconnected, then $q_R\geq 2$.  So to get $q_R=1$, we
must have a connected subquiver on which the monopole charges are
$(1,-1,0,\dots,0)$, with all charges vanishing outside this
subquiver.  Conversely, all these operators have $q_R=1$.  So
there are a total of $p(p+1)/2$ monopole operators of $q_R=1$. (An
exception to this counting is mentioned shortly.) This is the
dimension of the Lie group $SO(p+1)$, and we claim that is indeed
the symmetry associated with the monopole operators.

If there is a single balanced node, the symmetry group is clearly
$SO(2)$, as there is a single monopole operator. For any two
consecutive balanced nodes, there are three monopole operators. A
detailed computation of three point functions on the sphere would be
needed to show directly that these three operators define an $SO(3)$
algebra, as opposed to  $SO(2)\times SO(2)\times SO(2)$. In any
event, this follows from the orthosymplectic mirror symmetry
construction of \cite{Feng:2000eq}.

Once one knows that the symmetry for two consecutive balanced
nodes is $SO(3)$, one can reason by induction and build up an
$SO(p+1)$ symmetry for a chain of $p$ consecutive balanced nodes.
Suppose that this is so for some value of $p$. Adding a $p+1^{th}$
balanced node adds $p+1$ monopole operators, which must transform
in a $p+1$-dimensional representation of the group $SO(p+1)$ that
is already present. This representation must be non-trivial,
because of the hypothesis that two adjacent balanced nodes (one of
which is the new one) generates $SO(3)$ symmetry. Hence it must be
the irreducible $p+1$-dimensional representation, which combines
with the $SO(p+1)$ from the first $p-1$ nodes to generate
$SO(p+2)$.

In one important case, the above counting needs modification.
Suppose that the gauge symmetry of one of our balanced nodes is
$SO(2)$. Such a node can only appear at the end of a quiver, as
there is no way to divide its $Sp(2)$ flavor symmetry.  The group
$SO(2)\cong U(1)$ is abelian, so in a theory with $SO(2)$ gauge
symmetry, there is a classical symmetry of the Coulomb branch, the
shift of the dual photon. Moreover, as $SO(2)$ has no Weyl group,
the elements
\begin{equation}\label{delf}\begin{pmatrix}0&1\\ -1 & 0
\end{pmatrix},~~\begin{pmatrix}0&-1\\ 1 & 0
\end{pmatrix}\end{equation}
of $\frak{so}(2)$ are not conjugate and correspond to distinct
homomorphisms $\frak{u}(1)\to\frak{so}(2)$.  Hence there are two
different monopole operators of $q_R=1$, roughly with positive or
negative monopole charge, at a balanced $SO(2)$ node. The group
associated with the classical symmetry and the two monopole
operators is $SO(3)$. Indeed, $SO(2)$ with $Sp(2)$ flavor symmetry
is equivalent to $U(1)$ with $n_f=2$, and so was one of the basic
examples in section \ref{hyperex}.

Now we claim that given a chain of $p$ balanced orthogonal and
symplectic nodes, of which the first has $SO(2)$ gauge symmetry,
the monopole operators and the classical symmetry of the Coulomb
branch generate an $SO(p+2)$ symmetry.  We have already
established the case of $p=1$.  Proceeding by induction in $p$,
adding a $p+1^{th}$ balanced node adds now $p+2$ monopole
operators, transforming non-trivially under $SO(p+2)$, and so
extending the symmetry to $SO(p+3)$.

Similarly, an orthosymplectic chain of $p$ balanced nodes, with
$SO(2)$ at each end, gives $SO(p+3)$ symmetry.

\subsubsection{Some Significant Examples}\label{somex}
\begin{figure}
  \begin{center}
    \includegraphics[width=5.5in]{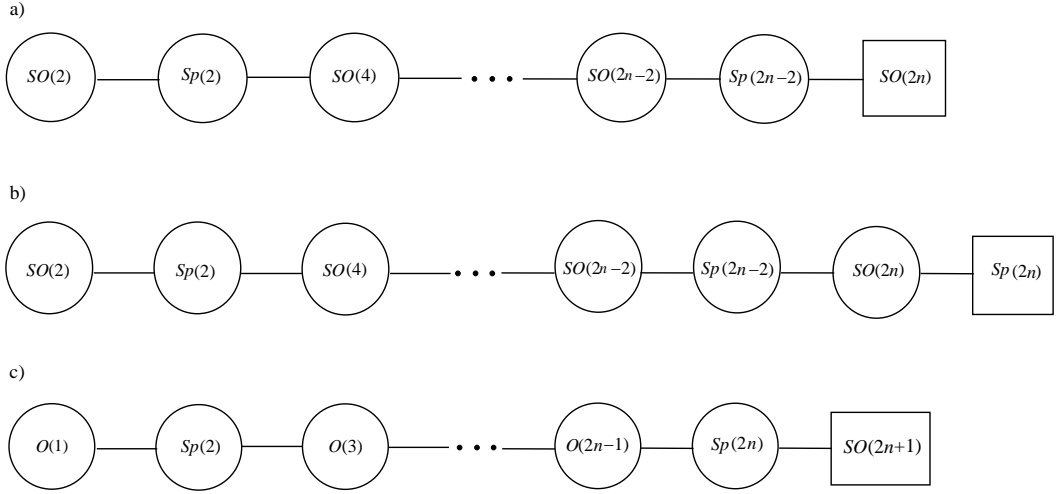}
  \end{center}
\caption{\small (a) A balanced chain
 ending with flavor symmetry
$SO(2n)$. (b) A similar balanced chain ending with flavor symmetry
$Sp(2n)$.  (c) A bad quiver  whose Higgs branch is the nilpotent
cone of $SO(2n+1)$. }
  \label{Fig54}\end{figure}

We will illustrate this idea with some significant examples.  In
fig. \ref{Fig54}(a), we show a balanced orthosymplectic quiver with
fundamental matter multiplets only at the last node.  The sequence
of groups is $SO(2)-Sp(2)-SO(4)-Sp(4)-\cdots -Sp(2n-2)$.  The flavor
symmetry at the last node is $SO(2n)$, and this is the classical
symmetry of the Higgs branch.
 The symmetry of the Coulomb branch resulting from a chain
of $2n-2$ balanced orthogonal and symplectic groups beginning with
$SO(2)$ is also $SO(2n)$.  A short computation shows that the
complex dimension of either the Higgs or the Coulomb branch is
$2n(n-1)$, which is the dimension of the nilpotent cone $\N$ of
the self-dual group $SO(2n)$.  These facts suggest that the IR
limit of this quiver describes $T(SO(2n))$. Indeed, it can be
shown along the lines of \cite{Feng:2000eq} and our arguments in
section \ref{examples} that this quiver is self-mirror and
describes the $S$-dual of Dirichlet boundary conditions for
$SO(2n)$.  We defer these matters to section \ref{bcos}.
Moreover, one can show directly by adapting\footnote{The Higgs
branch of $SO(2k)$ gauge theory coupled to fundamental
hypermultiplets with flavor symmetry $Sp(4k-2)$ is the same as the
Higgs branch of $O(2k-1)$ gauge symmetry with fundamental
hypermultiplets of the same flavor symmetry.  One can make this
substitution for all of the orthogonal nodes in fig.
\ref{Fig54}(a) or (b).  After doing so, the equivalence of the
Higgs branches of these quivers to the nilpotent cones of $SO(2n)$
and $Sp(2n)$ is a special case of the results of \cite{KobakSwann}
(see also a brief summary in \cite{KobakSwann2}).} the arguments
of \cite{KobakSwann} that the Higgs branch of this quiver is the
nilpotent cone of $SO(2n)$.

In fig. \ref{Fig54}(b), we show a similar balanced quiver but
continued one step farther. Now, the symmetry of the Higgs branch
is $Sp(2n)$ rather than $SO(2n)$.  The symmetry of the Coulomb
branch -- derived from $2n-1$ balanced nodes, the first group
being $SO(2)$ -- is $SO(2n+1)$.  This is the dual group of
$Sp(2n)$.  The Higgs and Coulomb branches have the right
dimensions to match the nilpotent cones $\N$ and $\N^\vee$ of
$T(Sp(2n))$.  Indeed, as we explain in section \ref{bcos}, this
theory describes the dual of Dirichlet boundary conditions for
$Sp(2n)$.  Again, the Higgs branch can be analyzed by adapting
arguments in \cite{KobakSwann} and does coincide with the
nilpotent cone of $Sp(2n)$.

As $SO(2n+1)$ is the dual group of $Sp(2n)$, the theory
$T(SO(2n+1))$ is the mirror of $T(Sp(2n))$.  It would be nice to
have a direct construction of $T(SO(2n+1))$, as opposed to its
mirror.  However, it seems that there is no good quiver that flows
in the IR to $T(SO(2n+1))$.  The quiver of fig. \ref{Fig54}(c),
based on the sequence $O(1)-Sp(2)-O(3)-Sp(4)-\dots$,  is shown in
\cite{KobakSwann} to have a Higgs branch that
coincides\footnote{\label{zurk} For this result to hold classically,
it is essential that the orthogonal gauge groups in the quiver are
ordinary orthogonal groups $O(k)$, as indicated, rather than special
orthogonal groups $SO(k)$.  See section \ref{irflow}.} with the
nilpotent cone of $SO(2n+1)$. Moreover, its Coulomb branch has the
same dimension as the nilpotent cone of $Sp(2n)$. However, this is
not a good quiver, as the symplectic nodes are imbalanced. So it
does not have a standard IR limit.  We argue in section \ref{bcos}
that this quiver theory flows to $T(SO(2n+1))$ in the infrared, but
that the $R$-symmetry that is realized in the infrared is not the
one that one sees in the underlying gauge theory.

\subsubsection{Infrared Flow Of $T(SO(3))$}\label{irflow}

For $SO(3)$, we can argue this directly. First we ignore the $O(1)$
gauge symmetry of the quiver and consider an $Sp(2)$ (or $SU(2)$)
theory with flavor symmetry $SO(4)$. Classically, the Higgs branch
$\H$ consists of two copies\footnote{\label{zry} See the analysis of
eqn. (3.4) in \cite{Seiberg:1994aj}, where this Higgs branch arises
in the study of a related four-dimensional model. From the point of
view of \cite{Seiberg:1996bs}, the Higgs branch is the one-instanton
moduli space on $\R^4$ with structure group the flavor group
$SO(4)$, and has two components because the instanton can be in
either factor of $SO(4)\cong SU(2)\times SU(2)$.} of the $\tx A_1$
singularity $\R^4/\Z_2$. The Coulomb branch of the model is
$\C=(\R^3\times S^1)/\Z_2$ \cite{Seiberg:1996nz}. Here $\Z_2$ acts
as a reflection of both $\R^3$ and the circle $S^1$. $\C$ has two
singularities (coming from the fixed points of the reflection of
$S^1$), each of which is an $\tx A_1$ singularity. At each
singularity of $\C$, $\C$ meets precisely one of the two components
of $\H$. (This structure was found in \cite{Seiberg:1994aj} in a
related four-dimensional model.  It is true in three dimensions for
similar reasons, as we explain below.)

Let $\M$ be the moduli space of vacua of the theory.  Near either
singularity of $\C$, $\M$ looks like two intersecting $\tx A_1$
singularities -- one being $\C$, and the other being the relevant
component of $\H$.  Either of these intersections gives the familiar
picture of $T(SU(2))$, or equivalently $T(SO(3))$.  However, the
$R$-symmetry that is part of the superconformal structure of
$T(SO(3))$ is not the microscopic $R$-symmetry of the underlying
$Sp(2)$ gauge theory. The microscopic $R$-symmetry is of course an
exact symmetry of $\C=(\R^3\times S^1)/\Z_2$, and comes from the
rotation of $\R^3$. Near either of the $\tx A_1$ singularities of
$\C$, the $SO(3)$ symmetry of $\C$ is enhanced to
$SO(4)/\Z_2=SO(3)_1\times SO(3)_2$.  Of these two factors, one of
them, say $SO(3)_1$, is the superconformal $R$-symmetry, and the
other, say $SO(3)_2$, is the expected $SO(3)$ global symmetry that
acts on the Coulomb branch of $T(SO(3))$.  The microscopic
$R$-symmetry is a diagonal subgroup of $SO(3)_1\times SO(3)_2$. (The
structure is the same as we described in footnote \ref{sasaki} for
the free vector multiplet, and similar to what we will find in
section \ref{splitting} for a certain splitting process involving
branes.)

The $Sp(2)$ gauge theory with $SO(4)$ flavor symmetry can flow to
$T(SO(3))$ in two different ways, since we have to pick one of the
singularities of the Coulomb branch $\C$.   The bad quiver of fig.
\ref{Fig54}(c) actually does not have this ambiguity, since the
gauge group is not quite $Sp(2)$ but $O(1)\times Sp(2)=\Z_2\times
Sp(2)$.  Here the $\Z_2$ factor exchanges the two components of $\H$
and the two singularities of $\C$.  Thus, after gauging this extra
$\Z_2$, there is only one singularity at which the bad quiver flows
to $T(SO(3))$.

To explain the claim about the action of $O(1)$, we note the
following. BPS monopoles of the $Sp(2)$ gauge symmetry appear in
this theory as instantons. In an instanton field, each
hypermultiplet flavor has a zero mode. The effective action of the
instanton field is roughly $\exp(i\phi)q_1q_2q_3q_4$, where $\phi$
is the dual photon, and $q_i$ is a fermion of the $i^{th}$ real
hypermultiplet.  This effective action has $SO(4)$ flavor symmetry,
but it does not have $O(4)$ flavor symmetry.  However, the
disconnected component of $O(4)$ is a symmetry if combined with a
$\pi$ shift in $\phi$.  Let $\Theta$ be  the product of a $\pi$
shift of $\phi$ and an element ${\rm diag}(-1,1,1,1)$ of $O(4)$.

To determine the action of $\Theta$, we note that $\phi$
parametrizes the $S^1$ in $\C=(\R^3\times S^1)/\Z_2$.  Hence a
$\pi$ shift of $\phi$ exchanges the two singularities of $\C$.
Also, classically, an $O(4)$ transformation of determinant $-1$
exchanges the two components of $\H$.  (See again the discussion
of eqn. (3.4) in \cite{Seiberg:1994aj}.  Alternatively, in the
instanton interpretation mentioned in footnote \ref{zry}, the two
components correspond to the two factors of $SO(4)\cong
SU(2)\times SU(2)$, which are exchanged by a reflection in
$O(4)$.) Thus, $\Theta$ exchanges the two singularities of $\C$
and the two components of $\H$ that meet these two singularities.
After dividing by $\Theta$, this bad quiver gauge theory has only
one singular point with a flow to $T(SO(3))$.

To complete the story for this bad quiver, we simply note that the
non-trivial element of the $O(1)$ gauge symmetry of this quiver
acts as $-1$ on just one of the four real hypermultiplets, so it
indeed corresponds to $\Theta$.

We will not make a similarly detailed analysis of the bad quivers
of higher rank.  We just note that for any $k$, the $Sp(2k)$
theory with $O(4k)$ flavor symmetry has many properties in common
with the example just described: the Higgs branch has two
components, exchanged by a flavor transformation of determinant
$-1$, and meeting the Coulomb branch on different loci.

\subsection{Unitary Quivers With Orthosymplectic Groups At The
End}\label{uniorth}

\begin{figure}
  \begin{center}
    \includegraphics[width=4.5in]{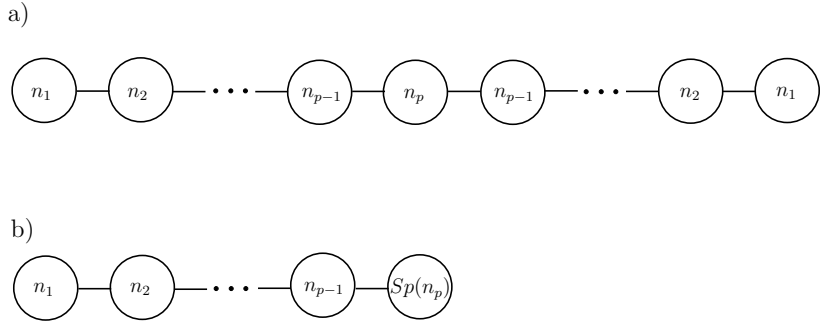}
  \end{center}
\caption{\small (a) A symmetric quiver of unitary groups. (A
circle labeled by an integer $k$ represents a $U(k)$ gauge group;
other groups will be indicated explicitly.)  (b) Its $\Z_2$
orbifold gives a quiver of unitary groups ending in an orthogonal
or symplectic group.   The symplectic case is depicted. }
  \label{Fig55}\end{figure}

Now we consider the situation of fig. \ref{Fig53}(c): a linear
quiver of $P-1$ unitary gauge groups followed by a single
orthogonal or symplectic node. Like orthosymplectic quivers, these
can arise as $\Z_2$ orbifolds of a quiver of unitary groups. In
this case, the $\Z_2$ symmetry must exchange the two ends of the
quiver (fig. \ref{Fig55}).

The orbifold interpretation leads us to expect that there is a
very simple inequality for the total $R$-charge. Just as in
section \ref{quivers}, the $R$-charge $q_R$ of a monopole operator
of charges $a_{i,k}$ is a sum of separate contributions $q_R^+$
and $q_R^-$  from those $a$'s that are positive or negative,
respectively:
\begin{equation}\label{pfulo}q_R=q_R^++q_R^-.\end{equation}

Each can be bounded as before. Indeed, the analog of eqn.
(\ref{plowb}) is the assertion that
\begin{equation}\label{plowbox} q_R^+\geq \sum_{i=1}^{P}\frac{ e_i}{2} \sum_{k|a_{i,k}\geq 0}
a_{i,k} +\sum_{k|a_{1,k}\geq 0} a_{1,k}(n_1-k)+\sum_{k|a_{s,k}\geq
0}^{n_s}a_{s,k}-\frac{1}{2}\sum_{k|a_{P,k}\geq
0}a_{P,k}\end{equation} for any choice of $s$ with $1\leq s\leq
P$. $q_R^-$ obeys a similar inequality with contributions from
negative $a_{i,k}$.

Setting $s=P$, we see immediately that if all $e_i$ are
nonnegative, then a monopole operator with nonzero charge at the
orthogonal or symplectic node has $q_R^\pm\geq 1/2$ and hence
$q_R\geq 1$.  This inequality also holds if the charges vanish at
the last node, in view of our previous results for unitary
quivers.

Now let us analyze the monopole operators of $q_R=1$ and hence the
symmetry of the Coulomb branch.  As usual, to get $q_R=1$,
monopole charges must vanish at any node with $e>0$.  So it
suffices to consider a quiver of $P$ nodes with $e_i=0$ at each
node.

Monopole operators whose charge vanishes at the $P^{th}$ node will
generate an $SU(P)$ symmetry, the usual result for a chain of $P-1$
balanced unitary nodes.  Now let us consider monopole operators that
have nonzero charge at the $P^{th}$ node.  The inequality
(\ref{plowbox}) with $s=P$ implies that to get $q_R=1$, the charge
at the $P^{th}$ node must be $(1,-1,0,\dots,0)$.  Moreover
$q_R^\pm\geq 1/2$, so to get $q_R=1$, we need $q_R^+=q_R^-=1/2$.
Considering the inequality with arbitrary $s$, we find that to get
$q_R^\pm=1/2$, the positive monopole charges at the $s^{th}$ node
must be $(1,0,0,\dots,0)$, up to permutation, and likewise the
negative charges must be of the same form, up to permutation.
Furthermore, by a familiar argument, the subquiver supporting the
positive charges must be connected, and likewise the subquiver
supporting the negative charges must be connected.  Conversely, when
all these conditions are imposed, we do get a monopole operator of
$q_R=1$.

Postponing for the moment some exceptions associated with $U(1)$
and $SO(2)$, the conditions just described give $P\times P=P^2$
monopole operators of $q_R=1$ with charge at the $P^{th}$ node.
How these operators transform under $SU(P)$ is completely
determined by how they transform under the maximal torus of
$SU(P)$, which is generated by the classical symmetries of the
Coulomb branch.  The nonzero weights that arise are differences
between weights of the fundamental representation and weights of
its conjugate (associated with positive and negative $a$'s,
respectively), so the monopole operators transform as the sum of
the adjoint representation of $SU(P)$ and a one-dimensional
trivial representation.

This is enough to ensure that the symmetry of the Coulomb branch
is (locally) $SU(P)\times SU(P)\times U(1)$.

Two exceptions should be pointed out. If a balanced $U(1)$ node is
present,  there is no room for a monopole operator to have charges
$(1,-1,0,\dots,0)$ at that node. A balanced $U(1)$ node must be at
the left end of the quiver, so the only $q_R=1$ monopole operator
that is removed by this limitation is the one that has both positive
and negative charges at every node. The absence of this operator
reduces the symmetry group to $SU(P) \times SU(P)$.

The other exception arises if the orthogonal or symplectic group
at the $P^{th}$ node is $SO(2)$.  The only balanced linear quiver
with this property involves the sequence of groups $U(1)-SO(2)$
for $P=2$.  The fact that $SO(2)$ is abelian results in an
enhancement of the symmetry of the Coulomb branch from
$SU(2)\times SU(2)$ (as suggested by the generic analysis) to
$Sp(4)$.  The Coulomb branch of this model has hyper-Kahler
dimension 2, and we suspect that it is isomorphic to $\BC^4/\Z_2$.

The hidden symmetries of the Coulomb branch that are associated with
monopole operators can be seen using mirror symmetry if the special
node is symplectic. They are classical symmetries of the mirror
quivers, which we describe next. Mirror symmetry for these quivers
was analyzed in \cite{Feng:2000eq}, and will be considered in
section \ref{morex}.

\def\CC{{\frak C}}

\subsection{Bifurcated Quivers}\label{bifurc}

So far all our results, both in section \ref{quivertheories} and
here, have involved linear quivers.  But some of the results have
close analogs for quivers of other types.  These analogs will be
important in the rest of the paper, when we include orbifolds and
orientifolds.

We start with the basic question of understanding a general quiver
of unitary gauge groups with only balanced nodes. Requiring that
every node of a unitary quiver is balanced is actually quite
restrictive, and there is a nice classification of such quivers.
Consider an arbitrary graph $\Gamma$ in which any two nodes are
connected by at most one line.\footnote{The argument will show
that this condition can be omitted, since without it the Cartan
matrix cannot be positive definite.} Let $N$ be the set of nodes,
and let $E$ be the set of edges, that is, the set of pairs of
points in $N$ that are connected by a line. A quiver and its
associated gauge theory are defined as follows. To every node
labeled by $p_i\in N$ we attach a positive integer $n_i$, the rank
of the group $U(n_i)$ that we attach to that node. For each edge
connecting points $p_i$ and $p_j$, we attach a bifundamental
hypermultiplet of $U(n_i)\times U(n_j)$. Finally, we assign $m_i$
fundamental hypermultiplets to the $i^{th}$ node. Of course, the
$m_i$ must be non-negative.

The condition for every node to be balanced is that
\begin{equation} 2 n_i - \sum_{j|(i,j)\in E} n_j = m_i
\end{equation} for all $i$.
It is convenient to express this condition in terms of the Cartan
matrix $\CC$ of the graph $\Gamma$.  $\CC$ is a matrix that acts on
a vector space $V$ that has a basis element $v_i$ for each node
$p_i\in N$.  We give $V$ a metric $(~,~)$ in which the $v_i$ are
orthonormal. In that basis, the Cartan matrix is
$\CC_{ij}=2\delta_{ij}-e_{ij}$, where $e_{ij}$ is 1 if nodes $p_i$
and $p_j$ are connected by a link, and zero otherwise.  The Cartan
matrix is a discrete version of the one-dimensional Laplace operator
$\Delta=-d^2/dx^2$, and like $\Delta$, it is real and symmetric.

In terms of the Cartan matrix, setting $\eurm n=\sum_i n_iv_i$ and
$\eurm m=\sum_i m_iv_i$, the condition for every node to be balanced
reads
\begin{equation}\label{turxo} \CC\eurm n=\eurm m.\end{equation}

Like any real, symmetric matrix, $\CC$ can be diagonalized with real
eigenvalues. $\CC$ shares with $\Delta$ the property that its
eigenvector $\eurm q$ with the lowest eigenvalue is unique up to a
scalar multiple and can be chosen to have all entries positive,
$\eurm q=\sum_iq_iv_i$ with all $q_i>0$. Suppose that $\CC\eurm
q=\lambda\eurm q$.  We have $\lambda(\eurm q,\eurm n)=(\CC\eurm
q,\eurm n)=(\eurm q,\CC\eurm n)=(\eurm q,\eurm m)$. Here $\eurm q $
and $\eurm n$ have positive coefficients, and $\eurm m$ has
non-negative coefficients, so the inner products involved are
positive, except that $(\eurm q,\eurm m)=0$ if $\eurm m=0$. We
deduce that $\lambda\geq 0$ and if $\lambda=0$ then $\eurm m=0$.

Since $\lambda$ was defined as the smallest eigenvalue of $\CC$, if
$\lambda>0$ then $\CC$ is positive definite. The graphs $\Gamma$
with positive-definite Cartan matrix are nothing else than the
${\textsf{ADE}}$ Dynkin diagrams.

If the smallest eigenvalue of $\CC$ is zero, then $\Gamma$ is the
extended Dynkin diagram of a group of $\textsf{ADE}$ type. Including
this case does not add much, because the requirement that $\eurm
m=0$ means that one $U(1)$ subgroup of the gauge group is decoupled.
We may as well ungauge the extended node of the Dynkin diagram,
reducing to the case that $\Gamma$ is a Dynkin diagram rather than
an extended Dynkin diagram.  This procedure gives a convenient
example of a balanced quiver for every $\textsf{ADE}$ diagram: start
with the extended Dynkin diagram with $\eurm m=0$ and with $\eurm n$
annihilated by $\CC$, and ungauge the extended node.

So we have shown that unitary quivers with all nodes balanced are
associated with $\textsf{ADE}$ Dynkin diagrams, and conversely
that for every choice of an $\textsf{ADE}$ diagram, there are
balanced quivers.

If we let $Q$ be a  quiver  based on a graph $\Gamma$ of type $G$,
where $G$ is any $\textsf{ADE}$ group, it is natural to suspect that
the Coulomb branch of the gauge theory associated with $Q$ (assuming
that it has a standard IR limit) has $G$ symmetry.  This would
generalize what we found in section \ref{quivertheories} for a
quiver of type $\textsf{A}$. Actually, the result for any $G$
essentially follows from the case of type $\textsf{A}$. For any
$\Gamma$, the construction of section \ref{quivertheories}
associates an $SU(2)$ symmetry of the Coulomb branch to every
balanced node. Moreover, for two balanced nodes that are not
adjacent, the two $SU(2)$'s commute, and for any two adjacent nodes,
the two $SU(2)$'s fit into an $SU(3)$ symmetry. By the usual
structure theory of Lie groups, it follows that the $SU(2)$'s
associated with all of the nodes generate together a group of type
$G$.

We define the excess of a quiver by $\eurm e=\CC\eurm n-\eurm m$. A
further conjecture along the lines of section \ref{quivertheories}
would assert that if we are given a quiver of unitary groups in
which each node is good, in the sense that the coefficients of
$\eurm e=\sum_ie_iv_i$ are all nonnegative, then the whole quiver is
good in the sense that all monopole operators satisfy $q_R\geq 1$.
Given this, we would hope that gauge theories associated with good
quivers flow to standard IR limits.
 The symmetry of the Coulomb branch would
then presumably be the product of simply-laced Lie groups
corresponding to the various balanced subquivers, times a $U(1)$
factor for each node with positive $e$.  We are not in a position
to prove this general conjecture.

\begin{figure}
  \begin{center}
    \includegraphics[width=2.5in]{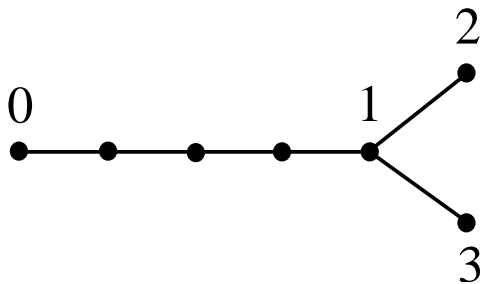}
  \end{center}
\caption{\small A ``bifurcated''  or $\tx D_n$ quiver.  Some nodes
that play an important role in the text are labeled.  The
bifurcation occurs at node 1, the short branches consist of nodes
2 and 3, and the outermost node of the long branch is labeled 0.}
  \label{Fig56}\end{figure}

In this subsection, we focus on quivers in the shape of a
$\textsf{D}_n$ diagram, or ``bifurcated'' quivers, as in fig.
\ref{Fig56}. We will assume that all nodes are good and we aim to
show that the whole quiver is good. The charge $q_R$ of a monopole
operator can be manipulated as usual: the contributions from
positive and negative charges can be separated, and one can restrict
to connected subquivers with nonzero charge at every node. The
contribution from negative charges equals the contribution from an
identical set of positive charges.

Let us consider the contribution to $q_R$ from the positive monopole
charges.  It is  the sum of three kinds of term.  At each node we
have contributions $\Delta_i=\frac{ e_i}{2} \sum_{k=1}^{n_i}
a_{i,k}$ and  $A_i=\sum_{k=1}^{n_i} a_{i,k} (2 n_i - 2k +1)$ (the
$a_i$ are arranged here in nondecreasing order). At each link there
is a contribution $B_{ij}=- \frac{1}{2} \sum_{k=1}^{n_i}
\sum_{t=1}^{n_{j}} \,2\,\,\mathrm{min}(a_{i,k},a_{j,t})$.  Our study
of linear quivers was based on inequalities such as
\begin{equation}\label{lowbo} B_{ij}\geq  -\sum_{k=1}^{n_i} a_{i,k} (n_i - k ) -
\sum_{t=1}^{n_{j}} a_{j,t} (n_j - t +1). \end{equation} The right
hand side is comparable to $-\frac{1}{2}A_i-\frac{1}{2}A_j$,
enabling us to bound  $\sum_{ij}B_{ij}$ in terms of $\sum_iA_i$.
This worked only because there were at most two $B$ terms
contributing at each node. As a result, we had a useful set of
inequalities for $q_R$ involving a sum of positive terms.

If we blindly use this method on a bifurcated quiver, the $A$ at
the node with three neighbors would not be sufficient to cancel
negative contributions from the three corresponding $B$ terms,
leaving a net negative contribution to the inequality. We need a
new trick, which fortunately is quite simple. As in the figure,
let us label the node with three neighbors as $i=1$, and the nodes
at the ends of the short branches of the quiver as $i=2$ and
$i=3$. Without loss of generality, assume $n_2\geq n_3$. Roughly,
we want to bound $B_{12}+B_{13}$ relative to  $-(\frac{1}{2} A_1 +
A_2 + A_3)$, after which there will still be a positive
contribution $\frac{1}{2} A_1$ to help in canceling the
contribution from the $B$'s on the long branch of the quiver.

Let us introduce a fictitious node labeled $c$ with $n_c=n_2+n_3$,
and define the sequence of charges $a_{c,t}$, $1\leq t\leq n_c$ at
this node by the following: $a_{c,t}=a_{2,t}$ if $t\leq n_2-n_3$,
while otherwise $a_{c,n_2-n_3+2t-1} =a_{3,t}$  and
$a_{c,n_2-n_3+2t} =a_{2,n_2-n_3+t}$. This definition ensures that
every $a_{2,t}$ and $a_{3,t}$ appears precisely once in the
sequence $a_{c,t}$. Hence we can write
\begin{equation} B_{1c}=B_{12}+B_{13} =- \frac{1}{2}
\sum_{k=1}^{n_1} \sum_{t=1}^{n_2+n_3}
\,2\,\,\mathrm{min}(a_{1,k},a_{c,t})
\end{equation}
Even though the  $a_{c,t}$ are not ordered, the method used to
derive the basic inequality (\ref{lowb}) or (\ref{ozowb}) is
perfectly valid, and gives
\begin{equation}\label{helpful} B_{1c}\geq  -\sum_{k=1}^{n_1} a_{1,k} (n_1 - k +1 ) -
\sum_{t=1}^{n_2+n_3} a_{c,t} (n_2+n_3 - t ). \end{equation} The
first term is familiar, and will be canceled against
$\frac{1}{2}A_1$. The second term receives contributions from both
the $2$ and $3$ nodes, which need to be disentangled:
\begin{align}\notag
-\sum_{t=1}^{n_2+n_3} a_{c,t} (n_2+n_3 - t) =&
-\sum_{t=1}^{n_2-n_3} a_{2,t} (n_2+n_3 - t) -\sum_{t=1}^{n_3}
a_{3,t} (n_2+n_3 - (n_2-n_3+2t-1))\\ &-\sum_{t=n_2-n_3+1}^{n_2}
a_{2,t} (n_2+n_3 - (2t - n_2+n_3)).
\end{align}
Finally, we have
\begin{equation}\label{firstone} B_{1c}\geq  -\sum_{k=1}^{n_1} a_{1,k} (n_1 - k +1 ) -
\sum_{t=1}^{n_3} a_{3,t} (2n_3 - 2t+1 )- \sum_{t=1}^{n_2} a_{2,t}
(2n_2 - 2t), \end{equation} as desired. If we had exchanged $2t$
and $2t-1$ in the definition of $a_{c,t}$, alternating the two
kind of charges the opposite way, we would have
\begin{equation}\label{secondone} B_{1c}\geq  -\sum_{k=1}^{n_1} a_{i,k} (n_i - k +1 ) -
\sum_{t=1}^{n_3} a_{3,t} (2n_3 - 2t )- \sum_{t=1}^{n_2} a_{2,t}
(2n_2 - 2t+1), \end{equation} Starting instead with the mirror
image of (\ref{helpful}),
\begin{equation}\label{elpful} B_{1c}\geq  -\sum_{k=1}^{n_1} a_{1,k} (n_1 - k ) -
\sum_{t=1}^{n_2+n_3} a_{c,t} (n_2+n_3 - t+1 ), \end{equation} a
similar argument gives
\begin{equation}\label{thirdone}
B_{1c}\geq  -\sum_{k=1}^{n_1} a_{i,k} (n_i - k ) -
\sum_{t=1}^{n_3} a_{3,t} (2n_3 - 2t+2 )- \sum_{t=1}^{n_2} a_{2,t}
(2n_2 - 2t+1), \end{equation}

If $j$ runs over the set $N$ of all nodes and as in the figure
$i=0$ labels the outermost node of the long branch, we have from
(\ref{firstone}) and (\ref{lowbo}), summed along the chain as in
section \ref{quivers}, the simple inequality
\begin{equation}\label{goodone}q_R\geq \sum_{j\in N} \Delta_j +
\sum_{k=1}^{n_2}a_{2,k}+\sum_{k=1}^{n_{0}}a_{0,k}(n_{0}-k).\end{equation}
For a quiver with all $e_i>0$, so that $\Delta_i\geq 0$, this
inequality implies that  $q_R \geq 1$, unless the charges at node
2 vanish. But if those charges vanish, then we reduce to a quiver
of type $\textsf{A}$, and again $q_R\geq 1$ by virtue of our
analysis in section \ref{quivertheories}.  We also learn from
(\ref{goodone}) and its analog for type $\textsf{A}$ that to get
$q_R=1$, the $\Delta_j$ must vanish, implying that $e_j=0$ at any
node with nonzero charges. As usual then, in a good quiver, to
analyze monopole operators of $q_R=1$, we can omit nodes with
$e_j>0$ and consider a connected balanced quiver -- one in which
all nodes are balanced.

A balanced quiver of type $\textsf{D}_n$ has monopole operators
that generate a $\textsf{D}_n$ symmetry of the Coulomb branch, by
virtue of an argument given earlier for all $\textsf{ADE}$
quivers.  It remains to show that there are no other monopole
operators of $q_R=1$. A useful fact is that, as in section
\ref{quivertheories}, to get $q_R=1$, the set of nodes at which
the charges are nonzero must be connected; otherwise each
connected component contributes at least 1 to $q_R$. Inequality
(\ref{goodone}) implies that to get $q_R=1$, the charge at the $2$
node, if not zero, must be elementary: $a_{2,k}=(0,0,\dots,0,1)$.
It is then also true that $a_{0,k}=(0,0,\dots,0,1)$. The modified
inequality
\begin{equation} \label{dneed}q_R\geq \sum \Delta_i +
\sum_{k=1}^{n_3}a_{3,k}+\sum_{k=1}^{n_{0}}a_{0,k}(n_{0}-k)\end{equation}
derived from (\ref{secondone})  shows that similarly
$a_{3,k}=(0,0,\dots,0,1)$. Then the third inequality
(\ref{thirdone})  can similarly be used to show that
\begin{equation}q_R\geq \sum \Delta_i -
\sum_{k=1}^{n_3}a_{3,k}+\sum_{k=1}^{n_s}a_{s,k}+\sum_{k=1}^{n_{0}}a_{0,k}(n_{0}-k).\end{equation}
for any $s$ in the long chain, implying that $\sum_ka_{s,k}$ must
be no greater than 2. Hence $a_{s,k}=(0,0,\dots,0,2)$ or
$a_{s,k}=(0,0,\dots,0,1,1)$. At this point it is straightforward,
though tedious, to show that any charge $(0,0,\dots,0,2)$ leads to
a monopole of $q_R>1$, and that  monopole operators with $q_R=1$
have at most one connected sequence of nodes with charges of the
form $a_{s,k}=(0,0,\dots,1,1)$ starting at the node $i=1$, with
other nonzero charges being of the form $a_{s,k}=(0,0,\dots,0,1)$.
What we have enumerated here are the positive roots of
$\textsf{D}_n$, so this is the symmetry generated by the monopole
operators.

\subsubsection{Orthosymplectic Quivers Of Type $\tx
D$}\label{orthod}

We have learned in section \ref{orquiv} that orthosymplectic quivers
have the same expression for the $R$-charge as unitary quivers once
the $R$-charge is expressed in terms of excesses, ranks, and
monopole charges. Apart from some exceptions involving $SO(2)$, the
number of monopole operators of $q_R=1$ in an orthosymplectic quiver
is one-half of what it is for the corresponding unitary quiver.

So we can apply our results  for a unitary quiver of type $\tx D_n$
to an orthosymplectic quiver of the same type. Let us start from the
case with no $SO(2)$ node involved. The total number of $q_R=1$
monopoles in the unitary case was $\mathrm{dim}(SO(2n))-
\mathrm{rk}(SO(2n)) = 2 n(n-1)$. Hence we expect $n(n-1)$ monopoles
in the orthosymplectic case. Monopole operators with charges
supported  on an ${\tx A}_{n-1}$ subquiver generate an $SO(n)$
symmetry. It is possible to argue by induction that the full
symmetry is actually $SO(n) \times SO(n)$.

If any endpoint of the quiver is an $SO(2)$ group, the symmetry is
enhanced. In particular an $SO(2)$ node at the end of the long
branch leads, for $n>4$, to  $SO(n+1) \times SO(n+1)$ symmetry. An
$SO(2)$ node at the end of a short branch implies that the central
node is $Sp(2)$, and then the quiver must be a $\tx D_4$ quiver,
with three $SO(2)$ gauge groups. (In particular, this is
equivalent to the $n=4$ case of an $SO(2)$ group at the end of the
long branch.) It takes some patience to count all the $33$
monopoles hiding in the quiver. The monopole operators and
classical symmetries at each $SO(2)$ node give rise to an $SU(2)$
symmetry group. The remaining $27$ monopole operators are in the
representation  $(\3 ,\3 ,\3)$ of this subgroup; hence the full
group turns out to be $Sp(8)$. Notice that the Coulomb branch of
the theory has hyper-Kahler dimension $4$. This and the symmetry
group suggest that the Coulomb branch is simply $\BC^8/{\Bbb
Z}_2$.

\section{Orientifolds And Orbifolds}\label{orient}

Dirichlet and Neumann are the most obvious half-BPS boundary
conditions in  $\N=4$ super Yang-Mills theory.  Intermediate
between them are boundary conditions in which the gauge group $G$
is reduced to a subgroup $H$ along the boundary.  Vector
multiplets of $H$ obey Neumann boundary conditions, while the rest
of the $G$ vector multiplets obey Dirichlet.

A particular case of this which is intuitively obvious and natural
is the case that $H$ is the subgroup of $G$ that commutes with a
symmetry $\tau$ of $G$ of order 2.  A symmetry of order 2 is known
as an involution, and may be either an inner automorphism or an
outer automorphism.  Boundary conditions associated with an
involution can be obtained by a simple $\Z_2$ orbifold of $\N=4$
super Yang-Mills on $\R^4$. One simply divides by the reflection
$y\to-y$ of space, accompanied by the gauge transformation $\tau$.
Of course, this must be extended to the full $\N=4$ theory in a
supersymmetric fashion.  As explained in section 2.2 of
\cite{Gaiotto:2008sa}, if we decompose the Lie algebra of $G$ as
$\frak g=\frak h\oplus \frak h^\perp$, where $\frak h$ is the Lie
algebra of $H$ and $\frak h^\perp$ is its orthocomplement, and
write $\Phi^\pm$ for the projections of a field $\Phi$ to $\frak
h$ and $\frak h^\perp$, then the necessary conditions for $\vec X$
and $\vec Y$ are
\begin{equation}\label{zelf}\vec X^+(0)=0=\vec
Y^-(0).\end{equation} Many instances of such boundary conditions
can be implemented in string theory via orientifolds or orbifolds.

In this section, we will analyze, for $G=U(n)$, the $S$-duals of the
boundary conditions associated with involutions.  These examples
illustrate in an interesting way some of the ideas of the present
paper.  They are quite different from examples that we have
considered so far, but are rather tractable, partly because of their
realizations in string theory. Another reason to study the
$S$-duality of these examples is that one can compare to a
mathematical theory developed by Nadler \cite{Nadler}, though we
will only go part way in that direction in the present paper.

\subsection{Three Types Of Involution}\label{threetypes}

Let us first classify the possible involutions of $G=U(n)$.

An inner involution $\tau$ is simply conjugation by an element $h$
of $U(n)$ that obeys $h^2=1$.  Such an $h$ has $p$ eigenvalues 1
and $q$ eigenvalues $-1$, for some $p$ and $q$ with $p+q=n$. We
will loosely follow the notation of \cite{Nadler} and call this an
involution of Class III. The subgroup $H$ that commutes with a
Class III involution is $U(p)\times U(q)$.  If $p=q$, we say that
the involution is symmetric.

An outer involution $\tau$ is complex conjugation composed with an
inner involution.  In other words, $\tau$ acts by $g\to w\bar g
w^{-1}$, where $g\to \bar g$ is complex conjugation and $w$ is an
element of $G$.  There are two essentially different cases,
depending on whether $\tau^2$ equals 1 or $-1$ when acting on the
fundamental representation of $U(n)$.

If $\tau^2=1$, we can take $\tau$ to be simply $g\to \bar g$.  We
will call this a Class I involution.  It reduces the gauge
symmetry from $G=U(n)$ to $H=O(n)$.

It is only possible to have $\tau^2=-1$ if $n$ is even.  In that
case, to realize $\tau^2=-1$, we can take $\tau$ to be $g\to w\bar
gw^{-1}$, where $w$ is the direct sum of $n/2$ blocks of the form
\begin{equation}\label{twob}\begin{pmatrix} 0 & 1 \\ -1 & 0
\end{pmatrix}.\end{equation}
We call this a Class II involution.  It reduces the gauge symmetry
from $G=U(n)$ to $H=Sp(n)$.

\begin{figure}
  \begin{center}
    \includegraphics[width=4in]{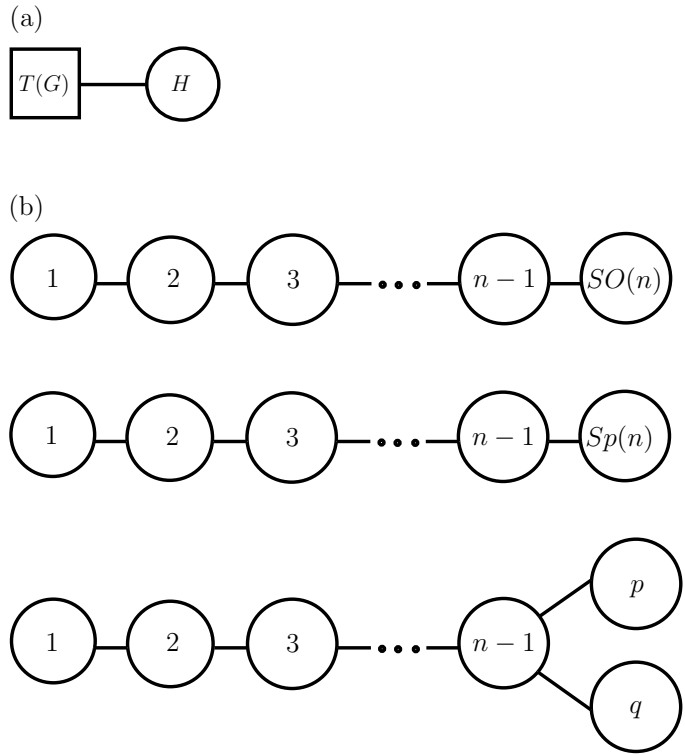}
  \end{center}
\caption{\small (a) Applying the ungauging procedure to find the
$S$-dual of a boundary condition defined by reducing the gauge
symmetry from $G$ to $H$ (with no Nahm pole or SCFT) means that we
gauge an $H$ subgroup of the $G$ symmetry of $T(G)$ and take the
IR limit. (b) In the present case, for $\tau$ of Class I, II, or
III, this procedure leads to the quivers shown here; a subgroup of
the symmetry of $T(U(n))$ has been gauged. Only Class I leads to a
good quiver.  Here and later, a node of a quiver labeled simply by
an integer $k$ represents a $U(k)$ gauge group; nodes that
represent other types of gauge group are labeled in more detail.}
  \label{Fig57}\end{figure}

Now let us begin to explore the $S$-duals of boundary conditions
associated with $\tau$ of Class I, II, or III. One of the most basic
questions is whether the $S$-dual has full $U(n)$ gauge symmetry and
so is obtained by coupling to a $U(n)$ invariant SCFT $\fB^\vee$. If
so, the mirror of $\fB^\vee$ is given by the familiar ungauging
procedure.  In the present case, this simply means (fig.
\ref{Fig57}(a)) that we gauge a subgroup $H$ of the symmetry group
$G$ of the Higgs branch of $T(U(n))$.  Since $T(U(n))$ can be
represented by a quiver, the same is true for the candidate
$\fB^\vee$, as indicated in fig. \ref{Fig57}(b).

However, of the three cases, only Class I leads to a good quiver.
For example, for the rightmost two nodes of the quiver of Class III
to be good, we need $n-1\geq 2p$ and $n-1\geq 2q$, which is
impossible since $n=p+q$.  Concerning Class II, we note that
according to (\ref{zano}), for an $Sp(n)$ node to be good, the
flavor symmetry of the fundamental hypermultiplets, when all other
gauge couplings are turned off, must be at least $SO(2n+2)$.
However, the Class II quiver in fig. \ref{Fig57}(b) only leads to a
flavor symmetry $SO(2n-2)$.

The Class I quiver in fig. \ref{Fig57}(b) has an $SO(n)$ node, and,
according to (\ref{kano}), for this node to be good, the flavor
symmetry of the fundamental hypermultiplets of $SO(n)$, when all
other gauge couplings are turned off, should be at least $Sp(2n-2)$.
That is precisely what we get from the Class I quiver in the figure.
So this quiver is good. In fact it is a balanced quiver of a type
considered in section \ref{uniorth}.  The symmetry of the Coulomb
branch is $SU(n)\times SU(n)$ (or $Sp(4)$ if $n=2$).

The Class I quiver has no Higgs branch.  To show this, we use the
fact that this quiver is the one considered in fig. \ref{Fig24},
which describes $T(SU(n))$, except that an $SO(n)$ subgroup of the
$SU(n)$ global symmetry of that quiver has been gauged.  As
explained in section \ref{unexdir}, the hyper-Kahler quotient of the
$T(SU(n))$ quiver is parametrized by an $n\times n$ matrix $M$ that
takes values in the nilpotent cone, or a deformation/resolution
thereof if FI parameters are turned on. For our purposes, it is
convenient to turn on complex FI parameters, so that the complex
equation obeyed by $M$ is deformed to (\ref{ortuf}), whose solutions
parametrize a deformation $\tilde \N$ of the nilpotent cone. We take
the FI parameters to be generic so that the eigenvalues of $M$ are
distinct.  To construct the Higgs branch of the Class I quiver in
the presence of the FI parameters, we need to take the hyper-Kahler
quotient of $\tilde \N$ by $SO(n)$. The complex moment map is the
antisymmetric part of $M$. $M$ acts on a vector space $V\cong
\BC^n$; the statement that it is symmetric means more invariantly
that it preserves a complex bilinear form (not a hermitian form) on
$V$.  The group of linear transformations of $V$ that preserve the
bilinear form is $SO(n)_\BC$. To show that the Higgs branch is
trivial, we must show that a symmetric matrix $M$ that obeys
(\ref{ortuf}) and in particular has distinct eigenvalues can be
diagonalized by an $SO(n)_\BC$ transformation. Indeed, since $M$ has
distinct eigenvalues, its eigenvectors furnish a complex basis for
$V$; as $M$ is symmetric, the eigenvectors are mutually orthogonal
with respect to the quadratic form, and we can choose them to be an
orthonormal basis.  So there is an orthonormal basis in which $M$ is
diagonal and thus we can diagonalize $M$ by an $SO(n)_\BC$
transformation.

The Coulomb branch $\C$ of the Class I quiver has  hyper-Kahler
dimension $n(n-1)/2+[n/2]$, where $[~~]$ denotes the integer part.
For $n=2$, the hyper-Kahler dimension is 2, and in view of the
$Sp(4)$ symmetry and absence of free hypermultiplets, it is
natural to suspect that $\C$ may be $\BC^4/\Z_2$. (Dividing
$\BC^4$ by $\Z_2$ preserves the $Sp(4)$ symmetry and projects out
the chiral operators of $q_R=1/2$.) This together with the absence
of a Higgs branch suggests that for $n=2$ the theory might be
simply a $\Z_2$ orbifold of a free theory. For $n>2$, we have not
been able to find a good candidate for $\C$, though because of its
large symmetry this might be possible.  It appears that for $n>2$
there is no good candidate for $\C$ as an orbifold of some
$\BC^{2s}$, suggesting that the SCFT is non-Gaussian despite the
absence of a Higgs branch.  This is not a familiar state of
affairs.  Known examples of non-trivial $\N=4$ SCFT's in three
dimensions generally have both Coulomb and Higgs branches.

For $\tau$ of Class II or III, the badness of the quiver indicates
that the dual of the orbifold  boundary condition has reduced
gauge symmetry. To understand the details, we will use a string
theory construction.

\subsection{String Theory Constructions}\label{orbior}

The gauge theory boundary conditions considered in this section
arise in four-dimensional gauge theory from a reflection $y\to -y$
together with a gauge transformation $\tau$ and suitable
transformations of other fields.   We want to understand how to
realize these boundary conditions in ten-dimensional string theory
with D3-branes that generate the $U(n)$ gauge symmetry interacting
with orbifold or orientifold five-planes that will generate the
boundary condition.

If $\tau$ is an outer automorphism of $U(n)$, we have to use an
orientifold five-plane.  The reason for this is that among the
symmetries of D3-branes, it is the reversal of the worldsheet
orientation of a string, often called $\Omega$, that acts on
$U(n)$ by an outer automorphism and maps the fundamental
representation of $U(n)$ to its complex conjugate.  Thus, to
relate Class I or Class II boundary conditions to a string theory
construction, we need to use some sort of orientifold.

For the same reason, if $\tau$ is an inner automorphism, we have
to use an orbifold operation -- constructed using a symmetry that
does not reverse the orientation of a string worldsheet.

Our basic constructions of boundary conditions use D5-branes that
fill dimensions 012456 and NS5-branes that fill dimensions 012789.
An orientifold fiveplane can preserve the same supersymmetry as a
parallel D5-brane, and a properly chosen $\Z_2$ orbifolding
operation can preserve the same supersymmetry as an NS5-brane
whose worldvolume coincides with its fixed point set.

Hence, we can represent Class I and Class II boundary conditions by
an orientifolding operation that reverses $y$ and $\vec Y$ (that is,
it acts as multiplication by $-1$ on $y=x^3$ and on $x^7,x^8,x^9$)
and leaves fixed the other coordinates -- so its fixed point set
coincides with the locus of a D5-brane in our constructions.  And
similarly, Class III boundary conditions come from a carefully
chosen orbifolding operation that reverses $y$ and $\vec X$ (that
is, one which acts as multiplication by $-1$ on $x^3,x^4,x^5,x^6$).
$S$-duality exchanges $\vec X$ and $\vec Y$, and as we will see,
although it does not merely exchange Class II and Class III, it maps
them to close cousins of one another.  On the other hand,
$S$-duality does not appear to map Class I to a perturbative
configuration.  This agrees with the fact that the dual of Class I
involves a non-trivial SCFT, since the Class I quiver in fig.
\ref{Fig57}(b) is good.

Now let us describe in more detail the relevant orbifolding and
orientifolding operations, all of which have been well analyzed in
the literature.  First of all, there are several different kinds
of orientifold fiveplane or O5-plane.  The most basic distinction
is according to whether the gauge symmetry of $n$ D5-branes
coincident with the O5-plane is $SO(2n)$ or $Sp(2n)$.  The former
case is called an O5$^-$-plane since it has a D5-brane charge of
$-1$.  The latter case has D5-brane charge $+1$ and is called an
O5$^+$-plane.

In our problem, we consider $n$ D3-branes with worldvolume
spanning directions 0123, and we obtain a boundary condition by
coupling to an O5-plane that spans directions 012456.  The
orientifolding operation gives a boundary condition in the
four-dimensional $U(n)$ gauge theory of the D3-branes, since it
reverses the $y=x^3$ direction.

In the case of a Class I boundary condition, which reduces $U(n)$
to $O(n)$ at the boundary, the flavor symmetry of boundary
hypermultiplets in the fundamental representation is a symplectic
group.  Boundary hypermultiplets come from coupling to D5-branes
supported on the orientifold plane, and the flavor symmetry is the
gauge symmetry of the D5-branes.  Symplectic gauge symmetry of
D5-branes corresponds to the case of an O5$^+$-plane, as asserted
in the last paragraph. So this is the orientifold that leads to a
Class I boundary condition.

Similarly, to get a Class II boundary condition, which reduces
$U(n)$ to $Sp(n)$ at the boundary, the flavor symmetry of
fundamental boundary hypermultiplets should be of orthogonal type.
So the appropriate orientifold plane is of type O5$^-$.

Finally, let us discuss the Class III boundary condition, which is
supposed to involve some sort of orbifold reflection of directions
3456, leaving fixed directions 012789.  Of course, we want to
define this orbifold in Type IIB superstring theory since we want
to do gauge theory on D3-branes.  In Type IIA superstring theory,
an orbifold that involves a simple reflection ${\cal I}_4$ of four
spatial coordinates preserves the same supersymmetry as an
NS5-brane supported on the orbifold fixed set.  But for Type IIB
this is not so.  (One way to see this is to observe that the
${\cal I}_4$ orbifold is chiral for Type IIB -- all unbroken
supersymmetries have the same six-dimensional chirality -- while
the worldvolume theory of a Type IIB NS5-brane is not chiral.)  To
preserve the supersymmetry of an NS5-brane, one must divide not by
${\cal I}_4$ but by ${\cal I}_4(-1)^{F_L}$, where $(-1)^{F_L}$ is
the operation that counts left-moving worldsheet fermions modulo
2.

\begin{table}
\caption{\small For each class of boundary condition derived from
an involution, we indicate here what type of string theory
orientifold or orbifold realizes it, and what is the corresponding
O5-brane or NS5-brane charge. }
\begin{center}
\begin{tabular}{c|c|c}
Class of Boundary Condition&Orientifold Or Orbifold&Five-brane
Charge
\\\hline
 I&O5$^+$ Orientifold & 1 \\ II &
O5$^-$ Orientifold &$-1$ \\ III & ${\cal I}_4(-1)^{F_L}$ Orbifold
& 0
\\
\end{tabular}
\end{center}\label{golbo}
\end{table}

The correspondence between involutions and string theory
constructions is summarized in Table \ref{golbo}. In the last column
of the table, we list the fivebrane charge of the appropriate
orientifold or orbifold fiveplane.  For orientifolds, this is the
D5-brane charge, while for the ${\cal I}_4(-1)^{F_L}$ orbifold, it
is the NS5-brane charge.  This NS5-brane charge is zero; it can be
computed as an integral at infinity of the $H$-field (the curvature
of the string theory $B$-field) but the orbifold $H$-field vanishes.

\subsection{Nonperturbative Duality}\label{nonper}

A nonperturbative duality relating some configurations of the type
just described was discovered some years ago
\cite{Kutasov:1995te,Sen:1996na}. Since the duality symmetry
$S:\tau\to-1/\tau$ exchanges D5-branes, which are related to the
first two rows in our table, with NS5-branes, which are related to
the third, we look for a transformation that relates a Class III
boundary condition to Class I or II.

The fivebrane charges make clear what the duality must be.  The
${\cal I}_4(-1)^{F_L}$ orbifold has zero fivebrane charge, while the
O5$^\pm$-planes have fivebrane charge $\pm 1$.  We can make
fivebrane charge 0 by combining an O5$^-$-plane with a single
D5-brane.  But there is no way to get to charge 0 by adding
fivebranes to an O5$^+$-plane, which already has positive charge.
(Adding anti D5-branes would of course break supersymmetry.)  So the
only reasonable conjecture is that the ${\cal I}_4(-1)^{F_L}$
orbifold is $S$-dual to an O5$^-$-plane plus one D5-brane.

Apart from the fact that the fivebrane charges and the unbroken
supersymmetries match, one of the original arguments for this
assertion is that the gauge symmetries match.  Quantization of the
twisted sector of the ${\cal I}_4(-1)^{F_L}$ orbifold gives a single
massless vector multiplet, with gauge group $U(1)$.  On the other
hand, a single D5-brane at an O5$^-$-plane has gauge symmetry
$SO(2)$, or equivalently $U(1)$.

We want to apply this duality to the gauge theory on the D3-branes.
In the field of the ${\cal I}(-1)^{F_L}$ orbifold, there are two
kinds of fractional D3-brane, depending on whether the generator of
the orbifold symmetry acts on the Chan-Paton bundle of the D3-brane
as multiplication by $+1$ or $-1$.  If we pick $p$ D3-branes of one
type and $q$ of the other type, with $n=p+q$, we get $U(n)$ gauge
symmetry broken at the boundary to $U(p)\times U(q)$. This is the
general Class III boundary condition.

There is no such choice to be made for the Class II orientifold:
we simply have in bulk $n$ D3-branes, with gauge symmetry reduced
from $U(n)$ to $Sp(n)$ at $y=0$.  Since there is no choice to be
made, we have a puzzle to resolve.  For what values of $p$ and
$q$, if any, is the orbifold dual to the orientifold?

This question actually has a very simple answer.  In the
orientifold, with $n$ being even, it is possible to give
expectation values to $\vec Y$, breaking $Sp(n)$ to $U(1)^{n/2}$,
and leaving no D3-branes at $\vec Y=0$ (that is, at
$x^7=x^8=x^9=0$). So
 in the $S$-dual orbifold, if there is one, it must similarly be possible
 to reduce the gauge group to $U(1)^{n/2}$ by displacing all D3-branes in $\vec X$ (recall
 that $S$-duality exchanges $\vec X$ and $\vec Y$) and leaving none at $\vec X=0$.
 This is possible precisely if
$p=q=n/2$; otherwise, there are ``fractional D3-branes'' that cannot
be removed from $\vec X=0$.  So this must be the right case for the
duality.

Let us look at this a little more closely, taking account of the
boundary condition (\ref{zelf}), which constrains $\vec X(0)$ and
$\vec Y(0)$.  For the orientifold, (\ref{zelf}) says that $\vec
Y(0)$ transforms in the adjoint representation of the unbroken
$Sp(n)$.  Its expectation value can indeed break $Sp(n)$ to the
maximal torus $U(1)^{n/2}$. For the orbifold, (\ref{zelf}) implies
that $\vec X(0)$ transforms in the bifundamental representation of
the unbroken $U(p)\times U(q)$. The case that $\vec X(0)$ can
break $U(p)\times U(q)$ to $U(1)^{n/2}$ is the symmetric case
$p=q=n/2$. So again that is the right case for this duality
between perturbative configurations. The unbroken groups
$U(1)^{n/2}$ that remain when the orbifold is perturbed in $\vec
X$ or the orientifold in $\vec Y$ are equivalent as subgroups of
$U(n)$, as one would expect from the $S$-duality between these two
configurations.

So we have our first case of using orbifolds and orientifolds to
answer the basic question from section \ref{threetypes}.  The
$S$-dual of the symmetric Class III boundary condition, with
$p=q=n/2$, is given by a Class II boundary condition supplemented
by boundary hypermultiplets.

\subsubsection{The General Case}\label{sumres}

We are obviously left with some questions.  (1) What is the
$S$-dual of a Class II boundary condition without boundary
hypermultiplets? (2) And what is the $S$-dual of a Class III
boundary condition with $p\not=q$?

We will show in sections \ref{zahm} and \ref{lastcase} that the
answers to these questions do not involve simple orbifold boundary
conditions, but involve Nahm poles:

(1)$'$ The $S$-dual of a Class II boundary condition that breaks
$U(n)$ to $Sp(n)$ (without boundary hypermultiplets) is a boundary
condition with a Nahm pole relative to the decomposition
$n=2+2+2+\dots 2$.  This breaks $U(n)$ to what we will call
$U(n/2)_2$.  Here $U(n/2)_2$ is a diagonal subgroup of
$U(n/2)\times U(n/2)\subset U(n)$.

(2)$'$ The $S$-dual of a Class III boundary condition that breaks
$U(n)$ to $U(p)\times U(q)$ with $n=p+q$ and  $p\geq q+2$ is a
boundary condition built from a Nahm pole and a further reduction
of gauge symmetry. The Nahm pole is associated with the
decomposition $n=(p-q)+1+1+\dots+1$ and commutes with $U(1)\times
U(2q)\subset U(n)$.  The boundary condition further reduces
$U(1)\times U(2q)$ to $H=Sp(2q)$.  If $p-q=1$, there is no Nahm
pole; the dual boundary condition simply reduces the symmetry from
$U(n)$ to $Sp(n-1)$.  In each of these cases, and in contrast to
the symmetric case $p=q$, there are no boundary hypermultiplets.

\begin{table}
\caption{\small   The first column lists the unbroken subgroups
$H$ in boundary conditions in $SU(n)$ gauge theory that are
defined by an involution $\tau$.  The second column lists the
unbroken gauge symmetry $\tilde H$ of the $S$-dual boundary
condition. The third column describes the Nahm pole, if any, that
is part of the reduction of the dual gauge group from $SU(n)$ to
$\tilde H$. The fourth column describes the matter system that is
coupled to $\tilde H$. (The hypermultiplets indicated are in the
fundamental representation of $Sp(n)$.)}
\begin{center}
\begin{tabular}{c|c|c|c}
$H$&$\tilde H$&Nahm Pole & Matter System
\\\hline
 $SO(n)$&$SU(n)$ & None &Non-trivial SCFT \\ $Sp(n)$ &
$SU(n/2)_2$ &$n=2+2+\dots+2$ & None\\ $S(U(n/2)\times U(n/2))$
&$Sp(n)$ & None &  Hypermultiplets
\\ $ S(U(p)\times
U(q)),~p>q$ & $Sp(2q)$ & $n=(p-q)+1+1+\dots+1 $& None
\\
\end{tabular}
\end{center}\label{obolbo}
\end{table}

These results are summarized in Table \ref{obolbo}. The table gives
the group $H$ that is left unbroken by an involution $\tau$, and the
construction of the dual boundary condition in terms of a Nahm pole,
a group $\tilde H$ that commutes with the Nahm pole, and a matter
system with $\tilde H$ symmetry. The table has been written for
$G=SU(n)$ rather than $U(n)$. This is accomplished by merely
dropping  central $U(1)$ factors (which in our constructions obey
Dirichlet boundary conditions on one side, and Neumann on the other)
from various entries.

Our table can be compared to the first three lines in Table 1 of
\cite{Nadler}, which refer to the group ${\tx A}_{n-1}=SU(n)$.  What
is called $\frak g_\R$ in the first column of that table is the Lie
algebra of a real group $G_\R$ whose maximal compact subgroup we
call $H$. (From our point of view, the data determining $G_\R$ are
the choice of compact gauge group $G$ and involution $\tau$.) What
is called $\frak h^\vee$ in the fourth column is the
complexification of the Lie algebra of what we call $\tilde H$. With
this translation, our table is in perfect agreement with that of
\cite{Nadler}.

To make clear why the two tables should match,  we will briefly
describe the problem treated in \cite{Nadler}, but restated in gauge
theory language. In effect, $G$ gauge theory is studied on a
four-manifold $M$ with boundary; the boundary condition is
determined by an involution $\tau$ of $G$, which reduces $G$ to a
subgroup $H$. Then 't Hooft operators that are supported on the
boundary are classified. It is shown that, although 't Hooft
operators in the interior of $M$ are classified by representations
of the dual group $G^\vee$, 't Hooft operators supported at the
boundary of $M$ are related to representations of a more mysterious
group $\tilde H$, whose origin is not obvious. For example, $\tilde
H$ is not the dual group of either $G$ or $H$. From our point of
view, $\tilde H$ is the subgroup of the dual gauge group $G^\vee$
that is gauged in the dual boundary condition. So Wilson operators
at the boundary are $\tilde H$-valued.

The last two columns in Table \ref{obolbo} involve matters that
have apparently not been explored yet in the mathematical
literature. The Nahm pole is plausibly related mathematically to
Arthur's $SL_2$.  What from our point of view is the matter system
that is part of the dual boundary condition might show up
mathematically in a precise study of the 't Hooft operators.

Some remarks about the $\textsf{C}$ and $\textsf{D}$ cases of the
table in \cite{Nadler} are made at the end of section
\ref{orthobif}.

\subsection{Nahm Poles}\label{zahm}

Let us return to the duality between (a) a boundary condition that
breaks $U(n)$ to $Sp(n)$, with coupling to a boundary
hypermultiplet, and (b) a boundary condition that breaks $U(n)$ to
$U(n/2)\times U(n/2)$.

We want to understand the dual of (a) without the boundary
hypermultiplet.  Our strategy will be to use the fact that it is
possible to preserve supersymmetry  while giving a bare mass to the
hypermultiplet.  We will determine what parameter in (b) corresponds
to the hypermultiplet bare mass, and then we will determine the
limit of (b) when the bare mass becomes large.  This will give us
the dual of breaking $U(n)$ to $Sp(n)$, without the hypermultiplet.

Before we introduce any perturbation, the boundary conditions obeyed
by $\vec X$ and $\vec Y$ in the presence of boundary hypermultiplets
are
\begin{align}\label{gelf}\notag \vec X^+(0)+\vec\mu_Z & = 0\\
                                 \vec Y^-(0) & = 0, \end{align}
where $\vec\mu_Z$ is the moment map for the space $Z$ that
parametrizes the boundary hypermultiplets.   (This condition
coincides with eqn. (\ref{zelf}), except that now we include the
hypermultiplets.)  As explained in sections 2.2.3 and 2.3.5 of
\cite{Gaiotto:2008sa}, it is possible to add central constants to
these boundary conditions, which become
\begin{align}\label{gielf}\notag \vec X^+(0)+\vec\mu_Z & = \vec v\\
                                 \vec Y^-(0) & = \vec w. \end{align}
Here $\vec v$ takes values in the center of $\frak h$, and $\vec
w$ takes values in the subspace of $\frak h^\perp$ that commutes
with $\frak h$.  (Moreover, the components of $\vec w$ must
commute with each other.)

The parameters $\vec v$ are FI parameters in a generalized sense,
transforming non-trivially under $SO(3)_X$ and trivially under
$SO(3)_Y$.  The parameters $\vec w$ are mass parameters,
transforming non-trivially under $SO(3)_Y$.  In enumerating
parameters, we also must include the FI parameters and mass
parameters of the boundary theory, if any.  In the present
discussion, the boundary theory consists of free hypermultiplets and
has only mass parameters.

Now in (a), $\frak h=\frak{sp}(n)$, which is a simple Lie algebra
with trivial center.  So there are no FI parameters.  There are
two mass parameters.  One arises because in (\ref{gelf}), we can
take $\vec Y^-(0)=\vec d \cdot 1_n$, where $1_n$ is the identity
$n\times n$ matrix.  The parameter $\vec d$ would be absent if we
took the underlying gauge group $G$ to be $SU(n)$ instead of
$U(n)$. (We use $U(n)$ because it arises more naturally from
branes.)  The second parameter is the hypermultiplet bare mass
$\eusm m$.

Dually in (b), there are no mass parameters since the condition that
$\vec w$ should commute with $\frak h$ forces $\vec w=0$; moreover,
this boundary condition has no boundary hypermultiplets.  However,
in (b) there are two FI parameters, since the center of
$H=U(n/2)\times U(n/2)$ has rank two.   Embedding $U(n/2)\times
U(n/2)$ in $U(n)$ in terms of $n/2\times n/2$ blocks
\begin{equation}\begin{pmatrix} * & 0 \\ 0 & *
\end{pmatrix},\end{equation}
we can  take the boundary condition on $\vec X$ to be
\begin{equation}\label{zolfox}\vec X(0)
=\begin{pmatrix} \vec c_1 \cdot 1_{n/2} & * \\ * & \vec c_2\cdot
1_{n/2}\end{pmatrix}. \end{equation} (There is no $\vec \mu_Z$ term
here as in this description there are no boundary hypermultiplets.)

So the two parameters $\vec c_1$ and $\vec c_2$ must match on the
other side the parameters $\vec d$ and $\eusm m$.  The matching is
easy to do because one parameter on each side involves the center
$U(1)$ of $G=U(n)$.  The boundary conditions that we are considering
do not couple the two factors of $G\sim U(1)\times SU(n)$, so we can
consistently remove the center of $G$, which means on one side
setting $\vec d=0$ and on the other side $\vec c_1=-\vec c_2$. So
the dual of the hypermultiplet bare mass is a boundary condition
\begin{equation}\label{ozolfox}\vec X(0)
=\begin{pmatrix} \vec {\eusm m} \cdot 1_{n/2} & * \\ * & -\vec{\eusm
m}\cdot 1_{n/2}\end{pmatrix}. \end{equation}

Now let us discuss how to preserve supersymmetry in the presence
of this boundary condition.  We take the vacuum at infinity   to
be given by $\vec X=\vec Y=0$.  Supersymmetry then requires that
$\vec Y$ vanishes everywhere, but $\vec X$ cannot vanish
identically in view of the boundary condition (\ref{ozolfox}).
Rather, we must look for a solution of Nahm's equations $d\vec
X/dy+\vec X\times \vec X=0$ that obeys the boundary condition and
has $\vec X$ vanishing at infinity.

Let us first discuss how to do this for $n=2$.  The general solution
of the $SU(2)$ Nahm equations on the half-line $y\geq 0$ with $\vec
X\to 0$ at infinity is
\begin{equation}\label{dolx}\vec X= f \frac{\vec t}{y+y_0}
f^{-1},\end{equation} with $y_0>0$ and $f\in SU(2)/\Z_2$; $\vec t$
are the usual $\frak{su}(2)$ generators. We must adjust the
parameters $f$ and $y_0$ to obey the boundary condition.  Without
essential loss of generality, take $\vec{\eusm m}=(0,0,\eusm m_3)$
and work in the usual basis in which $t_3$ is diagonal and $t_1,t_2$
are purely off-diagonal.  Then to obey (\ref{ozolfox}), we need to
take $f=1$ and $y_0=1/|{\eusm m}|$.

The limit as $\vec{\eusm m}\to\infty$ is now easily described.  In
this limit, $y_0\to 0$ and $\vec X(y)$ has an irreducible Nahm pole
at $y=0$.

Actually, for $n=2$ this is not really a new result.  The groups
$SU(2)$ and $Sp(2)$ coincide, so the boundary condition that reduces
$SU(n)$ to $Sp(n)$ just coincides, when $n=2$, with Neumann boundary
conditions.  We already know that Neumann boundary conditions are
dual to an irreducible Nahm pole, and this is what we have just
rediscovered from another point of view.

Now, however, we can immediately generalize to the case of any $n$.
($n$ must be even for the question about reduction from $U(n)$ to
$Sp(n)$ to make sense.)  To solve the $SU(n)$ Nahm equations with
the boundary condition (\ref{ozolfox}), we just take the tensor
product of the $SU(2)$ solution (\ref{dolx}) with the rank $n/2$
identity matrix $1_{n/2}$.  Then, taking $\vec{\eusm m}\to\infty$ as
before, $\vec X$ acquires a Nahm pole which is obtained simply by
taking the tensor product of the standard rank 2 Nahm pole with
$1_{n/2}$.

This Nahm pole corresponds to a decomposition $n=2+2+2+\dots+2$. It
breaks $U(n)$ to $U(n/2)_2$, a diagonal subgroup of $U(n/2)\times
U(n/2)\subset U(n)$.  What we have learned is that the dual of the
boundary condition defined by the involution that breaks $U(n)$ to
$Sp(n)$ is a boundary condition defined by a Nahm pole that breaks
$U(n)$ to $U(n/2)_2$.

\subsection{The Last Case}\label{lastcase}

\begin{figure}
  \begin{center}
    \includegraphics[width=5in]{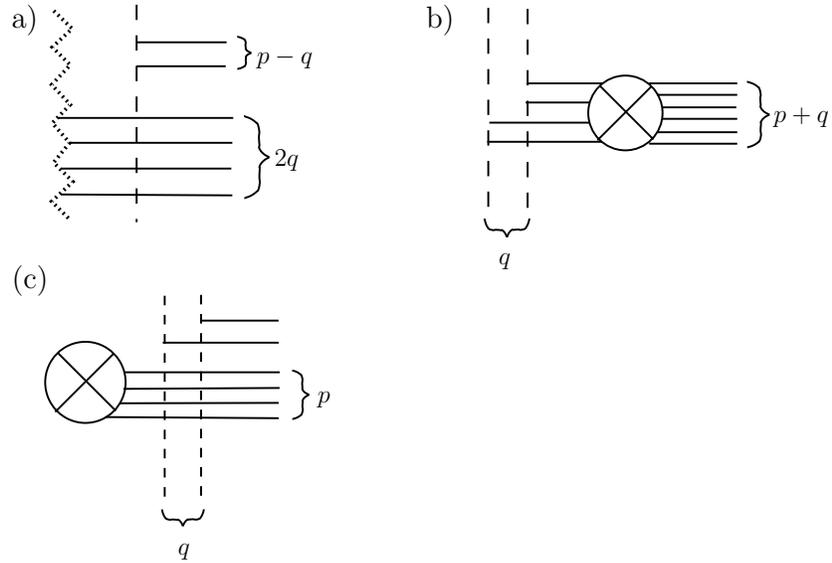}
  \end{center}
\caption{\small (a) A boundary condition in $U(p+q)$  gauge theory
(depicted here for $p=4$, $q=2$) consisting of a Nahm pole of rank
$p-q$ followed by an orientifold (represented by the vertical
jagged line) that reduces the gauge symmetry  to $Sp(2q)$. (b) A
brane configuration that is closely related to the $S$-dual of
(a).  $q$ D5-branes create a Nahm pole that is associated to the
decomposition $2q=2+2+\dots +2$.  In the limit that the separation
between these fivebranes vanishes, there is a global $U(q)$
symmetry; gauging this symmetry gives the $S$-dual of (a).  (c)
Instead, before gauging the symmetry, we can move the D5-branes
across the NS5-brane, arriving at this configuration.  Now if we
take the D5-branes to coincide and gauge the global $U(q)$
symmetry, we arrive at a gauge theory representation of the
$S$-dual of the original configuration (a).  This is a boundary
condition in $U(p+q)$ gauge theory in which a subgroup $U(p)\times
U(q)$ is gauged at the boundary.}
  \label{Fig57A}\end{figure}
To complete our explanation of Table \ref{obolbo}, we must justify
the last line, which describes the $S$-dual of a boundary condition
associated with an involution that breaks $SU(p+q)$ to $S(U(p)\times
U(q))$ for $p>q$.  One approach is to start with the third line of
the table, which says what happens for $p=q$, and ``flow'' to $q<p$
by giving suitable expectation values to the scalar fields $\vec X$
at infinity.

We will follow another approach, which is simple and possibly
illuminating.  In this, we will use an elementary relation between
Dirichlet and Neumann boundary conditions.  Suppose that, in any
gauge theory with gauge group $H$, we impose Dirichlet boundary
conditions, meaning that the gauge field $A$ and the generator
$\epsilon$ of a gauge transformation are both required to vanish at
the boundary. Then we get a theory in which $H$ acts as a global
symmetry at the boundary.  The global symmetry is just a gauge
transformation with $\epsilon$  constant, but not equal to 1, at the
boundary.  (See \cite{Gaiotto:2008sa}, section 2.2.2.) This
preserves the boundary condition $A=0$. As with any global symmetry,
we can seek to gauge this one. In the present case, this just means
that we allow the boundary value of $\epsilon$ to be non-constant
and thus arbitrary.  So we simply arrive at four-dimensional gauge
theory with a gauge parameter that is unrestricted at the boundary.
The gauge field is then also unconstrained at the boundary.  In
other words, gauging the global symmetry of Dirichlet boundary
conditions produces Neumann boundary conditions. We will call this
process the regauging trick.

The claim in the last row of the table is that if $\B$ is a
boundary condition defined by an involution $\tau$ that breaks
$SU(p+q)$ to $S(U(p)\times U(p))$, then the dual is a boundary
condition $\B^\vee$ consisting of a Nahm pole of rank $p-q$,
followed by a reduction of the structure group from $U(2q)$ to
$Sp(2q)$.  It is convenient to start with $\B^\vee$ and show that
the dual is $\B$.  $\B^\vee$ can be conveniently represented by a
configuration of $p+q$ D3-branes intersecting a D5-brane and
ending on an orientifold (fig. \ref{Fig57A}(a)).  As usual, to
work with branes we extend the symmetry from $SU(p+q)$ to
$U(p+q)$.  Later, we will factor out the central $U(1)$ from the
final statement.

 The $S$-dual
of the D5-brane in  fig. \ref{Fig57A}(a) is simply an NS5-brane.
As for the orientifold in the figure, it has no simple string
theory dual. If there were an additional D5-brane with no
D3-branes ending on it, there would be  a simple $S$-dual given by
the ${\cal I}_4\cdot (-1)^{F_L}$ orbifold. However, we learned in
section \ref{zahm} how to describe the dual of the boundary
condition due to the ``bare'' orientifold unaccompanied by an
extra D5-brane: it is given by a Nahm pole for the decomposition
$2q=2+2+\dots +2$, with gauging of the resulting $U(q)$ symmetry.
We can easily represent the Nahm pole by incorporating $q$
D5-branes with two D3-branes ending on each one, as in fig.
\ref{Fig57A}(b).  In this figure, a $U(q)$ global symmetry appears
if we take the separations between the D5-branes to vanish.  If we
gauge this symmetry, we arrive at the $S$-dual of the boundary
condition set by the bare orientifold. There is no convenient way
in a brane construction to gauge the $U(q)$ symmetry.  So we will
simply remember to gauge it at the end of the construction.

With this understanding, fig. \ref{Fig57A}(b) can be used to
construct the $S$-dual of fig. \ref{Fig57A}(a).  On the other
hand, we can make a standard brane manipulation.  We simply move
the D5-branes to the right of the NS5-brane, to arrive at fig.
\ref{Fig57A}(c).  Now only a single D3-brane ends on each
D5-brane, leaving $p$ D3-branes to end on the NS5-brane. The
collection of $q$ D5-branes therefore reduces the gauge symmetry
from $U(p+q)$ to $U(p)$, which then obeys Neumann boundary
conditions because of the NS5-brane.  From a field theory point of
view, this boundary condition admits a global $U(q)$ symmetry (the
commutant of the unbroken gauge group $U(p)$). The $U(q)$ symmetry
appears as a symmetry of the brane configuration if the D5-branes
are taken to be coincident. To construct the $S$-dual, we are now
supposed to gauge this global $U(q)$ symmetry.  At this stage, we
have a $U(p+q)$ gauge symmetry in the half-space $y\geq 0$, with a
subgroup $U(p)\times U(q)$ gauged at the boundary.

Thus the dual of  a boundary condition $\B$ with $U(p+q)$ reduced
to $U(p)\times U(q)$ at the boundary, for $p>q$, is a boundary
condition $\B^\vee$  with a Nahm pole reducing $U(p+q)$ to
$U(1)\times U(2q)$, which is then reduced to $Sp(2q)$ at the
boundary. Notice that in $\B$, the central $U(1)$ obeys Neumann
boundary conditions, while in $\B^\vee$ it obeys Dirichlet
boundary conditions.  Factoring out this central $U(1)$, we arrive
at the statement of the last line of Table \ref{obolbo} for gauge
group $SU(p+q)$.

A key step in this argument -- gauging the $U(q)$ global symmetry
at the very end of the process -- was essentially the regauging
trick in which Neumann boundary conditions (here for the subgroup
$H=U(q)$) can be obtained by gauging the global symmetry of
Dirichlet boundary conditions.

\subsection{More Elaborate Examples}\label{morex}
Once one understands the $S$-duality between the O5$^-$-plane and
the ${\cal I}_4\cdot (-1)^{F_L}$ orbifold, one can understand the
$S$-duals of more general boundary conditions made by combining
these with fivebranes. We will describe a few examples and compare
the results we get to results of the standard $T(SU(n))$
construction.

\begin{figure}
  \begin{center}
    \includegraphics[width=3.5in]{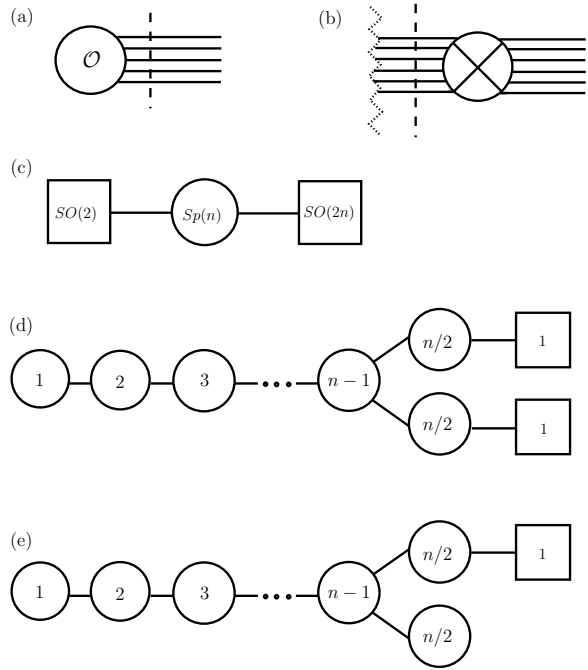}
  \end{center}
\caption{\small (a) A boundary condition in $U(n)$ gauge theory
created by a D5-brane together with an orbifold plane (which is
depicted by a circle containing the symbol $\cal O$). (b) The
$S$-dual, constructed from an NS5-brane,  D5-brane, and an
orientifold plane, represented by the jagged line. (The D5-brane
is generated from $S$-duality applied to the orbifold plane in
(a), while the NS5-brane is dual to the D5-brane in (a).) (c) A
quiver related to (b). It is balanced, so the Coulomb branch has
$U(1)$ symmetry. (d) The quiver description of the $S$-dual to (a)
that comes by using $T(SU(n))$.
 This should be mirror to the quiver in (c). (e) A boundary condition in which only one of the two
 $U(n/2)$ factors is coupled to a fundamental hypermultiplet leads to this quiver. }
  \label{Fig58}\end{figure}
We can start by considering the configuration in fig.
\ref{Fig58}(a), consisting of an the orbifold plane plus a
D5-brane. This adds a fundamental flavor to both $U(n/2)$ gauge
groups at the boundary. The $S$-dual configuration in fig.
\ref{Fig58}(b) involves $n$ D3-branes crossing an NS5-brane
followed by a D5-brane, and ending on an O5$^-$-plane. This
configuration produces a non-trivial SCFT, which arises as the IR
limit of a three-dimensional $Sp(n)$ gauge theory coupled to
fundamental hypermultiplets with flavor symmetry $SO(2n+2)$. This
gauge theory is balanced; hence the Coulomb branch has a hidden
$SO(2)$ symmetry. The bulk $U(n)$ gauge group is embedded in an
$SO(2n)$ subgroup of $SO(2n+2)$.

The $T(SU(n))$ recipe corresponds to the quiver in fig.
\ref{Fig58}(d).  This is  a balanced $\tx D_{n+1}$ type quiver,
with a hidden $SO(2n+2)$ symmetry in the Coulomb branch, and an
obvious $U(1)\cong SO(2)$ symmetry of the Higgs branch. The
quivers in (c) and (d) must be mirror and indeed we see that their
symmetries match.  Such mirror quivers have been considered before
in \cite{Hanany:1999sj}.

Alternatively, we can introduce a single flavor for one of the
$U(n/2)$ groups only. This makes sense as a half-BPS boundary
condition, though we cannot realize it by a brane construction. It
is entertaining to look at the $T(SU(n))$ prescription.  The
quiver in fig. \ref{Fig58}(e) has a single minimally unbalanced
node, and all other nodes are balanced. Though not a good quiver,
this quiver can be readily analyzed with the inequalities of
section \ref{bifurc}. The inequality (\ref{dneed}), with the
unbalanced node labeled as node 3, implies that a monopole
operator with charge at that node has $ q_R\geq 1/2$; moreover,
that value can indeed be achieved. Hence the quiver is ugly.
Omitting the unbalanced node, the symmetry group is $SU(n+1)$. The
unbalanced node has a classical $U(1)$ symmetry. The monopole
operators which previously extended $SU(n+1)\times U(1)$ to
$SO(2n+2)$ now have $q_R=1/2$ instead of $q_R=1$. They transform
in the antisymmetric tensor of $SU(n+1)$, and have charge $1$
under the classical $U(1)$ at the unbalanced node. The dimension
of the Coulomb branch is $n(n+1)/2$, which coincides with the
number of free $q_R=1/2$ hypermultiplets, so the theory is
completely free. The $U(n)$ gauge symmetry is embedded  in
 $SU(n+1) \times U(1)$. (How the center of $U(n)$ is embedded is not quite clear.)

\begin{figure}
  \begin{center}
    \includegraphics[width=3.5in]{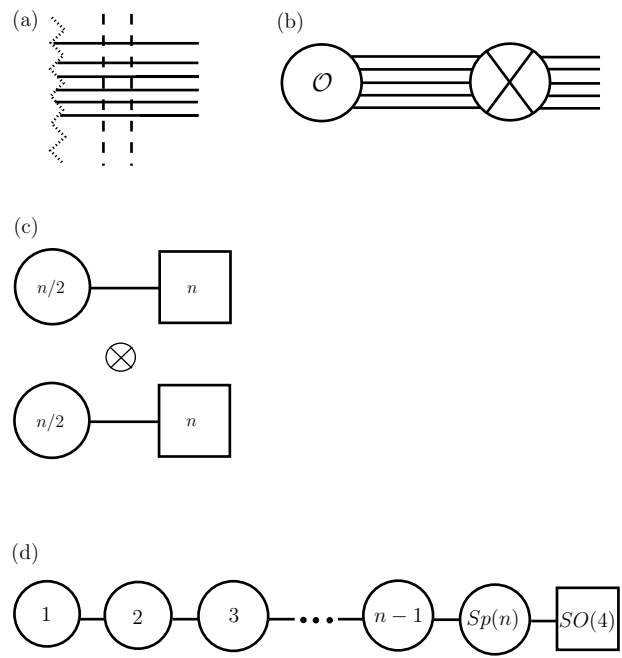}
  \end{center}
\caption{\small (a) The $Sp(n)$ orientifold plus two half
D5-branes. (b) The $S$-dual of (a), constructed from an orbifold
plane and a half NS5-brane. (c) The corresponding quiver. (d) The
result of applying the $T(SU(n))$ recipe to (a).}
  \label{Fig59}\end{figure}

One can add more matter fields coupled to $U(n/2)\times U(n/2)$ at
the boundary, but this does not seem to add many new ideas. Instead
we will do the opposite and add matter fields on the $Sp(n)$ side of
the duality. We begin with fig. \ref{Fig59}(a): $n$ D3-branes cross
two D5-branes and end on an O5$^-$-plane. This is a boundary at
which the $U(n)$ bulk gauge theory is broken to $Sp(n)$, and
fundamental hypermultiplets with $SO(4)$ flavor symmetry are coupled
to the surviving gauge group.

The $S$-dual brane configuration in fig. \ref{Fig59}(b) involves
an orbifold plane and a single NS5-brane. The  $4d$ $U(n)$ gauge
theory in the slab between the orbifold and the fivebrane is
broken to $U(n/2)\times U(n/2)$ by   the orbifold boundary
condition. At the NS5-brane, there are bifundamental
hypermultiplets coupling each of the $U(n/2)$ groups to the bulk
$U(n)$ gauge theory on the half space. In the infrared we are led
to a boundary condition in which the full $U(n)$ gauge theory is
preserved, and coupled diagonally to an SCFT which is the product
of two copies of a $U(n/2)$ three-dimensional gauge theory with
$n$ flavors, as in fig. \ref{Fig59}(c).

This SCFT has a hidden $SU(2) \times SU(2)$ symmetry on the
Coulomb branch due to the two independent balanced nodes. This
matches the $SO(4)$ flavor symmetry of the original boundary
condition.

It is also interesting to compare this to the $T(SU(n))$
prescription for the $S$-dual boundary condition.  The quiver is
depicted in fig. \ref{Fig59}(d), and is a balanced quiver with a
symplectic node at the end. We know from the monopole analysis
that we should expect an $SU(n) \times SU(n)$ hidden symmetry in
the Coulomb branch, which matches well the symmetry of the Higgs
branch of fig. \ref{Fig59}(c).

\begin{figure}
  \begin{center}
    \includegraphics[width=3.5in]{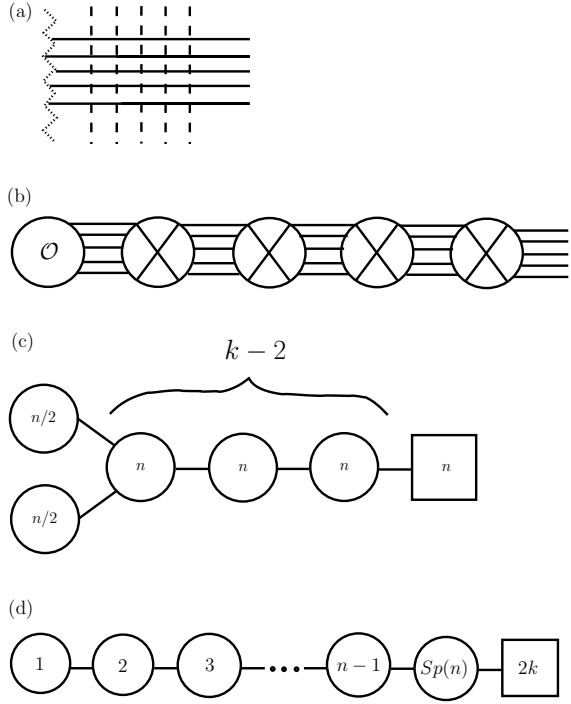}
  \end{center}
\caption{\small (a)  The analog of fig. \ref{Fig59}(a) with $k$
half D5-branes (sketched here for $k=5$) rather than 2.  (b-d) The
corresponding dual brane configurations and quivers. The main
difference is that the quiver in (c) is now connected and is of
type $\tx D$.  The quivers in (c) and (d) are still mirror. }
  \label{Fig60}\end{figure}
Finally we may consider a similar situation with $k>2$ flavors, as
in fig. \ref{Fig60}(a). The $S$-dual brane configuration in fig.
\ref{Fig60}(b) leads to a boundary condition in which the full
$U(n)$ gauge group at the boundary is coupled to the SCFT depicted
in fig. \ref{Fig60}(c). This SCFT is described by a balanced $\tx
D_k$ quiver. The monopole analysis predicts an $SO(2k)$ symmetry
of the Coulomb branch, which matches  the symmetry of the original
boundary condition. The $T(SU(n))$ prescription provides the
quiver in fig. \ref{Fig60}(d): only the unitary nodes are
balanced, and indeed we only expected an $SU(n)$ symmetry of the
Coulomb branch, not $SU(n) \times SU(n)$.

\section{Boundary Conditions For Orthogonal And Symplectic Gauge
Groups}\label{bcos} We want to extend some of the explicit
constructions of section \ref{examples} to orthogonal and
symplectic gauge groups. A well-known way to generate in string
theory an $\N=4$ super Yang-Mills gauge theory with orthogonal or
symplectic gauge groups is to consider D3-branes in the background
of an O3-plane. Brane constructions involving D3-branes and
fivebranes in the presence of an O3-plane have been introduced in
\cite{Feng:2000eq} as a tool to construct mirror pairs of
orthosymplectic linear quivers. We will review and clarify this
construction in the next subsection, and then adapt it to describe
$S$-duality of boundary conditions.

\subsection{Review of O3 Planes}
${\rm O}3$-planes in type IIB string theory come in four kinds,
distinguished by $\Z_2$ valued discrete fluxes of RR and NS
B-fields. The $S$-duality $SL(2,\Z)$ symmetry acts on the discrete
fluxes much like $SL(2,\Z)$ acts on the spin structures of a
two-torus. If the fluxes are zero, the orientifold plane, indicated
here as ${\rm O}3^-$, is invariant under $S$-duality and carries
$-1/4$ unit of D3-brane charge. $n$ D3-branes in the background of
an ${\rm O}3^-$-plane carry an $SO(2n)$ gauge theory.

\begin{table}
\caption{\small This table give the discrete RR and NS fluxes of
an O3-plane, its name, its D3-brane charge, the type of gauge
group is produces when combined with D3-branes, and its $S$-dual.}
\begin{center}
\begin{tabular}{c|c|c|c|c}
Fluxes&Name&D3-Brane Charge & Gauge Group & $S$-Dual
\\\hline
 $(0,0)$&O3$^-$ & $-1/4$ & $SO(2n)$ & O3$^-$\\$ (0,1)$ &
${\widetilde{\rm O}}3^-$ & $1/4$ & $SO(2n+1)$& O3$^+$\\
$(1,0)$ & ${\rm O}3^+$ &$1/4$ & $Sp(2n)$ & ${\widetilde{\rm
O}}3^-$
\\ $(1,1) $& ${\widetilde{\rm O}}3^+$ & $1/4 $& $Sp'(2n)$ & ${\widetilde{\rm O}}3^+$
\\
\end{tabular}
\end{center}\label{zolbo}
\end{table}

Adding half a D3-brane changes the RR discrete flux to 1 and the
D3-brane charge to $+1/4$. The resulting gauge group is $SO(2n+1)$.
This $\widetilde{{\rm O}}3^-$-plane is invariant under
$T:\tau\to\tau+1$ and transforms under $S:\tau\to-1/\tau$ to an
object ${\rm O}3^+$ with NS discrete flux. $n$ D3-branes in the
background of an ${\rm O}3^+$-plane carry an $Sp(2n)$ gauge group.
Finally a $T$ transformation on ${\rm O}3^+$ adds half a unit of
theta-angle to the $Sp(2n)$ gauge theory, leading to an object
called $\widetilde{{\rm O}}3^+$. Whenever we label the ${\rm
O}3$-planes by the corresponding gauge group, we will label this as
$Sp'(2n)$.  (The prime means the following: the theta-angle of the
$Sp'(2n)$ gauge theory differs by $\pi$ from the theta-angle of the
underlying Type IIB superstring theory.  In the other cases, the two
angles are equal.)

The most obvious question for us is what fivebrane configurations
will lead to simple boundary conditions and domain walls in the
presence of ${\rm O}3$-planes. Roughly speaking one can introduce
``half'' D5 and NS5-branes, which lead to simple field theory
constructions, similar to those for unitary groups. But there are
important subtleties.

An NS5-brane is  defined simply by a conformal field theory,
although not one which is known explicitly. If we take the
NS5-brane to have worldvolume in the $012789$ directions and to be
localized at $y=x^4=x^5=x^6=0$, then the SCFT is invariant under
reflection of directions $456789$, so one can construct an
orientifold SCFT. This describes the {\rm O}3-plane interacting
with what is usually called a half NS5-brane.  All fields and
couplings in this orientifold spacetime are obtained by applying a
$\Z_2$ projection to whatever one has before orientifolding.

At large distances along the $y$ direction, the target spacetime
of the orientifold must resemble one of the four flat space ${\rm
O}3$-planes. Because of the $H$-field of the half NS5-brane, the
NS flux will jump across the NS5-brane and the type of ${\rm
O}3$-plane will be actually different at large positive or large
negative $y$.  This means that if the gauge group is of
orthogonal type on one side of the NS5-brane, it will be of
symplectic type on the other.

Before orientifolding, if D3-branes end on an NS5-brane from both
sides, one  gets bifundamental hypermultiplets of the appropriate
$U(n)\times U(m)$ group.  In the present situation, $U(n)\times
U(m)$ is projected to $SO(n)\times Sp(m)$ (or $Sp(n)\times SO(m)$);
the bifundamental representation of $SO(n)\times Sp(m)$ is
pseudoreal, making it possible to define the $\Z_2$ projection for
the hypermultiplets. It follows from what we have just explained
that constructions based on half NS5-branes will generally produce
the sort of orthosymplectic quivers considered in section
\ref{orquiv}.

Given the existence of a half NS5-brane, $S$-duality implies the
existence of a ``half'' D5-brane across which the RR flux of the
O3-planes jumps. This will correspond to a domain wall between
$SO(2n)$ and $SO(2n+1)$ gauge theories, or between $Sp(2n)$ and
$Sp'(2n)$. The two cases are fundamentally different, however. In
the $Sp(2n)$ case, the half D5-brane can be constructed explicitly
from free fields, and all its properties can be calculated. In the
orthogonal case, there is no explicit construction, and this will
lead to some unusual properties.  The difference can be explained
as follows.

In general, before introducing the O3-plane, we can consider for any
positive integer $k$ a system of $k$ D5-branes of worldvolume
$012456$. Their gauge symmetry is $U(k)$. To make an orientifold
projection of this object, we need to choose an outer involution
$\tilde \tau$ of the Chan-Paton bundle that squares to $\pm 1$. An
outer involution is one which acts by complex conjugation times
conjugation by an element of $U(k)$. Similarly to define the
orientifold projection for D3-branes, one picks  an outer involution
$\tau$ squaring to $\pm 1$. The D3-brane gauge symmetry is
orthogonal or symplectic for $\tau^2=1$ or $-1$. The flavor symmetry
of hypermultiplets arising from $3-5$ strings is the gauge symmetry
of the D5-branes and is likewise orthogonal or symplectic for
$\tilde \tau^2=1$ or $-1$. As before, the D3-D5 strings admit an
orientifold projection only if one group is orthogonal and one is
symplectic, so we need $\tau^2=-\tilde\tau^2$.  (Of course one can
also give a conformal field theory explanation of this fact as in
\cite{Gimon:1996rq}.)

To get a half D5-brane, we want $k=1$, but this is  possible only if
$\tilde\tau^2=1$ and hence $\tau^2=-1$. So it is possible only for
an O3-plane of $Sp$ type. Making the orientifold projection of the
usual D3-D5 system leads to a single fundamental hypermultiplet of
$Sp(2n)$, with flavor symmetry $O(1)$. The jumping of the RR flux
across the half D5-brane means that the theta-angle must jump by
$\pi$ in crossing it. This is consistent with the $\Z_2$ anomaly of
three-dimensional $Sp(2n)$ gauge theory with an odd number of real
hypermultiplets: the anomaly can be cancelled by a half-integral
Chern-Simons term, or equivalently in our context by letting the
four-dimensional theta-angle jump by $\pi$.

If $\tilde \tau^2=-1$, $k$ must be even. After taking the $\Z_2$
projection, the D5-brane charge $k/2$ is an integer. The object
with smallest D5-brane charge has $k=2$ and is called a full
D5-brane.

We call the $k=2$ object a full D5-brane for either sign of
$\tilde\tau^2$.  However, for $\tilde \tau^2=1$, the full D5-brane
is trivially a direct sum of two half D5-branes.

For $\tilde\tau^2=-1$, the zero mode of the worldvolume scalar
field $\Phi$ which parametrizes the relative motion of the two
original D5-branes along $y$ is projected out by the orientifold.
The relative motion of two D5-branes is described by fields valued
in the adjoint representation of $SU(2)$, so let us omit the
diagonal part of $\Phi$ and consider only the adjoint-valued part.
We also set $\vec x=(x^4,x^5,x^6)$. The orientifold projection on
$\Phi$ is
\begin{equation}\label{orproj}\Phi(-\vec x)=-\Phi(\vec
x),\end{equation} ensuring that $\Phi$ has no zero mode.

There is no free field construction of a half D5-brane in the
presence of an ${\rm O}3$-plane of orthogonal type, but still, as
already noted, $S$-duality implies its existence. In crossing such
an object, since the RR flux jumps, the gauge group jumps between
$SO(2n)$ and $SO(2m+1)$ for some $n,m$. Actually it is possible to
show that this object must exist without making use of $S$-duality.
We use the fact that in Type IIB superstring theory, without any
orientifolding, one can have a supersymmetric configuration
consisting of a D5-brane with different numbers $p$ and $q$ of
D3-branes ending on the two sides. The gauge group jumps from $U(p)$
to $U(q)$, with a Nahm pole of rank $|p-q|$, but with no extra
degrees of freedom supported at the intersection. (See section 3.4.4
of \cite{Gaiotto:2008sa} for more detail.) The orthogonal-type
orientifold projection of this configuration has $SO(p)$ gauge
symmetry on one side, and $SO(q)$ on the other. It still has no
degrees of freedom supported at the intersection, since the
projection of nothing is nothing. It is only consistent if $p-q$ is
odd, because that is the  case that the Nahm pole is real.  So this
type of D-brane configuration exists even though it cannot be
constructed with free two-dimensional fields.

At this stage, for intersection with an O3-plane of orthogonal gauge
symmetry, we have two superficially similar configurations with the
same D5-brane charge, namely a full D5-brane or a pair of parallel
half D5-branes (fig. \ref{Fig61}). If the number of half D3-branes
on the left and right of the figure is $p$ (corresponding to $SO(p)$
gauge symmetry), then the number inside must differ from $p$ by an
odd number.  For reasons that will become clear, we have chosen the
number inside to be $p+1$.

\subsubsection{Splitting And $R$-Symmetry}\label{splitting}

\begin{figure}
  \begin{center}
    \includegraphics[width=3.5in]{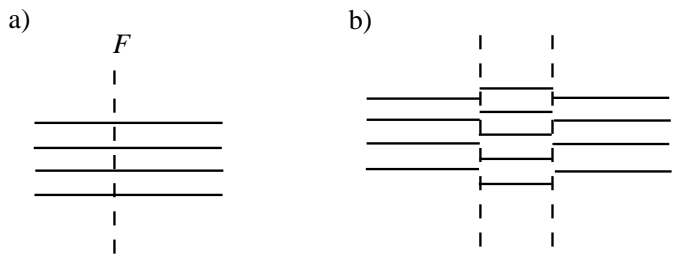}
  \end{center}
\caption{\small (a) D3-branes parallel to an O3-plane of
orthogonal type (not drawn) intersecting a full D5-brane.  Here
and later, a vertical dotted line labeled by the letter $F$
represents a full D5-brane, while an unlabeled vertical dotted
line represents a half D5-brane. (b) The same O3-plane
intersecting a pair of parallel half D5-branes.}
  \label{Fig61}\end{figure}
Consider a full D5-brane that intersects an O3-plane of orthogonal
type? Can we ``split'' the full D5-brane into a pair of half
D5-branes by giving an expectation value to $\Phi$? The orientifold
projection (\ref{orproj}) makes it impossible to give $\Phi$ a
constant expectation value, which we usually think of as the way to
separate two branes. What saves the day is that the full D5-brane
has an $Sp(2)\cong SU(2)$ gauge symmetry. To preserve the
supersymmetry of the orientifold plane, we need not take $\Phi$ to
be constant.  It is enough to obey the Bogomolny equations of
$SU(2)$:
\begin{equation}\label{bogeq} F=\star D\Phi.\end{equation}
The usual hedgehog solution of monopole charge 1 on $\R^3$ is of
the form
\begin{equation}\label{nogeq}\Phi=f(r) \vec x\cdot\vec
t,\end{equation} where $\vec t$ are the $\frak{su}(2)$ generators
and  $r=|\vec x|$ (and the gauge field is given by an analogous
ansatz). Moreover, $f(r)\sim L/r$ for large $r$ for some constant
$L$, so that $|\Phi|$ approaches $L$ at infinity.

This solution obeys the orientifold condition (\ref{orproj}).  At
large $r$, it describes two D5-branes separated by an amount $L$,
just as in fig. \ref{Fig61}(b). (For small $r$, inside the core of
the monopole, the solution is more complicated; the size of the
core is of order $1/L$ and so is negligible for large $L$.  This
is consistent with the semiclassical picture of fig.
\ref{Fig61}(b) in which $L$ is much larger than the string scale
and the core is not seen.) Interpreting the solution on
$\R^3/\Z_2$ rather than $\R^3$, the monopole charge is $1/2$. This
represents one extra half D3-brane stretched between the two
D5-branes in the figure, in addition to those already present in
fig. \ref{Fig61}(a). This explains the fact that in fig.
\ref{Fig61}(b), there is precisely one extra half D3-brane between
the two half D5-branes.

So the two configurations can be deformed into each other.  But
there is a very important point to be made about the symmetries.
In fig. \ref{Fig61}(a), there is an $SU(2)_X$ symmetry that
rotates $\vec x$. (The group that acts faithfully on $\vec x$ is
of course $SO(3)_X$.)  There is also an $SU(2)$ gauge symmetry of
the D5-branes, which is realized in the D3-brane theory as a
flavor symmetry $SU(2)_F$.

On the other hand, in the separated configuration of fig.
\ref{Fig61}(b), we see only one combination of these symmetries --
the rotation symmetry $SU(2)'_X$ that acts on $\vec x$. (Again, it
is the quotient $SO(3)'_X$ that acts faithfully on $\vec x$.)  The
second symmetry has been lost. The reason that we have given
$SU(2)'_X$  a different name from the $SU(2)_X$ rotation symmetry
of fig. \ref{Fig61}(a) is that actually they do not coincide even
in the limit $L\to 0$.   The hedgehog solution is not invariant
under either the rotation symmetry $SU(2)_X$ or the gauge symmetry
$SU(2)_F$, but only under a diagonal subgroup.  It therefore is
this diagonal subgroup that is the symmetry of the separated
configuration with two half D5-branes and so corresponds to
$SU(2)'_X$.

Going back to the configuration with the full D5-brane, from the
point of view of the $SO(p)$ gauge theory on the D3-brane, the low
energy physics is described by the coupling of the bulk
four-dimensional gauge fields to a three-dimensional SCFT.  This
SCFT is a free field theory that describes the bifundamental
hypermultiplet $H$.  The bosonic components of $H$ transform as
$(1/2,1/2)$ under $SU(2)_X\times SU(2)_F$.  They also have conformal
dimension 1/2, telling us that they must transform with spin 1/2
under the $R$ symmetry that is part of the superconformal algebra.
Since $SU(2)_F$ commutes with supersymmetry while $SU(2)_X$ is an
$R$-symmetry, the candidate $R$ symmetries are $SU(2)_X$ or a
diagonal subgroup  of $SU(2)_X\times SU(2)_F$. The condition that
$H$ must transform with spin 1/2 tells us that the superconformal
$R$-symmetry in the low energy limit of fig. \ref{Fig61}(a) is
actually $SU(2)_X$.

 The split configuration has a mass scale $L^{-1}$. We can
recover from it a superconformal field theory by going to a low
energy limit in which the separation of the half D5-branes is
unimportant and we recover the physics of the full D5-brane.
However, in view of what we have said, the $R$-symmetry $SU(2)'_X$
that is visible in the split configuration is not the
superconformal $R$-symmetry of the IR limit.

In a sense, this should come as no surprise. The split
configuration  does not obey the usual constraint that linking
numbers should be nondecreasing from left to right. The linking
numbers in the orthosymplectic case are defined for half branes in
essentially the same way that they are defined in the unitary case
for full branes: the linking number of a half fivebrane is the
number  of half fivebranes of the opposite kind to its left plus
the jump in the half D3-brane charge across it, including the
D3-brane charge of the O3-planes. From left to right of fig.
\ref{Fig61}(b), the linking numbers are $1$ and $-1$.  So
naturally, the ultraviolet and infrared $R$-symmetries are
different.

Even though the configuration of fig. \ref{Fig61}(b) violates the
linking number constraint, the above analysis implies that as the
separation between the two half D5-branes is taken to zero, there
is a smooth limit to the unsplit configuration of fig.
\ref{Fig61}(b). This is rather special in that, as shown via
Nahm's equations in section 3.5.1 of \cite{Gaiotto:2008sa}, a
generic D3-D5 configuration that violates the linking number
constraint does not have a similar natural limit to an unsplit
configuration when the D5 separations are taken to zero. This is
shown in \cite{Gaiotto:2008sa} for unitary groups (that is,
without the O3-plane) and is also true for generic configurations
that violate the linking number constraint in the presence of the
O3-plane.  But evidently (and as one can verify from Nahm's
equations), it is not true when the violation of the linking
number constraint comes only by splitting full D5-branes.

Violating the usual linking number constraint means that, under
$S$-duality, we should expect to encounter bad quiver gauge
theories. We will now give a simple example.  We define a full
NS5-brane to be the $S$-dual of a full D5-brane.  We want to
determine the IR dynamics at the intersection of an O3-plane with
a full NS5-brane, by splitting the full NS5-brane to a pair of
half NS5-branes and constructing a quiver.  $S$-duality will
enable us to determine exactly what is the correct quiver.

\begin{figure}
  \begin{center}
    \includegraphics[width=4.5in]{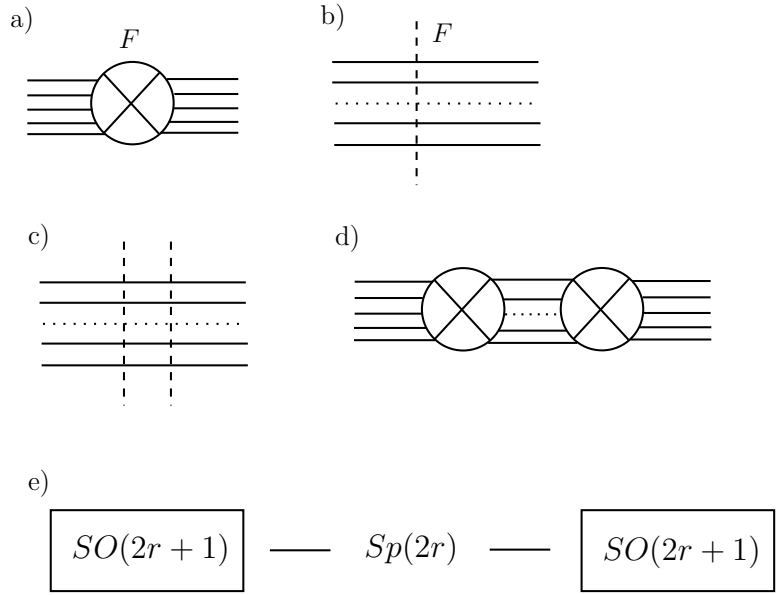}
  \end{center}
\caption{\small (a) A full NS5-brane intersecting an O3-plane of
type $SO(2r+1)$ (sketched here for $r=2$). $S$-duality leads to
(b), in which the O3-plane is now of symplectic type. Splitting
the full D5-brane leads to (c), and finally $S$-duality brings us
back to (d).  The theory in the central slab is $Sp'(2r)$.  The
three-dimensional limit is described by the good quiver shown in
(e).  In (b) and (c), and below, the horizontal dotted line
represents a half unit of D3-brane charge carried by an O3-plane
of symplectic type.  This dotted line  contributes to linking
numbers (though not to gauge symmetries) and drawing it lets us
distinguish visually between O3-planes of orthogonal or symplectic
type. By contrast, in these figures, an O3-plane of orthogonal
type is not explicitly drawn.}
  \label{Fig62}\end{figure}
First we consider a full NS5-brane intersecting an $\tilde{\rm
O}3^-$-plane with gauge group $SO(2r+1)$ for some $r$ (fig.
\ref{Fig62}). The $S$-dual is a full D5-brane intersecting an
O3$^+$-plane with gauge group $Sp(2r)$.  This can be
straightforwardly split to two perturbative half D5-branes, with
gauge group $Sp'(2r)$ between them. Then applying $S$-duality again,
we find that the original configuration with the full NS5-brane is
dual to a system of two half NS5-branes with $Sp'(2r)$ in the
central slab.  The difference between $Sp(2r)$ and $Sp'(2r)$ is
unimportant in the low energy three-dimensional limit, so that limit
gives  an $Sp(2r)$ gauge theory with flavor symmetry $SO(4r+2)$,
coming from bifundamentals at the two ends. This is a good and in
fact balanced theory, in the sense of section \ref{osymp}, so the
infrared and ultraviolet $R$-symmetries agree, consistent with the
fact that in this case the splitting of the D5-brane was
straightforward. Moreover, since the theory is balanced, the Coulomb
branch has an $SO(2)$ symmetry in the infrared, which is dual to the
flavor symmetry of two half D5-branes.

The splitting of a full NS5-brane intersecting an $\tilde{\rm
O}3^+$-plane is similar, since the dual D5-brane again intersects
an O3-plane of symplectic type and can be split straightforwardly.
The gauge theory describing the full NS5-brane turns out to be a
balanced  $SO(2r+1)$ gauge theory  with flavor symmetry $Sp(4r)$.
Again the ultraviolet and infrared $R$-symmetries match and the
Coulomb branch has an $SO(2)$ symmetry in the infrared.

\begin{figure}
  \begin{center}
    \includegraphics[width=3.5in]{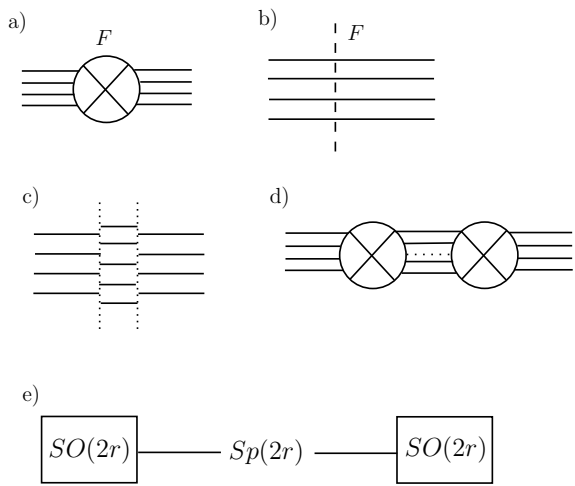}
  \end{center}
\caption{\small (a) A full NS5-brane intersecting an O3-plane of
type $SO(2r)$ (again sketched for $r=2$). $S$-duality leads again
to (b), and splitting the full D5-brane and applying $S$-duality
again leads to (c) and then (d).  But now, as the O3-plane in (b)
is of orthogonal type, the $R$-symmetry is modified in this
process. Accordingly, the quiver in (e), which describes the
three-dimensional limit of the configuration in (d), is now a bad
one.  }
  \label{Fig63}\end{figure}
The other two cases behave differently.  A full NS5-brane
intersecting an O3$^-$-plane is shown in fig. \ref{Fig63}(a). In
carrying out the usual duality operations, we have to split a full
D5-brane interacting with an O3-plane of orthogonal type, so we
expect the $R$-symmetry to be modified. Indeed, the gauge theory
description turns out to be an $Sp(2r)$ theory with flavor symmetry
$SO(4r)$.  This theory is bad, by the criterion of section
\ref{osymp}, so it is indeed impossible to match the ultraviolet and
infrared $R$-symmetries. In the infrared, the Coulomb branch is
supposed to have an $Sp(2)$ global symmetry, matching the flavor
group of the full D5-brane.  But as the gauge theory is bad, we
cannot see this symmetry by studying monopole operators.

The last case of a full NS5-brane intersecting O3$^+$-plane is
similar.  It is dual to an $SO(2r+2)$ theory with flavor symmetry
$Sp(4r)$.  This is a bad theory according to section
\ref{orthosymp}, consistent with the fact that the $R$-symmetry is
modified in splitting the full D5-brane.

\subsubsection{Monopoles And Orientifolds}\label{monor}

Depending on the type of orientifold projection, only a  net even
or odd number of half D3-branes can end on a half D5-brane.  From
the viewpoint of the D3-brane theory, this reflects the reality
property of the Nahm pole.  We want to investigate the matter from
the point of view of the D5-brane theory.

The endpoint of a half D3-brane looks like a singular monopole of
charge 1 in the $U(1)$ gauge theory of the D5-brane.  So to use
D5-brane field theory, we will omit the points where D3-branes
end, and discuss the topology of the orientifold projection on a
large two-sphere $S$  given by $r={\rm constant}$.

\def\LL{{\mathcal L}}
Let $\LL\to S$ be any line bundle. Let $\pi:S\to S$ be the
antipodal map $\vec x\to -\vec x$. There exists an antiunitary
isomorphism $\phi$ between $\LL$ and $\pi^*(\LL)$, because $\LL$
is completely classified topologically by its first Chern class
$c_1(\cal L)$, which is odd under both $\pi^*$ and complex
conjugation.  There is no natural choice of $\phi$, but $\phi^2$
is independent of the choice and equals $1$ or $-1$, depending
only on the topology of $\LL$. The formula is in fact that
$\phi^2=(-1)^{c_1(\LL)}$.

Now we can refine the criterion for when a half D5-brane exists.
The definition of the orientifold projection requires a choice of
antiunitary isomorphism $\tilde\tau$ from the Chan-Paton gauge
bundle of the D5-brane to itself.  For a single half D5-brane,
this bundle is a line bundle $\LL$, and in view of what is said in
the last paragraph, we have $\tilde\tau^2=(-1)^{c_1({\LL})}$. If
the D3-brane gauge theory is symplectic, we want $\tilde\tau^2=1$,
so in this case $c_1(\LL)$ must be even.  Since a half D3-brane
ending on a half D5-brane carries magnetic charge, this result
means that only a net even number of half D3-branes can end on a
single half D5-brane, as we already know.

If the D3-brane gauge theory is orthogonal, we want
$\tilde\tau^2=-1$, so $c_1(\LL)$ must be odd.  Hence only a net
odd number of half D3-branes can end on the half D5-brane, as we
also already know.

\begin{figure}
  \begin{center}
    \includegraphics[width=4in]{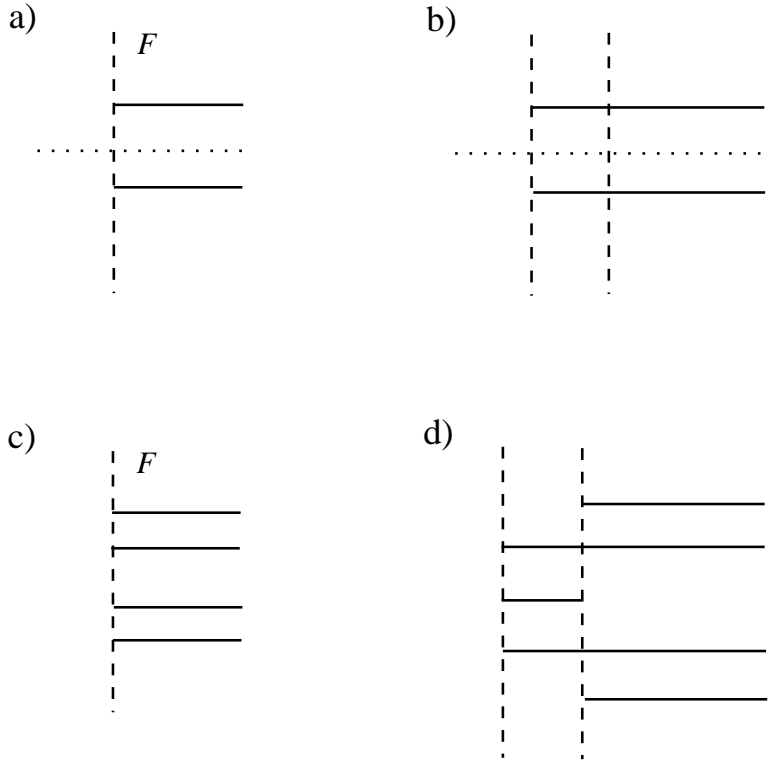}
  \end{center}
\caption{\small (a) An O3-plane of symplectic type intersecting a
full D5-brane, on which $4s+2$ half D3-branes end, drawn here for
$s=0$. (b) The gauge bundle at infinity in the last figure is
$\LL\oplus \LL$, so after splitting the full D5-branes, one might
expect $2s+1$ half D3-branes to end on each half D5-brane.
Instead, solving the Bogomolny equations leads to the situation
drawn here, the numbers (from left to right) being $2s+2$ and
$2s$. (c) An O3-plane of orthogonal type intersecting a full
D5-brane on which $4s$ half D3-branes end, drawn here for $s=1$.
(d) Upon splitting the D5-brane, an extra D3-brane fragment is
created via the Bogomolny equations, with the result that a net
odd number of half D3-branes ends on each half D5-brane. }
  \label{Fig63a}\end{figure}

Now, taking the gauge theory of the D3-branes to be symplectic, we
want to consider a problem of $4s+2$ half D3-branes ending on a
full D5-brane (fig. \ref{Fig63a}(a)).  The D5-brane gauge bundle
$E\to S$ is now a rank two bundle with $c_1(E)=4s+2$. It is
possible for such a configuration to have an $SU(2)$ global
symmetry which we call $SU(2)_F$.  For this, we require
$E=\LL\oplus \LL$, where $\LL$ is a line bundle with
$c_1(\LL)=2s+1$. The orientifold projection now has an unusual
property. The antiunitary isomorphism $\tilde\tau:E\to\pi^*E$ must
square to 1, even though any antiunitary isomorphism $\phi:\LL\to
\pi^*\LL$ obeys $\phi^2=-1$. So we pick
\begin{equation}\label{normox}\tilde\tau=\begin{pmatrix} 0 &
\phi\\ -\phi & 0 \end{pmatrix}, \end{equation} still leaving
$SU(2)_F$ symmetry.

Now we would like to split the D5-brane to a pair of half
D5-branes. Because of the form of $\tilde\tau$, we are in a
situation similar to what happened when we tried to split a full
D5-brane in the presence of an orthogonal O3-plane.  The zero mode
of the relevant adjoint-valued field $\Phi$ is projected out, but
the splitting is possible anyway with the help of  an 't
Hooft-Polyakov monopole. The result is that the split
configuration  exists, but it is not true, as one might expect
from the choice $E=\LL\oplus \LL$ with $c_1(\LL)=2s+1$, that
$2s+1$ half D3-branes end on each half D5-brane. Rather, the 't
Hooft-Polyakov monopole gives an extra half D3-brane between the
two half D5-branes, so from left to right the net numbers are
$2s+2$ and $2s$, as in fig. \ref{Fig63a}(b). As before,
$SU(2)_X\times SU(2)_F$ is broken to a diagonal subgroup.

This has a variant (fig. \ref{Fig63a}(c)) for a full D5-brane
interacting with an O3-plane of orthogonal type.  We recall that
this is obtained by orientifolding a configuration with a pair of
D5-branes intersecting the O3-plane.  Suppose that, before the
orientifolding, a net of $4s$ D3-branes end on the pair of
D5-branes.  We can take the D5-brane gauge bundle $E\to S$ to be
$E=\LL\oplus \LL$, $c_1(\LL)=2s$, giving a configuration again
with flavor symmetry $SU(2)_F$.  The orientifold projection must
obey $\tilde\tau^2=-1$, and since $c_1(\LL)$ is even, ensuring
that $\phi^2=1$, we can accomplish this with $\tilde\tau$ of the
same form as in (\ref{normox}).  With the help of the Bogomolny
equations, we can again split the full D5-brane, once again
breaking $SU(2)_X\times SU(2)_F$ to a diagonal subgroup and
arriving at fig. \ref{Fig63a}(d).

\subsection{Simple Boundary Conditions From Branes}
The basic idea of section \ref{examples} was to realize boundary
conditions in $U(n)$ gauge theory in terms of D3-branes ending on
a system of fivebranes, and use the properties of the branes to
determine the action of $S$-duality.  We have developed the tools
we need to do the same for orthogonal and symplectic groups, by
adding an O3-plane to the previous constructions.

$n$ D3-branes ending on a single NS5-brane gives Neumann boundary
conditions in $U(n)$ gauge theory.  So, applying an orientifold
projection, $n$ half D3-branes ending on a single half NS5-brane
gives Neumann boundary conditions in $SO(n)$ or $Sp(n)$ gauge
theory.\footnote{A variant is the case that an $Sp'$ theory ends
on a half NS5-brane, with a single half D3-brane on the other
side. There is then a real fundamental hypermultiplet of $Sp(n)$
at the boundary.}

The dual of Neumann boundary condition then corresponds to a
configuration in which all D3-branes end on a single half D5-brane.
This must correspond to a regular Nahm pole, as otherwise upon
solving Nahm's equations, it would support a moduli space of vacua.
If the bulk gauge group is $Sp(2n)$, the regular $\frak{su}(2)$
embedding corresponds to the $2n$-dimensional irreducible
representation of $SU(2)$, and similarly for $SO(2n+1)$ it
corresponds to the $2n+1$-dimensional irreducible representation. On
the other hand, the regular embedding of $SO(2n)$ corresponds to the
decomposition $2n=(2n-1)+1$. This is consistent with the fact that
on the other side of the half D5-brane the orientifold is of the
$\widetilde{\rm O}3^-$ type, which supports odd orthogonal gauge
symmetry. Only $2n-1$ of the $2n$ half D3-branes stop at the half
D5-brane, leaving $O(1)$ on the other side.

More general Nahm poles can be produced by combining several half
D5-branes. Consider an $SO(n)$ bulk gauge theory. We have learned
that at a single half D5-brane only an odd number of half
D3-branes can end. This produces naturally any Nahm pole of odd
rank. We can split any Nahm pole in which every summand is odd
dimensional into a sequence of elementary Nahm poles, ordered by
increasing dimension, in complete parallel with the unitary
construction. $S$-duals of such boundary conditions are trivially
found. The corresponding set of half NS5-branes gives rise to a
good linear  orthosymplectic  quiver. The quiver has balanced
nodes whenever successive  summands of the Nahm pole have the same
dimension, and the enhanced orthogonal symmetries of the Coulomb
branch match the subgroup of $SO(n)$ which commutes with the Nahm
pole.

From a field theory point of view, this is not the end of the
story. In the most general $\frak{su}(2)$ embedding in
$\frak{so}(n)$, it is possible to have an even number  of summands
of the same even dimension. If there are $2k$ such summands, the
commutant of this embedding contains a factor of $Sp(2k)$.  This
is the symmetry group associated to $k$ full D5-branes; hence it
is natural to suspect that full D5-branes will be needed to
describe these even rank Nahm poles. The Nahm pole can indeed be
realized by orientifold projection of $2k$ poles of rank $2d$ in
the unitary gauge theory.

To describe the $S$-dual configuration via gauge theory, we need
to split the D5-branes to half D5-branes (whose duals are half
NS5-branes that have a simple gauge theory interpretation).   This
splitting, however, involves the process of fig. \ref{Fig63a}(c,d)
in which a half D3-brane is created and the half D5-branes violate
the linking number constraint.  After the splitting, an odd number
of half D3-branes end on each half D5-brane, but the $R$-symmetry
no longer coincides with the $R$-symmetry of the unsplit
configuration. Once the configuration has been split in this way,
it is straightforward to identify its $S$-dual as an
orthosymplectic quiver.  The only drawback is that this quiver
will be a bad quiver, with a bad node corresponding to each
unusually ordered pair of half D5-branes.

We can proceed in much the same fashion for symplectic gauge
theories. Now the simplest Nahm poles to describe are the ones of
even rank. Those are easily mapped to half D5-branes on which an
even numbers of half D3-branes end, with the gauge group changing
from $Sp(2n)$ to $Sp'(2m)$ or viceversa. As usual a Nahm pole with
several even summands can be decomposed into a sequence of
elementary Nahm poles of increasing rank, and then $S$-duality is
straightforward. The result is a good orthosymplectic quiver.
Again the commutant of the $\frak{su}(2)$ embedding has orthogonal
factors for every set of summands of the same dimension. This will
map to sequences of consecutive balanced nodes in the $S$-dual
quiver gauge theory.

Just as in the case of an orthogonal gauge group, the full story
is more complicated. In the most general $\frak{su}(2)$ embedding
in $\frak{sp}(2n)$, it is possible to have a pair of summands of
the same odd dimension $d=2s+1$. The simplest example of this is
Dirichlet boundary conditions, where all  summands are of
dimension 1. We have learned in section \ref{monor} that we can
realize two Nahm poles of the same odd dimension with a full
D5-brane using the orientifold projection (\ref{normox}). We have
also learned how to split such a full D5-brane, with the help of
the Bogomolny equations. This leads to a configuration like that
of fig. \ref{Fig63a}(b) in which, from left to right, $d+1$ half
D3-branes end on the first half D5-brane and $d-1$ on the second.
As before, this splitting modifies the $R$-symmetry.

$S$-duality is straightforward and leads to a linear
orthosymplectic quiver. The quiver, however, will have a bad node
for each pair of unusually ordered Nahm poles in the split
configuration.
\begin{figure}
  \begin{center}
    \includegraphics[width=4.5in]{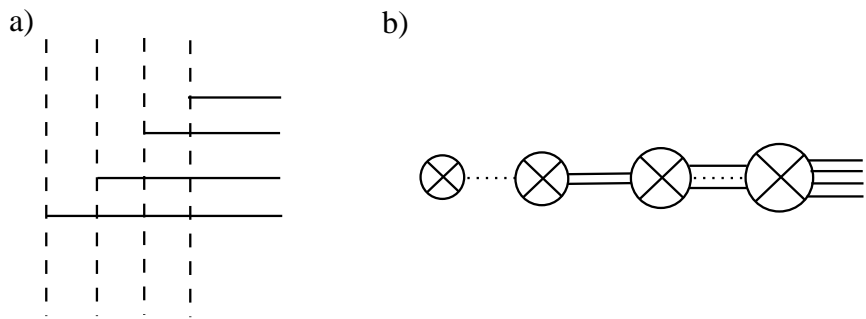}
  \end{center}
\caption{\small (a) Dirichlet boundary conditions in $SO(2n)$
gauge theory can be achieved by letting $2n$ half D3-branes end on
$2n$ half D5-branes (sketched here for $n=2$).  (b) The $S$-dual
configuration, which leads to the self-mirror quiver of fig.
\ref{Fig54}(a).}
  \label{Fig64}\end{figure}

\subsubsection{Quiver Representations For $T(G)$}\label{quivrep}

We are finally prepared to give a quiver description of $T(G)$ for
orthogonal and symplectic groups. The simplest example is
$T(SO(2n))$. The Dirichlet boundary condition for $SO(2n)$ is
realized by $2n$ half D3-branes, each ending on a separate half-D5
brane, as in fig. \ref{Fig64}(a). The $S$-dual of this brane
configuration is depicted in fig. \ref{Fig64}(b), and corresponds
to the good and perhaps even beautiful self-mirror quiver in fig.
\ref{Fig54}(a), which describes $T(SO(2n))$.

\begin{figure}
  \begin{center}
    \includegraphics[width=4.5in]{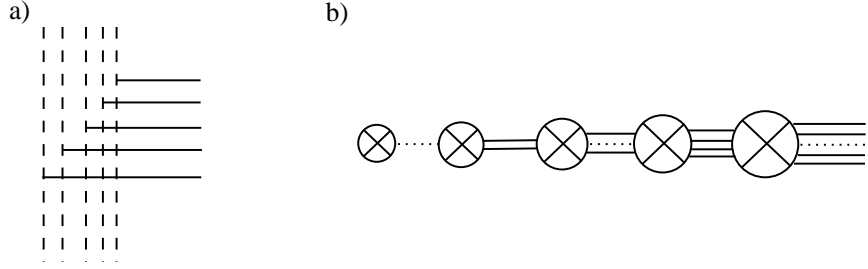}
  \end{center}
\caption{\small (a) A brane realization of Dirichlet boundary
conditions for $SO(2n+1)$, sketched here for $n=2$. (b) The
$S$-dual, which leads to the quiver of fig. \ref{Fig54}(b).}
  \label{Fig65}\end{figure}

One more half D3 and half D5-brane leads (fig. \ref{Fig65}) to a
Dirichlet boundary condition for $SO(2n+1)$, and to a slightly
longer quiver in fig. \ref{Fig54}(b). This is a good quiver
description of $T(Sp(2n))$.

\begin{figure}
  \begin{center}
    \includegraphics[width=5.5in]{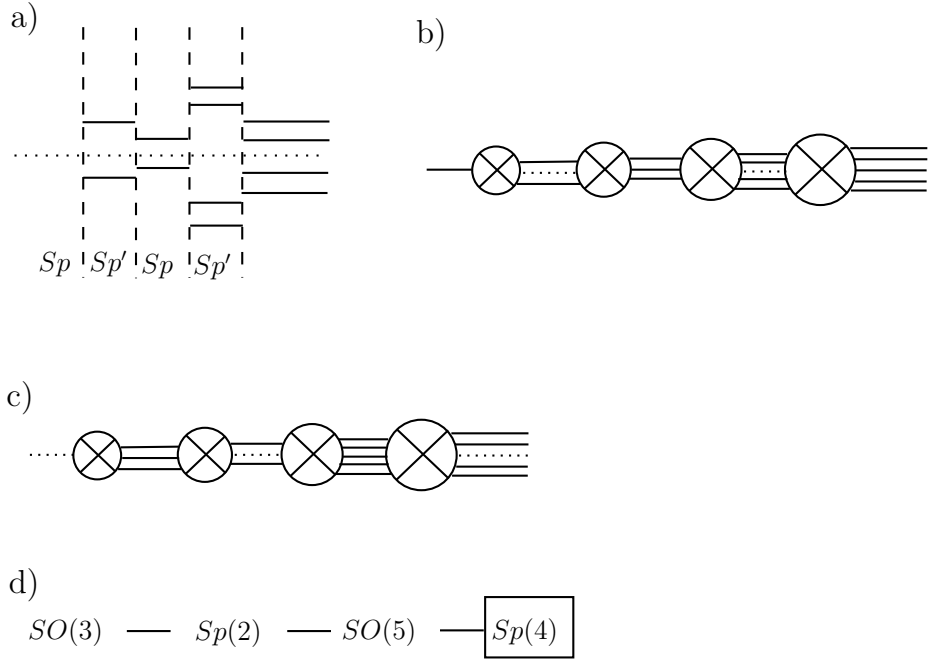}
  \end{center}
\caption{\small (a) A brane realization of the dual of Dirichlet
boundary conditions for $Sp(2n)$, sketched here for $n=2$. Between
the D5-branes, the gauge theories are alternately $Sp$ and $Sp'$
theories; that is, the $\theta$-angle jumps by $\pi$ in crossing
each half D5-brane. (b) The $S$-dual of (a), leading to the bad
quiver of fig. \ref{Fig54}(c) which describes $T(SO(2n+1))$. (c)
If we exchange all $Sp$ and $Sp'$ labels in (a), the dual looks
like this, and the gauge theory limit is the bad quiver in (d).
In (a), (b), and (c), linking numbers are alternately 2 and 0.   }
  \label{Fig66}\end{figure}

The theory $T(SO(2n+1))$ (which is also the mirror of $T(Sp(2n))$)
is the dual of Dirichlet boundary conditions for an $Sp(2n)$ gauge
group. Dirichlet boundary conditions are a very special case of an
even number of odd dimensional summands in the decomposition of
$2n$.  As in the examples just treated, we can realize Dirichlet
boundary conditions for $Sp(2n)$ by letting $2n$ half D3-branes end
on $2n$ half D5-branes.  But now, when we separate the half
D5-branes in $y$ in preparation for $S$-duality, the $R$-symmetry is
modified and several half D3-branes are added, to lead to the
configuration depicted in fig. \ref{Fig66}(a). The dual NS5-brane
configuration is depicted in fig. \ref{Fig66}(b), and the resulting
bad quiver in fig. \ref{Fig54}(c). This gives a description of
$T(SO(2n+1))$ which completely obfuscates the symmetries of the
Coulomb branch.  For $T(SO(3))$, this description was analyzed in
section \ref{irflow}.

Finally, we can ask for the $S$-dual of a Dirichlet boundary
condition for $Sp'(2n)$. The half D5-brane configuration is
essentially identical to the one in fig. \ref{Fig66}(a), but with
the labelling $Sp$ and $Sp'$ permuted. The $S$-dual boundary
condition involves the bad quiver in fig. \ref{Fig66}(c). The
resulting theory, which we could still call $T(Sp'(2n))$, is
apparently self-mirror, and would be useful to find the $S$-dual
of boundary conditions for an $Sp'$ gauge theory.

We can similarly give quiver descriptions of the various
$T^\rho_{\rho^\vee}(G)$ for orthogonal and symplectic gauge
groups. The simplest case, which does not require brane
manipulations, is $T_{\rho^\vee}(G)$, which is the $S$-dual of a
$\rho^\vee$ Nahm pole for $G^\vee$. The Nahm pole is built out of
half D5-branes by following the rules already formulated, and the
$S$-dual orthosymplectic quiver has the same structure as the
$T(G)$ quiver, but with missing nodes.

It is almost as easy to describe a general
$T^\rho_{\rho^\vee}(G)$: $\rho^\vee$ can be realized as a
configuration of  D5-branes in the $G^\vee$ duality frame, and
converted to an identical configuration of  NS5-branes in the $G$
duality frame. In some cases, splitting of these NS5-branes will
modify the $R$-symmetry, but  we do not need to split fully the
D5-brane configuration that generates $\rho$. It is convenient to
represent summands in $\rho$ of the ``correct'' dimension (odd for
orthogonal $G$, even for symplectic $G$) by a half D5-brane, and
pairs of summands of the ``wrong'' dimension by full D5-branes. To
order the branes properly so as to get a gauge theory description,
a half D5-brane with a Nahm pole of odd dimension $d$ needs to be
moved across $d$ half NS5-branes, and ends up representing a
single real flavor at the $d^{th}$ node of the quiver (which is
symplectic). A full D5-brane with two poles each of even dimension
$d$ also is moved across $d$ half NS5-branes and ends up as a full
flavor at the $d^{th}$ node, which now is orthogonal. The ranks of
the gauge groups at the nodes are then determined by following the
brane manipulations or more simply from the linking numbers of the
original half NS5-brane configuration.

The set of $T^\rho_{\rho^\vee}(G)$ includes all mirror pairs built
form linear orthosymplectic quivers.

\subsection{Examples of Interesting Boundary Conditions}

As in section \ref{examples}, we can use these methods to
understand $S$-duality for a much wider class of examples. We
consider some illustrative cases involving Neumann boundary
conditions with fundamental matter at the boundary.
\begin{figure}
  \begin{center}
    \includegraphics[width=5.5in]{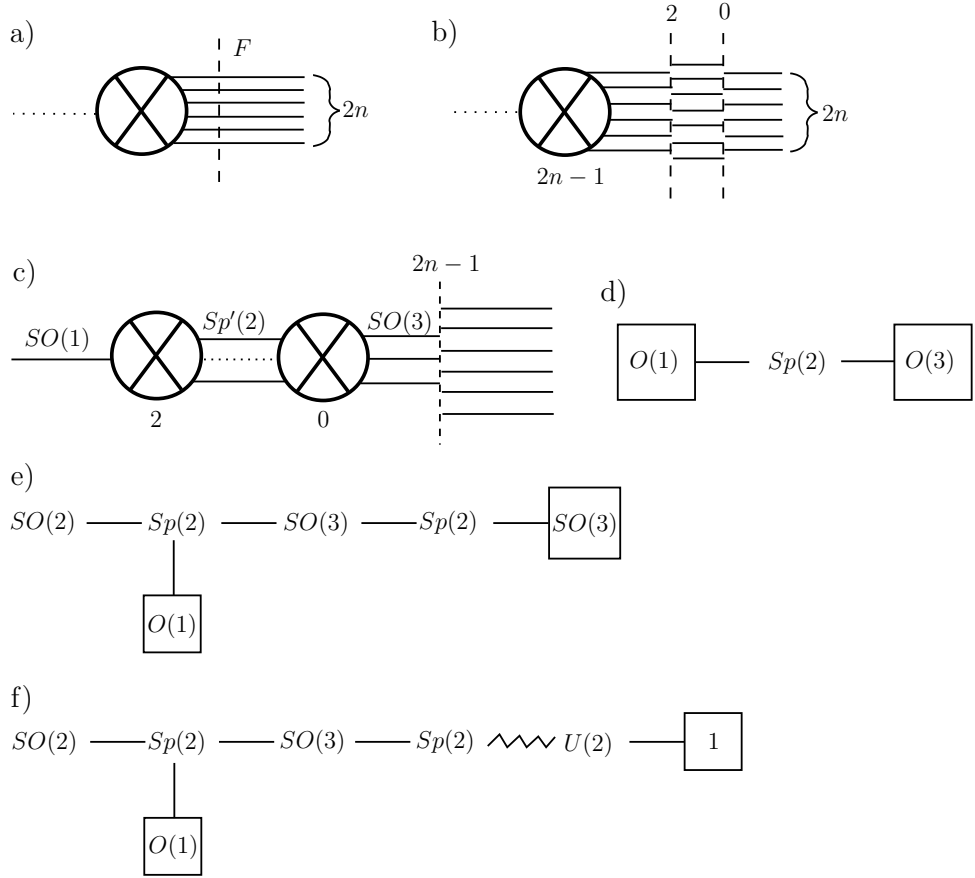}
  \end{center}
\caption{\small  (a) D3-branes intersecting a full D5-brane,
leading to $SO(2n)$ gauge theory coupled to a fundamental
hypermultiplet of flavor symmetry $Sp(2)$.  (b) Splitting the
D5-brane. Linking numbers are indicated. (c) The $S$-dual of (b).
The gauge symmetry is reduced to $SO(3)$ by a Nahm pole. (d) The
$SO(3)$ gauge symmetry near the boundary is coupled to the SCFT
that is generated by this quiver. This is a bad quiver
representing $T(SU(2))$. (e) The quiver representing
$T^\rho(SO(2n))$ for the decomposition $2n=(2n-3)+1+1+1$, sketched
here for $n=3$. (f) The result of diagonally coupling $SO(3)$
gauge fields to $T^\rho(SO(2n))\times T(SU(2))$ is this
``quiver,'' in which the jagged line represents the tensor product
of the fundamental representation of $Sp(2)$ and the fundamental
of $SO(3)$, viewed as a $U(2)$ representation.}
  \label{Fig67}\end{figure}

We begin with the brane configuration in fig. \ref{Fig67}(a). It
produces a  Neumann boundary condition in $SO(2n)$ gauge theory
coupled to a fundamental hypermultiplet with an $Sp(2)$ flavor
symmetry. $2n$ half D3-branes end on the NS5-brane, but the half
D3-brane charge of the O3-plane also jumps by $-1$ across the
NS5-brane, so its linking number is $2n-1$. To do $S$-duality, the
full D5-brane has to be split as in fig. \ref{Fig67}(b). The
resulting configuration has two half D5-branes of linking numbers
$2$ and $0$. The reordered $S$-dual configuration is shown in fig.
\ref{Fig67}(c). The $SO(2n)$ gauge group is broken to $SO(3)$ by a
Nahm pole of dimension $2n-3$, and it is coupled to the CFT
associated to the quiver in fig. \ref{Fig67}(d), which describes an
$Sp(2)$ gauge theory with $SO(4)$ flavor symmetry group. Notice that
one of the four flavors arises from the bifundamental hypermultiplet
at the leftmost NS5-brane, where the gauge theory jumps from
$Sp'(2)$ to $O(1)$. This bad quiver was analyzed in section
\ref{irflow}. Its low energy limit is $T(SO(3))=T(SU(2))$, though in
this flow the microscopic $R$-symmetry is not the one that is
relevant in the infrared.

We can check that our general prescription based on $T(SO(2n))$
agrees with this answer. We want to reproduce the original boundary
condition as the $S$-dual of the boundary condition in fig.
\ref{Fig67}(c). As a Nahm pole $\rho$ (related to the decomposition
$2n=(2n-3)+1+1+1$) is present, we need to use $T^{\rho}(SO(2n))$,
which is the infrared limit of the good quiver in fig.
\ref{Fig67}(e). We could couple this to the quiver in fig.
\ref{Fig67}(d) to produce a dual boundary condition, but as that
quiver is bad we would learn little from it. Alternatively we can
couple it to a more useful description of the theory: the usual
$T(SU(2))$ realization as $U(1)$ with two flavors. The price to pay
is that the resulting theory is not a quiver in the strict sense, as
the $SO(3)\sim SU(2)$ node is coupled  to hypermultiplets in both
the triplet and  the doublet of $SU(2)$. One can verify that
monopoles with $q_R=1/2$ exist in this quiver, with weights which
match the ones of an $SO(2n)\times Sp(2)$ bifundamental free
hypermultiplet. (To get all the monopoles, one needs to know that in
coupling $SO(3)$ to the $T(SU(2))$ quiver the gauge group is really
$U(2)$, not $SO(3) \times U(1)$.)

\begin{figure}
  \begin{center}
    \includegraphics[width=5.5in]{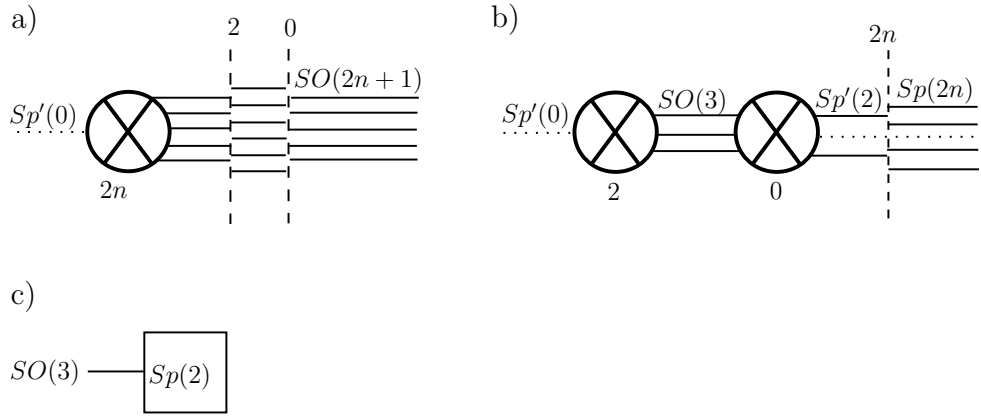}
  \end{center}
\caption{\small (a) Neumann boundary condition for $SO(2n+1)$ (in
the figure $n=2$) with addition of a full D5-brane, which has been
split to arrive at this picture. Linking numbers are indicated. The
$R$-symmetry has been modified by splitting D5-branes.  (b) The
$S$-dual configuration after standard reordering. (c) The bad quiver
representing the boundary SCFT. }
  \label{Fig68}\end{figure}
A similar brane configuration in fig. \ref{Fig68}(a) gives the same
boundary condition for $SO(2n+1)$. The construction of the $S$-dual
boundary condition is rather similar to the previous example; in
particular the linking number of the NS5-brane is $2n$ and those of
the two half D5-branes are again $2$ and $0$. The final result is
shown in fig. \ref{Fig68}(b) and corresponds to a boundary condition
for an $Sp(2n)$ gauge theory reduced to $Sp(2)$ by a Nahm pole of
dimension $2n-2$. The SCFT living at the boundary is depicted in
fig.  \ref{Fig68}(c): it is the IR limit of an $SO(3)$ gauge theory
with $Sp(2)$ flavor symmetry. This is again a bad quiver theory, but
it is also a very well known theory in disguise: $\N=8$ $SO(3)$
gauge theory. In the infrared, the moduli space of this theory is a
mixed branch: $\R^8/\Z_2$, with an $SO(8)$ $R$-symmetry group. Since
the $Sp(2)$ flavor symmetry is coupled to the bulk gauge fields,
only the Coulomb factor of the moduli space is really visible, and
corresponds to the moduli space of Nahm equations in fig.
\ref{Fig68}(a).

\begin{figure}
  \begin{center}
    \includegraphics[width=3.5in]{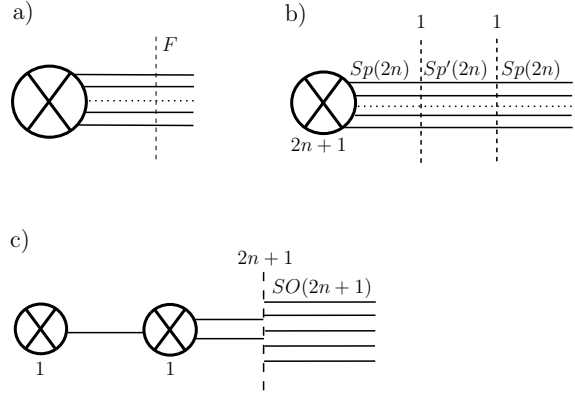}
  \end{center}
\caption{\small (a) $Sp(2n)$ gauge theory interacting with a full
D5-brane at the boundary.  (b) The split configuration, with
linking numbers indicated. (c) The $S$-dual. }
  \label{Fig69}\end{figure}
The third example, a Neumann boundary condition for $Sp(2n)$
together with a coupling to a fundamental hypermultiplet of flavor
symmetry $SO(2)$, is depicted in fig. \ref{Fig69}(a). The brane
manipulations are more elementary, as the full D5-brane is
equivalent to two half D5-branes, as in fig. \ref{Fig69}(b). The
$S$-dual configuration in fig. \ref{Fig69}(c) shows clearly that the
dual $SO(2n+1)$ gauge theory is broken to an $SO(2)$ subgroup by a
Nahm pole of dimension $2n-1$, and the $SO(2)$ subgroup is gauged at
the boundary.

\begin{figure}
  \begin{center}
    \includegraphics[width=5.5in]{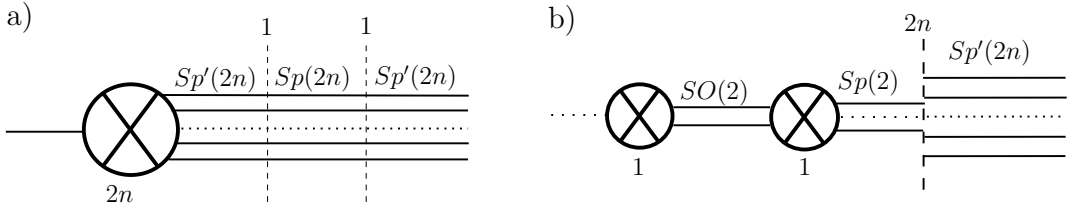}
  \end{center}
\caption{\small (a) $Sp'(2n)$ gauge theory coupled to fundamental
hypermultiplets with $SO(3)$ flavor symmetry. In the split
version, the hypermultiplets come  from both the two half
D5-branes and the boundary.  (b) The $S$-dual configuration. }
  \label{Fig70}\end{figure}
Finally the same brane configuration for an $Sp'(2n)$ gauge group
represents an $Sp'(2n)$ gauge theory coupled to fundamental
hypermultiplets with an $SO(3)$ flavor symmetry at the boundary,
because of the extra $O(1) \times Sp'(2)$ bifundamental
hypermultiplet at the location of the NS5-brane. The split brane
configuration and the $S$-dual brane configuration are shown in fig.
\ref{Fig70}. The dual $Sp'(2n)$ gauge theory is broken to $Sp(2)$ by
a Nahm pole of dimension $2n-2$ and coupled to an $SO(2)$ gauge
theory with $Sp(2)$ flavor symmetry group, which is nothing else but
$T(SU(2))$.

\subsection{O3-Planes With O5-Planes}\label{orthobif}

Now we will study  orthosymplectic gauge groups  realized by the
combination of an O3-plane and an O5-plane.  O5-planes were
described in  section \ref{orient}, where  we also studied an
${\cal I}_4(-1)^{F_L}$ orbifold fiveplane.  So in toto we have
three objects: the O3-plane with reflection of the coordinates
$456789$, the O5-plane with reflection of coordinates $3789$, and
the orbifold fiveplane with reflection of $3456$.  We will now
study models in which all three of these objects are present.  In
fact, as soon as one has any two of them, the third arrives for
free, since the product of any two of these reflection symmetries
is the third.

To some extent, we can $S$-dualize the product of these objects if
we know how to $S$-dualize them separately, since we can go to a
region in spacetime in which only one of the reflection symmetries
has a fixed point.  Though O5$^+$ does not have a  useful
$S$-dual, there is a useful duality
\cite{Kutasov:1995te,Sen:1996na} involving O5$^-$. The following
two objects are $S$-dual: (i) the combination of O5$^-$ with a
D5-brane; (ii) the orbifold fiveplane.

Now suppose that objects (i) and (ii) are both present. This gives
a configuration that is invariant under $S$-duality at least away
from the locus where the two objects meet.  That locus, which is
where $\vec x$ and $\vec y$ vanish, will be the locus of an
O3-plane.  The O5$^-$ creates orthogonal gauge symmetry for
D5-branes, so D3-branes intersecting it should have symplectic
gauge symmetry.  Hence the O3-plane is of symplectic type.  For
the overall configuration to be $S$-dual, the gauge group must be
$Sp'(2n)$ and so the O3-plane is of type $\tilde{\rm O}3^+$. We
therefore propose that the combination of the following objects is
$S$-dual:  an O5$^-$-plane together with a D5-brane; a orbifold
fiveplane; and an $\tilde{\rm O}3^+$-plane.

This statement implies the $S$-duality of a certain field theory
boundary condition.  This is a boundary condition in $Sp'(2n)$
gauge theory in which $Sp'(2n)$ couples to a fundamental
hypermultiplet and is broken at the boundary to $Sp(n)\times
Sp(n)$.  Of course, $n$ must be even.

\begin{figure}
  \begin{center}
    \includegraphics[width=5.5in]{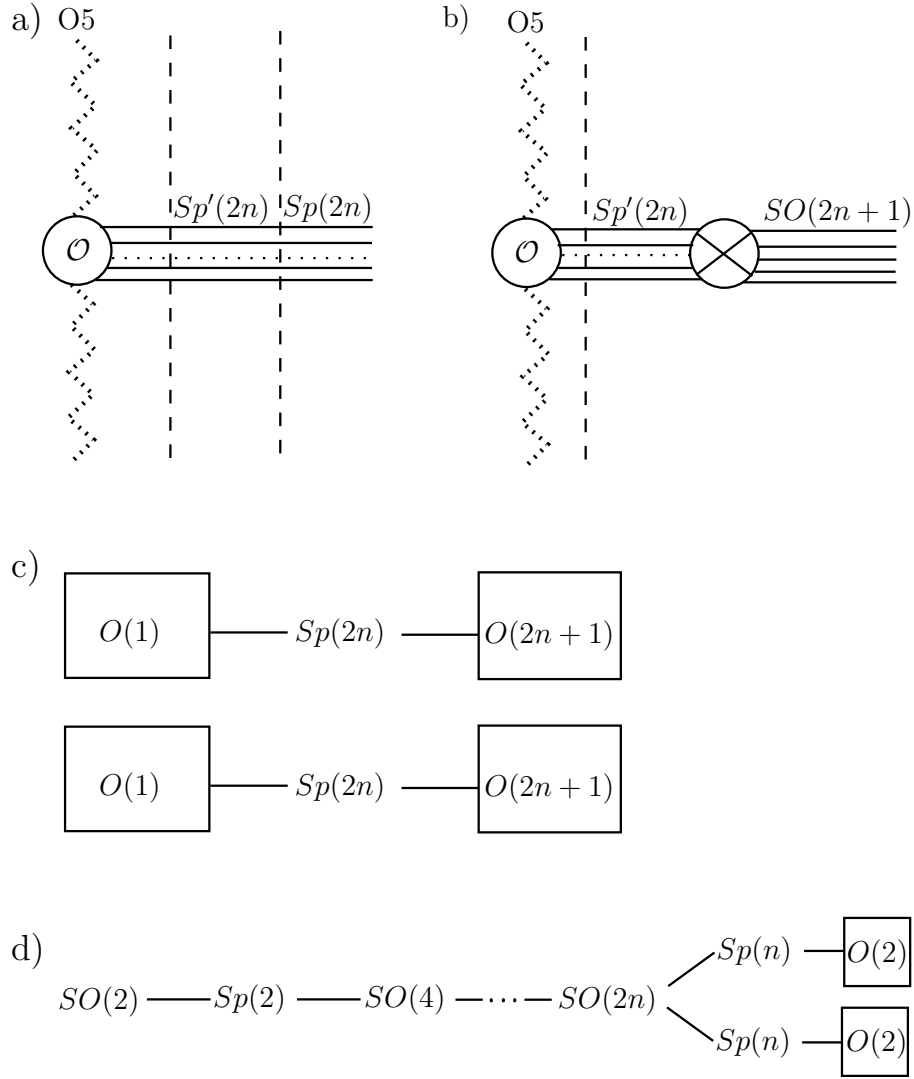}
  \end{center}
\caption{\small (a) A configuration involving a combination of O3
and O5 orientifold planes  and an orbifold plane, together with
two half D5-branes. (The circle containing the symbol $\cal O$
represents the orbifold plane.)  (b) $S$-duality converts the
rightmost half D5-brane to a half NS5-brane, while, as explained
in the text, the rest of the configuration is self-dual. (c) The
corresponding boundary CFT. (d) The mirror of the same CFT, built
through the use of $T(Sp(2n))$. }
  \label{Fig71}\end{figure}

Once we understand $S$-duality for this example, we can deduce how
$S$-duality must act for a very large class of additional
examples.  We will give an interesting and illustrative example,
using the knowledge gained in section \ref{bifurc}.  We add one
more half D5-brane to the system, so that it represents a boundary
condition for an $Sp(2n)$ gauge theory, broken to $Sp(n)\times
Sp(n)$, with each factor coupled to a hypermultiplet with $SO(2)$
flavor symmetry.  The global symmetry of the Higgs branch is
therefore $SO(2)\times SO(2)$.

The brane construction and its $S$-dual are depicted in fig.
\ref{Fig71}(a,b). The $S$-dual boundary condition for the
$SO(2n+1)$ gauge theory involves coupling with the product of two
copies of a certain boundary theory.  That theory is the IR limit
of $Sp(n)$ gauge theory coupled to hypermultiplets with $SO(2n+2)$
flavor symmetry, as depicted in fig. \ref{Fig71}(c). These
balanced theories each have an $SO(2)$ symmetry of the Coulomb
branch,  matching the $SO(2)\times SO(2)$ flavor symmetry of the
original boundary condition. We can reproduce the mirror of this
theory through $T(Sp(2n))$, as usual. The resulting composite
gauge theory is depicted in fig. \ref{Fig71}(d): it is a balanced
orthosymplectic quiver with the shape of a ${\tx D}_n$ Dynkin
diagram, and its Coulomb branch has $SO(2n+2) \times SO(2n+2)$
symmetry, as desired.
\begin{figure}
  \begin{center}
    \includegraphics[width=4.5in]{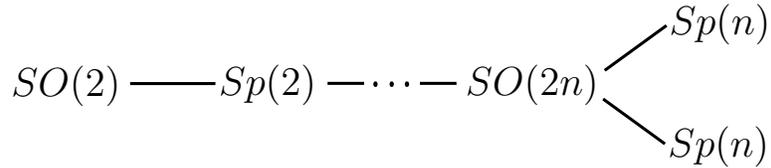}
  \end{center}
\caption{\small A bad quiver resulting from application of the
$T(Sp(2n))$ recipe. }
  \label{Fig71a}\end{figure}

Our starting point has been a self-dual configuration that describes
$Sp'(2n)$ broken at the boundary to $Sp(n)\times Sp(n)$ with a
boundary hypermultiplet.  Can we find the $S$-dual with the
hypermultiplet removed?  This question is superficially similar to
the one treated in section \ref{zahm}. One could try to imitate the
approach used there by adding an extra half D5-brane, as we have
done in fig. \ref{Fig70}(a), and then turning on a mass parameter to
remove the boundary hypermultiplets (which means, in terms of
branes, that the two half D5-branes  recombine into a full D5-brane
which is then displaced to large $\vec y$). Unfortunately, that mass
parameter is dual to the ``hidden'' FI parameter in fig.
\ref{Fig71}(c), which corresponds to the $SO(2)$ symmetry of the
Coulomb branch. That deformation is difficult to analyze, so we will
not follow this path.

Alternatively, we can use the standard $S$-duality prescription
involving $T(Sp(2n))$.  This gives rise to the bad quiver in fig.
\ref{Fig71a}, indicating that in the desired dual boundary
condition, the $SO(2n+1)$ dual gauge symmetry is broken.

We will just conjecture an answer based on the analogy with the
similar boundary conditions for the unitary case: a Nahm pole
related to the decomposition $2n+1= 2+2+\dots +2 +1$ together with
a gauging of the ``level two''  subgroup $Sp(n)_2$ which commutes
with the Nahm pole. We can extend this conjecture to a boundary
condition breaking $Sp(p+q)$ to $Sp(p) \times Sp(q)$ at the
boundary ($p>q$): the dual has a Nahm pole related to the
decomposition $2n+1= 2+2+\dots +2 +(p-q)$ together with a gauging
of the ``level two''  subgroup $Sp(q)_2$ which commutes with the
Nahm pole. This statement, which we suspect can be justified by
adapting the arguments of section \ref{lastcase}, is related to
the case CII in Table 1 of \cite{Nadler}. The case CI in that
table is a case in which the $T(G)$ prescription leads to a good
quiver, so there is full gauge symmetry at the boundary; the split
case of DI/DII is similar.

\section{$S$-Duality and Theta-Angle}\label{stheta}
It is natural to wonder if our construction of $S$-dual pairs of
boundary conditions extends to an action of\footnote{\label{wef}
When the gauge group is $G_2$ or $F_4$, the appropriate duality
group is not actually $SL(2,\Z)$, but a group generated by
$T:\tau\to \tau+1$, $S:\tau\to -1/n_{\frak g}\tau$, where
$n_{\frak g}$ is 2 for $G_2$ and 3 for $F_4$, as shown in
\cite{Argyres:2006qr}. For simplicity, we will refer to the
duality group as $SL(2,\Z)$.} $SL(2,\Z)$ on half BPS boundary
conditions corresponding to the full duality group of $\N=4$ SYM.
In this section, we will argue that the answer is positive, and
define a pair of abstract transformations $S$ and $T$ which
generate the duality group.

Before trying to do this, we should explain what exactly the
statement is supposed to mean.  Let us first review the case that
we have focused on so far: $\tau$ is imaginary and we want to know
how a boundary condition transforms under $S:\tau\to-1/\tau$.  We
start with a boundary condition in weakly coupled $G$ gauge
theory, and $S$ maps us to an equivalent configuration in strongly
coupled $G^\vee$ gauge theory.  This is not really illuminating,
however.  What we really mean by the dual boundary condition is a
boundary condition in {\it weakly} coupled $G^\vee$ gauge theory.
To find it, after acting with $S$, we have to continue the
$G^\vee$ boundary condition from strong coupling back to weak
coupling.  This makes sense because all boundary conditions that
we have studied in this paper can be defined (at $\theta=0$) for
any value of the gauge coupling $g$.    The ability to continue a
boundary condition in $g$ is built into what we mean in studying
the action of $S$ on boundary conditions.

This has an analog when we include $\theta$ and consider a more
general duality transformation $\gamma:\tau\to (a\tau+b)/(c\tau+d)$.
We start at ${\rm Im}\,\tau>>1$ and, say, ${\rm Re}\,\tau=0$.  After
acting with $\gamma$, we land in the strongly coupled region,
generically with a different value of $\theta=2\pi\,{\rm Re}\,\tau$.
Then we have to continue back to the starting point.  For this, we
have to restrict ourselves to boundary conditions that can be varied
with both ${\rm Im}\,\tau$ and ${\rm Re}\,\tau$ in a natural way.

Another way to describe the situation is this.  If one is given a
boundary condition that can be naturally continued as $\tau$
varies in the upper half-plane, then one can approach any cusp on
the real $\tau$ axis and ask what boundary condition is present in
that duality frame. So under these conditions it makes sense to
ask how $SL(2,\Z)$ acts on a boundary condition. If instead one is
given a boundary condition that is only defined for imaginary
$\tau$, then it only makes sense to ask what happens under
$\tau\to -1/\tau$.

While very general half-BPS boundary conditions allow a natural
variation of the gauge coupling $g^2=4\pi/{\rm Im}\,\tau$, only very
special ones admit a similar variation of $\theta$. For example,
consider a boundary condition with full $G$ gauge symmetry coupled
to a boundary SCFT $\fB$ with $\N=4$ supersymmetry. We suppose that
$G$ acts on the Higgs branch of $\fB$. A generalization of such a
boundary condition, preserving its full supersymmetry, is possible
\cite{Gaiotto:2008sa}
 precisely if the moment
map $\vec\mu$  satisfies the so-called fundamental identity: the
complex moment map $\mu_\BC$ associated with any choice of an
$\N=2$ subalgebra must obey\footnote{See eqn. (3.57) of
\cite{Gaiotto:2008sd} for the fact that there can be a constant on
the right hand side of the fundamental identity.}
\begin{equation}\label{fundid} \Tr \,\mu_\BC^2={\rm constant}.\end{equation}
The same condition allows coupling of $\fB$ to a three-dimensional
gauge theory with Chern-Simons action.

Similarly the Coulomb branch of an $\N=4$ theory can be  coupled
to bulk gauge fields if the twisted moment map obeys the
fundamental identity. A perhaps surprising generalization
\cite{Hosomichi:2008jd} is that if we are given a pair of
theories, one with a $G$-action on the Higgs branch, and one with
a $G$-action on the Coulomb branch, both satisfying the
fundamental identity, their product can be coupled to a $G$ Chern
Simons gauge theory preserving the full $\N=4$ supersymmetry. By
contrast, if both theories have $G$ action on the same branch,
such a coupling is generally not possible, as the fundamental
identity is not additive in $\mu$.

A wide class of $G$-invariant $\N=4$ SCFTs that obey the
fundamental identity has appeared in this paper.  These are the
theories $T_{\rho^\vee}(G)$ for any $G$ and $\rho^\vee$. The Higgs
branch of such a theory is always a union of nilpotent orbits. The
complex moment map $\mu_\BC$ takes values in those orbits; hence
its quadratic Casimir vanishes, and the fundamental identity is
obeyed, with the constant being zero. The examples given in
\cite{Gaiotto:2008sd} are special cases of these. If one makes an
FI deformation of $T_{\rho^\vee}(G)$, smoothing the singularities
and preserving the $G$ symmetry, the nilpotent orbits are deformed
to semisimple ones, and the (\ref{fundid}) remains valid, now with
a possibly nonzero constant.

A more obvious example of a boundary condition that can be naturally
continued for all $\tau$ is  Dirichlet modified by a Nahm pole. With
such a boundary condition, the topological term $(\theta/8\pi^2)\int
\,Tr\,F\wedge F$ can be added to the action, preserving all
supersymmetry.  This gives a natural way to vary $\theta$.  So such
boundary conditions should lie on an $SL(2,\Z)$ orbit. Not
coincidentally, the duality transformation $S:\tau\to -1/\tau$
converts Dirichlet with a pole of type $\rho^\vee$ to a boundary
condition $\B$ associated with the theory $T_{\rho^\vee}(G)$. Since
this theory obeys the fundamental identity, the boundary condition
$\B$ can again be contained in an $SL(2,\Z)$ orbit.  Orbits of this
type are the only $SL(2,\Z)$ orbits of half-BPS boundary conditions
that we know about.

\subsection{Definition Of $S$ And $T$}\label{defst}

For simplicity, we will describe the action of $SL(2,\Z)$ in terms
of a transformation on the three-dimensional SCFT $\fB$ that lives
at the boundary. When the dual does not have full gauge symmetry
but contains a Nahm pole or a gauge group reduction, one must
adapt the following procedure along the lines of section
\ref{symbr}.

When we say that $\fB$ is a theory with $G$ symmetry, what we mean,
to be more exact, is that we are given a precise recipe to couple
$\fB$ to a background $\N=4$ supermultiplet with gauge group $G$.
Assuming the fundamental identity is obeyed, such a coupling can
include a Chern-Simons coupling. We define the $T$ operation as a
unit shift of the Chern-Simons coefficient.

To define an $S$ operation, we need to refine what we mean by
saying that $T(G)$ has $G\times G^\vee$ symmetry.  Again, we need
to specify a standard coupling of $T(G)$ to background $G\times
G^\vee$ vector multiplets.  We specify this coupling by asking
that it should be parity-symmetric.  (This excludes the
possibility of adding $\N=4$ Chern-Simons couplings, which
otherwise are possible since $T(G)$ obeys the fundamental
identity.)

We call the $G$ symmetry of $T(G)$ a direct action, and the
$G^\vee$ symmetry a twisted action. The terminology is motivated
by the idea that $G$ acts on the Higgs branch, which is
parameterized by hypermultiplets, while $G^\vee$ acts on the
Coulomb branch, which is parameterized by twisted hypermultiplets.

Now we are in position to define the $S$ action: a theory $\fB$
with a direct action of $G$ is mapped by $S$ to a composite theory
$T(G^\vee) \times_G \fB$ with a direct $G^\vee$ action. Here we
define $T(G^\vee) \times_G \fB$ as the result of gauging the
product of the twisted $G$ action on $T(G^\vee)$ and the direct
$G$ action on $\fB$. In general, there will be a Chern-Simons
action for $G$ implicit in the coupling to $\fB$; otherwise we add
a supersymmetric Yang-Mills coupling and then flow to the
infrared. This $S$ operation was essentially defined in section
\ref{genpres}; in the analysis of fig. \ref{Fig52}, we argued that
it satisfies $S^2=1$. We now plan to show\footnote{As mentioned in
footnote \ref{wef}, for gauge group $G_2$ or $F_4$, $S$ acts
differently on the upper half plane. Consequently, the appropriate
relation is not $(ST)^3=1$.  The argument that follows really
shows that any word in $S$ and $T$ that acts trivially on the
upper half plane acts trivially on the theory; in this form, it
applies also to $G_2 $ and $F_4$.} that
 $(ST)^3$ is
also $1$, so that the two transformations generate an $SL(2,\Z)$
duality group. (For $G=U(1)$, where everything is much more
elementary, the fact that $(ST)^3=1$ can be shown by a direct path
integral computation \cite{Witten:2003ya}.)

We would like to mimic the $S^2=1$ proof, which used the Janus
interpretation of $S$. Unfortunately the $\N=4$ Janus
configurations are relatively ``rigid'' (only certain paths in the
upper half plane are allowed), so that different $\N=4$ Janus
walls cannot be concatenated in a fashion which preserves $\N=4$
supersymmetry. On the other hand, it is possible to relax this
constraint at the price of reducing supersymmetry from $\N=4$ to
$\N=3$. Indeed, there is a relatively straightforward description
of an $\N=3$ Janus configuration which allows for a generic
$y$-dependence of $\tau$. We start with the fact
\cite{Kao:1992ig,KLL,Kapustin:1999ha,Gaiotto:2007qi} that for any
group $\hat G$ and hypermultiplet representation $R$ there is an
$\N=3$ action with a gauge coupling and a Chern-Simons coupling.
Both of these couplings depend on choices of invariant quadratic
forms on the Lie algebra, of which the first should be positive
definite. We take $\hat G$ to be the infinite dimensional group of
$G$-valued functions $g(y)$ of the real variable $y$, and we take
$R$ to be the twisted version of the adjoint representation
described in section 2.3.1 of \cite{Gaiotto:2008sa}. We pick
quadratic forms on the Lie algebra that depend on arbitrary
functions of $y$, as at the end of section 2.3.1 in that
reference. The result is an $\N=3$ Janus configuration with an
arbitrary $\tau(y)$.

We can represent the $S$ operation by a Janus domain wall which
interpolates from $\tau$ to $-1/\tau$. Similarly $T$ is a Janus
wall which interpolates between $\tau$ and $\tau+1$. So $(ST)^3$
comes from a succession of six Janus domain walls, at the end of
which we return to the initial value of the coupling  parameter
$\tau$. Each of the six domain walls preserves $\N=4$
supersymmetry, but the combination has only $\N=3$. As we flow to
the infrared, the details of the path are forgotten, and we only
remember the initial and final points of the path.  Since these
coincide, the infrared limit is a trivial domain wall, confirming
that $(ST)^3=1$. Notice that a similar temporary $\N=3$
deformation which flows to a fixed point with enhanced
supersymmetry has been used in \cite{Aharony:2008ug}.

\subsection{Effective Action For Interaction With a $(p,q)$
Fivebrane}\label{effacint}
 A long standing puzzle in string theory
has been to describe the intersection between $n$ D3-branes and a
$(p,q)$ fivebrane. As an illustration of our construction, we will
use it to resolve this puzzle. We start from a  configuration that
is already understood, $n$ D3-branes intersecting a single
NS5-brane, and apply a general $SL(2,\Z)$ transformation. To put
this in our framework, we use the folding trick to describe this
intersection as a boundary condition for  $U(n) \times U(n)$ gauge
theory.

An important fact is that the folding trick reverses orientation,
so it maps $T$ to $T^{-1}$ while preserving $S$. The initial
boundary condition consists of a parity-invariant coupling to a
bifundamental hypermultiplet of $U(n)\times U(n)$. The action of
$T^k$ gives Chern-Simons coefficients $(k, -k)$ for $U(n) \times
U(n)$. A single bifundamental hypermultiplet with these
Chern-Simons coefficients gives a basic solution of the
fundamental identity \cite{Gaiotto:2008sd}. This theory, which was
described in detail in the reference, describes the intersection
of $n$ D3-branes with a $(1,k)$ fivebrane.
\begin{figure}
  \begin{center}
    \includegraphics[width=6in]{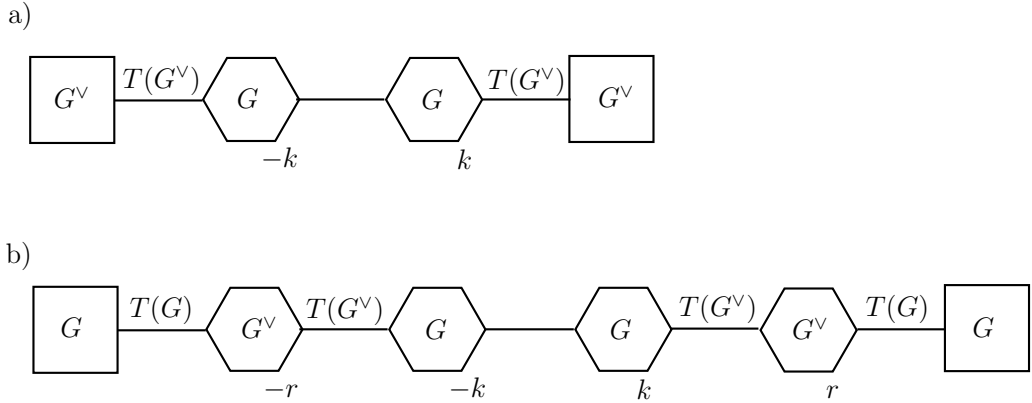}
  \end{center}
\caption{\small   Schematic representations of ``quivers'' in
which the hexagons represent Chern-Simons gauge theories, with
levels indicated by subscripts; the central edges  represent
ordinary bifundamental hypermultiplets, while other edges
represent couplings to $T(G)$ or $T(G^\vee)$, as labeled;  with
levels indicated by subscripts; and the squares indicate the
remaining flavor symmetries. }
  \label{Fig72}\end{figure}

An application of $S$ maps this theory to the ``quiver'' in fig.
\ref{Fig72}(a), which therefore describes the intersection of $n$
D3-branes and one $(k,1)$ fivebrane. A hexagon in the figure
represents a $U(n)$ gauge group with the Chern-Simons coefficient
indicated by the subscript.  Repeated action of $T$ and $S$
generates longer ``quivers'' of this type.  For example, the
result of $ST^rST^k$ is represented in fig. \ref{Fig72}(b).

Continuing in this way, we get a ``quiver'' describing the
interaction of $n$ D3-branes with a $(p,q)$ five-brane whenever $p$
and $q$ are relatively prime.  Even if $p$ and $q$ are not
relatively prime, but have greatest common divisor $k>1$, we can
make the same type of construction starting with $k$ NS5-branes. We
simply begin with the conventional linear  quiver of fig.
\ref{Fig40}(e), which describes the intersection of $n$ D3-branes
with a chain of $k$ NS five-branes, and then apply the above
operations.

We can generalize this slightly to the case that the numbers of
D3-branes on the two sides of the $(p,q)$ fivebranes are
different.  A special case is that there are D3-branes on only one
side, leading to a boundary condition in $U(n)$ gauge theory.  A
boundary condition involving a chain of NS5-branes with varying
linking numbers leads to the $SL(2,\Z)$ orbit containing the
theories $T_{\rho^\vee}(SU(n))$.

\subsection{$(p,q)$ Fivebranes And Fractional Chern-Simons
Couplings}\label{fractional}

We will conclude with an analysis of a single D3-brane ending on a
$(p,q)$ fivebrane. (Supersymmetry will play no important role and
is omitted.)  We begin with the case of a $(1,0)$ fivebrane. The
boundary theory is trivial; that is, the $U(1)$ gauge theory of
the D3-brane obeys Neumann boundary conditions, coupled to nothing
else.  We regard this trivial boundary theory as a theory with
$U(1)$ symmetry by introducing a background $U(1)$ gauge field $B$
whose couplings are zero.  Then we act with $T^k$, after which the
action for $B$ is a level $k$ Chern-Simons action $k\int B\wedge
dB/4\pi$.  In view of the description of $T(U(1))$ in section
\ref{tuone}, acting with $S$ means that we add a second $U(1)$
gauge field $A$ with coupling $\int A\wedge dB/2\pi$.  At this
point, then, the boundary action is
\begin{equation}\label{polyko}\frac{1}{2\pi}\int_{\partial
M}A\wedge dB +\frac{k}{4\pi}\int_{\partial M}B\wedge d
B.\end{equation}  This is the boundary action for a single
D3-brane ending on a $(k,1)$ fivebrane. In that application, $B$
is defined only on the boundary, but $A$ is defined in bulk (and
has a conventional bulk kinetic energy).

A somewhat inaccurate procedure that is frequently followed at
this stage is to treat $B$ as a linear field, ignoring the fact
that it may have quantized Dirac fluxes.  In this approximation,
one can perform a Gaussian integral over $B$, leading to a
boundary Chern-Simons coupling for $A$ that is not properly
quantized:
\begin{equation}\label{olyko} -\frac{1}{4\pi k}\int_{\partial M}
A\wedge dA.\end{equation} This is not really the right answer,
because in deriving it one has omitted the sum over fluxes of $B$.
Still, this computation sheds light on the sense in which  one
might  claim  \cite{Kitao:1998mf} that ending a D3-brane on a
$(k,1)$ five-brane induces a Chern-Simons coupling $-1/k$.

An action much like (\ref{polyko}) is often studied in relation to
the fractional quantum Hall effect. (For example, see eqn. (2.11)
in \cite{Wen}.)  In that context, $A$ is the ordinary
electromagnetic vector potential, and $B$ is an effective $U(1)$
gauge field induced by strong coupling effects in a
two-dimensional material. The couplings (\ref{polyko}) in that
context are supported on a defect in spacetime -- the world-volume
of the material -- rather than on a boundary.  In that context,
the effective Chern-Simons coefficient for $A$ is the quantum Hall
conductivity.   This conductivity is unaffected by the sum over
fluxes of $B$, so the computation leading to (\ref{olyko}) is a
valid way to explain the fractional quantum Hall effect.

On the other hand, we will get into trouble if we take
(\ref{olyko}) literally.  For example, consider a D3-brane
suspended between a $(1,0)$ fivebrane, which generates Neumann
boundary conditions, and a $(1,k)$ fivebrane.  (The configuration
is not supersymmetric, but that does not affect the point we are
about to make.)  The effective three-dimensional physics is given
by the action (\ref{polyko}), now understood in purely
three-dimensional terms. This theory is completely consistent, but
if we naively treat $B$ as a Gaussian field and integrate it out,
we will arrive at the theory (\ref{olyko}) which is inconsistent,
because the Chern-Simons coefficient is not properly quantized.
\bigskip

We would like to thank D. Ben-Zvi, E. Frenkel, D. Nadler, N.
Hitchin, and B. Webster for comments and M. Turansick for
assistance. Research of DG supported in part by DOE Grant
DE-FG02-90ER40542. Research of EW supported in part by NSF grant
PHY-0503584.

\bibliography{strings}{}
\bibliographystyle{JHEP-2}
\end{document}